\documentclass[12pt]{iopart}
\usepackage{graphicx}
\usepackage{iopams}
\usepackage[colorlinks,citecolor=blue]{hyperref}
\expandafter\let\csname equation*\endcsname\relax
\expandafter\let\csname endequation*\endcsname\relax
\usepackage{amstext, amsfonts, amssymb, amsmath}
\usepackage{bm}
\usepackage{bm}
\begin{document}

\title{Topological phononic metamaterials}

\author{Weiwei Zhu\footnote{These authors contributed equally}}
\address{Department of Physics, National University of Singapore, Singapore 117542, Singapore}
\author{Weiyin Deng\footnote{These authors contributed equally}}
\address{School of Physics and Optoelectronics, South China University of Technology, Guangzhou, Guangdong 510640, China}
\author{Yang Liu\footnote{These authors contributed equally}}
\address{Institute of Theoretical and Applied Physics, School of Physical Science and Technology \& Collaborative Innovation Center of Suzhou Nano Science and Technology, Soochow University, Suzhou 215006, China}
\author{Jiuyang Lu\footnote{These authors contributed equally}}
\address{School of Physics and Optoelectronics, South China University of Technology, Guangzhou, Guangdong 510640, China}
\author{Zhi-Kang Lin}
\address{Institute of Theoretical and Applied Physics, School of Physical Science and Technology \& Collaborative Innovation Center of Suzhou Nano Science and Technology, Soochow University, Suzhou 215006, China}
\author{Hai-Xiao Wang}
\address{College of Physics and Technology, Guangxi Normal University, Guilin 541004, China}
\author{Xueqin Huang\footnote{Corresponding author: phxqhuang@scut.edu.cn}}
\address{School of Physics and Optoelectronics, South China University of Technology, Guangzhou, Guangdong 510640, China}
\author{Jian-Hua Jiang\footnote{Corresponding author: jianhuajiang@suda.edu.cn}}
\address{Institute of Theoretical and Applied Physics, School of Physical Science and Technology \& Collaborative Innovation Center of Suzhou Nano Science and Technology, Soochow University, Suzhou 215006, China}
\author{Zhengyou Liu\footnote{Corresponding author: zyliu@whu.edu.cn}}
\address{Key Laboratory of Artificial Micro- and Nano-structures of Ministry of Education and School of Physics and Technology, Wuhan University, Wuhan 430072, China}
\address{Institute for Advanced Studies, Wuhan University, Wuhan 430072, China}

\begin{abstract}
The concept of topological energy bands and their manifestations have been demonstrated in condensed matter systems as a fantastic paradigm toward unprecedented physical phenomena and properties that are robust against disorders. Recent years, this paradigm was extended to phononic metamaterials (including mechanical and acoustic metamaterials), giving rise to the discovery of remarkable phenomena that were not observed elsewhere thanks to the extraordinary controllability and tunability of phononic metamaterials as well as versatile measuring techniques. These phenomena include, but not limited to, topological negative refraction, topological `sasers' (i.e., the phonon analog of lasers), higher-order topological insulating states, non-Abelian topological phases, higher-order Weyl semimetal phases, Majorana-like modes in Dirac vortex structures and fragile topological phases with spectral flows. Here we review the developments in the field of topological phononic metamaterials from both theoretical and experimental perspectives with emphasis on the underlying physics principles. To give a broad view of topological phononics, we also discuss the synergy with non-Hermitian effects and cover topics including synthetic dimensions, artificial gauge fields, Floquet topological acoustics, bulk topological transport, topological pumping, and topological active matters as well as potential applications, materials fabrications and measurements of topological phononic metamaterials. Finally, we discuss the challenges, opportunities and future developments in this intriguing field and its potential impact on physics and materials science.
\end{abstract}

\submitto{\RPP}
\maketitle
\tableofcontents
\clearpage

\section{Introduction}

Phononic metamaterials~\cite{science.289.1734liu,nat.mater.5.452fang}, including acoustic metamaterials~\cite{science.289.1734liu,nat.mater.5.452fang,phys.rev.e.70.055602li,phys.rev.lett.102.194301zhang,nat.rev.mater.1.16001cummer,sci.adv.1501595ma,appl.phys.lett.91.183518chen} and mechanical metamaterials~\cite{nat.rev.mater.2.17066bertoldi,adv.eng.mater.21.1800864surjadi,adv.mater.24.4782lee}, was proposed about three decades ago and have been developed comprehensively in both fundamental science and applications, from both theory and experiments. The concept of phononic crystals~\cite{phys.rev.b.62.5536psarobas,phys.rev.b.64.172301cleland,phys.rev.lett.88.104301yang,phys.rev.lett.93.024301yang,mater.today.12.34lu}, the motherboard of various phononic metamaterials, arises when one organize phononic structures in periodic patterns. By applying the Bloch theory on phononic waves in periodic structures, phononic band structures can be obtained. The merit here is that such phononic band structures are highly tunable if the phononic structures and their periodic patterns are manipulated. This tunability gives rise to  remarkable control over acoustic waves and mechanical waves, leading to extraordinary effects such as phononic band gaps, negative refraction, unconventional effective parameters, versatile control of phononic waves, as well as subwavelength imaging, trapping and waveguiding. After decades of study, phononic metamaterials became a material platform with fantastic controllability and reliable measurement techniques.

In the past years, topological phononic metamaterials have attracted tremendous research interest~\cite{commun.phys.1.97zhang,nat.rev.phys.1.281ma,Xin2020,nat.rev.mater.xue,ann.n.y.acad.sci.yves}. The field has been growing rapidly ever since it is born. Studies on topological phononic metamaterials aim to exploit topological physics---an important branch of condensed matter physics where the topological properties of energy bands play central roles~\cite{phys.rev.lett.62.2747zak,resta2007,phys.rev.lett.95.226801kane,phys.rev.lett.98.106803fu,rev.mod.phys.82.1959xiao,rev.mod.phys.82.3045hasan,rev.mod.phys.83.1057qi,nature.464.194moore,rev.mod.phys.90.015001armitage}---to explore new phenomena, materials, and their applications. The development of topological phononic metamaterials have brought many unprecedented effects into phononics and acoustics. For instance, the Zak phases and topological reflection, Weyl phonons and topological negative refraction, topological `sasers' (i.e., the phonon analog of lasers), non-Hermitian topological phononic metamaterials, topological Wannier cycles, bulk-defect correspondences, higher-order topological phases of phonons, non-Abelian topological states of phonons, phononic Dirac cones and double zero effective medium, to list but only a few. The field is still growing rapidly with lots of new ideas emerging in both fundamental and application aspects. The purpose of this review is to introduce the basic notions, models, pictures, and methods as well as the main progresses, opportunities, and challenges in this field so that readers from a broad range of disciplines can understand the main pages of this field. In this way, interdisciplinary studies and new ideas can continuously fuel researches in this direction.

The motivations of harnessing band topology in phononic metamaterials came from two main perspectives: First, phononic metamaterials are particularly appealing for the realization and observation of various topological phenomena. Thanks to their excellent tunability, topological phononic metamaterials can be designed into almost any geometry to realize various topological phases, ranging from valley- and spin-Hall phases to Weyl, Dirac, nodal line, higher-order, and fragile topological phases. Furthermore, mechanical systems provide extraordinary ability of detecting the bulk and boundary phononic states (both the wavefunctions and the spectra) with excellent position, wavevector, and frequency resolutions. Second, the developments of topological phononic metamaterials greatly enrich the study of metamaterials for unprecedented control of phononic waves and mechanical energies. These two motivations have been the main driving power for the development of topological phononic metamaterials, leading to researches ranging from the exploration of extreme topological phases that are hard to realize in condensed matter systems, to new ideas and phenomenon that have not yet been proposed in any other field, to cutting edge applications such as high-performance mechanical sensors, on-chip information processing based on integrated acousto-electronic systems, high-quality resonators, and robust frequency combs, etc. With these remarkable progresses and enormous research interest, the study on topological phononic metamaterials has been moving rapidly for years.

This review article is structured as follow: In Sec.~II. we introduce the basic aspects of phononic metamaterials, the fundamental concepts and methods for topological phononics as well as some phenomenological theories such as the Jackiw-Rebbi theory. In Sec.~III, we elaborate on various topological phononic metamaterials, forming four main categories: First, topological acoustic metamaterials including acoustic valley Hall, spin Hall, Chern insulators, and their generalizations, higher-order topological phases, topological semimetals, non-Abelian topological states, topological defects, and topological states protected by projective symmetries. Second, mechanical metamaterials at macroscopic scales such as spring mass systems. Third, micron-and nano-scale mechanical metamaterials that are appealing for on-chip applications. Fourth, non-Hermitian topological phononic systems that exhibit unconventional properties. In addition, we also discuss specific topics such as synthetic dimensions, artificial gauge fields, Floquet topological phononic systems, and bulk topological transport. Other topics covered here include Thouless pumping of phonons, topological active matters, acoustic skyrmions etc. For completeness, in Sec.~IV, we summarize the commonly used fabrication and measurement methods for the study of topological phononic metamaterials. In Sec.~V, we discuss the potential applications of topological phononic metamaterials. Finally, we give outlooks with discussions on future challenges and possible developments in Sec.~VI. Alongside with previous reviews~\cite{commun.phys.1.97zhang,nat.rev.phys.1.281ma,Xin2020,nat.rev.mater.xue,ann.n.y.acad.sci.yves}, this article intended to provide a systematic and the most up-to-date review on topological phononic metamaterials.

\section{Fundamental aspects}
\subsection{Phononic metamaterials}
\begin{figure}[htbp]
\flushright
    \includegraphics[width=0.85\linewidth]
    {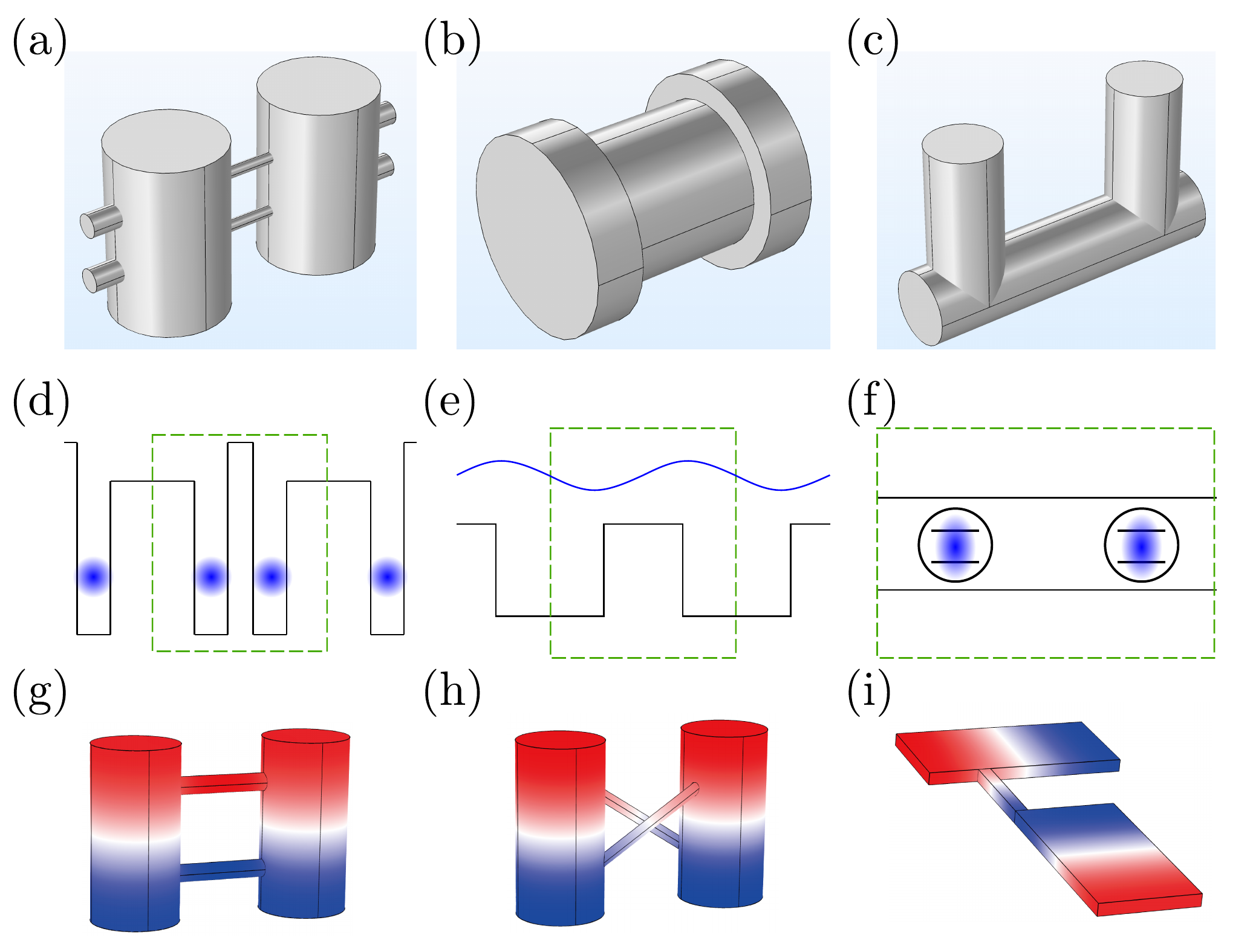}
    \caption{Three classes of topological phononic metamaterials. (a) Unit cell of coupled phononic cavities. (b) Unit cell of phononic crystals. (c) Unit cell of phononic locally resonant metameterials. (d) Tight-binding picture of electrons in periodic potential with deep well. (e) Free electron gases picture for electrons in periodic potential with shallow well or the energy of electrons is high. (f) Electrons are bounded by atoms and interact with each other by radiating light into waveguide. The unit cells are boxed by green boxes. (g) Positive coupling between two cavities with $p_z$ orbitals. (h) Negative coupling between two cavities with $p_z$ orbitals. (i) Negative coupling between one cavity with $p_x$ orbital and another with $p_y$ orbital.}
    \label{classification}
\end{figure}

Topological phononic metamaterials are periodically distributed arrays of artificial phononic atoms, which usually support robust topological edge states~\cite{commun.phys.1.97zhang,nat.rev.phys.1.281ma,Xin2020,nat.rev.mater.xue,ann.n.y.acad.sci.yves}. They are analogies of topological states for electrons in periodic potential. It is well known that the behavior of electrons scattered by periodic potential can be understood by different physical picture like tight-binding picture, free electron gas picture, etc~\cite{mrsbulletin_2001}. As the analogies, the topological phononic metamaterials can be classified into three classes corresponding to different physical pictures, based on the types of artificial atoms (resonators or scatterers) and the couplings style (near field couplings or far field couplings), as shown in Fig.~\ref{classification}. Figure~\ref{classification}(a) shows the first class of the phononic metamaterials, where the waves are confined in the coupled phononic cavities~\cite{phys.rev.lett.117.224301yang}. The confined waves in different cavities are coupled by near field through coupling tubes with small radius. The phononic waves in such system correspond to electrons in periodic deep well shown in Fig.~\ref{classification}(d) and can be described by the tight-binding
picture. In such picture, the waves were considered to be localized in the deep well. The electrons can hop between different wells by quantum tunneling and form pass bands. Figure~\ref{classification}(b) is the one-dimensional phononic crystal which is composed of AB layers with different cross-sectional areas~\cite{nat.phys.11.240xiao}. One plane wave will be scattered when it propagates from one layer to another layer. The phononic waves in such system correspond to electrons in periodic shallow well or electrons of high energy shown in Fig.~\ref{classification}(e) known as free electron gas picture. In such picture, the waves were considered to propagate and be scattered at the edges of potential wells. The destructive interference between forward and scattering waves forms the forbidden bands. Figure~\ref{classification}(c) is the locally resonant phononic metamaterial~\cite{science.289.1734liu,nat.phys.9.55lemoult}. The side branch tubes are local resonators. The local resonators can be excited by the propagating waves in the main waveguide and then leak waves into the main waveguide. Such system corresponds to a light-matter interaction system shown in Fig.~\ref{classification}(f), where the electrons are bounded by positively charged ions and interact with each other by radiating light. In such picture, the electrons in ground state can absorb photons from waveguide to excited states and the excited electrons can radiate photons into the waveguide. The interaction between forward propagating photons and radiating photons forms a band structure for quasi-particles as the hybridization of electrons and photons.

The three classes of topological phononic metamaterials have their own advantages. The first class of coupled phononic cavities provides a nice platform to realize topological states described by the tight-binding models. The couplings between cavities are realized by connecting tubes. Both the nearest-neighbor couplings and long-range couplings can be easily obtained. The coupling strength is tuned by changing the radius of the connecting tubes. Even negative couplings can be realized by connecting two local positions of two cavities with opposite field phase~\cite{nature555.342Serra-Garcia,phys.rev.B.101.161301zhu,phys.rev.lett.124.206601qi,nat.commun.12.1888gao}. Fig.~\ref{classification}(g) shows the positive coupling between two cavities with $p_z$ orbitals and Fig.~\ref{classification}(h) shows the negative coupling. Fig.~\ref{classification}(i) shows the negative coupling between one cavity with $p_x$ orbital and the other cavity with $p_y$ orbital. Besides, the negative coupling also can be realized by inceasing the length of coupling tubes~\cite{phys.rev.applied.14.024023chen}. The phononic crystal provides a nice platform to realize topological states in scattering (continuous) system. The topological properties of the system are determined by the symmetry of lattice, the symmetry of individual scatterer and the relative position between different scatterers. Fruitful topological phases have been obtained by only rotating the scatterers~\cite{phys.rev.lett.116.093901lu,nat.phys.15.582zhang}. Its feasibility and tunability have been widely proven both also in two-dimensional systems and three-dimensional systems. The locally resonant phononic metamaterial provides a nice platform to realize subwavelength topological states whose sizes are more compact. Besides, due to the couplings between cavities are far field, it also provides a nice platform to study topological states in open system. The topological properties of the system can be tuned by changing resonant frequency of the cavities or the relative position between the cavities.

\subsection{Conventional and higher-order topology} To introduce conventional topological insulators and higher-order topological insulators, we first introduce the concept of codimension, which is the difference between the dimensions of the system and the topological boundary states. The codimension of conventional topological insulators is one, i.e., $n$-dimensional system supports $n-1$-dimensional gapless boundary states. The Chern insulator~\cite{phys.rev.lett.61.2015haldane,phys.rev.b.74.235111thonhauser}, $Z_2$ topological insulator~\cite{phys.rev.lett.95.146802kane}, valley topological insulator~\cite{phys.rev.b.84.195444ding,phys.rev.lett.107.256801qiao,phys.rev.b.89.085429kim,npg.asia.mater.6.e147song,phys.rev.x.9.031021liu}, etc. belong to conventional topological insulators. The topological properties of conventional topological insulators are protected by bulk topology. Higher-order topological insulators are those states with codimension larger than one~\cite{phys.rev.lett.119.246402song,sci.adv.4.eaat0346schindler,phys.rev.b.97.205136khalaf,phys.rev.b.98.081110vanmiert,phys.rev.lett.120.026801ezawa,phys.rev.b.98.205129matsugatani,phys.rev.b.98.201114franca,phys.rev.lett.123.216803park,phys.rev.lett.124.036803chen,phys.rev.b.103.205123takahashi}. The two-dimensional higher-order topological insulator has gapped edge states but supports topological corners states. $n$-dimensional system supports $n-m$-dimensional ($m>1$) boundary states. The higher-order topological insulators can be classified into intrinsic and extrinsic ones~\cite{phys.rev.b.97.205135geier}. The intrinsic higher-order topological insulator is protected by bulk topology. Spatial symmetry plays an important role in it. The extrinsic higher-order topological insulator is protected by boundary states. The higher-order topology can be more general. Gapless states can support topological corner states in the bulk continuum~\cite{phys.rev.x.11.011016chen}. The higher-order Dirac/Weyl semimetals are also proposed as the higher-order topologies~\cite{phys.rev.b.102.094503zhang,phys.rev.lett.125.146401wang}.

\subsection{Topological band theory}
We have roughly introduced three classes of topological phononic metamaterials in previous part. Although they are corresponding to different physical pictures, all of them have periodic structures described by the periodic Hamiltonian
\begin{equation}
    \hat{H}(\mathbf{r})=\hat{H}(\mathbf{r}+\mathbf{R}),
\end{equation}
where $\mathbf{R}$ is the Bravais lattice vector. For the coupled phononic cavities, $\hat{H}(\mathbf{r})$ usually can be described by a sparse matrix, $H_{mn}=\langle m|\hat{H}(\mathbf{r})|n\rangle$. $H_{mn}$ is the near field couplings between the $m^{\text{th}}$ cavity and the $n^{\text{th}}$ cavity. For the phononic crystals, $\hat{H}(\mathbf{r})$ are usually described by partial differential operators~\cite{phys.rev.b.89.134302lu}, e.g. $-\rho^{-1}\kappa\nabla\cdot\nabla$, where $\nabla$ is the Nabla operator, $\rho$ is the mass density and $\kappa$ is the bulk modulus. For the locally resonant phononic metamaterials, $\hat{H}(\mathbf{r})$ usually can be described by a dense matrix~\cite{phys.rev.lett.121.124501zhu}, $H_{mn}=\langle m|\hat{H}(\mathbf{r})|n\rangle$. $H_{mn}$ is the far field couplings between the $m^{\text{th}}$ cavity and the $n^{\text{th}}$ cavity. The spectrum of the system can be obtained by solving the following eigen-equation,
\begin{equation}
    \hat{H}(\mathbf{r})\psi(\mathbf{r})=E\psi(\mathbf{r}).
\end{equation}
Due to $\hat{H}(\mathbf{r})$ is periodic in real space, the crystal momentum is a good quantum number. According to the Bloch theory, the eigenstates of the system satisfy the following expression,
\begin{equation}
    \psi_{n\mathbf k}(\mathbf{r})=
    u_{n\mathbf{k}}(\mathbf{r})e^{i\mathbf{k \cdot r}},
\end{equation}
where $u_{n\mathbf{k}}(\mathbf{r})$ is the periodic part of the Bloch wave function. The energy bands of the system can be obtained from the following equation,
\begin{equation}  H(\mathbf{k})u_n(\mathbf{k})=E_n(\mathbf{k})u_n(\mathbf{k}).
\end{equation}
Here $H(\mathbf{k})=e^{-i\mathbf{k}\cdot\mathbf{r}}\hat{H}(\mathbf{r})e^{i\mathbf{k}\cdot\mathbf{r}}$ is obtained from the Fourier transformation of the real space Hamiltonian. $H(\mathbf{k})$ or equivalently its eigenvalues $E_n(\mathbf{k})$ and eigenvectors $u_n(\mathbf{k})$ defines the band structure.

$E_n(\mathbf{k})$ also known as energy bands determine the most important transport properties of waves in the bulk, e.g., the waves are extended in the pass band and localized in the forbidden band. Recent studies on topological states pay more attention to the eigenvectors $u_n(\mathbf{k})$. Two systems with same energy bands $E_n(\mathbf{k})$ can have dramatically different topological properties, which are distinguished by various topological invariants determined by the global properties of the Bloch waves. For example, the usual Berry phase~\cite{berry1984} of the $n^{\text{th}}$ band can be obtained from the periodic part of the Bloch wave function by the definition
\begin{equation}\label{berry}
    \theta_{n}=\oint_{\mathcal{C}}d\mathbf{k}\cdot\langle u_{n}(\mathbf{k})|i\bigtriangledown_{\mathbf{k}}|u_{n}(\mathbf{k})\rangle,
\end{equation}
where $\mathcal{A}_{n}(\mathbf{k})=\langle u_{n}(\mathbf{k})|i\bigtriangledown_{\mathbf{k}}|u_{n}(\mathbf{k})\rangle$ is the Berry connection corresponding to a vector potential in momentum space and $\mathcal{C}$ is a closed path in momentum space. The $\mathcal{A}_{n}(\mathbf{k})$ is a gauge dependent quantity. We can see that the eigenvector has an arbitrary phase choice $|u'_{n}(\mathbf{k})\rangle=e^{i\phi_n(\mathbf{k})}|u_{n}(\mathbf{k})\rangle$ known as gauge transformation. After the transformation, the corresponding Berry connection is changed to $\mathcal{A}'_{n}(\mathbf{k})=\langle u'_{n}(\mathbf{k})|i\bigtriangledown_{\mathbf{k}}|u'_{n}(\mathbf{k})\rangle=\mathcal{A}_{n}(\mathbf{k})-\nabla_{\mathbf{k}}\phi_n(\mathbf{k})$. However the Berry phase $\theta_{n}$ is gauge independent and has physical meaning.  For one-dimensional model, $\mathcal{C}$ is the first Brillouin zone and $\theta_{n}$ is the Zak phase of the $n^{\text{th}}$ band. The Zak phase determines the Wannier centers of the system and corresponds to the polarization. Specially, when the system supports inversion symmetry, the Zak phase is quantized and only take values $0$ or $\pi$~\cite{phys.rev.lett.62.2747zak}.

From the Berry connection, we can define the Berry curvature $\mathbf{\Omega}_{n}(\mathbf{k})=\nabla\times \mathcal{A}_{n}(\mathbf{k})$. The Berry curvature is gauge independent and corresponds to magnetic field in momentum space. The well-known Chern number for two-dimensional system can be obtained by the integration of the Berry curvature in the whole first Brillouin zone,
\begin{equation}
    2\pi C_{n}=\int_{\mathrm{BZ
    }}d^2\mathbf{k} \cdot \mathbf{\Omega}_{n}(\mathbf{k}).
\end{equation}
The Chern number can be connected with the Berry phase by the following relation~\cite{adv.phys.64.227weng},
\begin{equation}\label{chern}
    \begin{split}
    2\pi C_{n}=&-\int_{-\pi}^{\pi}\int_{-\pi}^{\pi}dk_xdk_y(\partial_{k_x}\mathcal{A}_{n}^{y}-\partial_{k_y}\mathcal{A}_{n}^{x})\\
    =&\int_{-\pi}^{\pi}dk_y\partial_{k_y}\left(\int_{-\pi}^{\pi}dk_x\mathcal{A}_{n}^{x}\right)\\
    =&\int_{-\pi}^{\pi}d\theta_{n}(k_y).
    \end{split}
\end{equation}
Here $\theta_{n}(k_y)=\int_{-\pi}^{\pi}dk_x\mathcal{A}_{n}^{x}$ is the Berry phase calculated from the one-dimensional integration of $\mathcal{A}_{n}^{x}(k_x,k_y)$ along the $k_x$ direction for each fixed $k_y$. We notice the Chern number is just the winding of the Berry phase $\theta_{n}(k_y)$ as a function of $k_y$. Such expression has more physical meaning. We know that the Berry phase $\theta_{n}(k_y)$ determines the Wannier center of one-dimensional system and also determines the topological edge states. $\theta_{n}(k_y)$ is also called the Wannier bands. The nonzero winding of the Berry phase means the gapless topological edge states. According to the bulk-boundary correspondence, the Chern number determines the number of topological chiral edge states. One important property of topological edge states is that they are robust against disorders. Such property can be used to design robust topological devices.

\begin{figure}[htbp]
\flushright
    \includegraphics[width=0.85\linewidth]{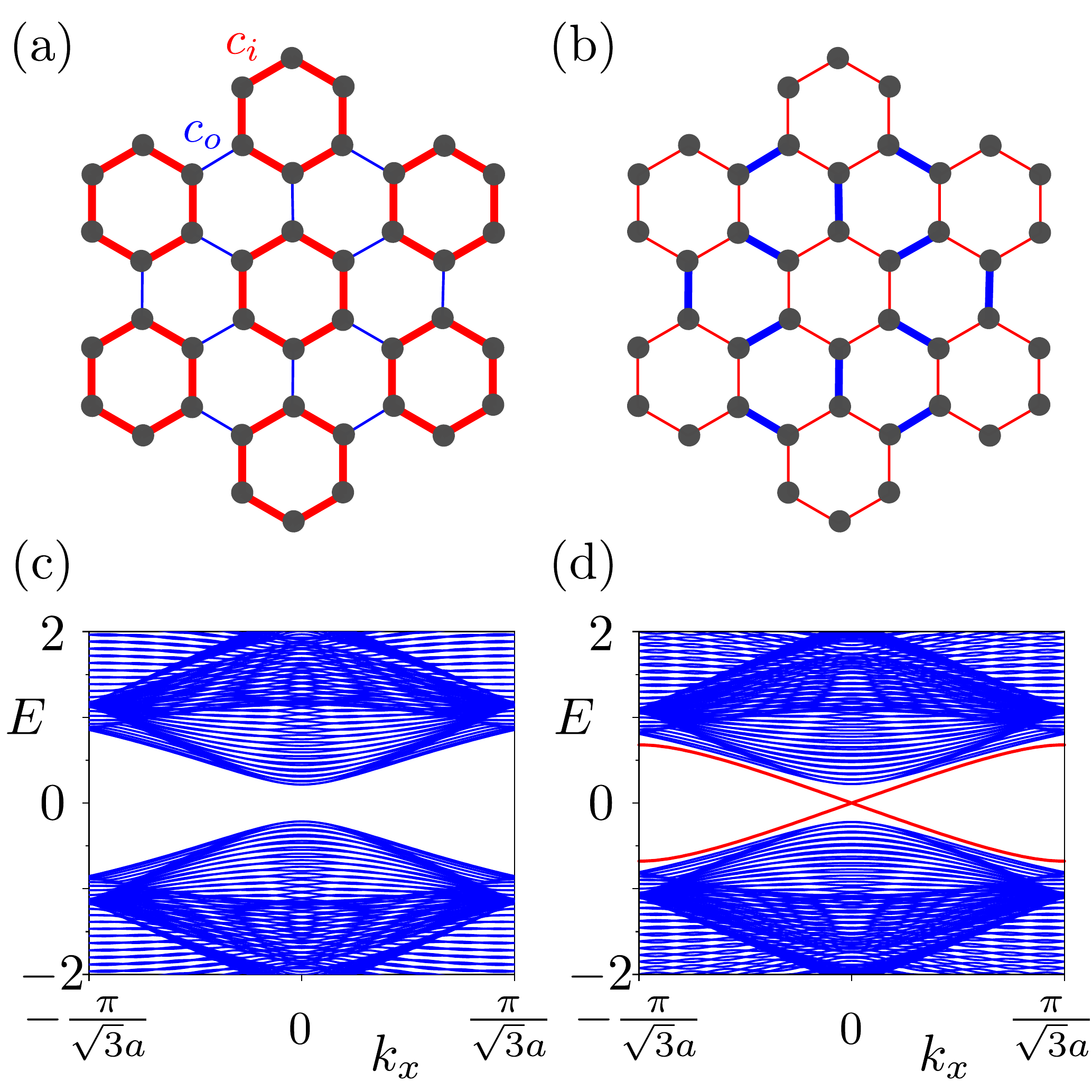}
    \caption{The breathing honeycomb lattice. (a) The topological trivial model, where $c_i>c_o$. (b) The topological nontrivial model, where $c_i<c_o$. (c) The energy band of strip structure for the topological trivial model. $c_i=1.2,c_o=1$. (d) The energy band for the topological nontrivial model. $c_i=1,c_o=1.2$. The edge states are colored by red.}
    \label{dimerisedhoneycomb}
\end{figure}

Although the definitions of the Berry phase in Eq.~\ref{berry} and the Chern number in Eq.~\ref{chern} are clear, they are still hard to be calculated practically. The reason is that the derivative of $|u_{n}(\mathbf{k})\rangle$ to $\mathbf{k}$ requires $|u_{n}(\mathbf{k})\rangle$ to be continuous with $\mathbf{k}$, which is usually hard to realize in calculation. Systems with multiple bands increase the calculation complexity further. Besides, the Berry connection is a gauge dependent quantity, which also brings difficulties in numerical calculation. Latter, we will introduce the Wilson loop approach, which is compatible with numerical implementation to calculate the topological invariants. Symmetry indicators are another way to distinguish topological states. Before that, we first introduce a concrete model to show the difference between topological trivial states and nontrivial states. Latter we will also use this model to show the power of the Wilson loop approach and symmetry indicators. The model is the breathing honeycomb lattice shown in Fig.~\ref{dimerisedhoneycomb}(a) and is described by the Hamiltonian~\cite{sci.rep.6.24347wu}
\begin{equation}\label{eq1}
  H=\sum\limits_{\langle i,j\rangle}c_{ij}a_i^\dag a_j,
\end{equation}
where $a_i^\dag$ ($a_i$) is the creation (annihilation) operator for a particle at the $i$th site, and $c_{ij}$ is the nearest neighbor coupling. Each unit cell contains six sites. The couplings in one unit cell are $c_{ij}=c_\mathrm{int}$ and the couplings between unit cells are $c_{ij}=c_\mathrm{ext}$. The model only contains the real nearest-neighbor coupling, which is friendly to experiment in phononics. As we show below, such simple model can describe a first-order topological insulator or a second-order topological insulator, depending on the parameters. The topological properties of the model can be understood by Wilson loop or symmetry indicators, so that it can be treated as a standard example to introduce basic concepts for topological phononics.

By doing a Fourier transformation to real-space Hamiltonian in Eq.~\ref{eq1}, we obtain the Hamiltonian in momentum space as
\begin{equation}\label{eq2}
  H(\mathbf{k})=\left(\begin{array}{cccccc}
0&c_\mathrm{int}&0&c_\mathrm{ext}e^{i\mathbf{k}\cdot\mathbf{a}_1}&0&c_\mathrm{int}\\
c_\mathrm{int}&0&c_\mathrm{int}&0&c_\mathrm{ext}e^{i\mathbf{k}\cdot\mathbf{a}_2}&0\\
0&c_\mathrm{int}&0&c_\mathrm{int}&0&c_\mathrm{ext}e^{i\mathbf{k}\cdot\mathbf{a}_3}\\
c_\mathrm{ext}e^{-i\mathbf{k}\cdot\mathbf{a}_1}&0&c_\mathrm{int}&0&c_\mathrm{int}&0\\
0&c_\mathrm{ext}e^{-i\mathbf{k}\cdot\mathbf{a}_2}&0&c_\mathrm{int}&0&c_\mathrm{int}\\
c_\mathrm{int}&0&c_\mathrm{ext}e^{-i\mathbf{k}\cdot\mathbf{a}_3}&0&c_\mathrm{int}&0
\end{array}
  \right),
\end{equation}
where $\mathbf{k}=(k_x,k_y)$ is the Bloch momentum, $\mathbf{a}_1=(0,a)$, $\mathbf{a}_2=(\sqrt{3}a/2,a/2)$,  $\mathbf{a}_3=(\sqrt{3}a/2,-a/2)$, and $a$ is the lattice constant. The spectrum of the system can be obtained by solving the eigen equation,
\begin{equation}\label{eq3}
  H(k)u_{n}(\mathbf{k})=E_{n}(\mathbf{k})u_{n}(\mathbf{k}),
\end{equation}
where $u_{n}(\mathbf{k})$ is the periodic part of the Bloch wave function, and $n$ is the index of the band.

Two configurations are studied, one is the topological trivial case with $c_\mathrm{int}>c_\mathrm{ext}$ shown in Fig.~\ref{dimerisedhoneycomb}(a), and the other is the topological nontrivial case with $c_\mathrm{int}<c_\mathrm{ext}$ shown in Fig.~\ref{dimerisedhoneycomb}(b). The spectrum of the strip structure with periodic boundary condition along the $x$ direction and open boundary condition along the $y$ direction is studied. For the topological trivial case, there is no states in the band gap as shown in Fig.~\ref{dimerisedhoneycomb}(c), while for the topological nontrivial case, there are topological edge states in band gap as shown in Fig.~\ref{dimerisedhoneycomb}(d). Note that each edge supports a pair of counter-propagating edge states. Besides, the system is a weak topological insulator in the sense that the topological edge states are gapless only in the zigzag edge but become gapped for other edge boundaries.

\subsection{Wilson loop} Wilson loop approach is a nice tool to calculate the topological invariant, which is gauge invariant and compatible with numerical implementations~\cite{adv.phys.64.227weng}. The method is quite general and can be used to describe both tight-binding models and continuous models~\cite{new.j.phys.21.093029wang}. It is also convenient to use it to describe multi-band system like introduced in Fig.~\ref{dimerisedhoneycomb}. Here we first provide the concept of the Wilson loop and then apply it to describe the topological properties of the breathing honeycomb lattice discussed above. Considering a matrix $M^{k_i,k_{i+1}}$, whose elements are obtained from

\begin{equation}
 M_{nn'}^{k_i,k_{i+1}}= \langle u_{n,k_i}|u_{n',k_{i+1}}\rangle,
\end{equation}
where $n$ and $n'$ are band indices covering all the $N$ bands below the concerned band gap, and $k_i$ and $k_{i+1}$ are two closer points on a loop in Brillouin zone. The Wilson loop can be obtained from the matrix product of a class of $M^{k_i,k_{i+1}}$,

\begin{equation}
    W=\prod_{i=1}^{N}M^{k_i,k_{i+1}},
\end{equation}
where $k_1,k_2,\ldots,k_{N}$ form a closed loop in the Brillouin zone and $k_{i+N}=k_{i}$. For one-dimensional problem, the Berry phases of each band are given by

\begin{equation}
    \theta_n\equiv i\log{w_n},
\end{equation}
where $w_n$ is the $n^{\text{th}}$ eigenvalue of $W$. For higher-dimensional system, we can consider the Berry phase along one direction $k_\parallel$ as a function of $k_\perp$, i.e., $\theta_n(k_\perp)$. For example, the well-knownn Chern number in two-dimensional system can be obtained by the winding of $\theta_n(k_y)$ as introduced in Eq.~\ref{chern}.

\begin{figure}[htbp]
\flushright
    \includegraphics[width=0.85\linewidth]{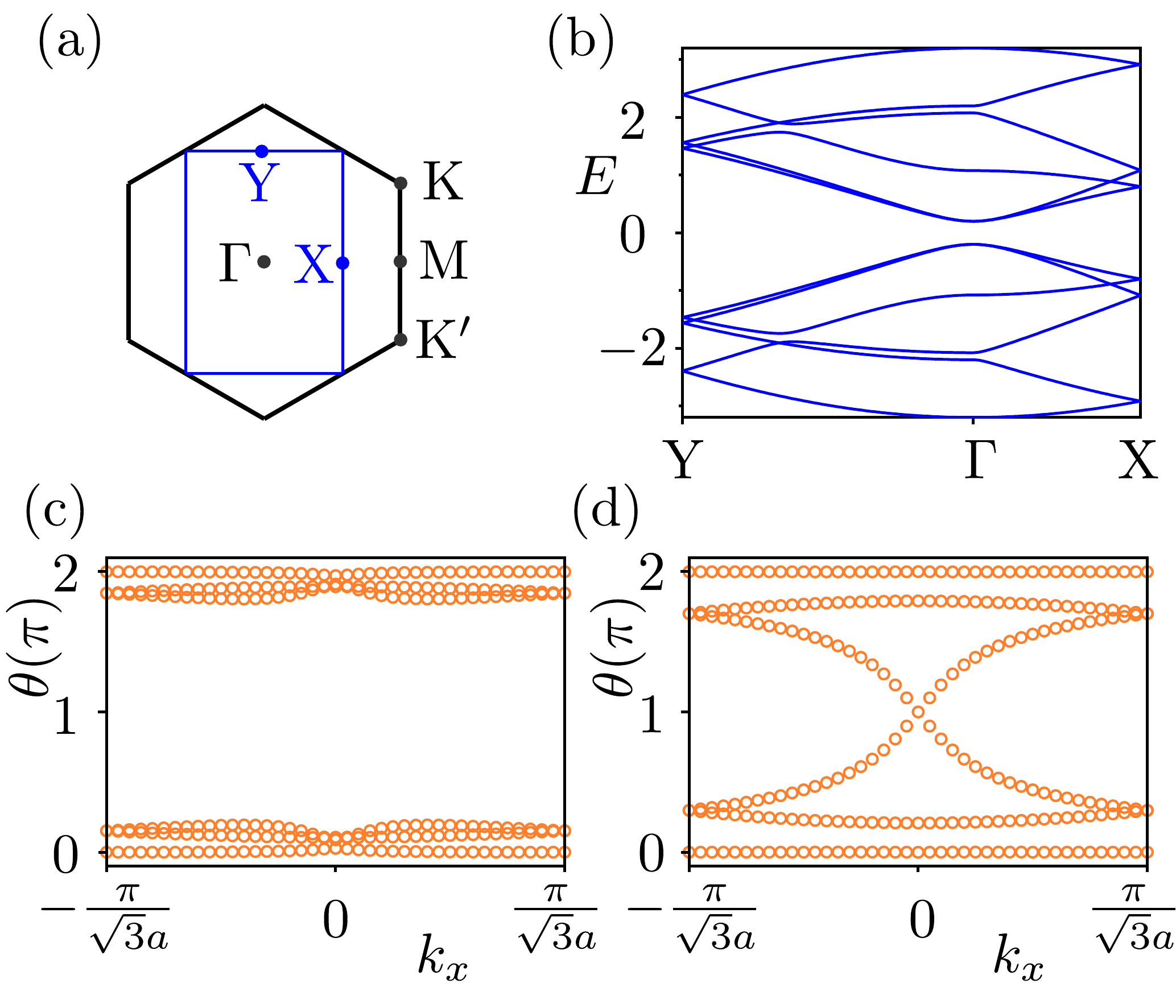}
    \caption{Bulk bands and the Berry phase $\theta_n(k_x)$ as function of $k_x$ (also known as Wannier bands). (a) The first Brillouin zone colored by black and the reduced Brillouin zone colored by blue. (b) The bulk bands in the reduced Brillouin zone. (c) The Wannier bands of the lower six bands for the topological trivial model. $c_i=1.2,c_o=1$. (d) The Wannier bands of the lower six bands for the topological nontrivial model. $c_i=1,c_o=1.2$.}
    \label{wilson}
\end{figure}

Next we apply the Wilson loop approach to study the topological properties of the breathing honeycomb lattice. The first Brillouin zone of the breathing honeycomb lattice is a regular hexagon shown in Fig.~\ref{wilson}(a). $\Gamma$, $\mathrm{M}$, $\mathrm{K}$ and $\mathrm{K}'$ are the high symmetric momentum points. From Fig.~\ref{dimerisedhoneycomb}, we notice that the unit cell is enlarged when considering the strip structure along the zigzag direction. The enlarged unit cell corresponds to a reduced first Brillouin zone which is shown in Fig.~\ref{wilson}(a) by blue color. To describe its topological properties, we need to consider the bulk bands of the enlarged unit cell. The original unit cell contains six sites so that there are six bulk bands. When considering the enlarged unit cell, the bulk bands are fold and there are totally twelve bulk bands with six below the zero gap and six up to the zero gap as shown in Fig.~\ref{wilson}(b). The Berry phases for the lower six bands are calculated by the Wilson loop approach. The results for the topological trivial model and topological nontrivial model are shown in Figs.~\ref{wilson}(c) and ~\ref{wilson}(d), respectively. We notice for the topological trivial model, the Berry phases for all six bands are located around 0 or $2\pi$, which can be continuously transformed to 0 or $2\pi$, thus is just an atomic insulator. However, for the topological nontrivial model, the Berry phases of the two bands are crossing around $\pi$, which is corresponding to a nonzero polarization and is the topological origin of the topological edge states in Fig.~\ref{dimerisedhoneycomb}(d). The Berry phase crossing and the topological edge states here have similar behavior to the topological insulators with time-reversal symmetry. The topological edge states in Fig.~\ref{dimerisedhoneycomb}(d) are double degenerate. However, they are different. Here the Wannier bands in Fig.~\ref{wilson}(d) are still gapped while the Wannier bands of topological insulator are gapless. The reason is that here the topological breathing honeycomb lattice is a weak topological insulator and the topological edge states only exist along specific directions. In addition, the topological edge states in Fig.~\ref{dimerisedhoneycomb}(d) do not connect with the bulk states.

\subsection{Symmetry indicators} Above we have shown that the topological states can be described by the Wannier bands calculated from the Wilson loop approach. However, the calculation is still inconvenient. Fu and Kane have shown one simple way to calculate the topological invariant of topological insulators with inversion symmetry, which only consider the inversion symmetry eigenvalues at high symmetric momentum points~\cite{phys.rev.b.76.045302fu}. This method has inspired the development of symmetry indicators of band topology, which can be used to deal with various symmetries and different topological states~\cite{j.phys.condens.matter.32.263001po,annu.rev.condens.matter.phys.12.225cano}. The development of symmetry indicators enables the diagnosis of topological materials~\cite{nature.566.486tang,nature.566.475zhang} and further inspires new concepts, such as higher-order topological insulators and fragile topological insulators~\cite{phys.rev.lett.121.126402po}.

Atomic insulators with crystalline symmetry have symmetric and exponentially localized Wannier functions. Given a crystal symmetry, the band representation for all possible atomic insulators at high symmetric momentum points can be obtained by placing different orbitals at different maximal Wyckoff positions. Insulators with specific symmetry can be classified into different topological phases by comparing their symmetry eigenvalues at high symmetric momentum points with atomic insulators. The insulators, which can be treated as sum of atomic insulators, are classified into normal insulators. Those can be smoothly deformed into a linear superposition of atomic insulator with some coefficients being negative are classified into fragile topological insulators. The conventional topological insulators are more stable, and they can not be transformed into atomic insulators without breaking symmetry or closing band gap, so that can not be expressed as a linear superposition of atomic insulators. Due to the existence of symmetry, the normal insulator can be further subdivided into different topological phases. For example, the atomic insulators with Wannier function at boundary of unit cell are topologically different from the atomic insulators with Wannier functions at the center of unit cell. In this sense, we also can treat them as topological nontrivial states although they usually do not support gapless topological edge states. Those topological phases can be roughly distinguished by checking the symmetry eigenvalues at high symmetric momentum points~\cite{j.phys.condens.matter.32.263001po,annu.rev.condens.matter.phys.12.225cano}.

\begin{figure}[htbp]
\flushright
    \includegraphics[width=0.85\linewidth]{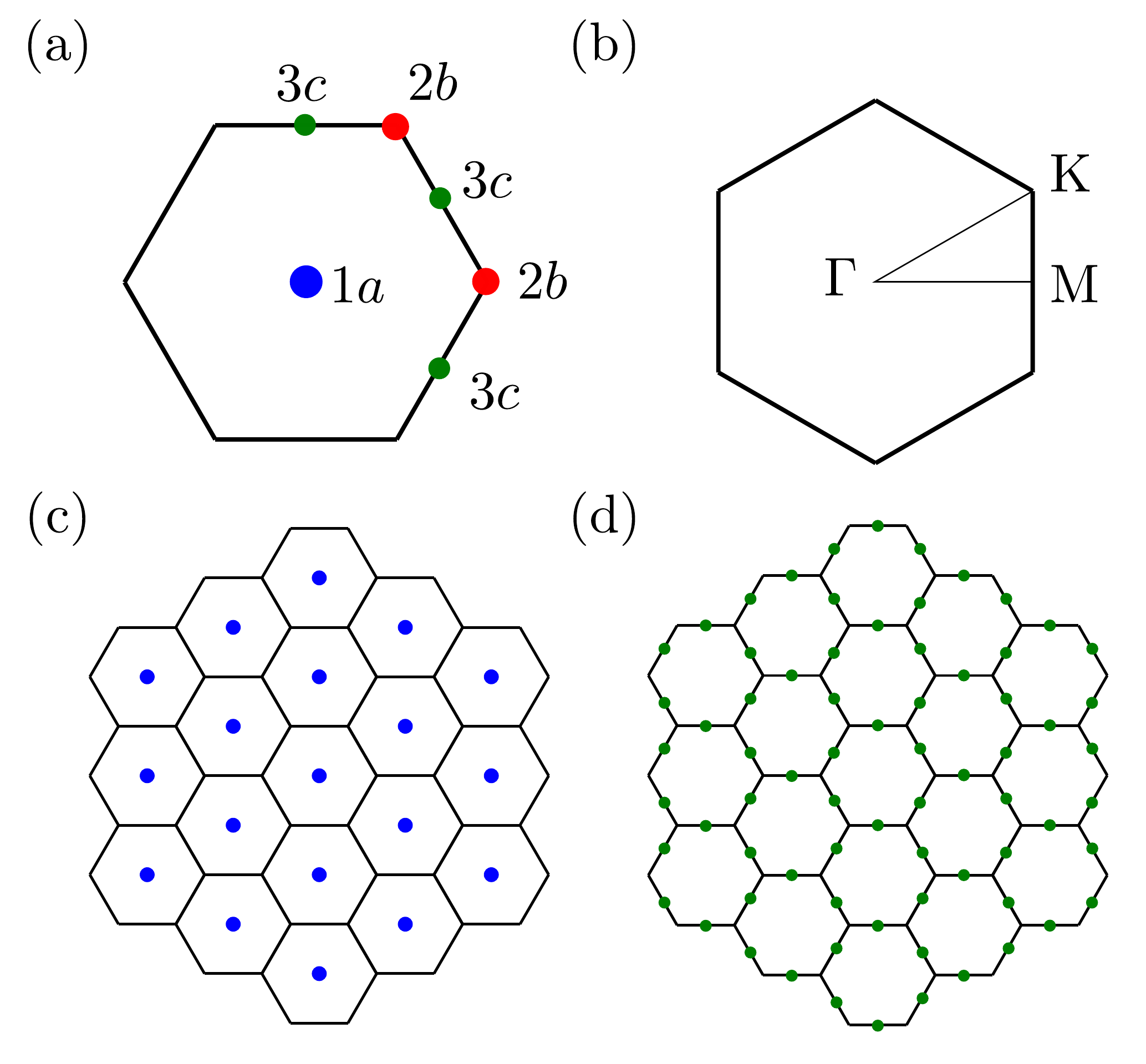}
    \caption{Charge distribution for breathing honeycomb lattice. (a) Maximal Wyckoff positions for breathing honeycomb lattice. (b) The Brillouin zone. (c) Charge distribution for the trivial case with $c_i>c_o$. (d) Charge distribution for the nontrivial case with $c_i<c_o$.}
    \label{symmetry}
\end{figure}

We still use the breathing honeycomb lattice shown in Fig.~\ref{dimerisedhoneycomb} as an example. The Hamiltonian in Eq.~\ref{eq2} supports six fold rotation symmetry
\begin{equation}
    C_6H(\mathbf{k})C_6^{\dag}=H(R_6\mathbf{k}),
\end{equation}
where $C_6$ is a six by six matrix with elements $C_{m,\mod(m+1,6)}=1$ and $R_6$ is a matrix that rotates the crystal momentum by $2\pi/6$ as
\begin{equation}
  R_6=\left(\begin{array}{cc}
        \cos(2\pi/6) & \sin(2\pi/6) \\
        -\sin(2\pi/6) & \cos(2\pi/6)
      \end{array}\right).
\end{equation}
The maximal Wyckoff positions for the breathing honeycomb lattice are shown in Fig.~\ref{symmetry}(a). There are one orbital
at the center of the unit cell $1a$ whose little group is $C_6$, two orbitals at position $2b$ whose little group is $C_3$, and three orbitals at position $3c$ whose little group is $C_2$. In momentum space, the first Brillouin zone is shown in Fig.~\ref{symmetry}(b). $\Gamma$, M and K are three high symmetric momentum points. The Bloch Hamiltonian supports $C_6$ symmetry at $\Gamma$, $C_2$ symmetry at M and $C_3$ symmetry at K. We consider the lowest three bands so there are three orbitals per unit cell. There are three possible atomic insulators which can be equivalent to it: the first one is built by placing three orbitals at position $3c$, the second by placing one orbital at $1a$ and two orbitals at $2b$, and the third by placing three orbitals at $1a$. The $C_2$ symmetry eigenvalues for $c_i>c_o$ are $-1,+1,+1$ at $\Gamma$ point and $-1,+1,+1$ at M point. It is equivalent to one $p$ orbital and two $s$ orbitals at $1a$. So the charge is distributed at the center of the unit cell, as shown in Fig.~\ref{symmetry}(c). Such phase is topologically trivial. The $C_2$ symmetry eigenvalues for $c_i<c_o$ are $-1,-1,-1$ at $\Gamma$ point and $-1,+1,+1$ at M point. It is equivalent to three $p$ orbitals at $3c$. The charge distribution is shown in Fig.~\ref{symmetry}(d). We notice there are charges on the edges and corners. They contribute to edge states and corner states. So the breathing honeycomb lattice with $c_i<c_o$ is still an atomic insulator, but it is topologically nontrivial with $c_i>c_o$.

\subsection{Floquet topological insulators and non-Hermitian topological insulators} Topological states have been extended to periodically driven system, which are known as Floquet topological insulators~\cite{phys.rev.lett.106.220402jiang,phys.rev.b.87.201109tong,phys.rev.x.3.031005rudner,phys.rev.b.95.195155morimoto,phys.rev.b.79.081406oka,phys.rev.b.82.235114kitagawa,phys.rev.b.84.235108kitagawa,nat.commun.8.13940hubener,nat.phys.16.38mciver,nat.phys.16.1058wintersperger,nat.rev.phys.2.229rudner}. Different from the static system, the Hamiltonian of a Floquet system is time dependent and periodic in time $H(t)=H(t+T)$, where $T$ is the driving period. Although the energy is not a conserved quantity any more, the quasienergy is conserved. We seek quasistationary solutions that have the form $|\psi(\mathbf{k},t)\rangle=e^{-i\varepsilon(\mathbf{k})t}|\phi(\mathbf{k},t)\rangle$, where $|\phi(\mathbf{k},t)\rangle$ is periodic in time with period $T$. The quasienergy band $\varepsilon(\mathbf{k})$ can be obtained from the eigenequation,
\begin{equation}\label{floquet}
  U_{T}(\mathbf{k})|\phi(\mathbf{k},t)\rangle=e^{-i\varepsilon(\mathbf{k})T}|\phi(\mathbf{k},t)\rangle,
\end{equation}
where $U_{T}(\mathbf{k})\equiv\mathcal{T}\exp[-i\int_{0}^{T}H(\mathbf{k},\tau)d\tau]$ is the Floquet operator, and $\mathcal{T}$ denotes the time-ordering operator.

At the beginning, Floquet engineering was used as a method to tune the topological properties of band structures. A normal insulator can be driven into topological nontrivial states by Floquet engineering~\cite{phys.rev.let.106.220402jiang}. Most Floquet topological states can be described by the topological invariants as the static case. The Floquet system can be treated as a static system with effective Hamiltonian $H_{\text{eff}}(\mathbf{k})\equiv\frac{i}{T}\log U_{T}(\mathbf{k})$. Later studies show that Floquet system can support topological states without static counterpart. From Eq.~\ref{floquet}, we notice $U_{T}(\mathbf{k})$ is an unitary matrix so that $\varepsilon(\mathbf{k})$ is periodic in frequency domain with period $2\pi/T$. It means the Floquet system can support fully gapless states~\cite{phys.rev.b.79.081406oka}. The unitary $U_{T}(\mathbf{k})$ is topologically nontrivial if it can not be continuously transformed into an identity matrix. Besides, Floquet system also supports topological states that are described by nontrivial topology of time evolution operator $U(\mathbf{k},t)\equiv\mathcal{T}\exp[-i\int_{0}^{t}H(\mathbf{k},\tau)d\tau]$. The most famous example is the anomalous Floquet topological insulator, which supports chiral edge states with zero Chern number~\cite{phys.rev.x.3.031005rudner}. Anomalous Floquet higher-order topological insulator is another example, which supports topological corner states with vanishing dipole and quadrupole polarizations~\cite{phys.rev.b.99.045441bomantara,phys.rev.b.100.085138rodriguez-vega,phys.rev.lett.124.057001hu,phys.rev.lett.124.216601huang,phys.rev.b.103.L041402zhu,phys.rev.b.104.l020302zhu}. We notice the Floquet operator $U_{T}(\mathbf{k})$ is unitary matrix which is similar to the scattering matrix. Their eigen-equations are equivalent to each other. So the Floquet topological states can be mimicked by scattering systems~\cite{phys.rev.lett.110.203904liang,phys.rev.b.89.075113pasek,phys.rev.x.5.011012hu,nat.commun.7.11619gao,phys.rev.b.95.205413delplace,phys.rev.lett.124.253601afzal}. Phononic coupled ring resonator system is a nice platform to simulate Floquet topological insulator, where the scattering process plays the role of time. Besides, the coupled waveguide system can also be used to mimic Floquet phenomena, where the propagation along specific direction in space plays the role of time~\cite{nature.496.196rechtsman,phys.rev.lett.117.013902leykam,nat.commun.8.13918mukherjee,nat.commun.9.4209mukherjee,nat.commun.8.13756maczewsky,nat.mater.19.855maczewsky,science.368.856Mukherjee}.

Non-Hermitian topology is another direction that attracts a lot of attention recently. Non-Hermitian band theory is widely different from the Hermitian one~\cite{nat.phys.14.11el-ganainy,adv.phys.69.249ashida,rev.mod.phys.93.015005bergholtz,nat.rev.phys.4.745ding}. First, the bulk spectrum of non-Hermitian system highly depends on the boundary condition. It breaks the usual bulk-boundary correspondence~\cite{phys.rev.lett.121.086803yao,phys.rev.b.99.201103lee}. Second, the bulk spectrum of non-Hermitian system is typically complex so that different kinds of band gaps can be defined: a non-Hermitian Hamiltonian is defined to have a point gap if and only if its complex-energy bands do not cross a reference point in the complex-energy plane; and a non-Hermitian Hamiltonian is defined to have a line gap if and only if its complex-energy bands do not cross a reference line in the complex-energy plane~\cite{phys.rev.x.8.031079gong,phys.rev.x.9.041015kawabata}. While the line gap topology can be transformed to a Hermitian counterpart without closing the band gap, the point gap is unique to non-Hermitian system. A topological nontrivial point gap with nonzero winding number contributes to the non-Hermitian skin effect~\cite{phys.rev.lett.125.126402zhang,phys.rev.lett.123.066404yokomizo}, causing all the bulk states localized under open boundary condition~\cite{phys.rev.lett.121.086803yao,phys.rev.lett.121.026808kunst,phys.rev.x.8.041031mcdonald,phys.rev.b.99.245116kunst,phys.rev.research.1.023013longhi,phys.rev.b.99.081103jin,phys.rev.lett.123.170401song,phys.rev.b.99.201103lee,phys.rev.lett.124.056802borgnia,phys.rev.b.100.035102deng,phys.rev.b.100.054301jiang}. While the non-Hermitian skin effect itself has topological origin from point gap, it also affects the usual band topology for line gap. The usual bulk-boundary correspondence does not work any more for non-Hermitian system with non-Hermitian skin effect. Non-Bloch topological band theory has been developed to recover the bulk-boundary correspondence for non-Hermitian system~\cite{phys.rev.lett.121.136802yao,phys.rev.lett.123.066404yokomizo,phys.rev.b.100.165430imura,phys.rev.b.101.195147kawabata,phys.rev.lett.124.066602longhi,phys.rev.lett.125.118001scheibner,phys.rev.b.101.045415zhang,phys.rev.lett.125.226402yang,phys.rev.research.2.013280zhu,pnas.117.29561ananya,nat.phys.16.747helbig,nat.phys.16.761xiao,phys.rev.b.103.195414zhu,phys.rev.b.103.L041404zhou,phys.rev.b.105.045422wu}. Except the non-Hermitian skin effect, other non-Hermitian topological states like non-Hermitian Chern bands~\cite{phys.rev.lett.121.136802yao,phys.rev.b.98.165148kawabata,phys.rev.b.98.245130chen,phys.rev.b.100.081104hirsbrunner,phys.rev.b.105.075128xiao}, hybrid skin-topological modes~\cite{phys.rev.lett.123.016805lee,phys.rev.lett.124.250402li,phys.rev.b.104.L161117ghorashi,phys.rev.b.104.L161116ghorashi,phys.rev.b.106.035425zhu,phys.rev.lett.128.223903li}, critical non-Hermitian skin effect~\cite{nat.commun.11.5491li} and second-order non-Hermitian skin effect~\cite{phys.rev.b.102.205118kawabata,phys.rev.b.102.241202okugawa,phys.rev.b.103.045420fu,phys.rev.b.104.L121101kim} have been widely studied.

\subsection{Phenomenological theories: Jackiw-Rebbi theory}
Dirac equation proposed to behave the relativistic and quantum mechanical electrons, now is mainly applied in condensed matter physics where it is widely extended to describe the low-energy excitations. The solution of Dirac equation induces the problem of negative energy modes, because physical electrons only carry positive energy. Based on this contradiction, Dirac speculated the existence of anti-electrons which was demonstrated by the experimental discovery of positrons in 1932~\cite{science.76.238anderson}. The original Dirac equation is written as follows,
\begin{equation}
(\boldsymbol{\alpha}\cdot\boldsymbol{p}+\beta{m})\Psi=i\frac{\partial \Psi}{\partial t}.
  \label{dirac equation}
\end{equation}
Here, coefficients $\boldsymbol{\alpha}$ and $\beta$ are matrices, which is a significant feature of the Dirac equation. $\boldsymbol{p}$ is the momentum operator with $\boldsymbol{p}=i\nabla$, and $m$ represents the mass term. The mass term of the normal Dirac equation is homogenous and position-independent. However, the defects or disturbances in the system change the constant mass term. These inhomogenous mass terms arising from defects or disturbances can seem like an effective phonon field implemented on the system.

In 1976, Jackiw and Rebbi obtained a soliton state with zero energy by solving the Dirac equation based on a soliton-monopole system when the Fermi field is present~\cite{phys.rev.d.13.3398jackiw,phys.scr.2012.014005jackiw}. This means that the inhomogenous and position-dependent mass term can give rise to some intriguing phenomena, as the soliton states emerge on the domain wall between two distinct lattices with opposite mass terms. These soliton states as a significant feature of Jackiw-Rebbi model can be realized in acoustic systems through proper mapping.

\begin{figure}[htbp]
\flushright
    \includegraphics[width=0.9\linewidth]{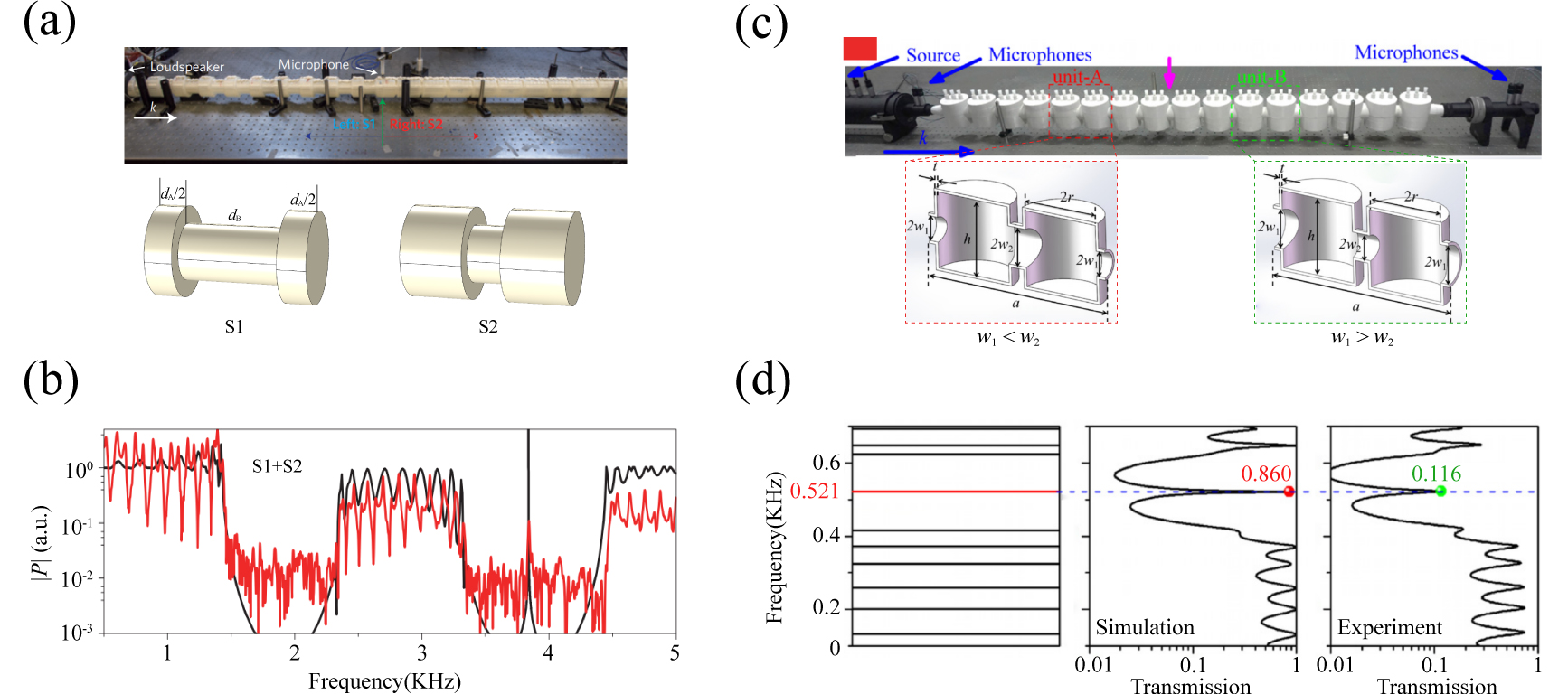}
    \caption{Jackiw-Rebbi domain wall in one-dimensional acoustic systems. (a) Configuration of the domain wall structure and its constituent unit cells S1 and S2. Green arrow in (a) marks the position of domain wall. (b) Acoustic pressure response spectrum at the domain wall labeled in (a). The soliton state has a strong response at the second gap. (c) Experimental set-up. The sample is an acoustic waveguide consisting of two kinds of phononic crystals: unit cell A and unit cell B. Pink arrow in (c) denotes the domain wall where a soliton state emerges between unit cells A and B. (d) Energy spectrum and its corresponding transmission spectrum for the sample in (c). The soliton state labeled as red line occurs in the energy gap. A transmission peak (green circle) indicating the existence of a soliton state is measured in experiment. Figures
   adopted from: (a) and (b), Ref.~\cite{nat.phys.11.240xiao}; (c) and (d), Ref.~\cite{appl.phys.lett.113.203501li}}
    \label{JRdminone-dimensional}
\end{figure}

Several recent papers reveal the method of realizing the acoustic analogues of Jackiw-Rebbi model. For instance, Xiao {\sl et al.} design a sandwich-like cylindrical waveguide structure~\cite{nat.phys.11.240xiao}: each acoustic unit cell has two wider tubes (tube-A) with a narrower tube (tube-B) inserted in (Fig.~\ref{JRdminone-dimensional}(a)). They demand that the longitudinal length of each unit cell is a constant 8.5cm, which means the adjustable parameter is only the length difference $\triangle d=(d_\text{A}-d_\text{B} )/2$ between tube-A and tube-B. By redistributing the lengths of tube-A and tube-B, they realize a band inversion of the second gap at the critical point $\triangle d$=0.49cm, where unit cells at its both sides have opposite effective mass terms. A $\mathbb{Z} _2$ symmetric geometry phase called as Zak phase~\cite{phys.rev.lett.62.2747zak}---the integral of Berry connection, is an excellent tool to distinguish topological properties between unit cells with opposite mass terms. As shown in Fig.~\ref{JRdminone-dimensional}(a), they combine two topologically different unit cells (S1 and S2) with opposite mass terms together to obtain a domain wall structure. As the statement of Jackiw-Rebbi theory, there is a soliton state localized on the domain wall. Such a soliton state is successfully observed by Xiao {\sl et al.} in the experiment and its acoustic response is shown in Fig.~\ref{JRdminone-dimensional}(b). In another work, Li {\sl et al.} propose a one-dimensional periodic acoustic system in exact analogy with the Su-Schrieffer-Heeger model~\cite{phys.rev.b.22.2099su,appl.phys.lett.113.203501li}, which formerly depicts two stable arrays of carbon atoms in polyethylene. The unit cell of the acoustic system is composed of two resonators and two junction tubes connecting them. The effective opposite mass term can be obtained through exchanging the radii of two junction tubes which control the intercell and intracell hoppings. A domain wall constructed through unit cells A and B with opposite mass terms is shown in Fig.~\ref{JRdminone-dimensional}(c). Similarly, there is an in-gap soliton state localized on the domain wall (Fig.~\ref{JRdminone-dimensional}(d)). This soliton state is confirmed by Li {\sl et al.} through transmission scanning and they verify the robustness of this soliton state. For mechanical systems, Vitelli {\sl et al.} also design a similar structure composed of a periodic arrangement of alternating massless rigid rotors, constrained to rotate about fixed pivot points, around an equilibrium angle $\bar{\theta}$ at odd-numbered sites and $\pi-\bar{\theta}$ at even-numbered sites~\cite{proc.natl.acad.sci.u.s.a.111.13004chen}. On the domain wall between rotors in right-leaning and left-leaning states (corresponds to distinct winding number in Ref.~\cite{nat.phys.10.39.kane}), a soliton state appears as expected. Both above works focus on one-dimensional domain wall systems and acquire the similar result that, a soliton state emerges on the domain wall predicted by Jakiw-Rebbi theory. And that prompts us to consider, what is the presentation of Jackiw-Rebbi model in a two-dimensional system.

\begin{figure}[htbp]
\flushright
    \includegraphics[width=0.9\linewidth]{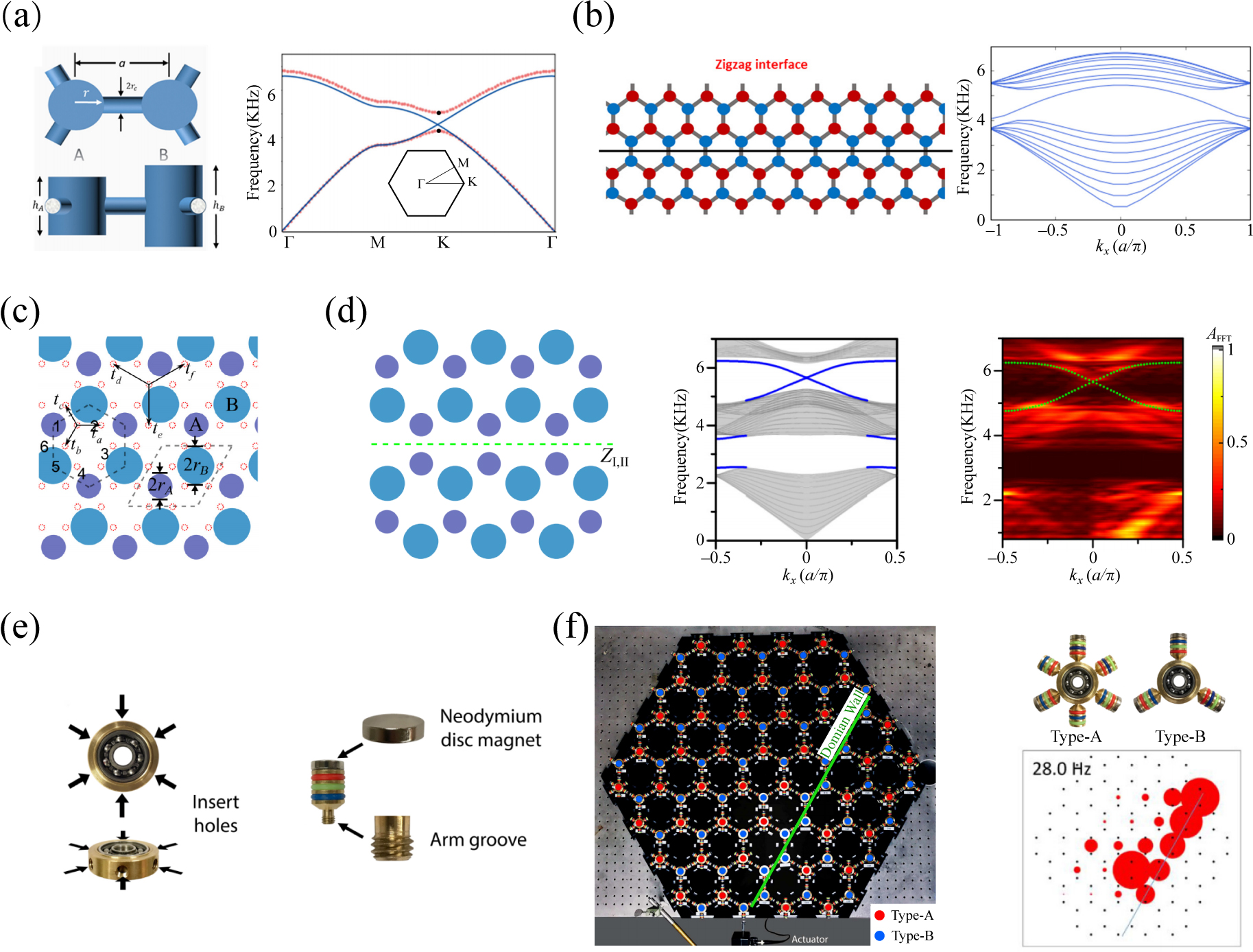}
    \caption{Jackiw-Rebbi domain wall in two-dimensional acoustic systems. (a) Configuration of the honeycomb acoustic unit cell and its corresponding energy spectrum with zero (blue curves) or nonzero (red curves) height difference $\Delta h$. In the case has nonzero $\Delta h$, an energy gap opens near the $K$ point in the Brillion zone. (b) Zigzag domain wall between two domains with opposite height difference $\Delta h=\pm0.15h$. A gapless edge state localized at the domain wall is shown in the gap of right spectrum. (c) Schematic of the acoustic crystals with two kinds of rigid rods embedded in the air background. (d) Schematic diagram of zigzag domain wall between domains I and II with opposite radius difference $\delta r=r_\text{A}-r_\text{B}$. The right panels are its calculated spectrum and measured spectra, where the green curves denote the edge dispersions. (e) Illustration of a spinner, with six insert holes to attach neodymium disc magnets. (f) Lattice configuration, with a domain wall (green line) divides the system into two distinct domains. The field distribution of the edge modes bound to the domain wall at 28Hz is presented in the right panel. Figures adopted from: (a) and (b), Ref.~\cite{j.appl.phys.123.091713yang}; (c) and (d), Ref.~\cite{phys.rev.appl.11.044086xie}; (e) and (f), Ref.~\cite{phys.rev.b.98.155138qian}}
    \label{JRdmintwo-dimensional}
\end{figure}

In one-dimensional systems, the behavior of Jackiw-Rebbi model is that, an in-gap soliton state is constrained to a zero-dimensional point called as the domain wall, and both sides of it have opposite mass terms. The simplest way to extend a one-dimensional system is that periodically arrange a one-dimensional system in other dimensions. Naturally, the 0D domain wall also expends its dimensions accordingly. There are numerous relevant works to present the properties of Jackiw-Rebbi model in two-dimensional systems. Zhang {\sl {\sl et al.}} design a periodic acoustic cavity system consisting of acoustic resonators~\cite{j.appl.phys.123.091713yang} connected by thin cylindrical coupling waveguides as shown in Fig.~\ref{JRdmintwo-dimensional}(a). Its energy spectrum exhibits linear dispersion near the $K$ and $K^\prime$ point, which is described by Dirac equation (Eq.~\ref{dirac equation}), named as Dirac cone. Through adjusting the height difference $\Delta h$ of cavities, they break the inversion symmetry of the system and successfully open a gap to introduce a topological index, valley Chern number, obtained by integrating Berry curvature around the valley. The opening of gap arises from the effective opposite mass terms performed around the intersection of Dirac cone~\cite{proc.natl.acad.sci.u.s.a.110.10546zhang}. As a result, valley Chern numbers in $K$ and $K^\prime$ valley host opposite signs, and their signs change as the sign of $\Delta h$ changes. When unit cells with distinct valley Hall phases are assembled to form a domain wall, the soliton state elaborated by Jackiw-Rebbi theory leads to gapless states propagating along the boundary, as shown in Fig.~\ref{JRdmintwo-dimensional}(b). In addition to the intuitive acoustic cavity configuration, acoustic scatterers and mechanical structures are also excellent platforms to realize the same physics. For example, Liu {\sl {\sl et al.}} proposed two kinds of valley Hall insulators based on triangular~\cite{nat.phys.13.369lu} and circular~\cite{phys.rev.appl.11.044086xie} (Fig.~\ref{JRdmintwo-dimensional}(c)) phononic scatterers, respectively. In the triangular scatterer structure, the valley Hall topological transition occurs when the rotation angle of triangular scatterer changes over $0^{\circ}$. Besides, in the circular scatterer system, it depends on the radius difference between two kinds of circular scatterers. Consequently, these two works observe the gapless chiral edge states transmitting along zigzag boundary line between two domains with opposite effective mass terms. Dispersion of edge modes is shown in Fig.~\ref{JRdmintwo-dimensional}(d). In another work, Prodan {\sl et al.} establish a mechanical counterpart based on magnetically coupled spinners~\cite{phys.rev.b.98.155138qian}---stainless steels with six arm grooves to attach neodymium disc magnet (Fig.~\ref{JRdmintwo-dimensional}(e)). And they observe the valley Hall edge modes stemming from the interaction of soliton states localized along the domain wall as shown in Fig.~\ref{JRdmintwo-dimensional}(f). There are other methods~\cite{phys.rev.lett.116.093901lu,phys.rev.appl.9.014001liu,nat.commun.11.762tian} to realize the Jackiw-Rebbi model in two-dimensional systems. These above works are enough to conclude that, Jackiw-Rebbi models are intuitively characterized by the establishment of a zero-energy soliton state and derived hinge states in lower dimensions than that of system, even though they usually appear at a finite frequency instead of zero energy in acoustic lattices. These states predicted by Jackiw-Rebbi theory are one kind of low-dimensional projections of the bulk properties termed as bulk-boundary correspondence. This correspondence is a useful tool to characterize the topology or polarization properties of materials.

Finally, we briefly discuss the analogues of Majorana fermions in acoustic systems. Complex fields $\Psi$ in Dirac equation describe charged particles and their anti-particles with opposite charges, such as electrons and positrons. In nature, there are also some neutral particles which are their anti-particles, as photons, phonons and gravitons etc. All of these are integer-spin bosons, and they are described by a real field $\Psi$.

\begin{figure}[htbp]
\flushright
    \includegraphics[width=0.9\linewidth]{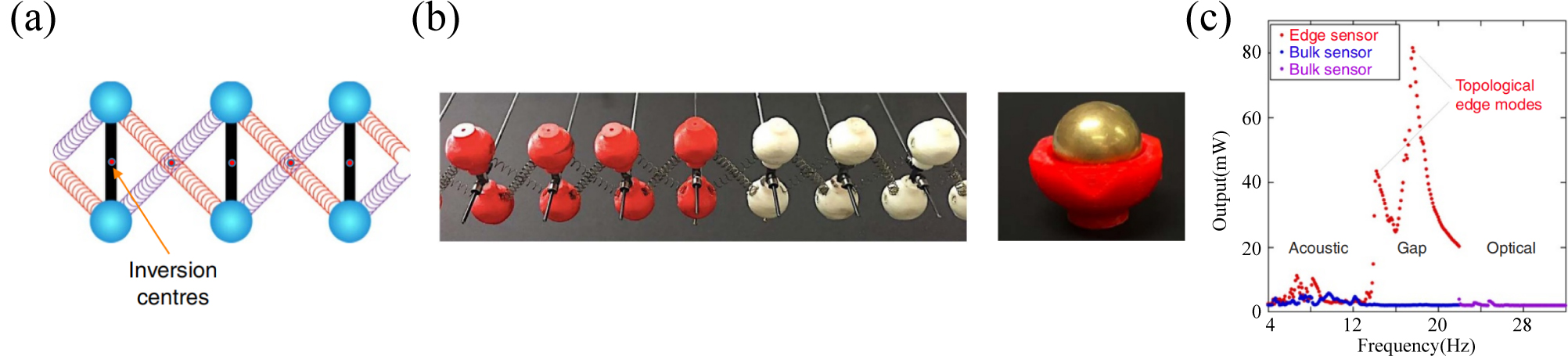}
    \caption{Majorana edge modes in the mechanical system. (a) Schematic diagram of dimer chain. (b) Experimental set-up. The experiment model consists of red and white dimers connected via springs. There is one such spring on one diagonal, while on the other there are two springs. The configuration of springs is switched between the red and white dimers. Each dimer consists of two brass balls encapsulated in plastic shells as shown in the right insert. (c) Measured phonon response. The split of the signal of edge mode is due to the nonlinear of the system and slight symmetry break. Figures adopted from Ref.~\cite{nat.commun.8.14587prodan}}
    \label{majoranainmechanic}
\end{figure}

Spin-1/2 neutral particles, whose anti-particles are themselves, are termed Majorana fermions playing an extremely vigorous role in the superconductor realm~\cite{science.364.1255jack}. The discussion about Majorana fermions needs the constraints of particle-hole symmetry protecting the superconductor. In the work of Prodan {\sl et al.}, they design a mechanical metamaterial~\cite{nat.commun.8.14587prodan} consisting of a periodic array of dimers connected via two springs as shown in Fig.~\ref{majoranainmechanic}(a). For each dimer, there are two motion degrees of freedom: one is translation and another one is rotation. The particle-hole symmetry of this system stems from the distinct behaviors of the translational and rotational degrees of freedom under inversion operation. Together with the intrinsic time-reversal symmetry, the particle-hole symmetry characterizes the BDI symmetry class. In one-dimensional fermionic topological superconductors from BDI symmetry class, the topological edge excitations are named Majorana fermions. Because, these edge modes are their self-mirror under the charge conjugation, like the Majorana fermion is its own antiparticle. E. Prodan {\sl et al.} measure the topologically protected edge modes in the experiment platform shown in Figs.~\ref{majoranainmechanic}(b) and~\ref{majoranainmechanic}(c), where the domain wall is established from the exchange of the number of springs on two diagonals for one side of unit cells. As shown in Fig.~\ref{majoranainmechanic}(b), they denote the distinct mechanical blocks with the red and white colors on dimers. The measured topological edge modes in Fig.~\ref{majoranainmechanic}(c) are invariant under a charge conjugation. For instance, the motion of these Majorana edge modes remains unchanged when the translational and rotational degrees of freedom are exchanged and follow with time-reversal operation. This is a significant feature to distinguish them from ordinary edge modes. This fascinating work gives a method to construct the mechanical analogues of Majorana fermions from BDI topological class, which inspired lots of researches in this direction. What is more, the majorana-like zero mode, reflected by the Dirac vortex originated from the Jackiw-Rossi mechanism, has been observed in the Kekul$\acute{e}$-distorded mechanical~\cite{Adv.mater.31.1904386chen} and acoustic~\cite{PRL.123.196601gao} metamaterials.

\section{Topological phononic metamaterials}

\subsection{Valley and spin Hall phononic crystals}
We first introduce the topological phononic crystals built on valley and pseudospin degrees of freedom, in which time-reversal symmetry remains invariant. Since these types of phononic crystals do not require time-reversal-breaking interactions, such as measurable responses to external magnetic fields or circulating fluid flows, they are easily achievable in experiment, and attracted considerable attention in the early days of exploring topological phononic crystals~\cite{science.349.47susstrunk,phys.rev.lett.116.093901lu,nat.commun.7.13368peng,nat.phys.12.1124he,nat.phys.13.369lu,Phys.Rev.Lett.120.116802lu,Phys.Rev.Lett.118.084303zhang}. Different from the various pseudospins recently proposed to mimic the quantum spin Hall effect~\cite{phys.rev.lett.95.226801kane,science314.1757bernevig,Science318.766konig}, valley is a long-stand concept in solid band theory~\cite{Phys.Rev.105.1933herring}, and originally introduced to indicate the local extrema appearing in band structures. Considering that achieving topology lies chiefly on the bands with the aggregation of Berry curvature, which is generally inversely proportional to width of the band gap, valleys are exactly the positions where nontrivial topology could occur. Since the wave motions in periodic systems are marked by Bloch momentum, valleys are also used to labeling topological states. In two-dimensional systems, a feasible way to obtain valley states is to break the Dirac cone degeneracy~\cite{Phys.Rev.B86.035141mei}. The Dirac cone in two-dimensional systems is typically represented by $k_x \sigma_x+k_y \sigma_y$, where the Dirac velocity is set to one without loss of generality and $\sigma_{i=x,y,z}$ are Pauli matrices. Since the Hamiltonian only exploits two of the three Pauli matrices, to open a gap in the Dirac cone we can introduce a perturbation term of the rest Pauli matrix, $m\sigma_z$, where $m$ is mass term tuning the band gap between the upper and lower valley states. Although the Dirac cone can appear at general positions of the Brillouin zone without any symmetry protection~\cite{phys.rev.b.89.134302lu}, the Dirac cone at the high symmetric position of the Brillouin zone is preferable to construct the valley states, which facilitates the calculation (no need to search the general positions in the Brillouin zone) and makes the valleys separate in the Brillouin zone as far away as possible (leaving enough space for regulation of the valley topology). Integration of the Berry curvature (localized around the valley point) over one valley region returns to a non-zero topological quantity, dubbed as valley Chern number. Since the time-reversal symmetry is intact, an integration of the Berry curvature over a whole band is always zero, and there always exists another valley in the same band carrying opposite valley Chern number.

\begin{figure}[htbp]
    \flushright
    \includegraphics[width=0.85\linewidth]{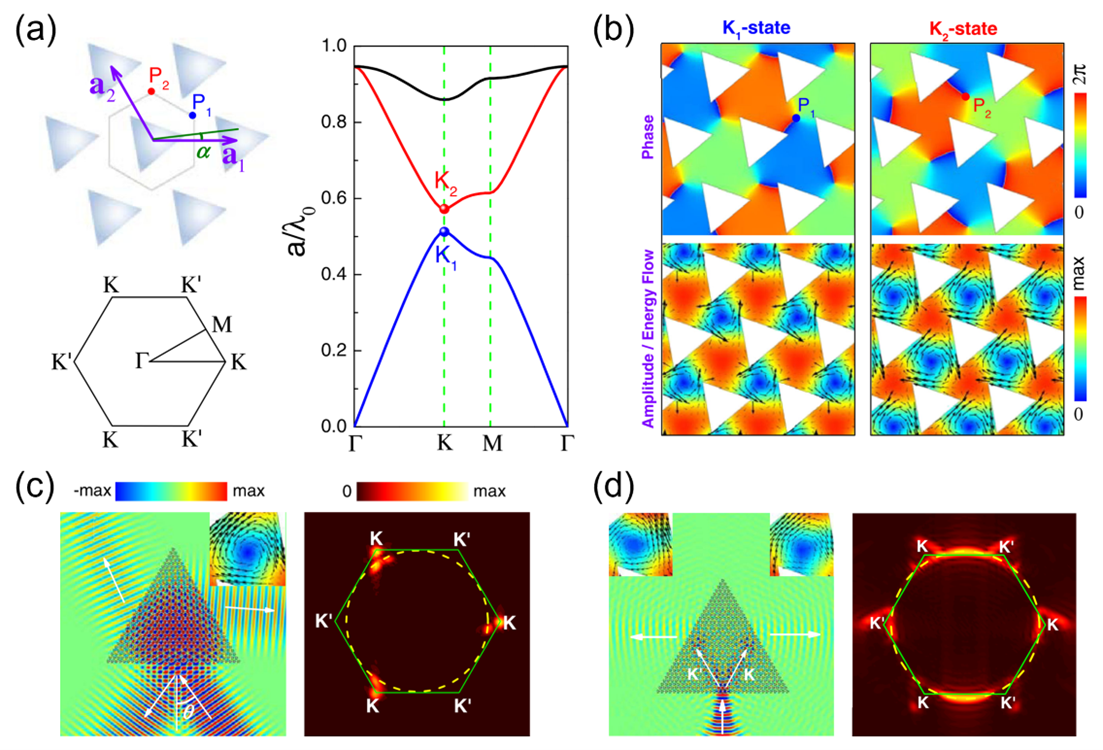}
    \caption{(a) Schematics of a hexagonal phononic crystals made of regular triangular rods. The $P_1$ and $P_2$ indicate the positions with $C_3$ symmetry. The band structure corresponding to a rotation angle $\alpha=10^{\circ}$. (b) Pressure field distributions for the valley states $K_1$ and $K_2$, where the top and bottom panels display (by color) the phase and amplitude patterns, respectively. The additional arrows in the bottom panels indicate the corresponding time averaged Poynting vectors. (c) Selection rules exemplified by the $K_2$ valley state, which is generated by a Gaussian beam incident onto a finite phononic crystals obliquely. The inset enlarges a vortex inside the phononic crystals. (d) The pressure distribution stimulated by a narrow Gaussian beam from the bottom, where the insets amplify the anticlockwise and clockwise vortices extracted from the left- and right-going beams inside the phononic crystals. The right panels in (c) and (d) display the corresponding Fourier spectra in momentum space, where the green solid and yellow dashed lines indicate the hexagonal first Brillouin zone and the isofrequency contour of free space, respectively. Figures adopted from~\cite{phys.rev.lett.116.093901lu}.}
    \label{jiuyang_Picture1}
\end{figure}

The valley phononic crystals was first built in a hexagonal lattice consisting of regular triangular rods as its scatterers~\cite{phys.rev.lett.116.093901lu}, where the lattice and the scatterer have the symmetry of point groups $C_{6v}$ and $C_{3v}$, respectively. The rotation of the scatterer [denoted by $\alpha$ in Fig.~\ref{jiuyang_Picture1}(a)] can cause a mismatch of the mirror symmetries between the lattice and the scatterers, which leads to a reduction of the symmetry of the phononic crystals. Specifically, the point group associated to the corner points ($K$ and $K^\prime$) of the hexagonal first Brillouin zone reduces from point group $C_{3v}$ for $\alpha=0^{\circ}$ to point group $C_3$ for $\alpha=10^{\circ}$, corresponding to a transition from Dirac cone dispersions to gapped valley dispersions [Fig.~\ref{jiuyang_Picture1}(a)]. Here the valley states at the $K$ point are focused, and those at the inequivalent K point can be deduced from the time-reversal symmetry. Demonstrated by the distributions of the phase, $\phi$, and the time-averaged Poynting vectors, $\langle \boldsymbol{S} \rangle \propto \bigtriangledown \phi$, of acoustic fields (upper panels and bottom panels in Fig.~\ref{jiuyang_Picture1}(b)), the valley states have peculiar vortex features, where the vortices are of reversed chirality for the valley states of lower and higher frequencies ($K_1$ and $K_2$), and are spatially separated at corners of the unit cell ($P_1$ and $P_2$). Therefore, the valley vortex feature, originating from the one-dimensional irreducible representations of the point group $C_3$, provides an obvious and measurable signal to detect the valley states. To characterize the quality of the vortex state, a quantity $Q=\int_{\rm{Cell}}({\bigtriangledown \times \langle {\boldsymbol{S}}\rangle})_z {\rm{d}}^2 r$ can be established~\cite{phys.rev.lett.116.093901lu}. This quantity is proportional to the acoustic angular momentum localized in one unit cell, $J_z=\int_{\rm{Cell}}{\langle{\bf{u}}\times \rho_0 {\bf u'} \rangle_z {\rm{d}}^2 r}$, where $u$ is the displacement of the vibration of acoustic field and $\rho_0$ is the mass density of the background. The $J_z$, as well as $Q$, acts as a continuous version of the phonon’s pseudospin angular momentum in atomic crystals~\cite{Phys.Rev.Lett.112.085503zhang,Phys.Rev.Lett.115.115502zhang}. The existence of nonvanishing $Q$ for the valley states implicate the intrinsic angular momenta of the valley states and the capability of rotating objects by transferring the angular momenta. Excitation of one single valley state in phononic crystals is critical to possible applications in acoustic manipulation. In Fig.~\ref{jiuyang_Picture1}(c), we show that the $K$ valley state can be excited by a Gaussian beam incidence, where the incident angle is chosen to satisfy the transversal momentum matching. Moreover, the time-reversal symmetry connected $K$ and $K^\prime$ valley states carry opposite chirality in the same frequency, which are used to realize a valley-chirality locked beam splitting~\cite{phys.rev.lett.116.093901lu,Phys.Rev.B95.174106ye} in analog to the valley Hall effect~\cite{Phys.Rev.Lett.108.196802xiao}. As shown in Fig.~\ref{jiuyang_Picture1}(d), under excitation by a narrow Gaussian beam, the acoustic wave in the phononic crystals spatially split into two branches, each of which has a vortex feature with opposite chirality. Based on this, we identify the phononic crystals with valley states reside in an acoustic valley Hall phase.

\begin{figure}[htbp]
    \flushright
    \includegraphics[width=0.85\linewidth]{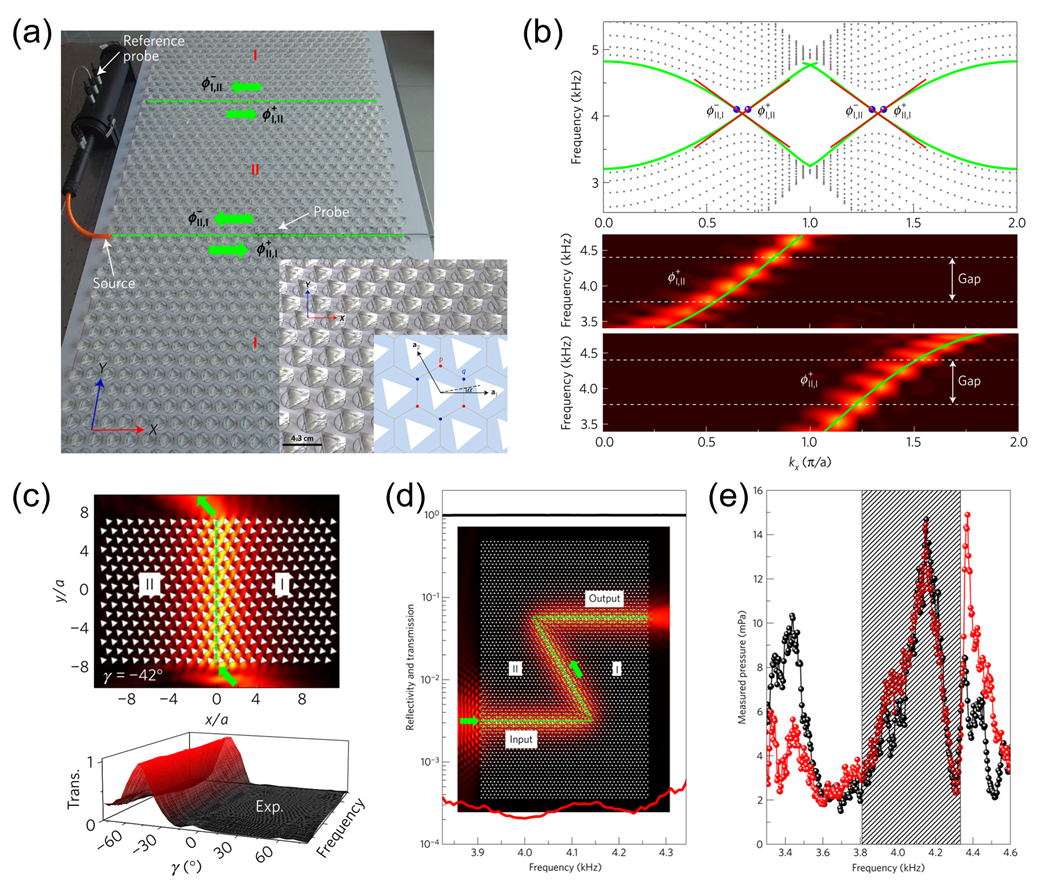}
    \caption{(a) Experimental setup to observe the topological valley Hall edge modes. The domains I and II represent the phononic crystals with $\alpha=-10^{\circ}$ and $10^{\circ}$, respectively. (b) Simulated (upper panel) and measured (lower panel) dispersions of the edge modes in the interfaces separating topologically distinct valley Hall phases. The simulation is performed by a superlattice structure containing two different horizontal interfaces. In each interface, one pair of time-reversal gapless valley edge are supported. (c) Valley-selective excitation of the valley Hall edge mode. Field distributions simulated for the sample with a vertical SC interface II–I, excited by the Gaussian beams at the incidence angles $\gamma =-42^{\circ}$ at 4.06 kHz. Lower panel: frequency-dependent angular spectra show the angular selectivity within the whole bulk gap. (d) and (e) Reflection immunity of the valley Hall edge modes from sharp corners. Black and red lines in (d) represent the calculated power transmission and reflection spectra for the two sharp turns in a zigzag path. Inset: field pattern simulated at 4.06 kHz. Black and red circles in (e) represent the measured transmitted pressure for the zigzag path and a straight channel sample, which are in good agreement in the bulk gap (shadow region). Figures adopted from~\cite{nat.phys.13.369lu}.}
    \label{jiuyang_Picture2}
\end{figure}

The topologically protected acoustic edges were later observed in interfaces between two distinct valley Hall phases~\cite{nat.phys.13.369lu}. The vortex chirality of the acoustic valley states can be switched by rotating the orientation of the triangular scatterers. Consider two phononic crystals consisting of scatterers rotated clockwise and counterclockwise by the same angle ($\pm \alpha$), they share the same symmetry of point group $C_3$ and are related by a horizontal mirror operation, which enables them identical valley dispersions yet reverses the chirality and position of the valley vortices. This leads to an inversion of the valley bands, if one focuses these properties of valley vortices. The phononic crystals thus possess two distinct valley Hall phases of opposite mass terms, valley Hall phase-I and valley Hall phase-II, corresponding to the rotation angles $\alpha$ belong to $(-60^{\circ},0)$ and $(0,60^{\circ})$, respectively~\cite{nat.phys.13.369lu}. The valley Chern number calculated for these two valley Hall phases are $C^K=1/2$ and $-1/2$, where the superscript denotes the $K$ valley~\cite{Phys.Rev.Lett.99.236809xiao}. As a result, the difference of $C^K$ across the interface between valley Hall phase-I and valley Hall phase-II is $\pm 1$, which, according to bulk-edge correspondence, leads to a valley induced edge state (of positive or negative group velocity) near the projection of $K$ valley. Similar situation occurs for the $K^\prime$, but $C^{K^\prime}=-C^K$ for the $K^\prime$ valley~\cite{phys.rev.lett.116.093901lu}. Such edge states are constructed in experiment with the setup shown in Fig.~\ref{jiuyang_Picture2}(a). A pair of counter-propagating valley edge states [green lines in the upper panel of Fig.~\ref{jiuyang_Picture2}(b)] are expected at interfaces, where the edge states $\phi_{({\rm{I}},{\rm{II}})}^\pm$ and $\phi_{({\rm{II}},{\rm{I}})}^\pm$ respectively label the gapless modes hosted by the interfaces I-II and II-I, along the $\pm x$ directions. Taking $\phi_{({\rm{II}},{\rm{I}})}^+$ as an example, its analytical expression is proportional to a spinor $(1,1)^T$ with respect to the valley states ($K_1$ and $K_2$) and exponentially decays away from the interface, which is solved by viewing valley Hall phase-I and valley Hall phase-II as continuum media characterized by the $K$ valley Hamiltonians with opposite mass terms~\cite{nat.phys.13.369lu}. The lower panel of Fig.~\ref{jiuyang_Picture2}(b) shows the measured spectra of the interface states by Fourier transforming the scanned acoustic pressure distributions along the interfaces, which agrees well with the simulation results. The valley edge states can not only be supported in the horizontal interface, but also the interface along an arbitrary direction (of angle $\theta$ with respect to the $+x$ direction). Keeping the left- and right-hand sides of the interfaces as valley Hall phases-II and I, the spinor factor of the edge state $\phi_{({\rm{II}},{\rm{I}})}^+$ becomes $(1,e^{i \theta})^T$. We show a vertical interface ($\theta=\pi/2$) in Fig.~\ref{jiuyang_Picture2}(c). To stimulate the valley edge state in that interface by an external Gaussian beam, the incident angel should fulfill a selection rule similar to that in Fig.~\ref{jiuyang_Picture1}(c). The selection rule of excitation reflects that the valley edge states are in nature composed of by the bulk valley states located at $K$ or $K^\prime$. To validate this phenomenon, we have measured transmissions validate the obliquely excitation. Note that a key feature of the valley edge state is its exponential decay away from the interface, and the decay length depends only proportionally on the mass-term difference, i.e., a constant related to the bulk properties. As shown in bottom panel of Fig.~\ref{jiuyang_Picture2}(c), the measured transmission is indeed frequency insensitive, in good agreement with the prediction based on the effective Hamiltonian~\cite{nat.phys.13.369lu}. The negligibly weak backscattering of the valley Hall phase edge states is demonstrated in a typical zigzag bending channel shown in Fig.~\ref{jiuyang_Picture2}(d), where the sound waves travel smoothly in the curved path despite suffering two sharp corners. Compared with the transmission of a straight channel of the same length, the reflection-immune edge states are experimentally verified, as shown in Fig.~\ref{jiuyang_Picture2}(e). Modifying the scatterers, valley Hall edge states can be more localized in the interface~\cite{Phys.Rev.Appl.9.034032zhang,j.appl.phys.123.091713yang}  and used in acoustic directional antennas~\cite{adv.mater.30.1803229.zhang} and lasing modes~\cite{nature.597.655hu}. The valley Hall edge states have been extended to phononic crystals of square lattice~\cite{Phys.Rev.B97.155124xia,Phys.Rev.Appl.12.024007zhu}, and further to heterostructures with Dirac materials inserted into the interface~\cite{nat.commun.11.3000wang}.

\begin{figure}[htbp]
    \flushright
    \includegraphics[width=0.85\linewidth]{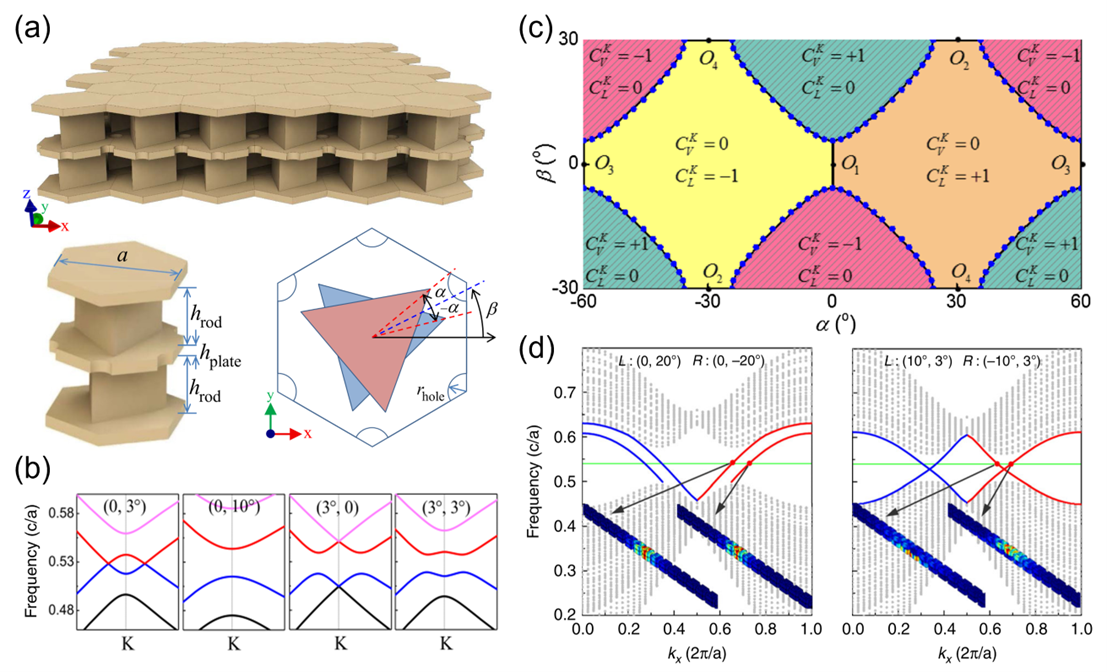}
    \caption{(a) Schematic of the coupled bilayer phononic crystals of hexagonal lattice. The angles $\alpha$ and $\beta$ together characterize the orientations of the triangular scatterers in both layers. (b) Local views of numerical dispersions for the bilayer phononic crystals with specified rod orientations. (c) Reduced phase diagram parametrized by the angles $\alpha$ and $\beta$. Different phases are labeled by two types of topological invariants $C_V^K$ and $C_L^K$, corresponding to valley and layer polarization. (d) Projected dispersions along an interface separating two topologically distinct valley Hall phases (left panel) and layer-valley Hall phases (right panel), where red and blue curves correspond to the edge modes projected by the $K$ and $K^\prime$ valleys. Figures adopted from~\cite{Phys.Rev.Lett.120.116802lu}.}
    \label{jiuyang_Picture3}
\end{figure}

Due to the lack of intrinsic degrees of freedom, such as spin and polarization, for acoustic waves, exploiting the layer degree of freedom~\cite{phys.rev.lett.106.156801zhang, proc.natl.acad.sci.u.s.a.110.10546zhang, nature.520.650ju} is of great significance and fruitful in phononic crystals. This approach becomes more feasible thanks to the developments in advanced three-dimensional printing technology. Beyond the single layer case, bilayer valley phononic crystals were proposed~\cite{Phys.Rev.Lett.120.116802lu}. As shown in Fig.~\ref{jiuyang_Picture3}(a), the bilayer phononic crystals consists of two layers of phononic crystals separated by a plate penetrated with a honeycomb array of circular holes, where the orientations of the triangular rods are characterized by the relative angle $\alpha$ and the common angle $\beta$. The combination of these angles enables the bilayer phononic crystals with various types of band structures, as exhibited in Fig.~\ref{jiuyang_Picture3}(b) in the vicinity of $K$ valley. The band structures, as well as their topological properties, can be captured by an effective Hamiltonian, which, in addition to the valley terms of single layer phononic crystals, possesses layer-dependent terms $\eta(\alpha s_z+\beta) \sigma_z-\bigtriangleup_c s_x$~\cite{Phys.Rev.Lett.120.116802lu}. The parameter $\eta$ depends on the spatial filling ratio of the scatterer rods, and $\bigtriangleup_c$ depends on the detailed geometry of the holes connecting the bilayers, both of which can be extracted from numerical simulations. For fixed parameters $\eta$ and $\bigtriangleup_c$, all the possible combinations of the angles $\alpha$ and $\beta$ leads to a reduced phase diagram as shown in Fig.~\ref{jiuyang_Picture3}(c). The phase diagram is separated by the straight and curved lines associated with ring and point degeneracies, respectively. The separated phase domains correspond to either nontrivial valley Hall phases (shaded regions) or nontrivial acoustic layer-valley Hall phases (unshaded regions), distinguished by the quantized topological invariant invariants $C_V^K$ and $C_L^K$~\cite{Phys.Rev.Lett.120.116802lu}. The former is a natural bilayer extension of the valley Chern number concerned in the monolayer system~\cite{nat.phys.13.369lu,Nat.Commun.8.1304wu,Phys.Rev.Lett.120.063902noh}, and the latter identifies the layer information and resembles that proposed for quantum spin Hall systems~\cite{Phys.Rev.Lett.95.136602sheng,phys.rev.lett.97.036808sheng}. The interfaces between these topologically nontrivial phases support acoustic edge states with different behaviors of the wave motions. Two cases are shown in figure 3(d). One is the interface between two valley Hall phases with the rod angles $\alpha=0^{\circ}$ and $\beta=\pm 10^{\circ}$. Crossing the interface $\bigtriangleup C_V^K=2$ and two edge modes (red curves) carrying positive group velocities appear near the $K$ valley, whose eigenfields disperse in both layers and indicate a mixing of layer pseudospins. For the interface between two layer-valley Hall phases with the rod angles $\alpha=\pm 10^{\circ}$ and $\beta= 3^{\circ}$, $\bigtriangleup C_L^K=2$ and in contrast to the valley Hall case the two edge modes carry different group velocities. In particular, the eigenfield concentrates dominantly in either the upper layer or the lower layer, that is a layer-pseudospin polarized transport. Therefore, assisted with the additional layer information, the valley edge modes can be either layer mixed or layer polarized, based on which even an interlayer converter has been conceived for possible application~\cite{Phys.Rev.Lett.120.116802lu}. The bilayer design late stimulates related research in square lattice~\cite{Phys.Rev.Appl.16.014058zhu}, and van der Waals metamaterials~\cite{Phys.Rev.B101.121103(R)dorrell}.

\begin{figure}[htbp]
    \flushright
    \includegraphics[width=0.85\linewidth]{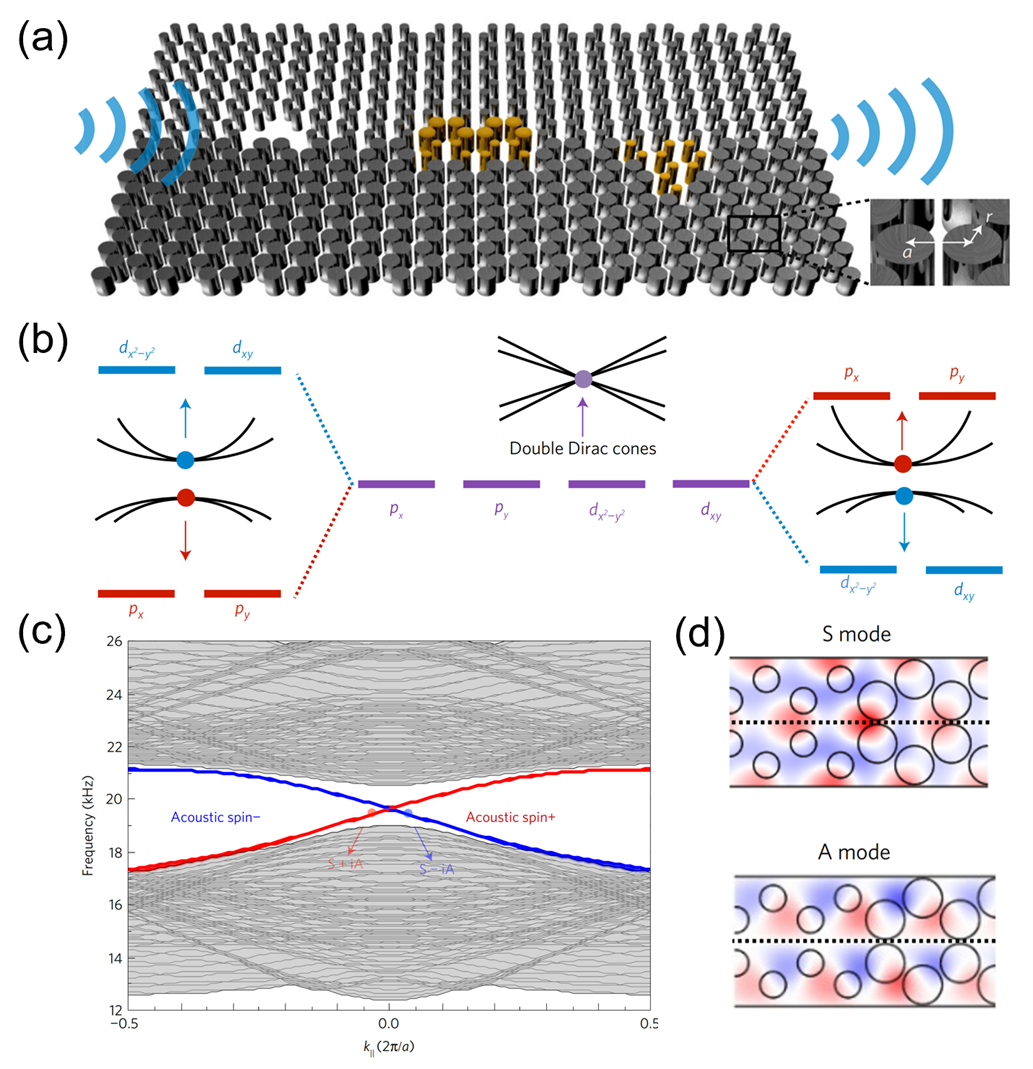}
    \caption{(a) Schematic of the phononic crystals constructed by a graphene array of stainless-steel rods with radii r. In the interface between two phononic crystals of rods with different filling ratios, defects, such as cavities, disorders, and bends, can be introduced to verify the robustness of the interface state propagation. The inset shows a zoom-in view of the rods of the phononic crystals. (b) Illustration of a band inversion process by tuning the filling ratio. At large and small filling ratios, the bands of the phononic crystals characterized by the band modes ($p$ and $d$) are inverted. The process of band gap closure and reopening indicates a topological phase transition, and at the critical point, an accidental double Dirac cone with a four-fold degeneracy is formed. (c) Calculated projected energy bands of a supercell consisting of two phononic crystals with inverted bands. The red and blue lines represent an acoustic pseudospin up and down edge states. The shadow regions represent the bulk bands. (d) Symmetric and anti-symmetric modes hybridized to constitute the edge states. Figures adopted from~\cite{nat.phys.12.1124he}.}
    \label{jiuyang_Picture4}
\end{figure}

\begin{figure}[htbp]
    \flushright
    \includegraphics[width=0.85\linewidth]{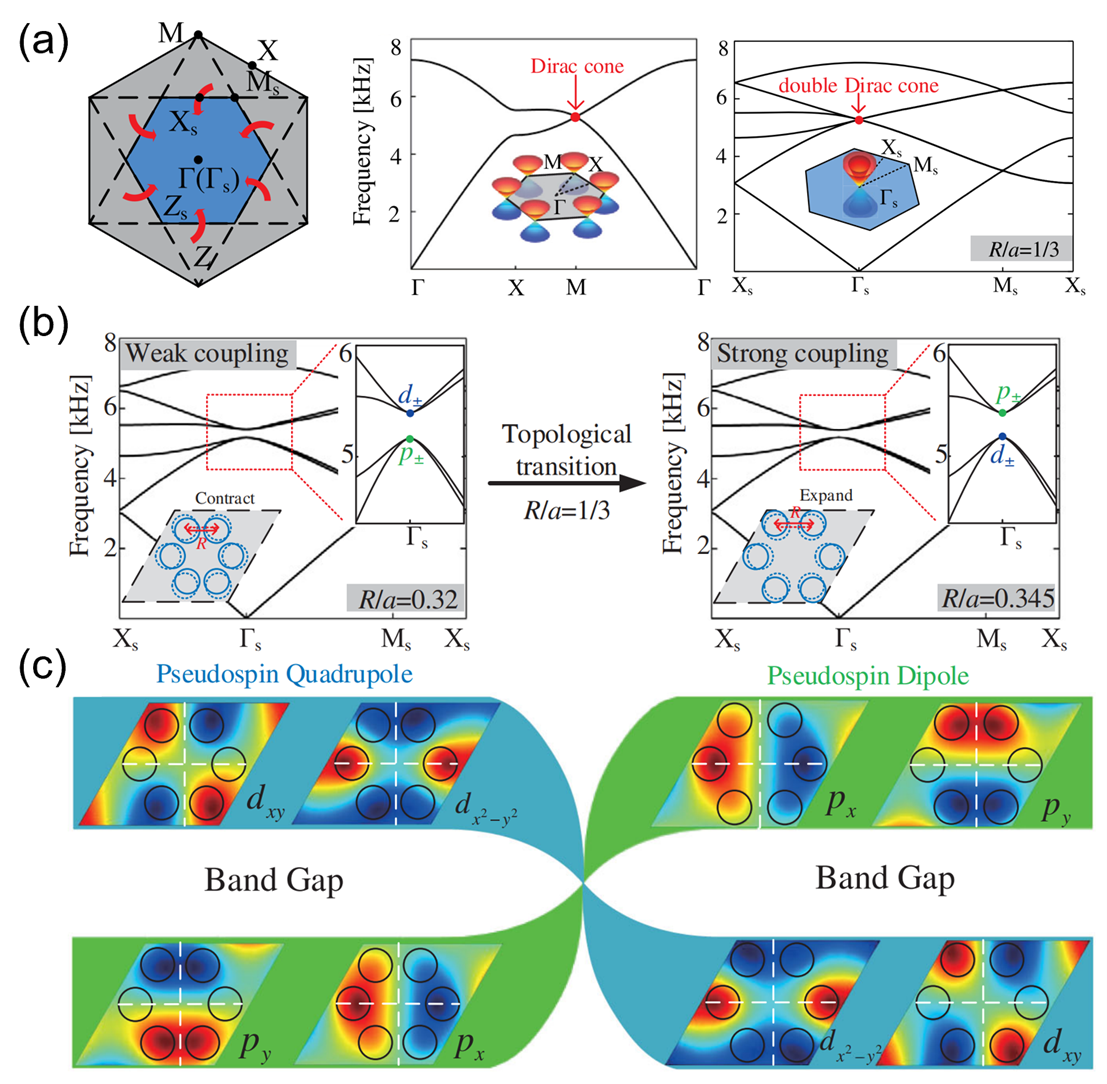}
    \caption{(a) Folding procedure to construct a double Dirac cone at the Brillouin zone center by expanding the primitive cell and folding two Dirac cones at the original Brillouin zone corners. Middle panel: dispersion relation of the lattice based on the original unit cell. (Inset: single Dirac cones at the Brillouin zone corners). Right panel: the final dispersion relation of the lattice based on the enlarged unit cell (Inset: double Dirac cone at the Brillouin zone center). (b) Topological transition as the coupling strength is tuned by contracting and expanding the six rods in one unit cell. Critical coupling appears at $R=a/3$ with double Dirac cone. (c) Topological modes inversion underlying the transition, where the dipole and quadruple modes act as the pseudospins. Figures adopted from~\cite{Phys.Rev.Lett.118.084303zhang}.}
    \label{jiuyang_Picture5}
\end{figure}

Another approach to increase adjustable degrees of freedom of classical waves is to access the topologically nontrivial band gap based on a degenerate state with higher degeneracy, such as the double Dirac cones. Following the approach, metamaterials, analogized to quantum spin Hall effect, were proposed for microwaves~\cite{Nat.Mater.12.233khanikaev,Phys.Rev.Lett.114.127401ma} and Lamb waves~\cite{nat.commun.6.8682mousavi}, where mixed polarizations of these vectorial fields are used to construct fermion-like pseudospins~\cite{Proc.NatlAcad.Sci.USA.113.4924he}. In contrast, for a scalar acoustic waves, the four-fold degeneracy of double Dirac cones cannot be achieved by degrees of freedom of intrinsic polarizations. To achieve the pseudospin states and the double Dirac cones in phononic crystals, accidental degeneracies of various Bloch modes were constructed by exploiting irreducible representations of systems with hexagonal lattice symmetry~\cite{nat.phys.12.1124he,Sci.Rep.6.32752wen}. The starting point is to realize the double Dirac cones in the Brillouin zone center via changing the filling ratio. As shown in Fig.~\ref{jiuyang_Picture4}(a), the phononic crystals is of a graphene lattice and its filling ratio is determined by the scatterers' radius r. At a critical filling ratio $r/a= 0.3928$ ($a$ is the lattice constant), an accidental double Dirac cone appears at the Brillouin zone center [Fig.~\ref{jiuyang_Picture4}(b)], whose dispersion can be anticipated from an effective Hamiltonian accidentally spanned by the basis $(p_x,p_y)$ and $(d_{x^2-y^2},d_{2xy})$, the two-fold irreducible representations of the point group $C_{6v}$~\cite{appl.phys.lett.105.014107li,Opt.Exp.23.12089li}. At a small or large filling ratio, the double Dirac cone is gapped by two two-fold degeneracy with the band modes being $(p_x,p_y)$ or $(d_{x^2-y^2},d_{2xy})$, where pseudospins for the bulk states can be achieved through hybridizing these modes as $p_\pm=p_x\pm i p_y$ and $d_\pm=d_{x^2-y^2}\pm id_{2xy}$ ($+$ and $–$ represent the pseudospin up and down, respectively). Therefore, by tuning the filling ratio, the energy band inversion occurs [Fig.~\ref{jiuyang_Picture4}(b)], leading to a topological transition of phononic crystals with $p$/$d$ pseudospins. Accordingly, in the interface composed by the phononic crystals with inverted bands, a pair of edge states is expected to exist and traverse the overlapped band gap, shown in Fig.~\ref{jiuyang_Picture4}(c). These edge states are localized at the interface and are recognized as $(S+iA)/(S-i A)$, where $S$ and $A$ are symmetric and anti-symmetric modes with respect to the middle line [dashed lines in Fig.~\ref{jiuyang_Picture4}(d)]. They can also be correlated to the pseudospins of the bulk states as $(S+i A)$/$(S-i A)=(p_+ + d_+)$/$(p_- - d_-)$, where the symmetric component $S$ can be represented as $S=(p_x + d_{x^2-y^2})/\sqrt{2}$ and the anti-symmetric component $A$ as $A=(p_y+d_{2xy})/\sqrt{2}$, as shown in Fig.~\ref{jiuyang_Picture4}(d). Hence, the two acoustic pseudospin edge states carry opposite pseudospin feature and propagate in opposite directions, implying a one-way spin-dependent propagation that is robust against various kinds of defect, such as cavities, disorders, and bends~\cite{nat.phys.12.1124he}. In the above works, the construction of double Dirac cones is based on the accidental degeneracy, which can be achieved in a more deterministic way of band folding~\cite{phys.rev.lett.114.223901wu,Phys.Rev.Lett.118.084303zhang,NewJ.Phys.19.075003yves}. Specifically, as shown in Fig.~\ref{jiuyang_Picture5}(a), by expanding the primitive unit cell of  phononic crystals of graphene lattice to enclose six rods, the Brillouin zone shrinks into a new one whose area is one third of the original Brillouin zone, and the Dirac cones locating at the corners of original Brillouin zone are folded to the $\Gamma$ point in the new Brillouin zone to form a double Dirac cone. To open a band gap in the double Dirac cone, new configurations of phononic crystals are created by contracting/expanding the metamolecule via concentrating/distracting the six rods. The gapped dispersions are shown in Fig.~\ref{jiuyang_Picture5}(b), with the pressure field distributions of the metamolecules depicted in Fig.~\ref{jiuyang_Picture5}(c). Owing to a multiple scattering inside the metamolecule, a pair of dipolar resonant states accompanied by a pair of quadrupolar resonant states are created, echoing the basis $p_{x/y}$ and $d_{(x^2-y^2)/{2xy}}$. Again, the band inversion caused by contracting and expanding the rods in one unit cell leads to counterpropagations of edge states with opposite pseudospin features. Note that the six-fold rotation symmetry plays an important role in the design of phononic crystals, as it allows two inequivalent two-dimensional modes as the cornerstones to build the pseudospins. With similar design schemes, acoustic analogues of quantum spin Hall effect have been realized in mechanical waves~\cite{nat.commun.9.3072yu,physrevx.8.031074miniaci}, mechanical systems~\cite{Phys.Rev.B97.060101zheng}, and recently extended to three-dimensional phononic crystals by layer-stacking structures~\cite{Nat.Commun.9.4555he,nat.commun.11.2318he}.

\subsection{Acoustic Chern insulator and spin-Chern insulator}
\begin{figure}[htbp]
    \flushright
    \includegraphics[width=0.85\linewidth]{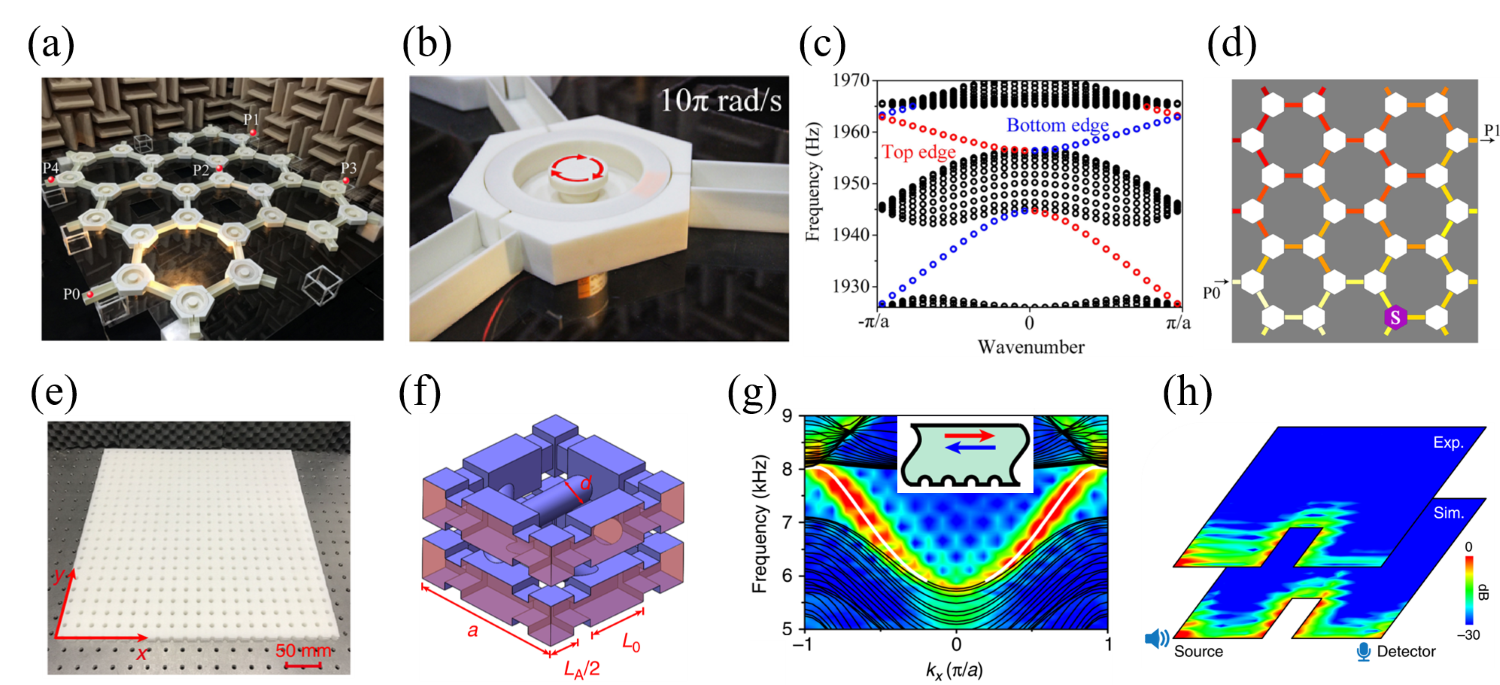}
    \caption{Acoustic Chern insulator and spin-Chern insulator. (a) Sample of the acoustic Chern insulator. (b) Sample of rotating three-port ring resonator with the chiral-structural rotor. The rotating rotor helps to generate a stable air flow, breaking the time-reversal symmetry of the system. (c) Band dispersion of chiral edge states for a ribbon (red) circles represent the gapless chiral edge modes at the bottom (top) boundary. (d) Measured pressure fields of anticlockwise chiral edge mode with a defect that a circulator denoted by S stops motion. The acoustic waves are input from port P0 and output to port P1. (e) Sample of the acoustic spin-Chern insulator. (f) Magnified side view of the sample. Green dotted box denotes the unit cell of the sample. (g) Measured band dispersion of helical edge states for a ribbon. Color maps denote the measured data, while white (black) lines represent the calculated edge (projected bulk) state dispersions. Inset shows the corresponding a pair of edge states counterpropagate along the top boundary. (h) Pressure fields for a sample possessing a rectangular defect, evidencing the backscattering robustness of the acoustic edge modes. (a)-(d) adopted from Ref.~\cite{phys.rev.lett.122.014302ding}, and (e)-(h) adopted from Ref.~\cite{nat.commun.11.3227deng}.}
    \label{weiyin.1.CI and SCI}
\end{figure}

According to the time-reversal symmetry, the two-dimensional topological insulators usually come in two different classes. The first breaks the time-reversal symmetry, and hosts the gapless chiral edge states. It hosts the quantum anomalous Hall effect and is commonly called Chern insulator, since its topological invariant is described by Chern number~\cite{phys.rev.lett.61.2015haldane,adv.phys.64.227weng}. For acoustic waves, circulating fluid flow is introduced to break the time-reversal symmetry~\cite{science343.516fleury}. Inspired by this, the acoustic Chern insulator is proposed in the acoustic metamaterial containing circulating fluids~\cite{nat.commun.6.8260khanikaev,newj.phys.17.053016ni,phys.rev.lett.114.114301yang}, and later realized in the resonator array with the chiral-structural rotors~\cite{phys.rev.lett.122.014302ding}. Figure~\ref{weiyin.1.CI and SCI}(a) shows the sample, i.e., a honeycomb lattice of ring resonators. Each resonator has a rotor rotating by a motor, generating a stable air flow and breaking the time-reversal symmetry, as shown in Fig.~\ref{weiyin.1.CI and SCI}(b). At this time, the bulk band hosts nontrivial topology with nonzero Chern number, and gives rise to the chiral edge states for an open boundary, as calculated in Fig.~\ref{weiyin.1.CI and SCI}(c). The measured pressure fields in Fig.~\ref{weiyin.1.CI and SCI}(d) show the anticlockwise chiral edge mode can transport across a defect where the circulator stops motion, demonstrating the robustness of the edge states.

The second class of two-dimensional topological insulator keeps the time-reversal symmetry, and hosts a pair of helical edge states. It hosts the quantum spin Hall effect in the electronic systems~\cite{phys.rev.lett.95.226801kane,science314.1757bernevig}, whose bulk band topology can be described by the $Z_2$ index~\cite{phys.rev.lett.95.146802kane} or the spin-Chern numbers equivalently~\cite{phys.rev.lett.97.036808sheng,chin.phys.b22.067201sheng}. The gapless feature of the helical edge states is guaranteed by the Kramer degeneracy originated from the spin-1/2 time-reversal symmetry. Different from the $Z_2$ index, spin-Chern numbers are well defined in the absence of any symmetries~\cite{phys.rev.b80.125327prodan}, thus are employed to identify pseudospin topological insulators~\cite{phys.rev.b82.165104li}, forming the concept of spin-Chern insulator~\cite{chin.phys.b22.067201sheng,phys.rev.b80.125327prodan,phys.rev.b82.165104li,phys.rev.b98.094310barlas,sci.rep.3.3435ezawa,phys.rev.lett.108.196806li,nat.commun.11.3227deng}. Whether the helical edge states in spin-Chern insulator are gapless or not, relates with the symmetry of system and the microstructure of sample boundary~\cite{phys.rev.lett.108.196806li}. Figure~\ref{weiyin.1.CI and SCI}(e) shows the sample of the acoustic spin-Chern insulator~\cite{nat.commun.11.3227deng}, which is a bilayer structure with a layer pseudospin degree of freedom. As shown in Fig.~\ref{weiyin.1.CI and SCI}(f), the chiral interlayer tubes induce the pseudospin-orbit coupling, giving rise to the nontrivial band topology described by nonzero spin-Chern numbers. The gapless helical edge states are calculated and observed in Fig.~\ref{weiyin.1.CI and SCI}(g). Owning to the pseudospin-momentum locking, they are robust against the backscattering induced by the defect, as shown in Fig.~\ref{weiyin.1.CI and SCI}(h).

\subsection{Weyl acoustic metamaterials}
\begin{figure}[htbp]
    \flushright
    \includegraphics[width=0.85\linewidth]{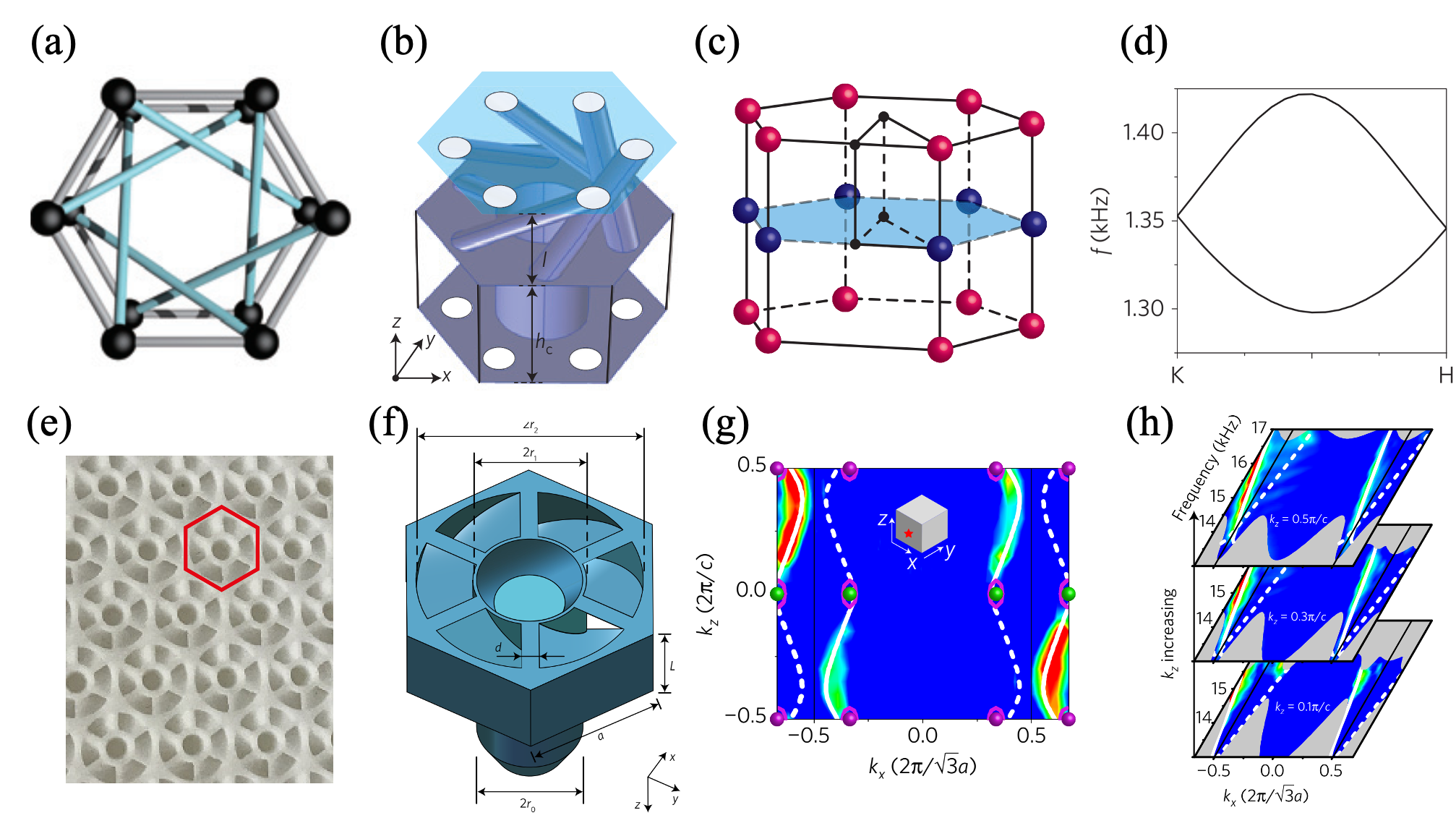}
    \caption{Weyl acoustic metamaterial with type-I Weyl point and Fermi-arc surface states. (a) Tight-binding model based on a layer-stacking strategy with chiral interlayer coupling, generating the Weyl points in momentum space. (b) Corresponding unit cell in the acoustic metamaterial. (c) Distribution of the Weyl points in the first Brillouin zone. Red (blue) spheres denote the Weyl points with topological charge +1 ($-1$). (d) Linear dispersion of the Weyl points along the $k_z$ direction. Type-I Weyl point is featured without overtilted feature. (e) Experimental sample of the acoustic metamaterial. (f) Unit cell of the sample, corresponding to the red hexagon in (e). (g) Equifrequency contours of the Fermi-arc surface states. Inset shows the detected XZ surface. (h) Dispersion of the surface states for three fixed $k_z$. In (f) and (g), color maps denote the experimental data, and the white solid and dashed lines represent the simulated results on the XZ and its opposite surfaces, respectively. (a)-(d) adopted from Ref.~\cite{nat.phys.11.920xiao}, and (e)-(h) adopted from Ref.~\cite{nat.phys.14.30li}.}
    \label{weiyin.3.type-I Weyl}
\end{figure}

Different from the topological insulators, topological semimetals are featured with band touching in momentum space, including zero-dimensional discrete points, one-dimensional continuum lines or two-dimensional surfaces~\cite{rev.mod.phys.90.015001armitage,nat.rev.mater.6.784hasan}. The band touching carries nontrivial topology and gives rise to boundary states. Conventional Weyl semimetals are crystals hosting type-I Weyl points---the twofold linear crossing points without overtilt in momentum space. The quasiparticle excitations around the type-I Weyl point satisfy the Weyl equation, thus behaves like Weyl fermion. The realization of Weyl fermions in lattice systems have been generalized from electronic systems to atomic systems (see for example~\cite{Phys.Rev.A.85.033640jiang,EPL.97.67004delplace}) and then to photonic (see for example~\cite{nat.photon.7.294lu,science.349.622lu,nat.commun.7.13038chen}) and phononic systems. In phononic crystal, the conventional Weyl semimetal was proposed~\cite{nat.phys.11.920xiao} and realized~\cite{nat.phys.14.30li,phys.rev.appl.10.014017ge} by a layer-stacking construction. The linear dispersion in two dimensions is easily obtained in graphene. Stacking the graphene along the $z$ direction with chiral interlayer coupling, as shown in Fig.~\ref{weiyin.3.type-I Weyl}(a), the linear dispersion also occurs in the $z$ direction, giving rise to the Weyl point. The effective Hamiltonian of Weyl point has the form as $H_\text{W}=A_x k_x \sigma_x+A_y k_y \sigma_y+A_z k_z \sigma_z$, where $\boldsymbol{k}$ is the wavevector, $\boldsymbol{\sigma}$ is the Pauli matrix, and $A_i$ with $i=x,y,z$ is the coefficient. Figure~\ref{weiyin.3.type-I Weyl}(b) shows the corresponding unit cell of the acoustic metamaterial. There exist four Weyl points in the first Brillouin zone, as shown in Fig.~\ref{weiyin.3.type-I Weyl}(c). The linear dispersion along the $k_z$ direction is calculated in Fig.~\ref{weiyin.3.type-I Weyl}(d). Robustness of the Weyl points has two aspects: the first is that the Weyl point is stable against weak perturbations owning to describing by all the three Pauli matrices; the second is that it hosts nontrivial topological charge described by Chern number $C=\text{sgn}(A_x A_y A_z )=\pm{1}$, behaving as the drain or source of Berry flux.

Owning to their topological nature, Weyl points always come in pairs with the zero summing charges, giving rise to Fermi-arc surface states and chiral anomaly. Unlike the Fermi-circle surface states in topological insulators, the Fermi-arc surface states are nonclosed and connect the Weyl points with opposite charges. Figures~\ref{weiyin.3.type-I Weyl}(e) and~\ref{weiyin.3.type-I Weyl}(f) show the sample and unit cell of the Weyl acoustic metamaterial in experiment, respectively~\cite{nat.phys.14.30li}. Fermi arcs, equifrequency contours of the surface dispersion, are observed in Fig.~\ref{weiyin.3.type-I Weyl}(g). The Fermi-arc surface states can also be revealed by the surface dispersions for fixed $k_z$, as shown in Fig.~\ref{weiyin.3.type-I Weyl}(h). These surface dispersions host the chiral surface states, attributed to the $k_z$-dependent Chern numbers, which are related with the charges of Weyl points. The equifrequency cut of the surface dispersions forms a trajectory connected a pair of Weyl points with opposite charges, i.e., Fermi arc. As such, the Fermi-arc surface states in the Weyl acoustic metamaterial exhibit topologically protected one-way propagation of acoustic waves in the presence of defects.

\begin{figure}[htbp]
    \flushright
    \includegraphics[width=0.85\linewidth]{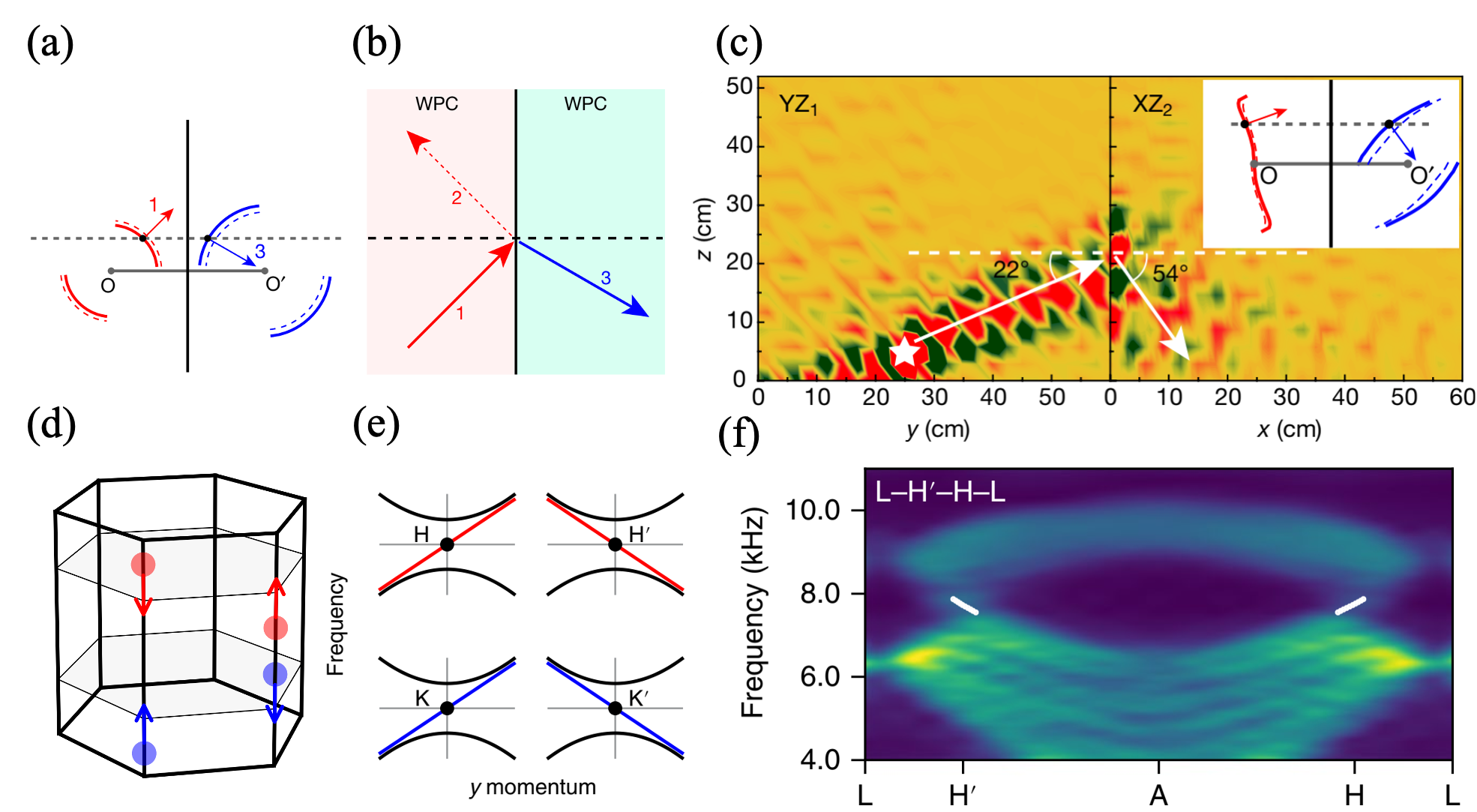}
    \caption{Topological refraction of the Fermi-arc surface states and zeroth Landau levels in the Weyl acoustic metamaterial. (a) Fermi arcs of two neighboring surfaces. Dashed curves denote the Fermi arcs at a slight larger frequency than the ones (solid curves). Arrows represent the directions of the group velocities. (b) Topological negative refraction at an interface between two neighboring surfaces. Due to the nonclosed property of Fermi arc, reflection of the surface states is forbidden. (c) Observation of the topological negative refraction of the Fermi-arc surface states in the Weyl acoustic metamaterial. Inset is the configuration of the Fermi arcs in the two neighboring surfaces. (d) Axial pseudomagnetic fields are constructed by shifting the Weyl points in momentum space. (e) Zeroth chiral Landau levels for the four Weyl points, owning to the axial pseudomagnetic fields. Red (blue) color denotes the topological charge +1 ($-$1) of the Weyl points. (f) Observation of the chiral Landau levels around the Weyl points at H' and H. Again, color maps denote the experimental data, and the white dots represent the simulated results. (a)-(c) adopted from Ref.~\cite{nature560.61he}, and (d)-(f) adopted from Ref.~\cite{nat.phys.15.357peri}.}
    \label{weiyin.4.Fermi arc and LL}
\end{figure}

The Fermi-arc surface states can give rise to topological negative refraction~\cite{nature560.61he}, providing a practical use of Fermi arcs. Figure~\ref{weiyin.4.Fermi arc and LL}(a) shows the Fermi arcs of two neighboring surfaces, which form an interface (black vertical line). Considering an incident wave along the direction of the red arrow (labelled 1), the refracted wave will propagate along the direction of the blue arrow (labelled 3), since the wavevector is conserved parallel to the interface. As shown in Fig.~\ref{weiyin.4.Fermi arc and LL}(b), this refraction is negative, because the refracted and incident waves are on the same side of the normal (dashed horizon line). More importantly, owing to the nonclosed nature of Fermi arcs, there are no modes for reflected wave, thus the reflection (dashed arrow labelled 2) is forbidden. Figure~\ref{weiyin.4.Fermi arc and LL}(c) shows the measured pressure fields on the surface of the Weyl acoustic metamaterial, evidencing the negative refraction without reflection of the Fermi-arc surface states.

Under a strong magnetic field, the Weyl point supports the chiral zeroth Landau level, which hosts one-way propagation related with the charge of Weyl point~\cite{nat.phys.15.357peri}. While adding a parallel electric field, the Weyl quasiparticles can be pumped through the zeroth Landau level between the Weyl points with opposite charges, leading to non-conservation of chiral currents, i.e., chiral anomaly, the other hallmark of Weyl semimetal. Since the acoustic wave is inert to the external magnetic field, constructing pseudomagnetic field is significant in acoustic metamaterials. By designing an effective space-dependent sublattice potential along the $x$ direction, the Weyl points are shifted along the $k_z$ or $-k_z$ direction, as shown in Fig.~\ref{weiyin.4.Fermi arc and LL}(d), axial pseudomagnetic fields are induced along the y direction in the Weyl acoustic metamaterial. Weyl points at $H$ and $K^\prime$ feel an opposite field to the ones at $H^\prime$ and $K$. Unlike the genuine magnetic field, the acoustic metamaterial keeps the time-reversal symmetry in the presence of the pseudomagnetic fields, giving rise to the chiral zeroth Landau levels illustrated in Fig.~\ref{weiyin.4.Fermi arc and LL}(e). For example, due to the pseudomagnetic fields with opposite directions acting on Weyl points at $H$ and $H^\prime$, the chiral Landau levels around them have opposite group velocities, although they host the same charge +1, as observed in Fig.~\ref{weiyin.4.Fermi arc and LL}(f).

\subsection{Other Weyl acoustic metamaterials}
\begin{figure}[htbp]
    \flushright
    \includegraphics[width=0.8\linewidth]{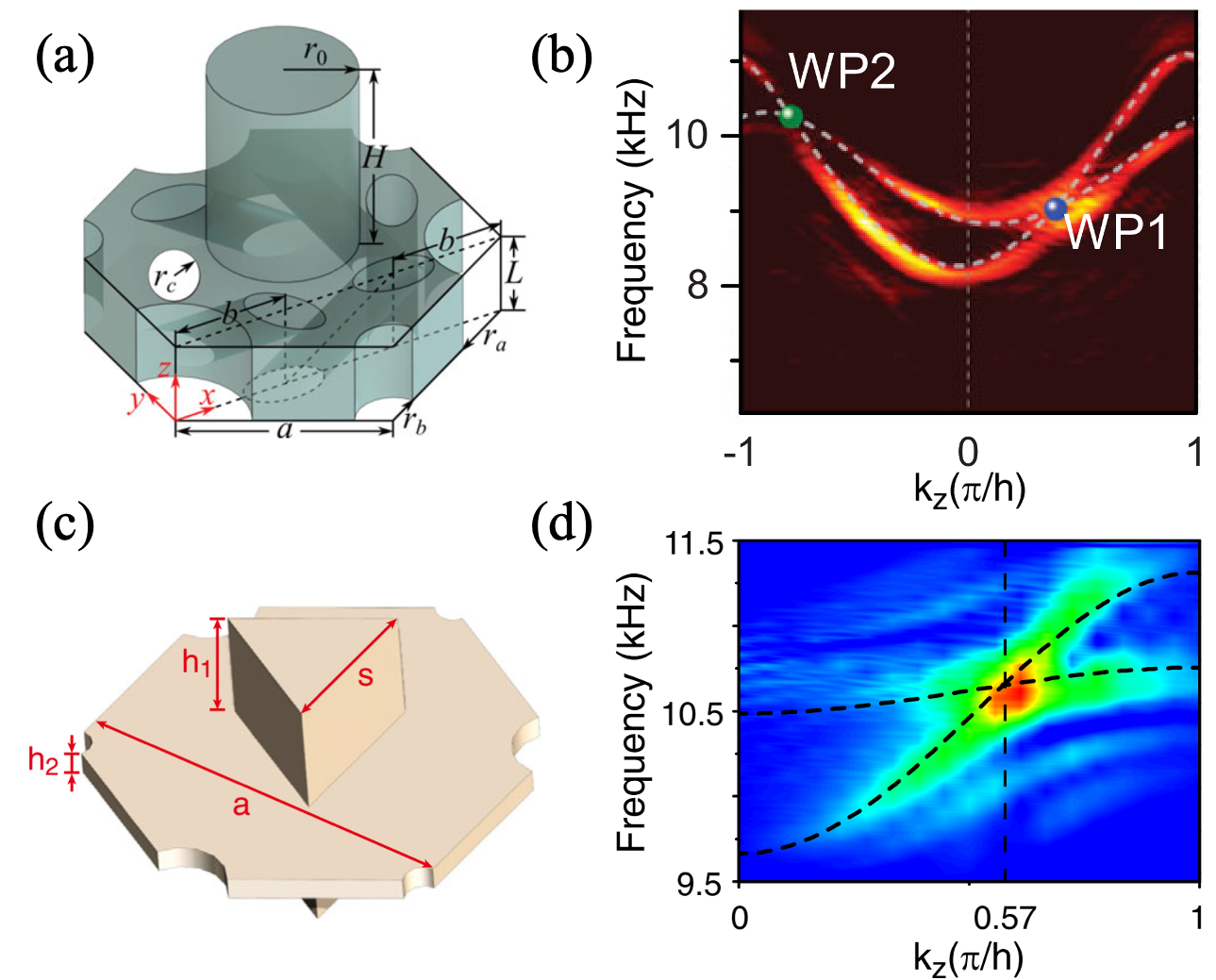}
    \caption{Weyl acoustic metamaterial with type-II Weyl points. (a) Unit cell of a type-II Weyl acoustic metamaterial. (b) Observation of the type-II Weyl points, which are not located at the same frequency. (c) Unit cell of the acoustic metamaterial with ideal type-II Weyl points. (d) Observation of the ideal type-II Weyl point. Four type-II Weyl points reside at the same frequency, guaranteed by the mirror and time-reversal symmetries. (a) and (b) adopted from Ref.~\cite{phys.rev.lett.122.104302xie}, and (c) and (d) adopted from Ref.~\cite{phys.rev.lett.124.206802huang}.}
    \label{weiyin.5.type-II Weyl}
\end{figure}

Unlike the Weyl fermions in high-energy physics, the quasiparticle excitations in crystals are not limited by the Poincaré symmetry, thus exhibit rich configurations with nontrivial topology in momentum space. The typical one is the type-II Weyl point, characterized by an overtilted cone dispersion~\cite{nature527.495soluyanov}. Although the topological charge of Weyl point is independent on the overtilted dispersion, this feature gives rise to unique properties different from the type-I Weyl point, such as the antichiral zeroth Landau level and anisotropic chiral anomaly. The type-II Weyl points have been predicted in the acoustic metamaterials with stacked hexagon and square structures~\cite{nat.phys.11.920xiao,phys.rev.lett.117.224301yang}, and later observed in experiments. Figures~\ref{weiyin.5.type-II Weyl}(a) and~\ref{weiyin.5.type-II Weyl}(b) show the unit cell and the observations of the type-II Weyl points in an acoustic metamaterial, respectively~\cite{phys.rev.lett.122.104302xie}. Due to lacking the relation by symmetry, the Weyl points are at different frequencies. Then the ideal type-II Weyl points, which are related by the mirror and time-reversal symmetries and thus reside at the same energy, have been realized in the acoustic metamaterial~\cite{phys.rev.lett.124.206802huang}, as seen in Figs.~\ref{weiyin.5.type-II Weyl}(c) and~\ref{weiyin.5.type-II Weyl}(d). By constructing the pseudomagnetic fields in this ideal type-II Weyl acoustic metamaterial, the antichiral zeroth Landau levels were observed~\cite{dengweiyin2022}.

\begin{figure}[htbp]
    \flushright
    \includegraphics[width=0.85\linewidth]{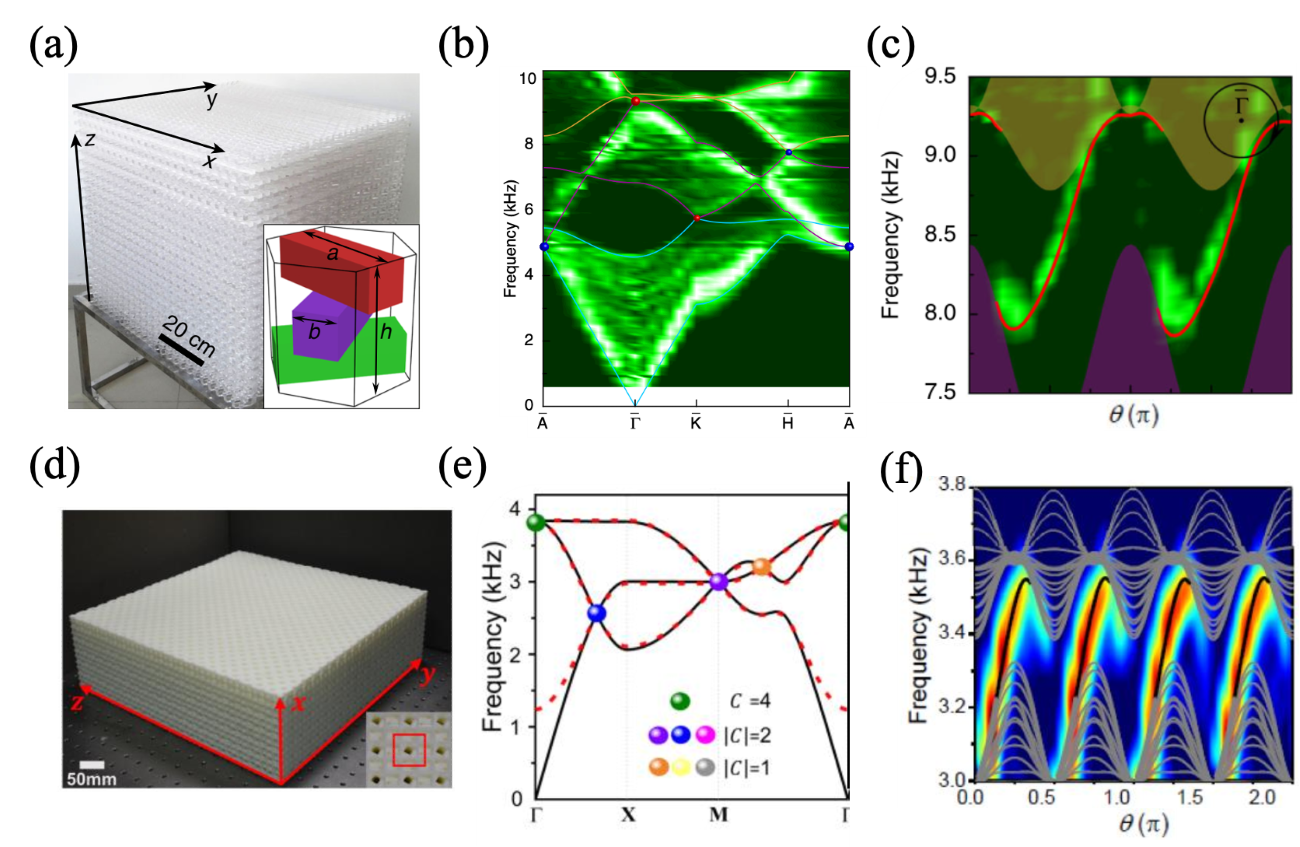}
    \caption{Quadratic and quadruple Weyl acoustic metamaterials. (a) Sample for the acoustic metamaterial with quadratic Weyl point. Inset shows the unit cell. (b) Projected bulk band dispersion along the high-symmetry lines. The projected quadratic Weyl points with charge +2 (red sphere) and -2 (blue sphere) are located at $\overline{\Gamma}$ and $\overline{\text{A}}$ points, respectively. (c) Surface dispersion along a loop centered at $\overline{\Gamma}$ point. Topological charge +2 of the quadratic Weyl point results in two chiral surface states with positive group velocity. (d) Sample for the acoustic metamaterial with quadruple Weyl point. Inset shows the enlarged view of the surface boundary with a unit cell (the red box). (e) Bulk dispersion along the high-symmetry lines. The blue sphere denotes the quadruple Weyl point. (f) Surface dispersion along a loop centered at the position of the quadruple Weyl point. Four chiral surface states with positive group velocity reveal the topological charge +4 of the quadruple Weyl point. (a)-(c) adopted from Ref.~\cite{nat.commun.11.1820he}, and (d)-(f) adopted from Ref.~\cite{arxiv:2203.14600luo}.}
    \label{weiyin.6.high charge Weyl}
\end{figure}

The second way to generalize the conventional Weyl point is changing the topological charge of the twofold degenerate point, which is no longer the linear dispersions in all directions, such as the quadratic and quadruple Weyl points~\cite{phys.rev.b98.214110chen,nat.commun.11.1820he,arxiv:2203.14600luo}. The Weyl point possessing high charge is usually protected by the spatial symmetries, to vanish the linear dispersion. Figure~\ref{weiyin.6.high charge Weyl}(a) shows the sample of acoustic metamaterial possessing a pair of quadratic Weyl points located at $\Gamma$ and A, together with the conventional Weyl points~\cite{nat.commun.11.1820he}. Quadratic Weyl points host charge $\pm{2}$, and featured with the quadratic dispersions in two directions and linear dispersion in the other one. Figure~\ref{weiyin.6.high charge Weyl}(b) shows the measured projected bulk band dispersion, exhibiting the nonlinear feature of the quadratic Weyl point. Two chiral surface states with positive group velocity in the surface dispersion along a loop centered at $\overline{\Gamma}$, as shown in Fig.~\ref{weiyin.6.high charge Weyl}(c), demonstrates the quadratic Weyl point possesses charge +2 at $\Gamma$. The quadruple Weyl point, which carries the highest charge of a twofold degenerate point with charge 4, has also been observed in an acoustic metamaterial with space group symmetry $P$432~\cite{arxiv:2203.14600luo}. Figures~\ref{weiyin.6.high charge Weyl}(d) and~\ref{weiyin.6.high charge Weyl}(e) show the sample and the bulk dispersion. The quadruple Weyl point exhibits the cubic dispersion in the (111) direction and quadratic dispersions in the other two directions. Similarly, as shown in Fig.~\ref{weiyin.6.high charge Weyl}(f), four chiral surface states with positive group velocity reveal the charge 4 of the quadruple Weyl point.

\begin{figure}[htbp]
    \flushright
    \includegraphics[width=0.85\linewidth]{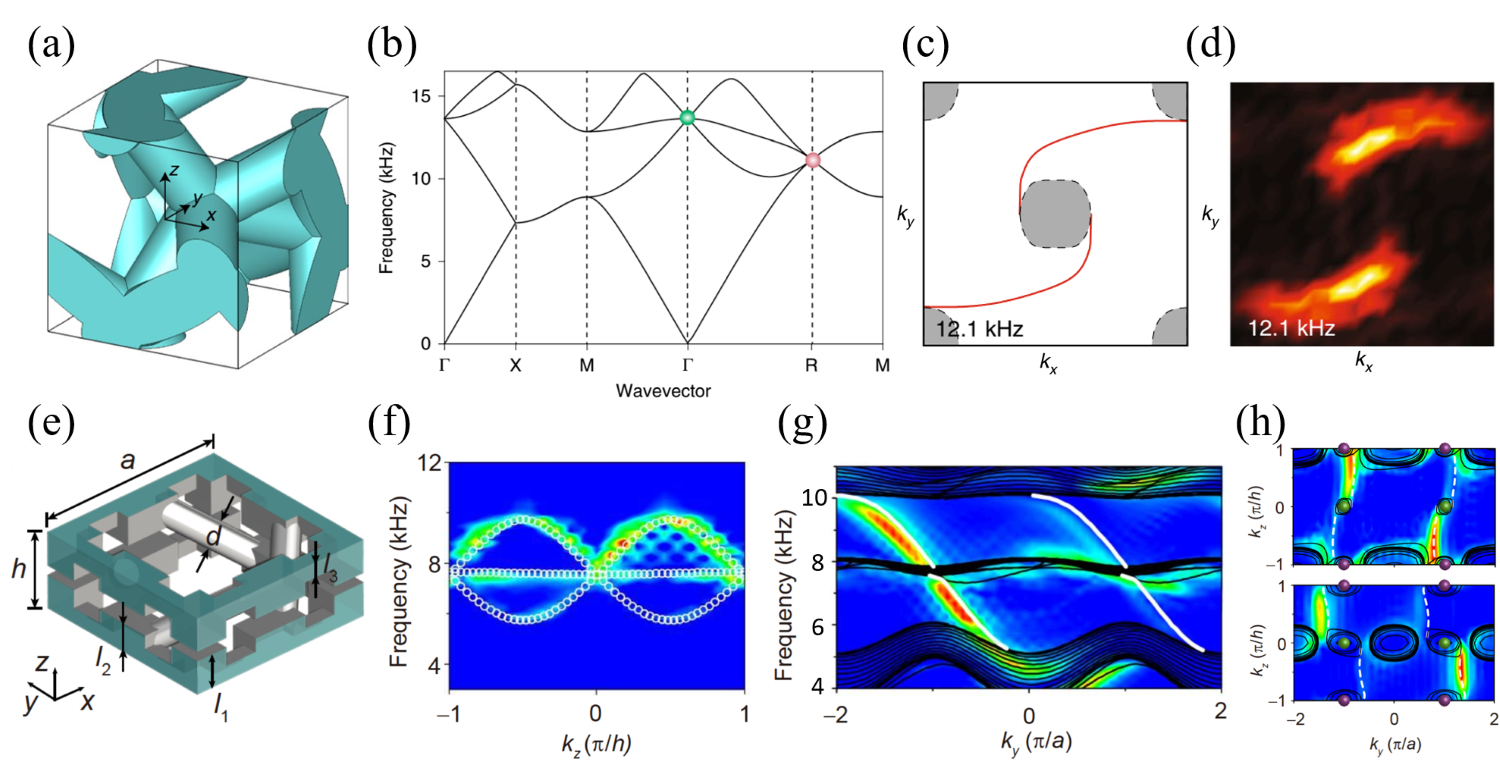}
    \caption{Spin-1 Weyl acoustic metamaterial. (a) Unit cell of an acoustic metamaterial with a spin-1 Weyl point and a charge-2 Dirac point. (b) Bulk dispersion of the corresponding acoustic metamaterial. A spin-1 Weyl point with threefold degeneracy (blue sphere) and a charge-2 Dirac point with fourfold degeneracy (red sphere) are located at $\Gamma$ and R, respectively. (c) and (d) Calculated and measured double-helicoid Fermi arcs connecting the projected spin-1 Weyl point and the charge-2 Dirac point. (e) Unit cell of an acoustic metamaterial with a pair of spin-1 Weyl points. (f) Observation of the pair of spin-1 Weyl points in the bulk dispersion. (g) Observation of surface dispersion with $k_z=0.5\pi/h$. The surface states emerge in the gaps between the first and second bands, as well as between the second and third bands. (h) Observation of double Fermi arcs at $f=7 \text{kHz}$ between the first and second bands  (top panel) and at $f=9 \text{kHz}$ between the second and third bands  (bottom panel). (a)-(d) adopted from Ref.~\cite{nat.phys.15.645yang}, and (e)-(h) adopted from Ref.~\cite{sci.china-phys.mech.astron.63.287032deng}.}
    \label{weiyin.7.spin-1 Weyl}
\end{figure}

The third way is changing the low-energy excitation from the spin-1/2 quasiparticle to the other spin one, accompanied by changing the degeneracy of the touching point. Spin-1 Weyl point is the typical one, which is the threefold degenerate point touched by two linear dispersions and an additional flat band, and carries charges ($-2$,0,2) or (2,0,$-2$) for the three bulk bands. The low-energy excitation has the form $\boldsymbol{k}\cdot \boldsymbol{S}$, where $\boldsymbol{S}$ is the vector of the spin-1 Gell-Mann matrix, rather than the spin-1/2 Pauli matrix, thus behaves as the spin-1 quasiparticle. Due to describing by only part of Gell-Mann matrices, spin-1 Weyl point is usually protected by external symmetries. Figure~\ref{weiyin.7.spin-1 Weyl}(a) is the unit cell of an acoustic metamaterial with space group symmetry $P$213~\cite{nat.phys.15.645yang}. As shown in Fig.~\ref{weiyin.7.spin-1 Weyl}(b), the acoustic metamaterial possesses a spin-1 Weyl point coexisting with a charge-2 Dirac point. Charge 2 of these two points leads to the double Fermi arcs in the surface Brillouin zone, as calculated and measured in Figs.~\ref{weiyin.7.spin-1 Weyl}(c) and~\ref{weiyin.7.spin-1 Weyl}(d), respectively. The spin-1 Weyl semimetal exclusively possessing spin-1 Weyl points can be realized in an acoustic metamaterial by stacking the Lieb lattice layer~\cite{sci.china-phys.mech.astron.63.287032deng}, with the unit cell given in Fig.~\ref{weiyin.7.spin-1 Weyl}(e). A pair of spin-1 Weyl points, as shown by the bulk dispersion in Fig.~\ref{weiyin.7.spin-1 Weyl}(f), are connected by the chiral symmetries. Charges ($-2$,0,2) and (2,0,$-2$) for the three bulk bands result in the surface states shown in Fig.~\ref{weiyin.7.spin-1 Weyl}(g) and double Fermi arcs in Fig.~\ref{weiyin.7.spin-1 Weyl}(h) emerging between the first and second bands, as well as between the second and third bands. Along this generalization, spin-3/2 and spin-5/2 quasiparticles can be explored in acoustic metamaterials.

\begin{figure}[htbp]
    \flushright
    \includegraphics[width=0.85\linewidth]{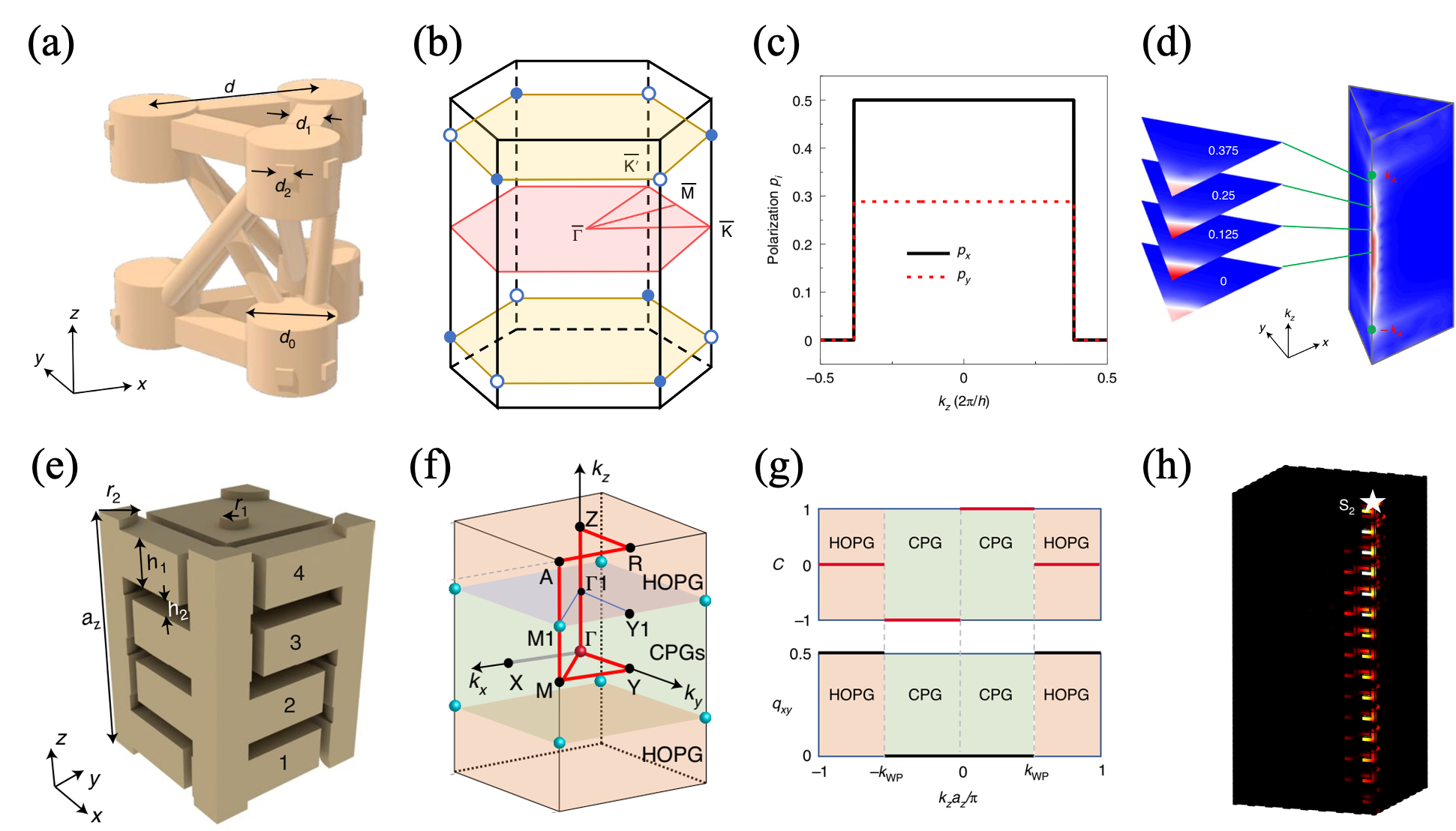}
    \caption{Higher-order Weyl acoustic metamaterials. (a) Unit cell of an acoustic metamaterial with two pair of higher-order Weyl points. (b) Distribution of the second-order Weyl points in the first Brillouin zone. Solid (hollow) spheres denote the Weyl points with topological charge $+1$ ($-1$). (c) Bulk polarization of the lowest band as a function of $k_z$. (d) Measured (left panel) and simulated (right panel) pressure fields, showing the hinge states resulting from the nonzero $k_z$-dependent bulk polarization. (e) Unit cell of an acoustic metamaterial with two higher-order Weyl points and a quadratic Weyl point. (f) Distribution of the Weyl points in the first Brillouin zone. Blue spheres denote the higher-order Weyl points, and red sphere denotes the quadratic Weyl point. (g) Chern number (top panel) and quadrupole index (bottom panel) as a function of $k_z$. (h) Measured pressure fields of the hinge states, attributed to the nonzero $k_z$-dependent quadrupole index. (a)-(d) adopted from Ref.~\cite{nat.mater.20.812wei}, and (e)-(h) adopted from Ref.~\cite{nat.mater.20.794luo}.}
    \label{weiyin.8.higher-order Weyl}
\end{figure}

Recently, the higher-order Weyl semimetals are realized in acoustic metamaterials. As a significant extension, the higher-order Weyl point not only hosts the nonzero topological charge, as the first-order feature, but also can be viewed as a transition point between second-order topological insulator and Chern insulator or topologically trivial insulator~\cite{phys.rev.lett.125.266804ghorashi,phys.rev.lett.125.146401wang}. Besides the two-dimensional Fermi-arc surface states, higher-order Weyl semimetal exhibits the one-dimensional hinge states, connecting the higher-order Weyl points. The higher-order Weyl semimetal was realized in an acoustic metamaterial by stacking the kogome lattice layer~\cite{nat.mater.20.812wei}, whose unit cell is shown in Fig.~\ref{weiyin.8.higher-order Weyl}(a). This system hosts the pure higher-order Weyl points of four, showing in Fig.~\ref{weiyin.8.higher-order Weyl}(b). In this case, the Chern number as a function of $k_z$ is always zero, but the bulk polarization as the higher-order topological invariant has distribution, as shown in Fig.~\ref{weiyin.8.higher-order Weyl}(c). The nonzero $k_z$-dependent bulk polarizations lead to the hinge states, as measured and calculated in Fig.~\ref{weiyin.8.higher-order Weyl}(d). The higher-order Weyl semimetal can also be realized in an acoustic metamaterial with a chiral screw symmetry $S_{4z}$~\cite{nat.mater.20.794luo}. Figure~\ref{weiyin.8.higher-order Weyl}(e) shows the unit cell. The acoustic metamaterial hosts two higher-order Weyl points with charge $-$1 and a quadratic Weyl point with charge +2, as depicted in Fig.~\ref{weiyin.8.higher-order Weyl}(f). The topological property of the acoustic metamaterial can be revealed by the two-dimensional topological index, Chern number as the first order topological invariant and quadrupole index, by considering $k_z$ as a parameter. As shown in Fig.~\ref{weiyin.8.higher-order Weyl}(g), the transition points both for the Chern number and quadrupole index are the higher-order Weyl points. Figure~\ref{weiyin.8.higher-order Weyl}(h) shows the measured pressure fields, demonstrating the hinge states, owning to the higher-order topology.

\subsection{Dirac acoustic metamaterials}
\begin{figure}[htbp]
    \flushright
    \includegraphics[width=0.85\linewidth]{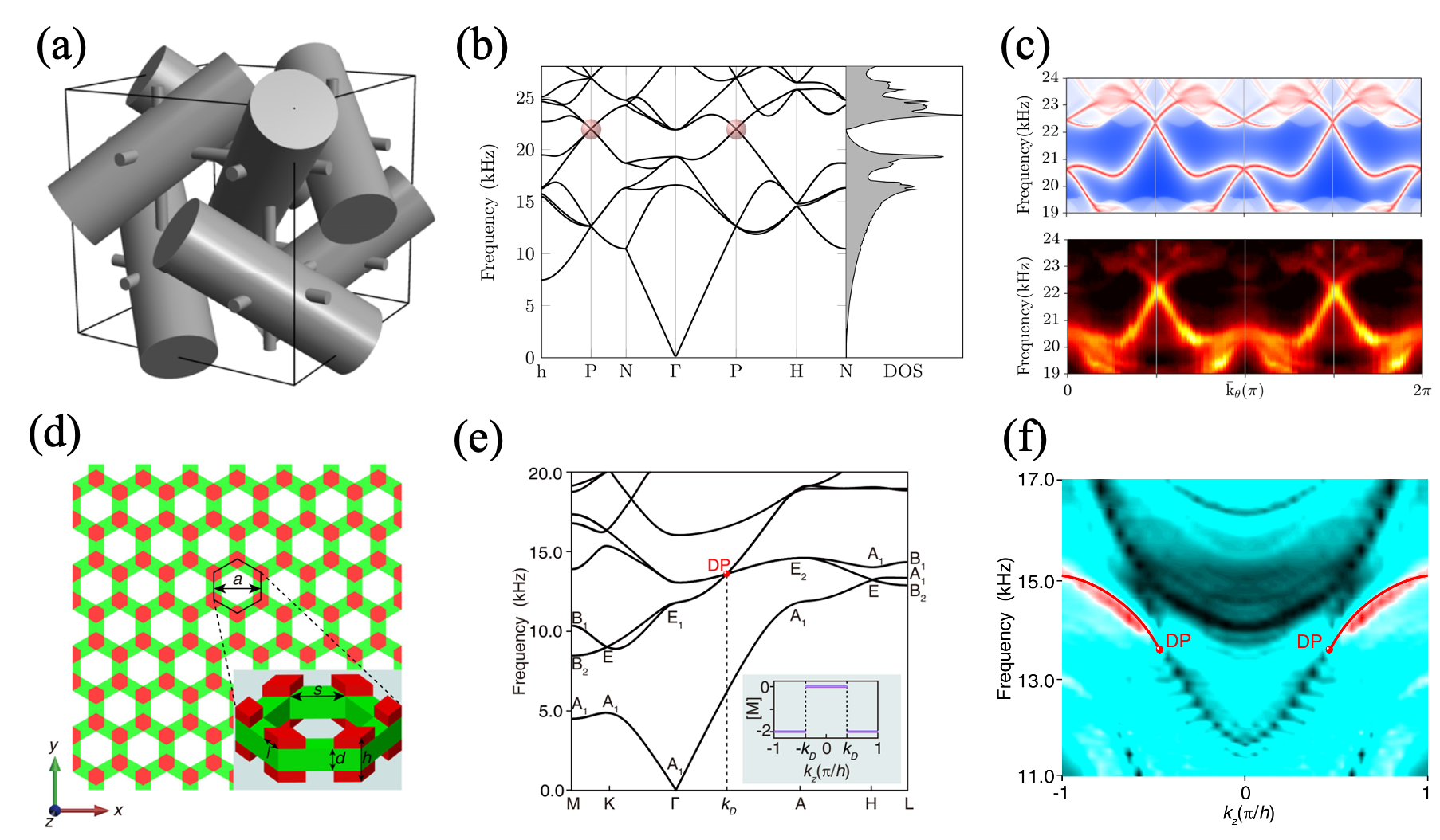}
    \caption{Dirac acoustic metamaterials. (a) Unit cell of an acoustic metamaterial hosting a pair of Dirac points with $Z_2$ charges. (b) Bulk dispersion along the high-symmetry lines. A Dirac point located at P point shows the four-fold degeneracy. (c) Calculated (top panel) and measured (bottom panel) surface dispersions along a loop around the projected Dirac points. (d) In-plane structure and unit cell (inset) of an acoustic metamaterial possessing a pair of higher-order Dirac points. (e) Bulk band dispersion along high-symmetry lines. Red sphere denotes the higher-order Dirac point. Inset shows a higher-order topological index as a function of $k_z$. (f) Measured (color) and calculated (red line) hinge dispersion, originated from the nonzero higher-order topological index. (a)-(c) adopted from Ref.~\cite{phys.rev.lett.124.104301cheng}, and (d)-(f) adopted from Ref.~\cite{phys.rev.lett.127.146601qiu}.}
    \label{weiyin.9.Dirac}
\end{figure}

Conventional three-dimensional Dirac point is a fourfold linear crossing point, in which the quasiparticle excitation behaves as Dirac fermion. The Dirac point is constituted by a pair of Weyl points with opposite charges, thus carries a charge described by $Z_2$ or spin-Chern number. The accidental Dirac point can be formed by band inversion~\cite{lightsci.appl.9.201xie}, while the symmetry-enforced Dirac point usually emerges in the acoustic metamaterial with nonsymmorphic space group~\cite{lightsci.appl.9.38cai,phys.rev.lett.124.104301cheng}. Figure~\ref{weiyin.9.Dirac}(a) shows the unit cell of an acoustic metamaterial with space group symmetry $Ia\overline{3}$, which hosts a pair of Dirac points~\cite{lightsci.appl.9.38cai,phys.rev.lett.124.104301cheng}. Figure~\ref{weiyin.9.Dirac}(b) is the calculated bulk dispersion, showing a Dirac point at P. The other one connected by the symmetry is at $-$P. Although the Fermi-arc surface states are usually not stable in Dirac semimetal~\cite{pnas115.8311le}, they are robust here protected by the combination of gride and time-reversal symmetries, as shown in Fig.~\ref{weiyin.9.Dirac}(c). Naturally, there exist the higher-order Dirac points, which also have been explored in acoustic metamaterials~\cite{phys.rev.lett.127.146601qiu,phys.rev.lett.128.115701xia}. Figures~\ref{weiyin.9.Dirac}(d) and~\ref{weiyin.9.Dirac}(e) show the sample and the bulk dispersion of a higher-order Dirac acoustic metamaterial, respectively. Similar to the Weyl case, the higher-order topology here can be revealed by the two-dimensional higher-order topological index, as calculated in the inset of Fig.~\ref{weiyin.9.Dirac}(e). Nontrivial higher-order topology gives rise to the hinge dispersion, as shown in Fig.~\ref{weiyin.9.Dirac}(f). In addition, there exist charge-2 Dirac point observed in acoustic metamaterials, which is constituted by two Weyl points with same charge~\cite{nat.phys.15.645yang}.

\subsection{Nodal line and nodal surface acoustic metamaterials}
\begin{figure}[htbp]
    \flushright
    \includegraphics[width=0.85\linewidth]{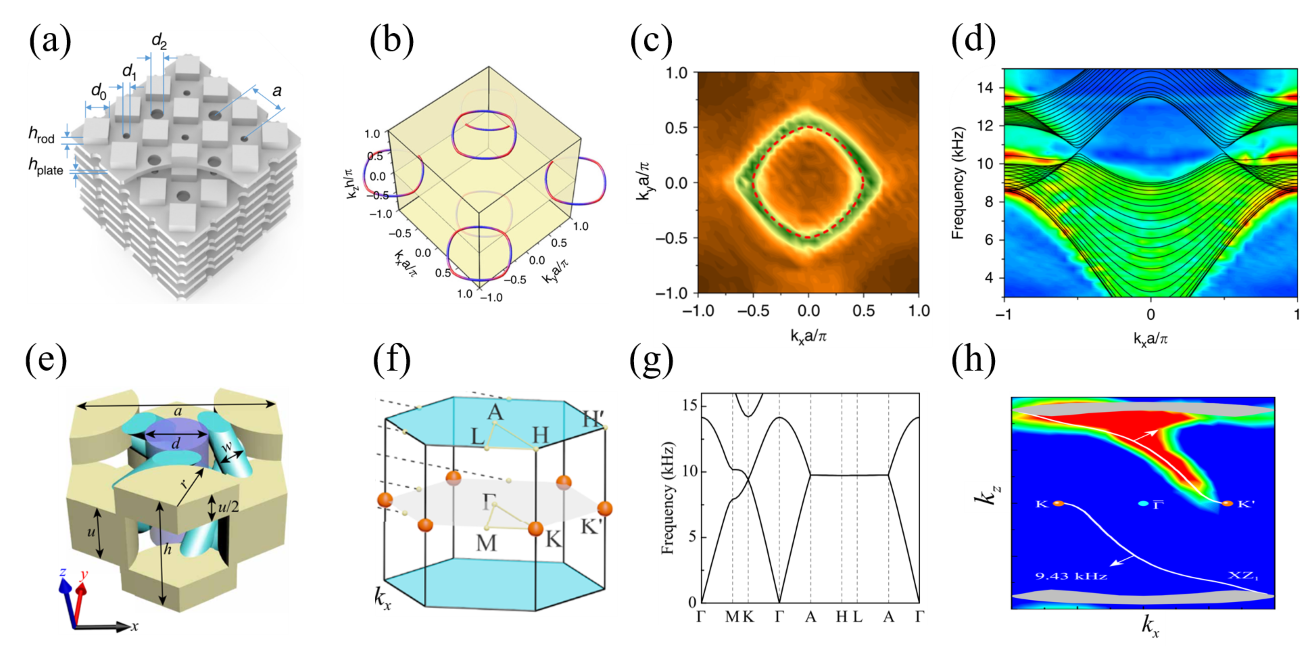}
    \caption{Nodal line and nodal surface acoustic metamaterials. (a) Schematics of an acoustic metamaterial sample hosting a nodal ring. (b) Distribution of simulated nodal rings in momentum space. (c) Measured nodal ring at the plane $k_z h=\pi$. Red dashed line is the calculated result. (d) Measured (color) and calculated (lines) surface dispersions for $k_y a=0$. Red lines show the drumhead surface states, originated from the nontrivial topology of the nodal ring described by $\pi$ Zak phase. (e) Unit cell of an acoustic metamaterial hosting a charged nodal surface with charge $-2$ and a pair of Weyl point with the same charge +1. (f) Distribution of the nodal surface (blue color) and Weyl point (orange spheres) in the first Brillouin zone. (g) Bulk dispersion along high-symmetry lines, showing the nodal surface at the plane $k_z h=\pi$. (h) Fermi arcs connecting the nodal surface and Weyl points. Bright color is the measured data, white curves and gray regions denote the calculated Fermi arcs and projected bulk bands. (a)-(d) adopted from Ref.~\cite{nat.commun.10.1769deng}, and (e)-(h) adopted from Ref.~\cite{sci.adv.6.eaav2360xiao}.}
    \label{weiyin.10.line and surface}
\end{figure}

Nodal line semimetal hosts the band touching of the one-dimensional continuum lines. Different from the nodal point possessing nonzero charge described by two-dimensional topological invariant, such as Chern number for Weyl point, nodal line usually carries nontrivial $\pi$ Zak phase, one-dimensional topological invariant, and give rise to the drumhead surface states~\cite{chin.phys.b25.117106fang}. The realization of nodal line in phononic system was first investigated theoretically by Xiong et al.~\cite{phys.rev.b.97.180101xiong} where nonsymmorphic crystalline symmetry is exploited to generate symmetry-protected, unavoidable phononic nodal lines. Later, phononic nodal lines were investigated in various platforms. Figure~\ref{weiyin.10.line and surface}(a) shows the acoustic metamaterial sample hosting a nodal ring in the first Brillouin zone~\cite{nat.commun.10.1769deng}, which is protected by a mirror symmetry. Figure~\ref{weiyin.10.line and surface}(b) shows the distribution of nodal ring in reciprocal space. Figure~\ref{weiyin.10.line and surface}(c) is the measured equifrequency contour, evidencing the existence of the nodal ring. By parameterizing $k_x$ and $k_y$, the Zak phase calculated in $k_z$ direction equals to $\pi$ or $0$, transiting at the projected nodal rings. Considering an open boundary along the $z$ direction, the end states resulted from the $\pi$ Zak phase form the flat drumhead surface states, as shown in Fig.~\ref{weiyin.10.line and surface}(d). Other geometric configurations, such as straight line~\cite{phys.rev.b100.041303qiu} and chain~\cite{phys.rev.appl.13.054080lu}, have also been realized in acoustic metamaterials.

When the band touching is the two-dimensional surface, i.e., nodal surface, the topological property returns to the case of nodal point~\cite{nat.commun.10.5185yang,sci.adv.6.eaav2360xiao}. Nodal surface carries a nonzero topological charge, thus behaves like the nodal point with the same charge, similar to the nodal line with a nonzero topological charge. Figure~\ref{weiyin.10.line and surface}(e) shows the unit cell of an acoustic metamaterial with the combination of time-reversal symmetry and a twofold screw symmetry along the $z$ direction. As shown in Fig.~\ref{weiyin.10.line and surface}(f), the system hosts a nodal surface with charge $-2$ and two Weyl points with the same charge +1. The bulk dispersion exhibiting the nodal surface is calculated in Fig.~\ref{weiyin.10.line and surface}(g). Two Fermi arcs connected the nodal surface and Weyl points are measured in Fig.~\ref{weiyin.10.line and surface}(h), as expected. Along this way, the nodal point may be extended to a nodal line degeneracy carrying the same charge in an acoustic metamaterial~\cite{phys.rev.b99.094206xiao}.

\subsection{Higher-order topological insulators}
Since 2018, an intriguing class of topological insulating phases called higher-order topological insulators, which go beyond the conventional bulk-edge correspondence, have attracted tremendous attention~\cite{nat.rev.phys.3.520xie,science.357.61benalcazar,phys.rev.b.102.125144xiong,chin.phys.lett.37.074302lin,phys.rev.lett.127.255501wei,phys.rev.lett.127.156401zheng,phys.rev.lett.129.125502yang,phys.rev.lett.129.084301liu,phys.rev.lett.125.255502yang,appl.phys.lett.117.151903meng,sci.bull.66.1959wu,j.appl.phys.127.075105wang,sci.bull.66.1740xu,phys.rev.lett.124.166804ren,sci.bull.67.488huang,phys.rev.lett.120.026801ezawa,nat.mater.18.108xue,nat.mater.18.113ni,phys.rev.b.98.045125ezawa,phys.rev.lett.126.156401zhang,phys.rev.lett.122.244301xue,adv.mater.31.1904682zhang,appl.phys.lett.117.113501yang,phys.rev.b.100.075120chen,nature555.342Serra-Garcia,nat.commun.11.65zhang,phys.rev.b102.035105lin,phys.rev.b.101.161116benalcazar,phys.rev.lett.124.206601qi,nat.commun.11.2108ni,phys.rev.lett.126.156801yang,nat.commun.10.5331zhang,EPL.137.15001lin}. In general, a $l$th order topological insulator in $n$ dimensional systems has topologically protected boundary states on the $n-l$ dimensional boundaries. For example, a two-dimensional higher-order topological insulator hosts gapped one-dimensional edge states and zero-dimensional corner states in the band gap. In this section, we discuss the recent progresses on phononic higher-order topological insulators.

\begin{figure}[htbp]
    \flushright
    \includegraphics[width=0.8\linewidth]{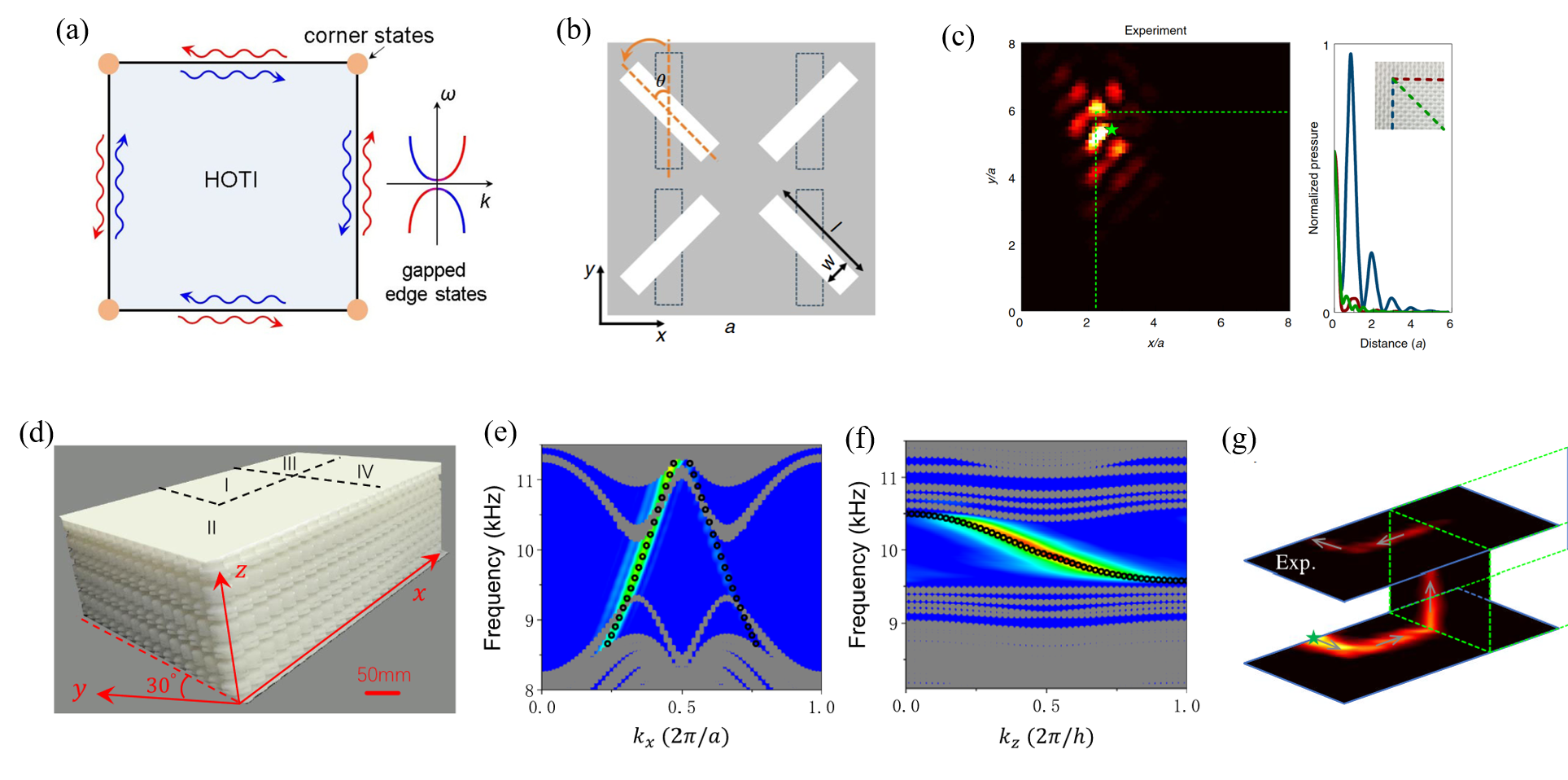}
    \caption{Higher-order topological insulators based on topological crystalline insulators. (a) Illustration of a two-dimensional higher-order topological insulator via Jackiw-Rebbi mechanism. (b) The unit cell of second-order topological acoustic metamaterials with nonsymmorphic symmetry. (c) The corner states as the emergence of Jackiw-Rebbi soliton.  (d) A photo of the printing phononic crystal sample with an L-shaped design. (e,f) Projected band dispersions along (e) the $k_x$ and (f) $k_z$ directions in the top surfaces. (h) Measured acoustic pressure field distributions for hinge transportation.(a) adopted from Ref.~\cite{phys.rev.b.102.125144xiong},  (b)-(c) adopted from Ref.~\cite{nat.phys.15.582zhang}, and (d)-(g) adopted from Ref.~\cite{phys.rev.lett.127.255501wei}.}
    \label{haixiao.TCI}
\end{figure}

The study of higher-order topological insulators stems from several aspects. First, with the progressive understanding of topological crystalline insulators (i.e., topological insulating phases protected by crystalline symmetries), it became clear that as the symmetry of the edge boundaries is lower than the bulk, the predicted topological edge states often become gapped---similar to the gapped edge states in electronic $Z_2$ topological insulators when the time-reversal symmetry is broken~\cite{Phys.Rev.B.83.205124jiang}. Surprisingly, the gapped edge states may by themselves become effectively lower-dimensional topological insulators and may support topologically protected, nontrivial corner or hinge states, yielding a dimensional hierarchy of topological boundary states~\cite{nat.phys.15.582zhang,nat.commun.11.2318he,phys.rev.b.102.125144xiong,chin.phys.lett.37.074302lin,phys.rev.lett.127.255501wei,phys.rev.lett.127.156401zheng,phys.rev.lett.129.125502yang,phys.rev.lett.122.086804liu,phys.rev.lett.125.255502yang,appl.phys.lett.117.151903meng,sci.bull.66.1959wu,j.appl.phys.127.075105wang,sci.bull.66.1740xu,phys.rev.lett.124.166804ren,sci.bull.67.488huang}. Figure~\ref{haixiao.TCI}(a) presents schematically an example: a two-dimensional higher-order topological insulator hosts both the gapped edge states and in-gap corner states. The gapped edge states originate from the bulk topological crystalline phase and the breaking of the protective crystalline symmetry at the edge boundaries. These gapped edge states can be described by massive one-dimensional Dirac equations. The switch of the sign of the Dirac mass at the adjacent edge boundaries then leads to the topological corner states according to the Jackiw-Rebbi mechanism.

A concrete example of the acoustic realization of such a two-dimensional higher-order topological insulator is given in Fig.~\ref{haixiao.TCI}(b). In this acoustic metamaterials, the topological transitions of the bulk and edges can be triggered independently by tuning a single geometric parameter $\theta$. To visualize the corner mode, a box-shaped supercell consisting of acoustic metamaterials with a distinct higher-order topology is constructed, of which the field pattern is displayed in Fig.~\ref{haixiao.TCI}(c). Such a proposal opening a new route towards tunable topological metamaterials and subsequently has been realized in the mechanical system~\cite{j.appl.phys.127.075105wang,sci.bull.66.1959wu}. Another common Jackiew-Rebbi mechanism induced higher-order topological insulator is the acoustic topological crystalline insulator (also known as the acoustic analog of the quantum spin-Hall insulator), in which the bulk topological index and the bulk-induced corner topological index can be defined via the symmetry-eigenvalues of Bloch functions at the high symmetry points ~\cite{phys.rev.b.99.245151Benalcazar}. When the acoustic topological crystalline insulator and normal insulator are placed in a supercell with specific symmetry, the nontrivial corner index difference gives rise to the corner states featured with pseudospin-momentum locking, leading to the higher-order topological quantum spin-Hall effect~\cite{chin.phys.lett.37.074302lin,phys.rev.lett.122.086804liu,phys.rev.lett.125.255502yang}. In addition, to overcome the shortage that the inducing topological corner mode emerges in very limited geometries, a general scheme based on Dirac vortices from aperiodic Kekul\'e modulation is also proposed~\cite{phys.rev.lett.126.226802wu}. By further extending to three dimensions, one may realize the hierarchies of topological insulators~\cite{nat.commun.11.2318he,appl.phys.lett.117.151903meng,phys.rev.lett.127.255501wei,sci.bull.66.1740xu} and Dirac points~\cite{phys.rev.lett.127.156401zheng,phys.rev.lett.129.125502yang}. For example, a three-dimensional acoustic analogy of topological insulator exhibits both two-dimensioanl gapless surface Dirac cone as first-order and one-dimensional gapless hinge Dirac dispersion as second-order topological insulators, supporting robust surface or hinge sound transport~\cite{nat.commun.11.2318he}. Such an idea can be further generalized to realize hinge propagation in three different directions~\cite{appl.phys.lett.117.151903meng,phys.rev.lett.127.255501wei,sci.bull.66.1740xu}. As depicted in Fig.~\ref{haixiao.TCI}(d), a three-dimensional sample consisting of three-dimensional topological phononic crystals is fabricated to illustrate the hinge states transporting along three different directions. By carefully designing onsite potential and the intracell and intercell layer coupling, different topological phases can be obtained in the stacked bilayer hexagonal lattice, which further enables the interface states. As shown in Figs.~\ref{haixiao.TCI}(e) and~\ref{haixiao.TCI}(f), the experimentally measured projected band dispersion along the $k_x$ and $k_z$ direction are observed, indicating the hinge transport along $x$ and $z$ directions, respectively. By further printing a sample that consisting of different phases, the one-way transport from hinge to hinge is demonstrated, of which a measured acoustic pressure field distribution is displayed in Fig.~\ref{haixiao.TCI}(g).

\begin{figure}[htbp]
    \flushright
    \includegraphics[width=0.8\linewidth]{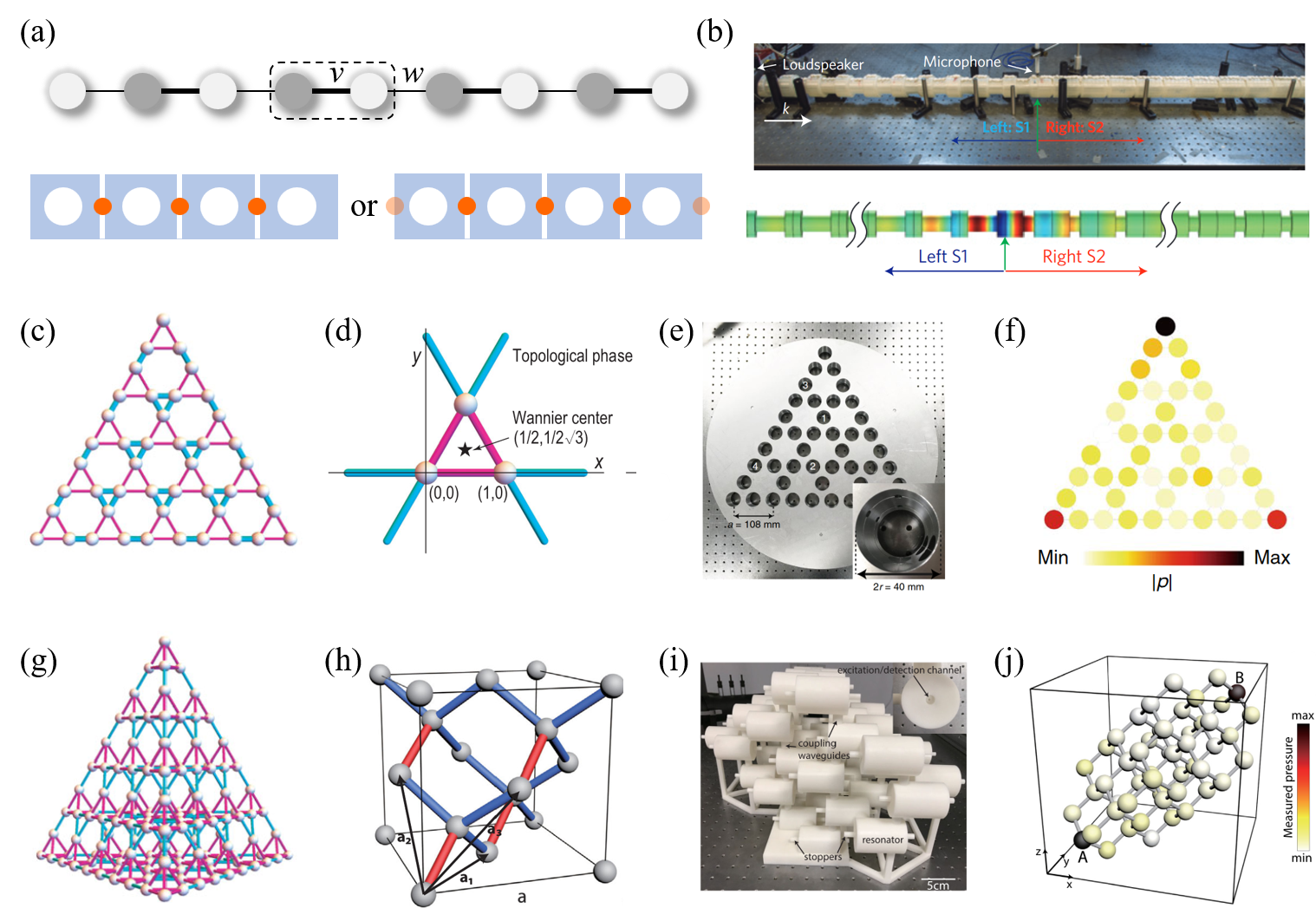}
    \caption{Wanniner-type topological insulators with dipole moments. (a) Schematic of Su-Schrieffer-Heeger model (upper panel), and the possible configurations of Wannier center for the Su-Schrieffer-Heeger model with nontrivial dipole moment (Lower panel). (b) A one-dimensional phononic crystal consisting of a nontrivial lattice and trivial lattices (upper panel) and the simulated field pattern of the topological model that localized at the interface between nontrivial and trivial lattices.  (c) Illustration of breathing kagome lattice with dimerized couplings. (d) The Wannier center configuration for breathing kagome lattice with nontrivial dipole moment, where there is a mismatch between Wannier center and lattice sites.  (e) A kagome lattice sample, and (f) the experimental density of states, which exhibits localization at corners. (g) Illustration of Pyrochlore lattice, which is the prototype of third-order Wannier-type topological insulators. (h)Schematic of the third-order Wannier-type topological insulator. (i) Photograph of the fabricated sample.  (j) Measured pressure maps at frequency of 2900Hz, corresponding to the corner modes.(b) adopted from Ref.~\cite{nat.phys.11.240xiao}. (c)-(d) and (g) adopted from Ref.~\cite{phys.rev.lett.120.026801ezawa}.(e)-(f) adopted from Ref.~\cite{nat.mater.18.108xue}. (h)-(j) adopted from Ref.~\cite{phys.rev.lett.122.244301xue}. }
    \label{haixiao.Wannier HOTI}
\end{figure}

The second origin of higher-order topological insulators is the polarization and Wannier center of crystalline insulators. In the modern theory of polarization, the bulk dipole moment is formulated through the Berry phase, which is related to the Wannier center. The dipole moment can be quantized by crystalline symmetries, such as mirror and rotation symmetries, making it eligible as a topological invariant. A typical example with quantized dipole moment is given by the well-known Su-Schrieffer-Heeger model [see the upper panel of Fig.~\ref{haixiao.Wannier HOTI}(a)], which describes one-dimensional spinless electronic system in a chain with staggered hopping amplitudes~\cite{phys.rev.lett.62.2747zak}.
When Su-Schrieffier-Heeger model has a nontrivial dipole moment, namely, the Wannier center configuration of the Su-Schrieffier-Heeger model differs from its lattice site configuration, the fractional charges emerge at boundaries that cannot be removed by any perturbations, as depicted in the lower panel of Fig.~\ref{haixiao.Wannier HOTI}(a). Although the above polarization theory was first proposed in solid physics, it still can be applied to the understanding of topological behavior in acoustics. Note that the dipole polarization should be replaced by the concept of the Wannier center, and the fractional boundary charge can be understood as due to the filling anomaly of the bulk and its manifestation at the boundaries~\cite{phys.rev.b.96.245115Benalcazar}. We remark that this picture holds not only in the tight-binding models but also in phononic systems which cannot be described by tight-binding models, as summarized in Ref.~\cite{EPL.137.15001lin}.

Inspired by the Su-Schrieffer-Heeger model, the first demonstration of nontrivial topology in acoustics was realized in 1D phononic crystals~\cite{nat.phys.11.240xiao}. An interface state emerges at the interface when two phononic crystals with different dipole moments are placed together [see Fig.~\ref{haixiao.Wannier HOTI}(b)]. Starting from the quantized dipole moment in one-dimensional systems, one can generalize it to higher-dimensional systems, of which have polarization forms similar to that of the dipole moment in one dimension. In the literature, topological crystalline insulators characterized by nontrivial dipole momentum are termed Wannier-type higher-order topological insulators~\cite{phys.rev.lett.120.026801ezawa,nat.mater.18.108xue,nat.mater.18.113ni,phys.rev.b.98.045125ezawa,phys.rev.lett.126.156401zhang,phys.rev.lett.122.244301xue,adv.mater.31.1904682zhang,phys.rev.lett.122.204301fan,commun.mater.2.62chen,appl.phys.lett.117.113501yang,phys.rev.b.100.075120chen,sci.adv.6.eaay4166Weiner,nat.commun.10.5331zhang,phys.rev.b.102.104113zheng,phys.rev.b.101.220102zhang,phys.rev.b.101.094107wakao}. A prototype of Wannier-type higher-order topological insulators is the kagome lattice with dimerized couplings[see Fig.~\ref{haixiao.Wannier HOTI}(c,d)], which theoretically proposed in a tight-binding model~\cite{phys.rev.lett.120.026801ezawa} and subsequently experimentally demonstrated in acoustic systems~\cite{nat.mater.18.108xue,nat.mater.18.113ni}[also see Fig.~\ref{haixiao.Wannier HOTI}(e,f)]. These topological phases are characterized by a topological invariant, defined as an integral of the Berry potential over the entire Brillouin zone. Since the dipole polarization is equivalent to the position of Wannier center [see Fig.~\ref{haixiao.Wannier HOTI}(c)], it is intuitively to distinguish the topological phase by depicting the Wannier center configuration~\cite{phys.rev.b.98.045125ezawa}. The mismatch between the lattice site and Wannier center gives rise to the higher-order topological corner states. By combining the valley degree of freedom, the emergence of corner states is tied to specific corners[see Fig.~\ref{haixiao.Wannier HOTI}(g-i)], results in the valley-selective corner states~\cite{phys.rev.lett.126.156401zhang}. To step further, one can extend the two-dimensional second-order topological insulator to the three-dimensional third-order topological insulator [see the tight-binding model in Fig.~\ref{haixiao.Wannier HOTI}(g)]. As shown in Fig.~\ref{haixiao.Wannier HOTI}(h), an anisotropic diamond lattice with cubic unit cell is proposed to mimic the third-order Wannier-type topological insulators. Experimentally, a third-order topological insulator with rhombohedronlike structure containing six rhombus-shaped surfaces [see Fig.~\ref{haixiao.Wannier HOTI}(i)] is carefully designed by stereo-lithography three dimension printing ~\cite{phys.rev.lett.122.244301xue}. Since a third-order topological insulator is characterized by the existence of topological corner states at certain corners of a finite three-dimension sample, Fig.~\ref{haixiao.Wannier HOTI}(j) also presents a typical corner state via measuring the local acoustic response at two corner resonators. Other schemes to third-order topological insulators also can be accessed by the acoustic pyrochlore lattice~\cite{sci.adv.6.eaay4166Weiner} and an Su-Schrieffer-Heeger-inspired cubic lattice~\cite{nat.commun.10.5331zhang,phys.rev.b.102.104113zheng}.

\begin{figure}[htbp]
    \flushright
    \includegraphics[width=0.8\linewidth]{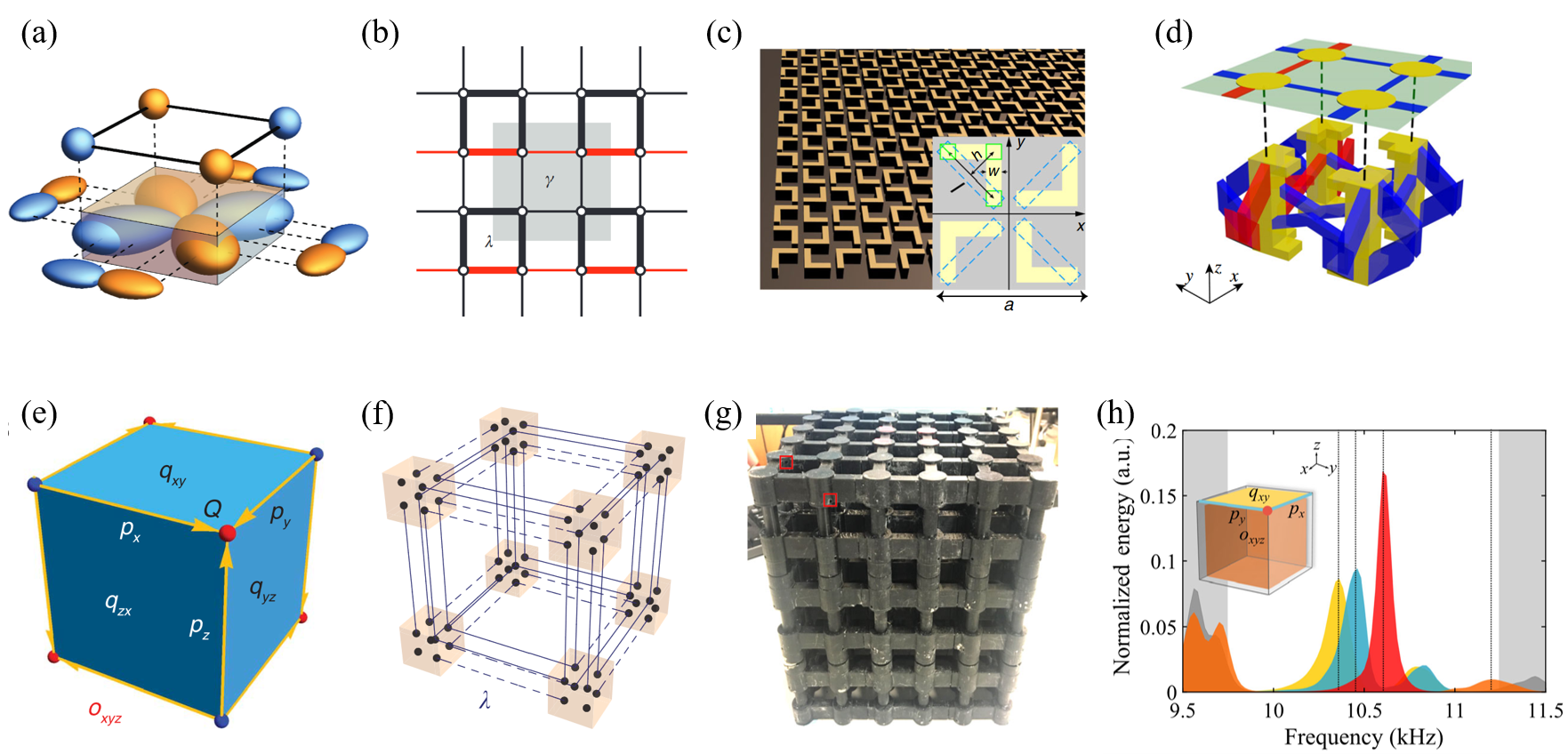}
    \caption{Quadrupole and octupole topological insulators. (a) Bulk quadrupole moment with its accompanying edge dipoles and corner charges. (b) A tight-binding model for quadrupole topological insulators. Thin (thick) lines refer to weak (strong) hoppings with strength $\gamma (\lambda)$, while the red (black) lines indicate a negative (positive) hopping amplitude.  (c) A two-dimensional acoustic metamaterial that realizes a quadrupole topological insulator based on symmetry approach. The inset shows the unit cell.  (d) Acoustic quadrupole topological insulator based on coupled acoustic resonators with different connectivity. (e) Bulk octupole moment $O_{xyz}$ and its boundary polarizations. (f) The tight-binding model for octupole topological insulators. (g) Acoustic octupole topological insulators that made of coupled resonators. (h) Normalized acoustic frequency-response spectra for the selective bulk (orange), surface (yellow), edge (blue) and corner sites (red) of a finite structure. (a)-(b) adopted from Ref.~\cite{nature555.342Serra-Garcia}.  (e)-(f) adopted from Ref.~\cite{science.357.61benalcazar}. (c) adopted from Ref.~\cite{nat.commun.11.65zhang}, (d) adopted from Ref.~\cite{phys.rev.lett.124.206601qi}, and (g)-(h) adopted from Ref.~\cite{nat.commun.11.2108ni}. }
    \label{haixiao.3.multipole HOTI}
\end{figure}

Although the dimensional hierarchy of the Jackiw-Rebbi mechanisms is an easy way to understand and construct higher-order topological insulators, historically, higher-order topological insulators were first found in the study of the multipole moment in crystalline insulators. In the modern theory of polarization, the bulk dipole moment is formulated through the Berry phase. However, the generalization to multipole moment leads more intricate formulation. In contrast to the Wannier-type higher-order topological insulators, multipole topological insulators have vanishing dipole moment but with quantized multipole moments~\cite{science.357.61benalcazar,phys.rev.b.96.245115Benalcazar,nature555.342Serra-Garcia,nat.commun.11.65zhang,phys.rev.lett.124.206601qi,nat.commun.11.2108ni} that are described by nested Wannier bands~\cite{science.357.61benalcazar,phys.rev.b.96.245115Benalcazar,phys.rev.b.101.161116benalcazar,phys.rev.b102.035105lin} and quantum operator formulations~\cite{Phys.Rev.B.100.245134kang,Phys.Rev.B.100.245135wheeler,Phys.Rev.B.100.245133ono}. For example, a quadruple topological insulator has vanishing dipole polarization but a finite quadrupole moment that protected by certain symmetries, as depicted in Fig.~\ref{haixiao.3.multipole HOTI}(a). Fig.~\ref{haixiao.3.multipole HOTI}(b) presents the tight-binding model for quadrupole topological insulators, which is a square lattice model with four sites in each unit cell with coexisting both positive and negative nearest-neighbor couplings~\cite{science.357.61benalcazar,phys.rev.b.96.245115Benalcazar}. The positive and negative couplings enable the $\pi-$flux per plaquette (the area of one unit cell), leading to two noncommutative reflection symmetries, which is necessary for emergency nontrivial quadrupole topology. To this end, a symmetry-based approach is introduced to realize quadrupole topological insulator~\cite{nat.commun.11.65zhang}, in which the quadrupole topology is protected by two noncommutative glide symmetries (see Fig.~\ref{haixiao.3.multipole HOTI}(c)). Another approach to acoustic quadrupole topological insulator is to generate both positive and negative hoppings directly by linking acoustic cavities with different connectivity according to the field morphologies of acoustic resonators ~\cite{phys.rev.lett.124.206601qi} (also see Fig.~\ref{haixiao.3.multipole HOTI}(d)).

Similar to the quadrupole topological insulator with quantized quadrupole moment, an octupole insulator has vanishing dipole polarization and quadrupole moment but a finite octupole moment (see Fig.~\ref{haixiao.3.multipole HOTI}(e)). The tight-binding model for octupole topological insulator use cubic lattice models with eight sites in each cell with coexisting positive and negative nearest shopping [also see Fig.~\ref{haixiao.3.multipole HOTI}(f)]~\cite{science.357.61benalcazar,phys.rev.b.96.245115Benalcazar}. Due to its complex tight-binding configurations, the three-dimensional octupole topology is even more challenging to be realized. Up till now, the only way to acoustic octupole topological insulator is to generate the positive and negative hoppings by linking resonators with different configurations~\cite{nat.commun.11.2108ni,nat.commun.11.2442xue}. A typical fabricated sample is given in Fig.~\ref{haixiao.3.multipole HOTI}(g). The manifestation of the octupole topological insulator can be identified by the emergence of the surface, edge, and corner states, respectively. As shown in Fig.~\ref{haixiao.3.multipole HOTI}(h), the measured acoustic frequency-response spectra for the selective bulk (orange), surface (yellow), edge (blue) and corner sites (red) of a finite structure provide the direct evidence for the acoustic octupole topological insulators.

Besides, it is also worthy to notice several new concepts that are related to the higher-order topological insulators. For example, by utilizing the richness classes of higher-order topologies, one may combine the conventional topology with higher-order topological phase in a single acoustic system, leading to the hybrid (or hierarchy) topological insulator~\cite{phys.rev.lett.126.156801yang,nat.commun.10.5331zhang}. In addition, an acoustic square-root topological insulator, whose topological property is inherited from the square of the Hamiltonian, is also demonstrated to host higher-order topology ~\cite{phys.rev.b.102.180102yan}. Last but not least, the gain-and-loss-induced higher-order topological insulator is also proposed to be implemented in sonic crystals~\cite{phys.rev.lett.122.195501zhang}. Subsequently, a higher-order topological insulator induced by deliberately introduced losses was experimentally realized in a phononic crystal~\cite{nat.commun.12.1888gao}.

\subsection{Non-Abelian topological states}
In this section, a brief introduction to the non-Abelian band topology and a short review of its recent experimental realizations in phononic metamaterials are presented. The non-Abelian band topology, which was first proposed by Wu {\it et al.} in 2019~\cite{science365.1273wu}, to some degree, complements and extends the conventional Abelian topological band theory. Let us first recall the Abelian band topology before we move on. As elucidated in other sections, the topological nature of band insulators, or degenerate band nodes, in general, can be characterized by, e.g., the Chern number, quantized Zak phase, and various winding numbers, which are usually defined within two subspaces spanned by “occupied” and “unoccupied” bands. These topological invariants or node charges are classified by Abelian groups so that they obey the additivity and commutativity of multiplication. A prototype example is that two Weyl points with opposite Chern charges tend to annihilate in collision.

Beyond the above paradigm, the non-Abelian band topology, as its name implies, exemplifies the non-Abelian topological charges of multiple bands, wherein the emergence and robustness of multi-band configurations cannot be explained by the conventional Abelian topological invariants. So far, the application of the non-Abelian band topology always demands two pivotal elements. First, the Hamiltonian of the system should enjoy the space-time inversion symmetry $I_{ST}=I_{S}T$, where $I_S$ denotes the spatial inversion $P$ (or $\pi$ rotation in two-dimensional spinless systems) and $T$ the time inversion. Second, multiple bands or gaps require to be taken into consideration simultaneously, rather than to be simply partitioned into two subspaces. The combined anti-unitary symmetry $I_{ST}$ acts as an on-site symmetry since it leaves every $k$ point fixed in momentum space. More importantly, $I_{ST}^{2}=1$ leads us to take a suitable basis so that $I_{ST}$ can be represented by the complex conjugation $K$, which yields
\begin{equation}
  I_{ST}H(k) I_{ST}^{-1}=H^* (k)=H(k);\quad I_{ST} |u(k)\rangle=|u(k)\rangle^*=|u(k)\rangle.
\end{equation}
In that case, the Hamiltonian as well as the corresponding eigenstates at each $k$ point is real-valued. The gauge freedom of eigenstates is reduced to $\pm 1$. The Hamiltonian at each $k$ is then encoded by an orthonormal rotation frame $\{u_i(k)\}$, which denotes a local geometric quantity of eigenstates. The way how the frame smoothly rotates  when the Hamiltonian evolves with $k$ contributes to the band topology in $I_{ST}$-symmetric systems. For example, the first Stiefel-Whitney number defined from the twist of real eigenstates is equivalent to the quantized Zak phase derived from the parallel transport based on complex eigenstates~\cite{CPB.28.117101Ahn}.

\begin{figure}[htbp]
    \flushright
    \includegraphics[width=0.85\linewidth]{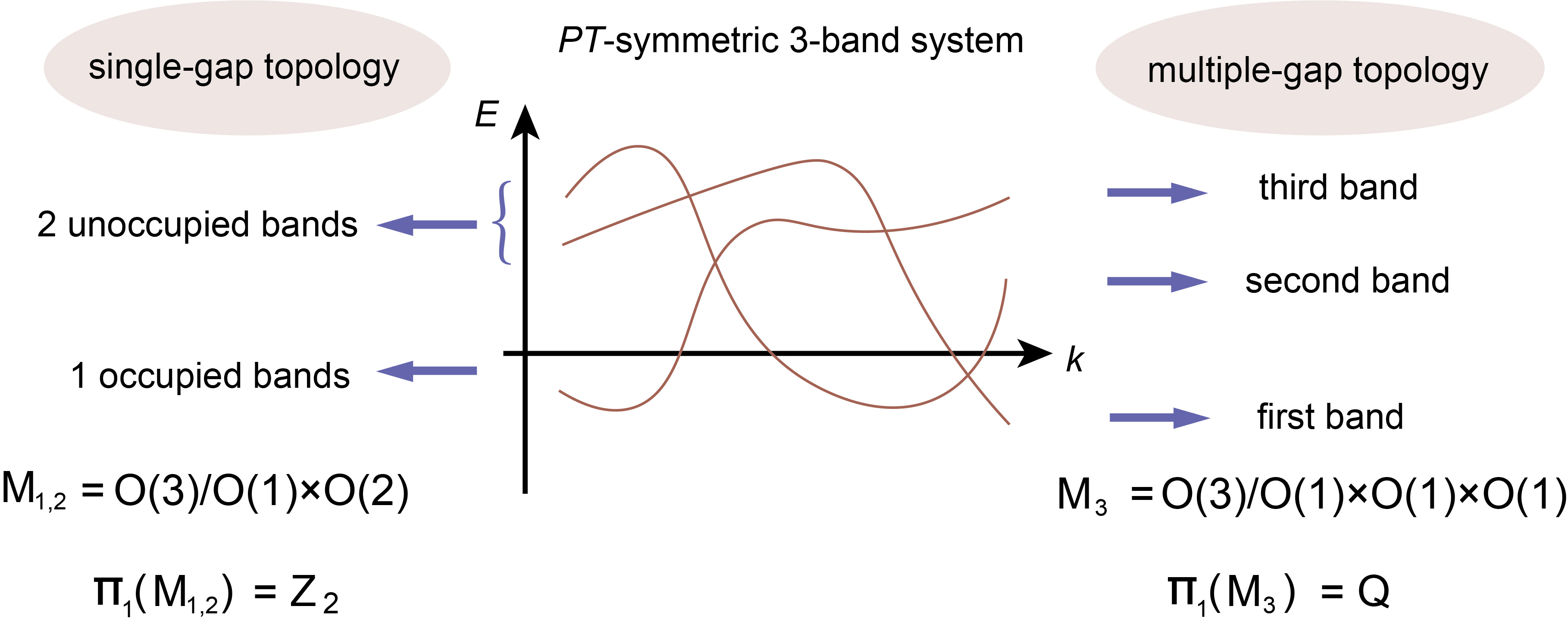}
    \caption{Single-gap and multiple-gap partitions of a $PT$-symmetric three-band system lead to Abelian and non-Abelian band topology, respectively. $M_{1,2}$ and $M_3$ are the Hamiltonian space while  $Z_2$ and $Q$ are their corresponding first-order homotopy groups.}
    \label{figure1}
\end{figure}

 The power of the multi-gap perspective is released when the order–parameter space of the Hamiltonian is involved. As illustrated in Fig.~\ref{figure1}, when we consider a $PT$-symmetric three-band spinless system, if we separate three bands into one occupied and two unoccupied bands, then the corresponding order–parameter space of the Hamiltonian, without degeneracy, is $M_{1,2}=O(3)/O(1)\times O(2)$, which is a real Grassmann manifold where $O(N)$ is the $N$-dimensional orthogonal group ~\cite{karoubi2008k}. According to the homotopy group theory, the first homotopy group of the real Grassmann manifold $M_{1,2}$ is $\pi_1(M_{1,2} )=Z_2$, which accounts for the $Z_2$ quantized Zak phase used in Abelian band topology~\cite{PRB.92.081201Fang,CPB.25.117106Fang,PRL.121.106403Ahn}.
%  accumulated along the non-contractible loops  or a closed loop encircling the band nodes in momentum space.

\begin{figure}[htbp]
    \flushright
    \includegraphics[width=0.85\linewidth]{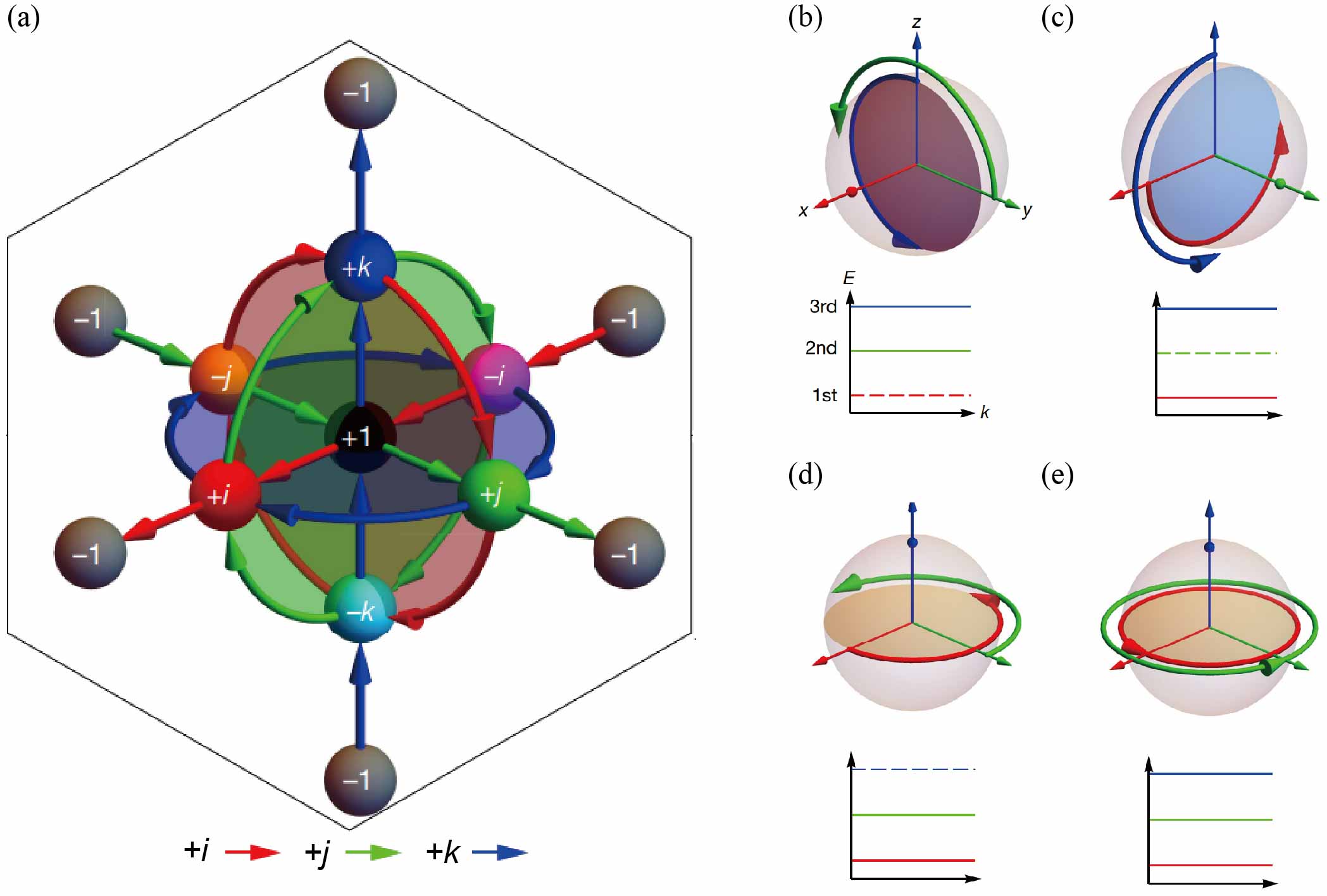}
    \caption{Graphical representation of
the quaternion group. (a) Multiplication rules of non-Abelian charges. (b)-(d) Non-Abelian charges of $i$,$j$ and $k$, represented by the $\pi$ rotation of the eigenstate frame around the first, second and third bands, respectively. (e) Non-Abelian charge $-1$ represented by $2\pi$ rotation of one eigenstate.  Figures adopted from Ref.~\cite{Nature.594.195Guo}.}
    \label{figure2}
\end{figure}

In comparison, when we consider the multiple gaps together (see the right panel in Fig.~\ref{figure1}), the order-parameter space of the Hamiltonian  is then altered to a real flag manifold~\cite{karoubi2008k}
\begin{equation}
  M_3=O(3)/O(1)\times O(1)\times O(1).
\end{equation}
The first homotopy group of the real flag manifold $M_3$ is
\begin{equation}
\pi_1(M_3)=Q=\{ \pm i,\pm j,\pm k,+1,-1 \},
\end{equation}
which is nothing but the non-Abelian quaternion group, the leading role in non-Abelian band topology. The associated non-Abelian group multiplications of the quaternion group are graphically represented in Fig.~\ref{figure2}(a), which shows that $i$, $j$ and $k$ satisfy $ij=k$, $jk=i$, $ki=j$ and $i^2=j^2=k^2=-1$. The topological charges of band gaps or band nodes in a $PT$-symmetric three-band spinless system can be characterized by these non-Abelian group elements. Taking the $PT$-symmetric three-band insulator system in one dimension as an example, the quaternion charges are defined by the rotation of the 3-eigenstate frame when $k$ runs from $-\pi$ to $\pi$ across the first Brillouin zone, as visualized in Figs.~\ref{figure2}(b)-\ref{figure2}(e). Specifically, the $\pi$ rotation of the frame around the first, the second and the third eigenstates induces the non-Abelian topological charges of $+i$, $+j$ and $+k$, respectively, for the global three bands. The $2\pi$ rotation of the frame around any one of the eigenstates constitutes the non-Abelian topological charge of $-1$. The identity charge $1$ indicates no rotation and thus corresponds to the trivial phase. The conjugate partners $-i$, $-j$ and $-k$ denote the inverse rotations. Remarkably, the elements of $Q$ are observable only up to conjugacy, which results from the sign freedom ±1 of real-valued eigenstates~\cite{NJP.24.053042Park}. The $PT$-symmetric three-band spinless system is the minimal model to encode the non-Abelian band topology.

In the wake of Wu’s pioneering work~\cite{science365.1273wu}, several theoretical~\cite{NP.16.1137Bouhon,PRX.9.021013Ahn,PRB.102.115135bouhon,PRB.101.195130Tiwari,kang2020non,NC.13.1Peng,NJP.24.053042Park,bouhon2022multi,Arxiv.jiang2022edge,PRResearch.3.043006Ezawa,slager2022floquet,wang2022experimental,chen2022non,peng2022multigap,lange2022topological} researches sprung up and have witnessed the rise of the non-Abelian band topology. For instance, the phonon spectra in layered silicates were predicted to host the non-Abelian physics~\cite{NC.13.1Peng}. By far, thanks to the designability of artificial materials, the non-Abelian band topology has been proposed, manipulated and probed in various metamaterials such as transmission line networks~\cite{Nature.594.195Guo,NC.12.1jiang}, phononic crystals~\cite{nat.phys.17.1239jiang,qiu2022minimal,jiang2022experimental} full-dielectric~\cite{ACSPhotonics.8.2746Park} and biaxial photonic crystals~\cite{PRL.125.053601,PRL.128.246601Wang,NC.12.1jiang,Light.10.1Wang,Arxiv.Wang} and spring-mass systems~\cite{Arxiv.park}. Among them, the first direct observation of the non-Abelian topological charges was achieved by Guo {\it et al.}~\cite{Nature.594.195Guo} in a one-dimensional $PT$-symmetric three-band insulator system using transmission line network metamaterials. The different non-Abelian quaternion charges, as represented in Figs.~\ref{figure2}(b)-\ref{figure2}(e), were extracted through the experimentally obtained eigenstate frame sphere.
\begin{figure}[htbp]
    \flushright
    \includegraphics[width=0.85\linewidth]{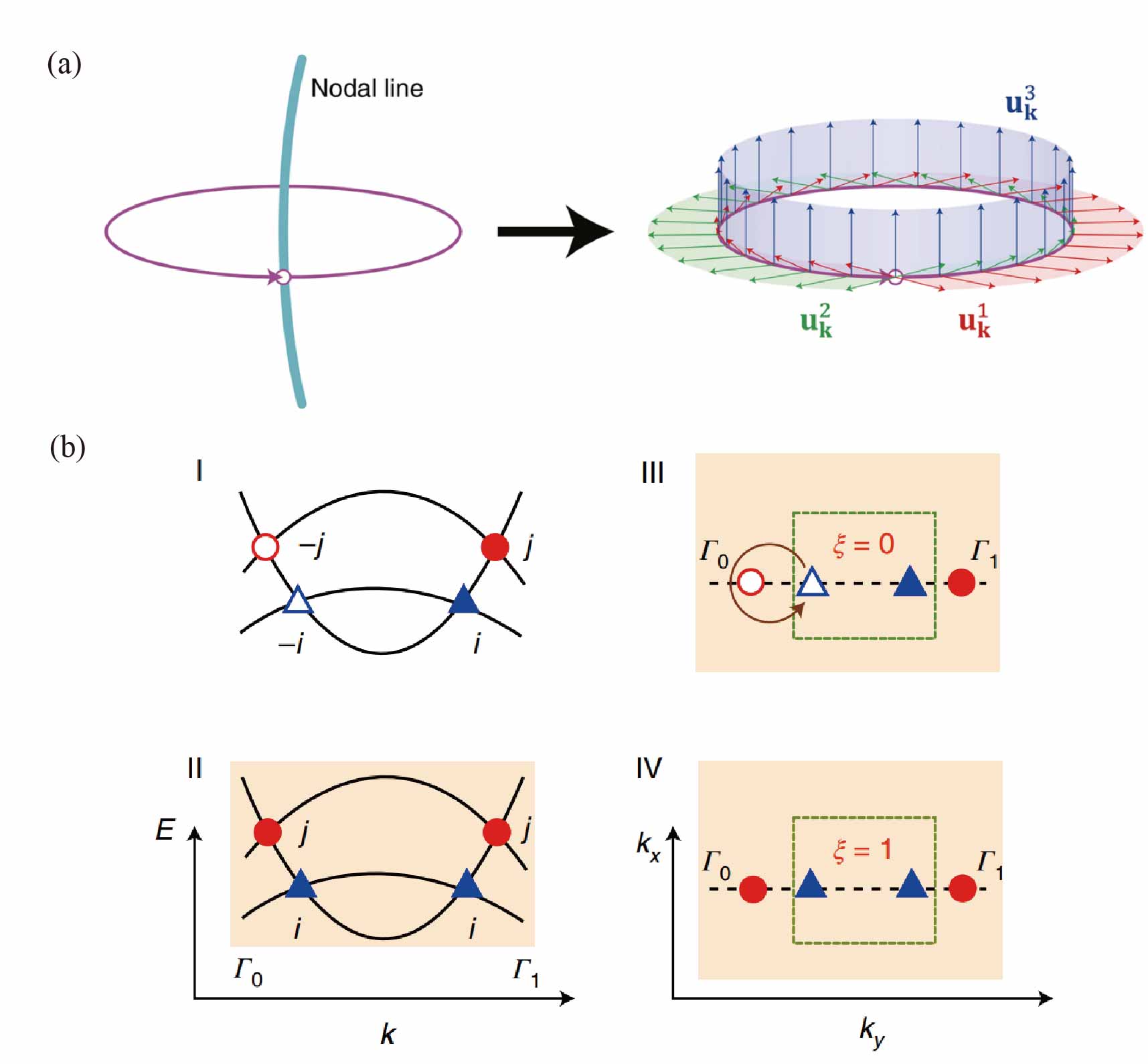}
    \caption{Non-Abelian characterization of Nodal systems. (a) Non-Abelian charges defined from the rotation of the eigenstate frame encircling a nodal line. (b) Illustration of the non-Abelian braiding and the topological Euler class in nodal structures. Figure (a) adopted from Ref.~\cite{Nanophotonics.11.2779Park}. Figure (b) adopted from Ref.~\cite{nat.phys.17.1239jiang}. }
    \label{nodal}
\end{figure}
Meanwhile, most researches about the non-Abelian band topology focus on three-band semimetal systems in two or three dimensions with entangled band nodes such as Weyl points, nodal lines, and nodal links. The possible configurations of semimetals are set strict constraints by non-Abelian charges, which are defined according to the rotation of the 3-eigenstate frame around a closed loop in momentum space encircling the band nodes, as sketched in Fig.~\ref{nodal}(a). Different configurations offer an excellent platform to explore the novel non-Abelian phenomena that are absent in one dimension, such as the non-Abelian braiding of topological charges and their multipath phase transitions. As shown in Fig.~\ref{nodal}(b), braiding a band node with charge $-i$ in the first gap around one with charge $-j$ in the second gap (III) changes their charges by a factor of $-1$, which is the nontrivial manifestation of the non-Abelian braiding, i.e., $(-j)(-i)(-j)^{-1}=i$ and $(-i)(-j)(-i)^{-1}=j$. After braiding, the two band nodes in each gap have the same charge (II). Such two nodes are topologically obstructed to annihilate, as the total charge is $i^2=j^2-1$, which is the key signature of a nontrivial braiding process. The non-zero integer-valued patch Euler class $\xi$ integrated over a Brillouin zone patch enclosing the pair of band nodes, defined as follows,
\begin{equation}
  \int_{k\in Brillouin\,zone\,patch}\langle{\nabla u_1(k)}|\times |\nabla u_2(k)\rangle,
\end{equation}
can then quantifies their stability~\cite{nat.phys.17.1239jiang}.
 \begin{figure}[htbp]
    \flushright
    \includegraphics[width=0.85\linewidth]{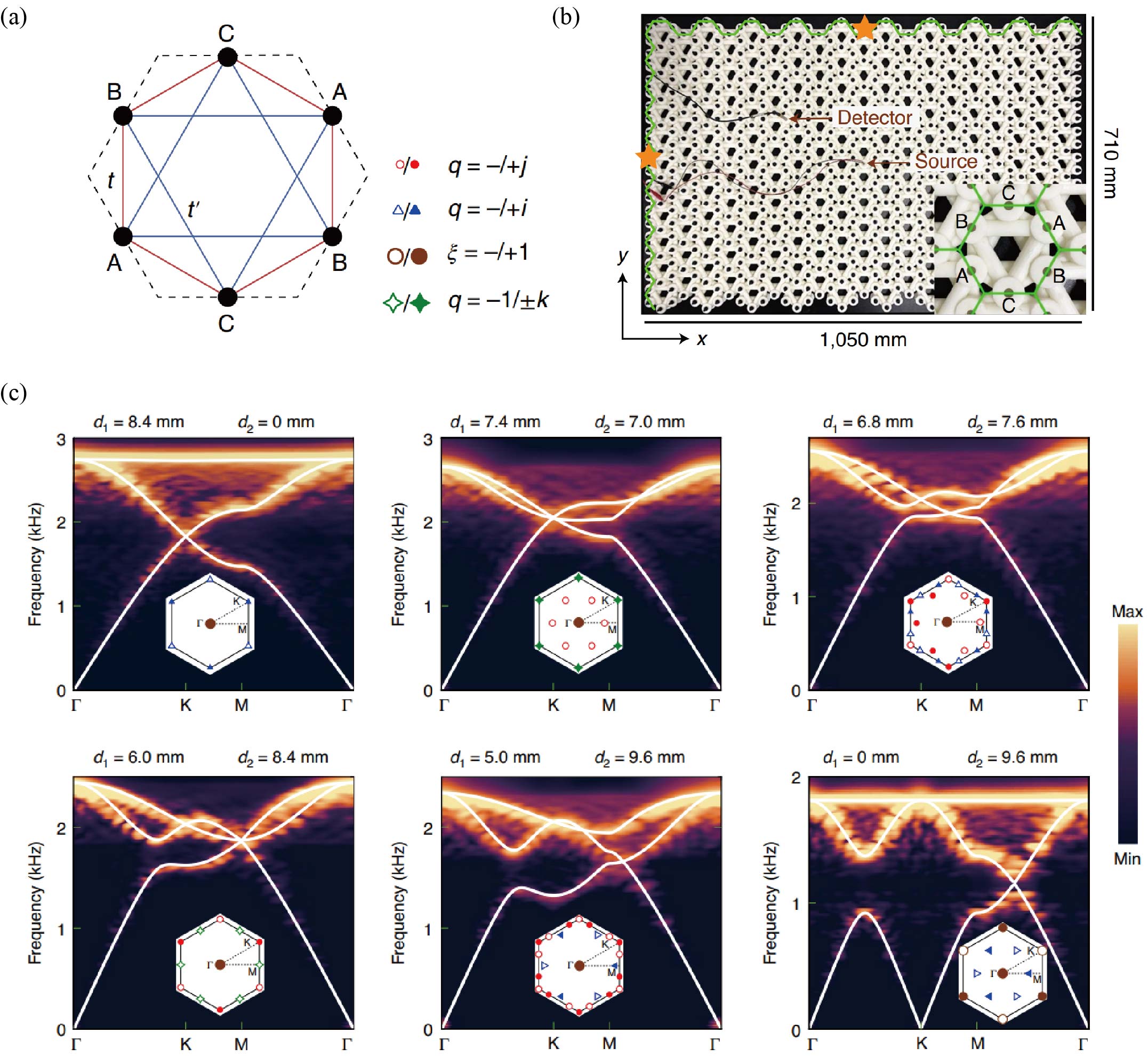}
    \caption{(a) A kagome tight-binding model with tunable nearest and next-nearest neighbor couplings. (b) Photograph of the experimental set-up for measurements. (c) Measured acoustic band structures for six geometric parameters with different nodal structures characterized by different non-Abelian charges. Figures adopted from Ref.~\cite{nat.phys.17.1239jiang}.}
    \label{jiang}
\end{figure}

The braiding and phase transitions of non-Abelian charged band nodes were first observed by Jiang {\it et al}. using tunable acoustic semimetals based on the kagome model~\cite{nat.phys.17.1239jiang}. The kagome model with only the nearest and next-nearest neighbor couplings [see Fig.~\ref{jiang}(a)] results in a two-dimensional $PT$-symmetric three-band semimetal. In experiments, the acoustic kagome lattices were 3D printed using photosensitive resin [see Fig.~\ref{jiang}(b)]. The sites of the kagome model are realized by cylindrical acoustic resonators. The nearest and next-nearest neighbor couplings are realized by horizontal tubes with different diameters that connect the resonators. Correspondingly, the first three $s$-orbital bands of the acoustic kagome lattice are always tangled together and the non-Abelian topological charges can be defined upon these band nodes. By adjusting geometric parameters of metamaterials, in other words, tuning the nearest and next-nearest neighbor couplings, the non-Abelian node charges are converted through the braiding of nodes in adjacent gaps, as well as their creation, annihilation, merging, transferring, and splitting. As implied in the evolution of the experimentally measured band dispersions in Fig.~\ref{jiang}(c), these detailed dispersions are associated with different non-Abelian topological charges. In each panel, the evolution tendency of node charges is denoted by the arrows in the inset. The labels of different charges are shown in Fig.~\ref{jiang}(a) for reference. Specifically, when the next-nearest neighbor couplings vanish [see the first panel of Fig.~\ref{jiang}(c)], Dirac points emerge at two $K$ points with charges $\pm i$ in the first band gap and a quadratic node with charge $-1$ emerges at the $\Gamma$ point. The quadratic node contributes to a nonzero patch Euler class of $\xi =1$, which remains during the tuning process, proving the topological robustness of the non-Abelian charge $-1$. Then, by continuously decreasing and increasing the nearest and next-nearest neighbor coupling strength, respectively, rich non-Abelian topological transitions occur. First, six pairs of Dirac nodes with charges $\pm j$, of which the total charge is 1, are created "in a vacuum” on the $\Gamma K$ and $\Gamma M$ lines in the second band gap. In the meantime, two halves of them head to $K$ and $M$ points, respectively. Once the six Dirac nodes arrive at $K$ points, linear triply-degenerate points emerge, with charges $\pm k$ since $(\pm i)(-j)=\mp k$ [see the second panel in Fig.~\ref{jiang}(c)]. Further tuning splits the triply-degenerate points and leads to  $\pm j$ charged Dirac nodes at $K$ points in the first gap and $\pm i$ charged Dirac nodes at $KM$ lines in the second gap [see the third panel in Fig.~\ref{jiang}(c)]. The $\pm i$ charged Dirac nodes continue to move and encounter the remaining six $-j$ charged Dirac nodes in the second gap at three $M$ points, which results in the formation of $-1$ charged linear triply-degenerate points [see the fourth panel in Fig.~\ref{jiang}(c)]. Later, the linear triply-degenerate points split again, giving rise to six $\pm i$ ($\pm j$) charged Dirac nodes at $\Gamma M$ ($KM$) lines in the first (second) gap. Finally, the quadratic nodes at $K$ points with charge $-1$ are created, forming the nonzero patch Euler classes of $\xi =\pm 1$. Notably, the linear triply-degenerate points are the unavoidable nodes for the braiding of two charges in adjacent gaps. Such triply-degenerate points are widely studied before and after the birth of the non-Abelian band topology~\cite{zhu2016triple,winkler2019topology,lenggenhager2021triple,park2021topological,feng2021triply,das2022mathbb,lenggenhager2022triple,lenggenhager2022universal}. The multi-gap bulk-boundary correspondences with both zigzag and armchair boundaries were also studied in experiments, which provide an indirect manifestation of the braiding of the band nodes, due to the fact that the Zak phase switches when passing through an odd number of Dirac nodes.

\begin{figure}[htbp]
    \flushright
    \includegraphics[width=0.85\linewidth]{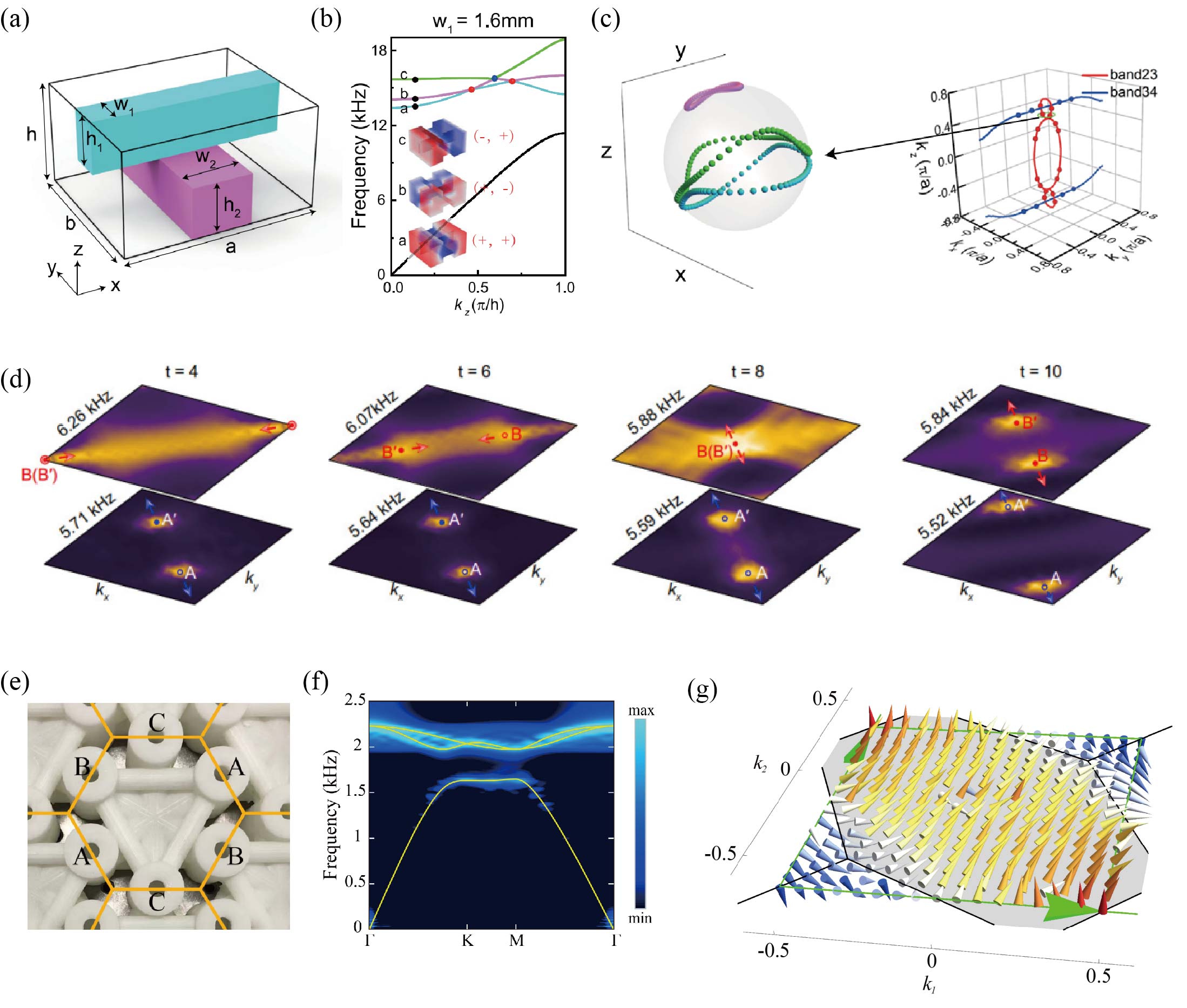}
    \caption{(a) The unit cell of the phononic crystal. (b) The corresponding band structures with three bands entangled together. (c) Non-Abelian charge $-1$ of the nodal link revelaed from the rotation of frame. Figures (a)-(c) adopted from Ref.~\cite{PRL.128.246601Wang}. (d) Experimental observation of non-abelian braiding of band nodes. Figure (d) adopted from Ref.~\cite{qiu2022minimal}. (e) The acoustic kagome lattice with third-next-neighbor couplings. (f) Gapped band structure. (g) Experimentally obtained meronic Skymion texture of the lowest band. Figures (e)-(g) adopted from Ref.~\cite{jiang2022experimental}.}
    \label{jiangliu}
\end{figure}

Soon after, Wang {\it et al.} gave the first experimental observation of the earring nodal links characterized by the non-Abelian quaternion charge of $-1$~\cite{PRL.128.246601Wang}. They fabricated a three-dimensional $PT$-symmetric phononic crystal based on a carefully designed $k\cdot p$ Hamiltonian. The phononic crystal is constructed layer-by-layer along the $z$-direction with a unit cell stacked by two twisted  cuboids [see Fig.~\ref{jiangliu}(a)]. The geometry has two mirror
symmetries and inversion symmetry which offers the opportunity that a nontrivial accidental triple degenerate node can emerge within three bands above the lowest band [see Fig.~\ref{jiangliu}(b)]. The rotation of the 3-eigenstate frame on the unit sphere along
the closed loop encircling the node shows that all three eigenstates rotate $2\pi$ around the $z$ axis [see Fig.~\ref{jiangliu}(c)], which denotes the key feature of the non-Abelian quaterion charge $-1$. By small perturbation, the triple degenerate node can reduce to
two kinds of earring nodal links where their link cannot be fulley gapped, which indicates the topological protection of quaterion charge $-1$. The earring nodal links were verified experimentally in band dispersions at different frequencies by performing three-dimensional Fourier transforms of the scanned fields. Compared to Jiang's work, Qiu {\it et al.} realized the similar nontrivial non-Abelian braiding process of band nodes in acoustic metamaterials based on a square lattice model ~\cite{qiu2022minimal}. However, taking a step forward, they provided the first experimental evidence of the stability of pair nodes of non-Abelian charges [see Fig.~\ref{jiangliu}(d)]. Moreover, the mirror eigenvalues were measured to characterize the  braiding process.

 The non-Abelian band topology is also generated to four-band $PT$-symmetric systems~\cite{NC.12.1jiang}, Floquet systems~\cite{slager2022floquet}, non-Hermitian territory~\cite{PRResearch.3.043006Ezawa,guo2022exceptional,konig2022braid}, nontrivial Euler insulators~\cite{ezawa2021topological,bouhon2022multi,zhao2022quantum,jiang2022experimental,guan2022landau} and so on. Notably, by introducing a nonzero second next-neighbor couplings in the acoustic Kagome lattice [see Fig.~\ref{jiangliu}(e)], the so-called Euler insulator is formed with the gapped lowest band [see Fig.~\ref{jiangliu}(f)] captured by a meronic Skymion texture [see Fig.~\ref{jiangliu}(g)]. Such texture is based on the orientation of real eigenstates in the two-dimensional Brillouin zone~\cite{jiang2022experimental}. Recent work reveals that edge state distributions of higher-order topological phases can be explained from a non-Abelian perspective~\cite{Arxiv.jiang2022edge}. We remark that a further study on non-Abelian band topology in both theory and experiments is deserved, as it is of fundamental importance using a global methodology to understand the novel topological phenomena which cannot be characterized by conventional band topology.

\subsection{Topological defects and bulk-defect correspondences}
\begin{figure}[htbp]
    \flushright
    \includegraphics[width=0.85\linewidth]{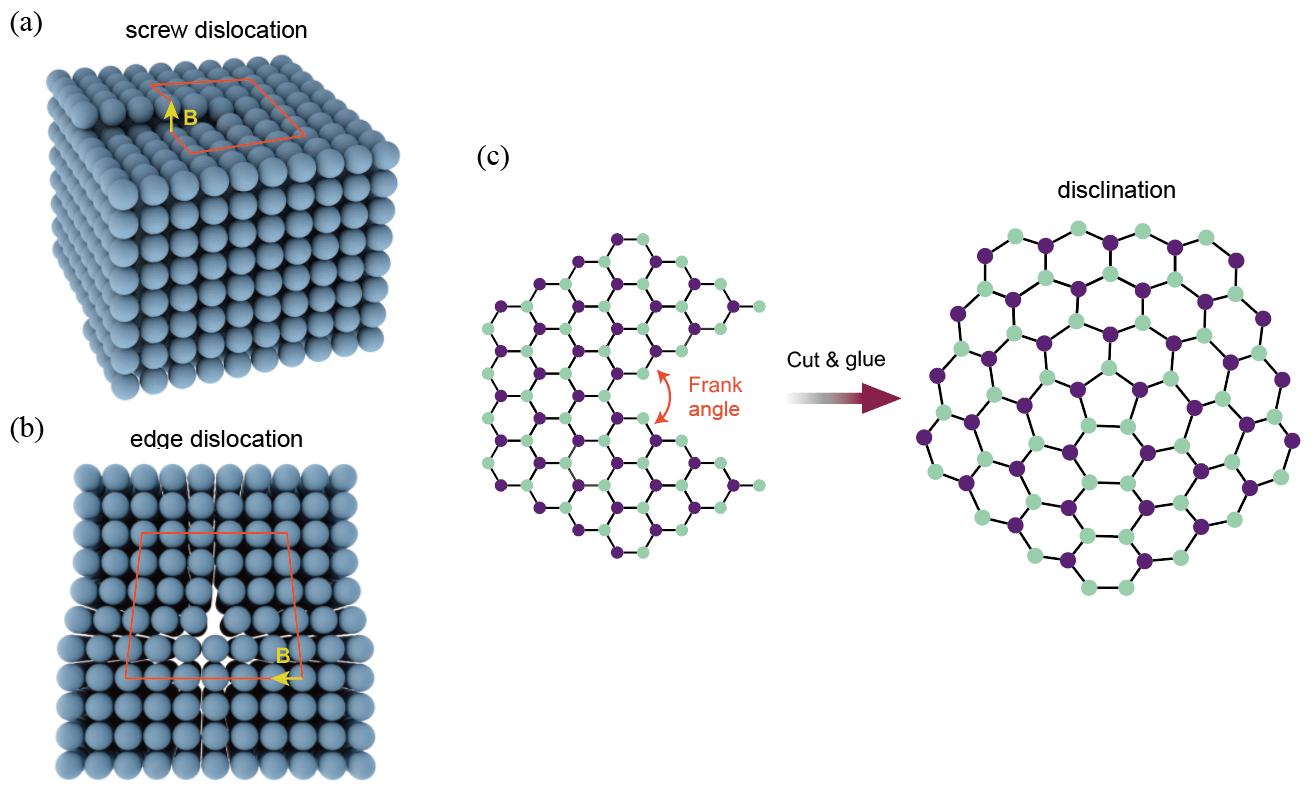}
    \caption{Illustration of a screw dislocation (a), an edge dislocation (b), and a disclination (c), respectively.}
    \label{defect}
\end{figure}

The bulk-boundary correspondence, stating that the nontrivial bulk band topology can be manifested as the existence of topologically protected lower-dimensional boundary states, is at the core of the topological band theory. Unprecedented progresses have been achieved in experimental verification of band topology by virtue of probing spectral signals at boundaries including surfaces, hinges or corners. Aside from boundaries which can be deemed as a sort of crystal defects due to the lattice destruction, there is a variety of defects that are embodied inside of crystals, such as vacancies, substitutions, dislocations and disclinations. They substantially exist in natural nanomaterials and nonequilibrium progresses~\cite{mermin1979topological,RMP.80.61Kleman}. These defects can be regarded as, to some extent, internal “boundaries” inside the bulk. A question is naturally remarked whether the bulk band topology can be revealed by these internal “boundaries”. One positive answer is provided by topological defects~\cite{Teo2010,ran2009one,de2014bulk,teo2017topological,li2020fractional,van2018dislocation,geier2021bulk,soto2020dislocation,kong2022topological,schindler2022topological,lin2022topological}, which include, for example, dislocations [see Figs.~\ref{defect}(a) and~\ref{defect}(b)] and disclinations [see Fig.~\ref{defect}(c)].

Topological defects are robust lattice distortion against local lattice deformation, distinguished from other defects by their stable topological charges such as Burger’s vectors and Frank angles defined in real space [see Fig.~\ref{defect}]. They provide an exceptional methodology to explore the topological band theory and extend the conventional bulk-boundary correspondence to the bulk-defect correspondence. The interplay between the real-space topological charges and the reciprocal-space band topology gives rise to the underlying principles. Such dual topology in both real- and reciprocal-space offers unprecedented opportunities in probing and inducing topological phenomena such as topological defect states and fractional spectral charges. Beside pieces of research in natural materials~\cite{hamasaki2017dislocation,nayak2019resolving}, most topological phenomena triggered by topological defects were demonstrated in, e.g., photonic, phononic, and electrical circuit systems, owing to the access to desirable lattice distortion in metamaterials. We here briefly review the experimental progress of topological phenomena induced by topological defects, particularly in phononic metamaterials. We proceed according to the classification of topological defects.

\begin{figure}[htbp]
    \flushright
    \includegraphics[width=0.75\linewidth]{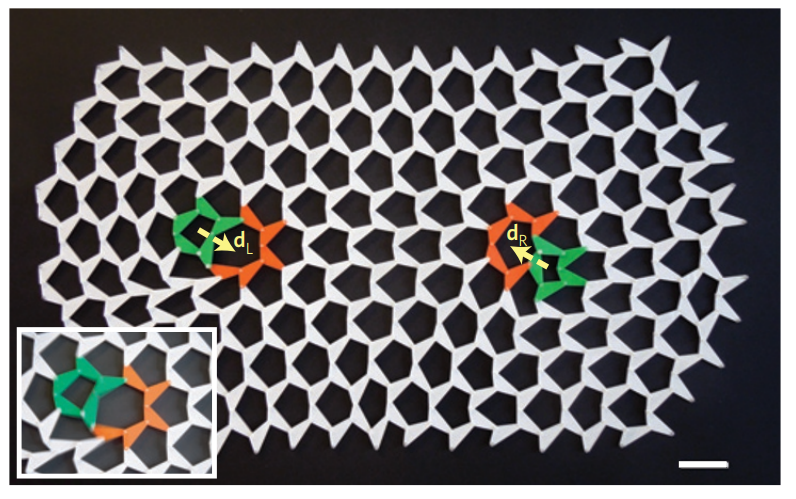}
    \caption{Topological mechanical modes localized at two dislocations in a plastic prototype of a deformed kagome network. Figure adopted from Ref.~\cite{NP.11.153Paulose}}
    \label{mechanical}
\end{figure}

An edge dislocation is characterized by the Burger’s vector, which acts as a zero-dimensional defect in two dimensions. Topological states bound to edge dislocations were earlier observed in mechanical metamaterials based on deformed kagome lattices~\cite{NP.11.153Paulose}, as shown in Fig.~\ref{mechanical}. The topological bound states arise from a delicate interplay between the momentum Berry phase and the Burger’s vertor in real space. Latter, topological zero-dimensional light-trapping on an edge dislocation is directly measured in photonic crystals by Li {\it et al.}~\cite{NC.9.1Li}, based on the topological edge states of Chern insulators. More recently, topological dislocation states were experimentally confirmed in a “weak” topological insulator constructed from stacked one-dimensional Su-Schrieffer-Heeger chains within an anisotropic magneto-mechanical system~\cite{PRApplied.14.064042Grinerg}. The first observation that partial edge dislocations can probe the higher-order band topology in two-dimensional and three-dimensional multipole insulators was implemented in circuit-based resonator arrays ~\cite{NC.13.1Yamada}. Besides, by introducing additional two synthetic translation dimensions, Chen {\it et al.} designed in two-dimensional dielectric photonic crystals a four-dimensional topological insulator phase with a nontrivial second Chern number~\cite{chen2021second}. The zero-dimensional dislocation states in non-synthetic space evolve into chiral gapless one-dimensional dislocation states in four-dimensional synthetic space due to the nontrivial second Chern number therein, which were also validated in experiments, These topological phenomena are yet to be explored in phononic systems.

\begin{figure}[htbp]
    \flushright
    \includegraphics[width=0.85\linewidth]{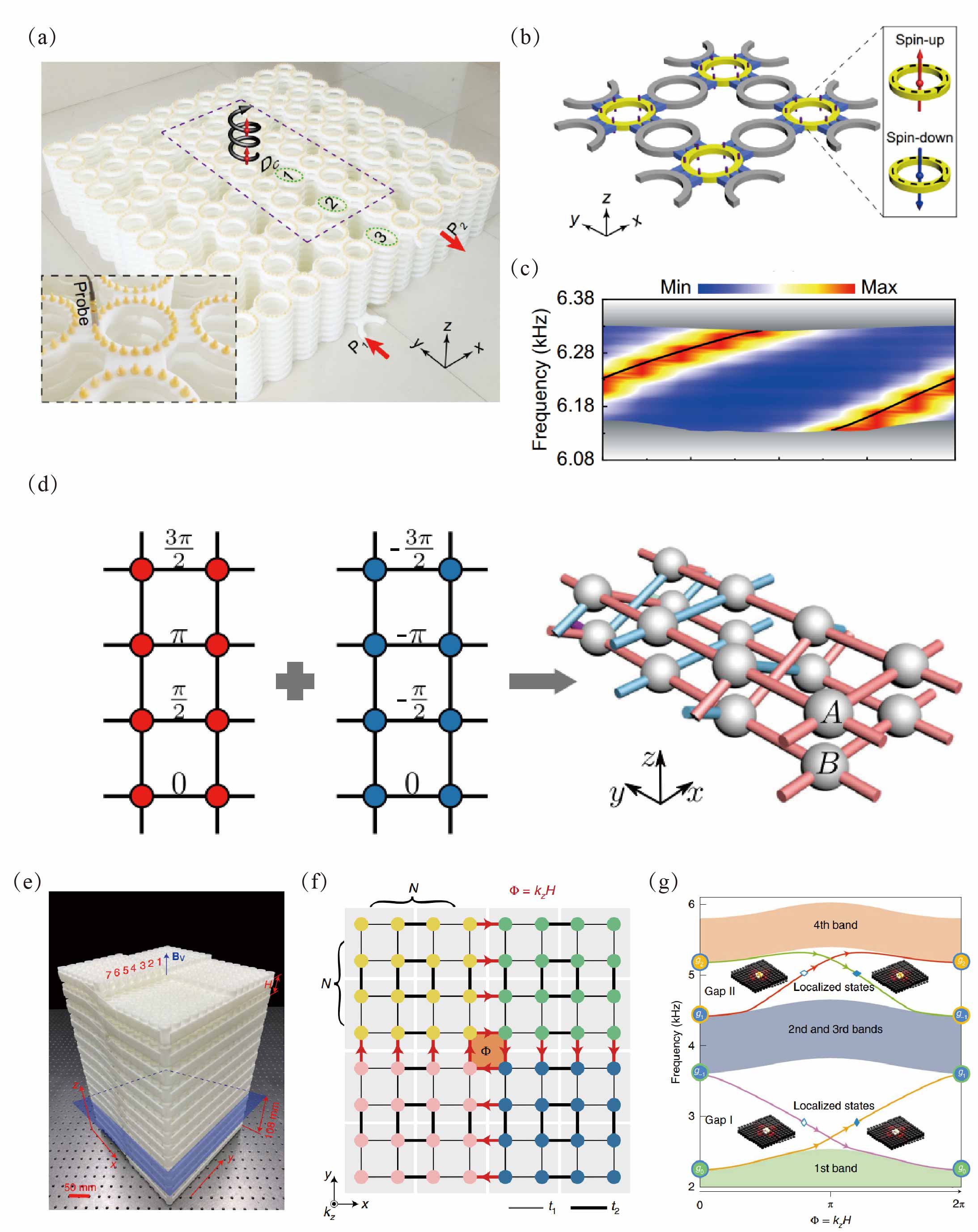}
    \caption{(a) Experimental three-dimensional phononic crystal with a dislocation. (b) Monolayer coupled ring-waveguides structure used for stacking the three-dimensional. The inset shows the pseudospins according to the circulating directions of sound. (c) Experimentally measured dispersion of the one-dimensional dislocation states. Figures (a)-(c) adopted from Ref.~\cite{NC.13.1Ye}. (d) Two square lattices with different magnetic flux per plaquette is stacked to form a two-dimensional topological insulator. Red (blue) colors denote positive (negative) couplings. Figure (d) adopted from Ref.~\cite{PRL.127.214301Xue}. (e) Experimental setup with a step screw dislocation to realize topological Wannier cycles. (f) An effective two-dimensional model after Fourier transform along the $z$-direction, which exhibits a local gauge flux $k_z$ ranging from $0~2\pi$. (g) Acoustic  spectra versus $k_z$ with localized states traveling two band gaps. Figures (e)-(g) adopted from Ref.~\cite{nat.mater.21.430lin}.}
    \label{liulin}
\end{figure}

As a matter of fact, topological defects first come to play in topological materials since Ran {\it et al.} in 2009 made the first prediction that in weak three-dimensional topological insulators~\cite{ran2009one}, the screw dislocation line can act the same role as boundaries to host one-dimensional topological gapless helical boundary states. A screw dislocation characterized by the Burger’s vector is a translation-breaking defect where the lattice planes follow spirally along the dislocation line. The emergence of topological dislocation states is jointly dictated, as follows, by the weak topological invariants $v_i$ of occupied bands and the Burger’s vector $\boldsymbol{B}$ of the dislocation,
\begin{equation}
 \frac{1}{2}\,\boldsymbol{B}\cdot\sum_{i}v_i\boldsymbol{G}_i = \pi\,(mod\,2\pi),
\end{equation}
where $\boldsymbol{G}_i$ are the basis vectors of the reciprocal lattice. The helical spin-polarized dislocation states were recently observed by Ye {\it et al.}~\cite{NC.13.1Ye} and Xue {\it et al.}~\cite{PRL.127.214301Xue}, respectively, in acoustic weak three-dimensional topological insulators implemented by coupled resonators [see Figs.~\ref{liulin}(a)-\ref{liulin}(c) from Ye's work]. In a different manner, the former work is based on stacked layers, while the latter utilizes a mathematical technology of pseudospin transformation [see Fig.~\ref{liulin}(d)]. More recently, a photonic realization of dislocation states with a synthetic dimension has also been achieved in a coupled optical waveguide array~\cite{Nature.609.931Lustig}.

In the meantime, a unique mechanism of the screw dislocation termed topological Wannier cycles that are irrelevant of weak topological invariants were first proposed and experimentally demonstrated by Lin {\it et al.}~\cite{nat.mater.21.430lin}, in an acoustic higher-order topological insulator based on the stacked two-dimensional Su-Schrieffer-Heeger model [see Fig.~\ref{liulin}(e)]. In this work, the screw dislocation plays the role of artificial local gauge flux insertion [see Fig.~\ref{liulin}(f)]. The disconnected four bands composed of Wannier orbitals with different rotation symmetry representations evolve cyclically into each other versus the local gauge flux, which is manifested as spectral flows traveling multiple band gaps and topological boundary states localized at the dislocation [see Fig.~\ref{liulin}(g)]. Topological Wannier cycles provide an alternative way in experiments to probe the higher-order band topology when the spectral signals at corners are unavailable in the absence of chiral symmetry.

Now let us turn to disclinations. Disclinations are zero- and one-dimensional topological defects in two- and three-dimensions, respectively, which, distinct from the breaking of translation symmetry of dislocations, disrupt the rotation symmetry of the pristine lattice by “adding” or “subtracting” a fan-shaped sector and then applying the cut-and-glue process [see Fig.~\ref{defect}(c)]. The real-space topology in disclinations is characterized by the Frank angle. Topological responses induced by disclinations include fractional bound charges~\cite{PRB.101.115115Li}, localized zero-dimensional and one-dimensional boundary states, topological disclination pumping~\cite{PRA.106.L021502Xie}, and so on. Akin to the aforementioned topological phenomena at dislocations, the disclination-induced fractional charges are determined by dual-topology in both real and momentum space as well~\cite{PRB.101.115115Li}. Here, the real-space topology is the Frank angle while the momentum-space topology is the topological invariants of topological crystalline insulators. In experiments, fractional disclination charges were firstly demonstrated by Liu {\it et al.}~\cite{Nature.589.381Liu} and Peterson {\it et al.}~\cite{Nature.589.376Peterson} in two-dimensional photonic crystals and transmission line systems, respectively. The quantized fractional charges were detected by quantitatively measuring the total spectral density-of-states at each unit cell within a frequency range. Fractional charges can in turn inspire the probe of bulk crystalline topology. Wang {\it et al.} validated in experiments that, by introducing a pair of disclinations with opposite Frank angles, internal topological edge states can be acquired in photonic valley Hall systems. The disclination line connecting two disclinations works as an internal boundary that splits the lattice into two topologically distinct regions and yields the robust propagation of topological edge states terminated at two disclinations. Such internal topological edge states may pave the way for free-form topological waveguiding against inevitable fabrication errors. Based on a similar two-disclination photonic system, a fresh topological phenomenon called topological disclination pumping was proposed by Xie {\it et al.}~\cite{PRA.106.L021502Xie}. Finally, we remark that fractional charges also emerge in the Su-Schrieffer-Heeger model at the kinks and edges, together with the topological kink and edge modes, which were recently confirmed in photonic~\cite{arxiv:2203.00206liang} and phononic~\cite{arxiv:2208.11882ge} systems.

\begin{figure}[htbp]
    \flushright
    \includegraphics[width=0.85\linewidth]{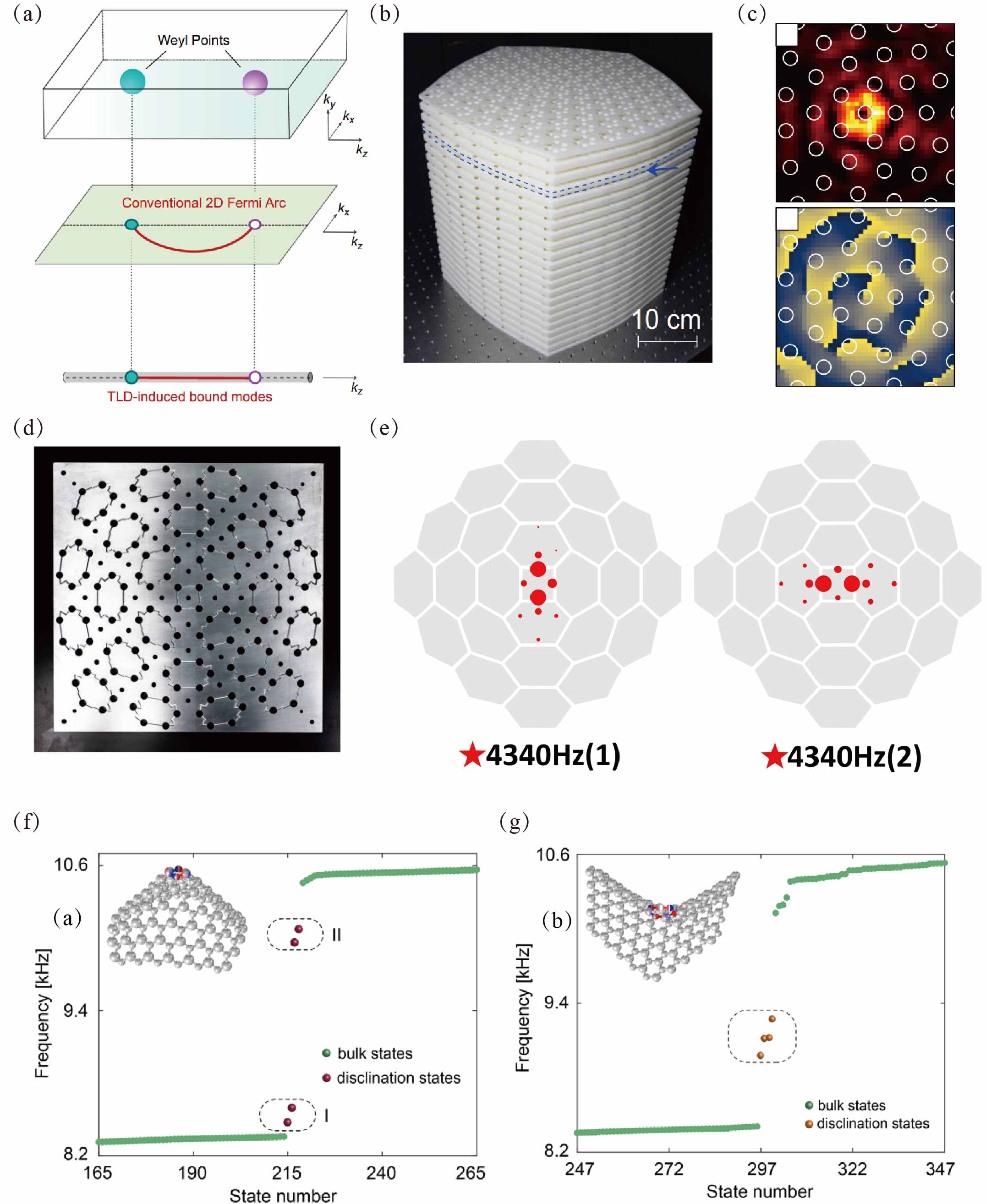}
    \caption{(a) Illustration of bound states in a Weyl semimetal with a disclination. The bound states exist between the projections of the Weyl points in the one-dimensional momentum space where the Chern number is nonzero. (b) Experimental phononic crystal with a disclination. Blue dashes indicate one of the layers. The arrow denotes one gap for acoustic probes. (c) Measured localized disclination modes and corresponding phase distributions. Figures (a)-(c) adopted from Ref.~\cite{NC.12.1Wang}. (d) Experimental chiral-symmetric acoustic lattice with coiled couplings. (e) Measured acoustic pressure distributions of two disclination states. Figures (d) and (e) adopted from Ref.~\cite{PRL.128.174301deng}. (f) and (g) Simulated acoustic eigenfrequencies around the topological band gap in the conic and hyperbolic lattices with a disclination, respectively. Figures (f) and (g) adopted from Ref.~\cite{PRL.129.154301chen}.}
    \label{weylchiralchen}
\end{figure}
Further on, Wang {\it et al.} again designed an acoustic three-dimensional Weyl semimetal with a one-dimensional disclination~\cite{NC.12.1Wang}, as shown in Figs.~\ref{weylchiralchen}(a)-\ref{weylchiralchen}(c). They found that the one-dimensional disclination line states carrying orbital angular momentum emerge only at the $k_z$ range with nonzero Chern numbers. The underlying physics can be interpreted by the interplay between the disclination-induced localized gauge fluxes and the nonzero Chern number. The disclination-induced gauge flux is of non-Abelian nature, indicating that multiple defects can be braided to have different net fluxes which have yet to be explored in metamaterials.

Disclinations always yield non-planar surfaces in nanomaterials such as graphene and fullerenes, due to the minimization of the deformation energy. The effect of nonplanar or non-Euclidean geometries on topological phenomena induced by topological defects is less unexplored. Using the Kamada-Kawai layout algorithm and based on the valley Hall acoustic metamaterials, Chen {\it et al.} designed the disclination-induced conic and hyperbolic surfaces, which respectively exhibit  a positive and negative Gaussian curvature~\cite{PRL.129.154301chen}. Furthermore, they focused on the topological bands induced by $p$-orbitals. The emergent topological bound states arise from the interplay among the real-space topology, the non-Euclidean geometry, the bulk band topology, and the p-orbital physics. The bound states [see Figs.~\ref{weylchiralchen}(f) and~\ref{weylchiralchen}(g)] were confirmed by clear experimental evidence.  Chen's work extends the bulk-disclination correspondence to non-Euclidean geometries and p-orbital physics, which may shed light on topological phenomena for electrons in nanomaterials with  non-Euclidean geometries that are crucial in understanding the properties of these materials.

It is known that phononic metamaterials often break chiral symmetry that pins the topological disclination states at the mid-gap in topological crystalline insulators. Deng {\it et al.}~\cite{PRL.128.174301deng} figured out this issue in experiments by ingeniously designing an acoustic crystal based on the coiled coupled-cavity model that preserves chiral symmetry [see Fig.~\ref{weylchiralchen}(d)]. Wherein, two mid-gap states protected by chiral symmetry localized at the disclination were measured [see Fig.~\ref{weylchiralchen}(e)].

We remark that topological defects, in a broad sense, are comprised not only of dislocations and disclinations, but also of vortices, skyrmions, monopole excitations, and so on. The Dirac vortex, as mentioned in section 2.7, manifesting the vortex texture of the phase of the two-dimensional Dirac mass term, was extensively explored recently in metamaterials, based on honeycomb lattices with Kekul$\acute{e}$-distortion. The Majorana-like zero modes bound at vortices were observed in acoustic~\cite{PRL.123.196601gao}, mechanical~\cite{Adv.mater.31.1904386chen,ma2021nanomechanical}, photonic~\cite{menssen2020photonic,noh2020braiding} systems and were proposed to enhance laser performance~\cite{gao2020dirac,xi2021topological,yang2022topological,ma2021room}.

The rapid progress suggests the rising of a new research frontier in topological materials and physics. We anticipate that with the development of topological phononic  systems, more emergent topological phenomena at topological defects can be discovered and their impacts on applications in phononics can be developed as well.

\subsection{Acoustic topological insulators induced by projected symmetry}
\begin{figure}[htbp]
    \flushright
    \includegraphics[width=0.85\linewidth]{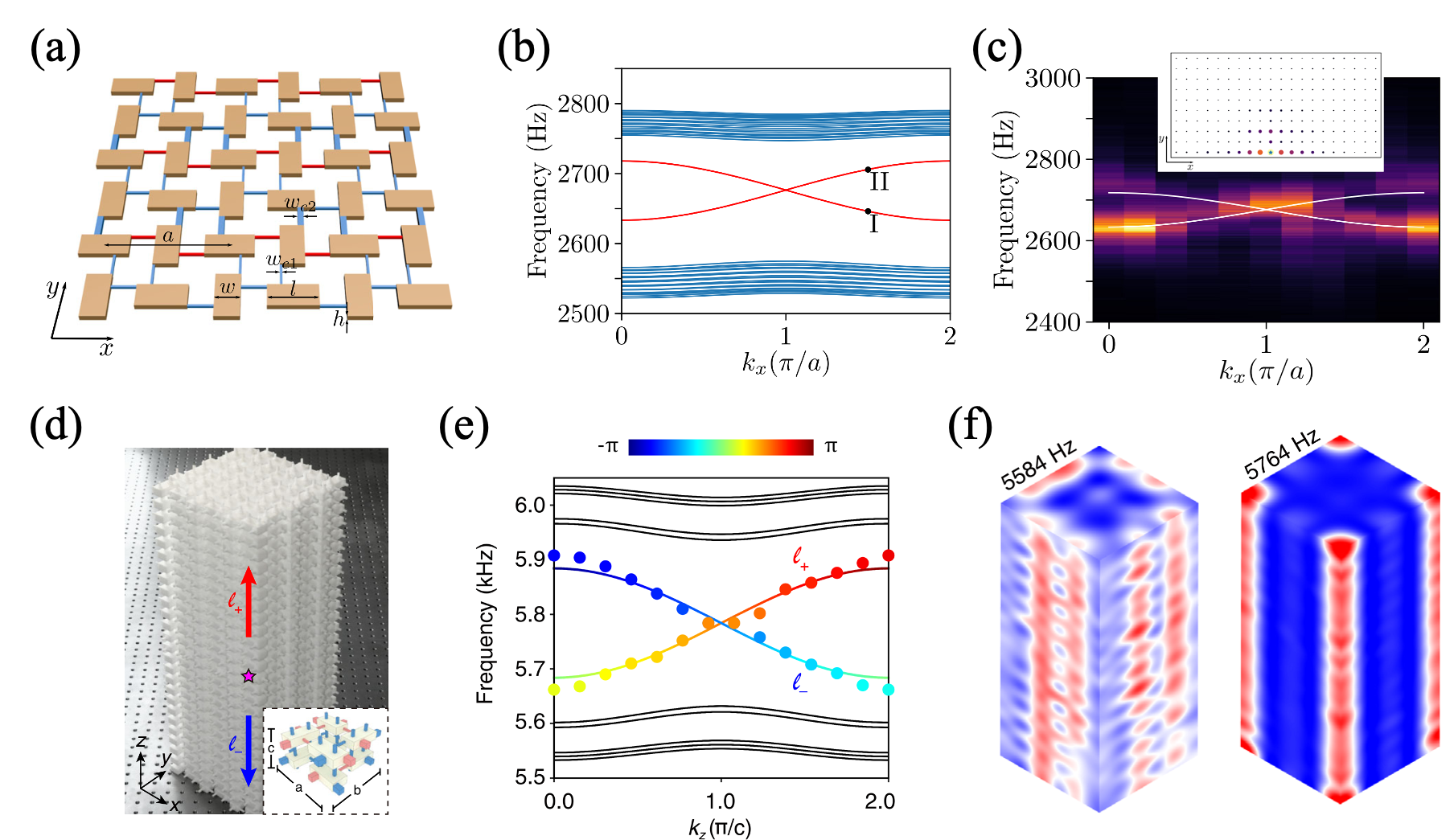}
    \caption{Acoustic Möbius topological insulators induced by projective symmetry. (a) Schematic of two-dimensional acoustic Möbius topological insulator, constructed by the resonators (orange) and tubes (blue and red). (b) Simulated projected dispersions for a ribbon with the open boundaries along the $y$ direction. (c) Measured edge dispersions. The white lines denote the simulated results, corresponding to the red lines in (b). Inset: Measured pressure fields. Both the color and size of the circles represent measured amplitudes. (d) Experimental sample of three-dimensional acoustic higher-order Möbius topological insulator. Inset shows the unit cell. (e) Measured (color dots) and calculated (color lines) hinge dispersions. Color scale represents the phase information of the projective translation eigenvalues. Black lines denote the calculated projected bulk and surface states. (f) Measured pressure fields for the surface (left panel) and hinge states (right panel), respectively. (a)-(c) adopted from Ref.~\cite{phys.rev.lett.128.116802xue}, and (d)-(f) adopted from Ref.~\cite{phys.rev.lett.128.116803li}.}
    \label{weiyin.2.Mobius TI}
\end{figure}

Owning to the gauge symmetry, the algebraic structure of crystal symmetries needs to be projectively represented, leading to new topological phases~\cite{phys.rev.lett.126.196402zhao,phys.rev.lett.127.076401shao,nat.commun.13.2215chen,phys.rev.b102.161117zhao}. A concrete example is the Möbius topological insulator induced by the projective translation symmetry in the presence of a $Z_2$ gauge field~\cite{phys.rev.b102.161117zhao}. Figure~\ref{weiyin.2.Mobius TI}(a) shows the schematic of a two-dimensional acoustic Möbius topological insulator~\cite{phys.rev.lett.128.116802xue}, in which the blue (red) tube has the positive (negative) hopping amplitude, giving rise to a $\pi$ gauge flux in a unit cell. Under this $Z_2$ gauge field, the staggered dimerization of hopping along the $y$ direction opens a bulk gap and generates nontrivial topology described by a Möbius $Z_2$ invariant related with the projective translation symmetry. As shown in Fig.~\ref{weiyin.2.Mobius TI}(b), the edge dispersions have the Möbius twist with a $4\pi$ periodicity, similar to the edge of a Möbius strip, and are completely detached from the bulk dispersions. The experimental evidences are shown in Fig.~\ref{weiyin.2.Mobius TI}(c), including the measured edge dispersions and the related pressure fields. Figure~\ref{weiyin.2.Mobius TI}(d) shows the acoustic sample of a three-dimensional higher-order Möbius topological insulator~\cite{phys.rev.lett.128.116803li}, by stacking the quadrupole topological insulator and keeping the projective translation symmetry along the $z$ direction. The measured hinge dispersions and hinge pressure fields give the smoking gun evidences, as shown in Figs.~\ref{weiyin.2.Mobius TI}(e) and~\ref{weiyin.2.Mobius TI}(f), respectively.

\subsection{Mechanical metamaterials at macroscopic scales}
As in acoustic systems, topological phenomena of phononic waves in mechanical systems~\cite{nat.commun.8.14201rocklin,Addit.Manuf.19.167Al-Ketan,J.Mech.Phys.Solids117.22chen,J.Mech.Phys.Solids122.54chen,PhysRevResearch.1.032047attig,science.367.6473patil,PhysRevB.101.024101wang,PhysRevLett.125.256802sirota} can be understood from the rich geometric phases of classical mechanics (Hannay 1985~\cite{j.phys.a-math.gen.18.221hannay}). Nevertheless, it gives insight when one examines the equation of motion in certain prototype topological mechanical systems.

We now present the formal consistency between the equation of motion for a generalized mass-spring model and the Schrödinger equation to see the mapping between classical systems and quantum systems. According to Ref.~\cite{nat.phys.12.621huber}, the equation of motion for a generalized mass-spring model is

\begin{equation}
   {\ddot{x}_i=-D_{ij} x_j +A_{ij} \dot{x}_j},
  \label{motion equation of mass-spring model}
\end{equation}
which follows Einstein’s summation convention. Here, $x_i (i\in N)$ represents one of the $N$ independent displacements and $\dot{x_i}$ is its time derivative. The mass term is merged into matrices $D_{ij}$ and $A_{ij}$. The real and symmetric positive-definite dynamical matrix $D_{ij}$ denotes the springs couplings, and the skew-symmetric matrix $A_{ij}$~\cite{proc.natl.acad.sci.u.s.a.113.E4767susstrunk} describes the non-dissipative coupling arising from velocity-dependent forces. The nonzero $A_{ij}$ only emerges in materials with nonreciprocal elements and plays a role similar to the Lorentz force of charged particles under a magnetic field. Symmetries about matrices $D_{ij}$ and $A_{ij}$ make the spring net a conservative (non-dissipative) system. Such mechanical systems are governed by Newton’s second law which is of second-order derivative in time. To bridge the classical mechanical systems to quantum systems (although, Eq.~\ref{motion equation of mass-spring model} is discrete and not enough to describe the continues mechanical systems), one can rewrite Eq.~\ref{motion equation of mass-spring model} in a similar form of Schrödinger equation as
\begin{equation}
  {i\frac{\partial}{\partial t}\left( \begin{matrix} \sqrt{D}^T{x} \\ i\dot{x}\end{matrix}\right)=\left( \begin{matrix} 0 & \sqrt{D}^T \\ \sqrt{D} & iA\end{matrix}\right)\left( \begin{matrix} \sqrt{D}^T{x} \\ i\dot{x}\end{matrix}\right)},
  \label{schrodinger equation of mass-spring model}
\end{equation}
which is of first-order derivative in time. Here, the positive square root of matrix $D$ can be obtained from its spectral decomposition. The Hamiltonian $H={\left( \begin{matrix} 0 & \sqrt{D}^T \\ \sqrt{D} & iA\end{matrix}\right)}$ is Hermitian due to the positive-defined matrix $D$ and skew-symmetric nature of matrix $A$. The vector $\psi=\left( \begin{matrix} \sqrt{D}^T{x} \\ i\dot{x}\end{matrix}\right)$ plays a role similar to the wavefunction $\psi$ in the Schrödinger equation.

The formulation of Eq.~\ref{schrodinger equation of mass-spring model} which is akin to tight-binding problems in quantum mechanics, drives us directly to implement topological theories on mechanical metamaterials. Although topological properties don’t rely on translation symmetry, topological theories are well presented in lattice systems where the Hamiltonian and eigenvectors take the Bloch forms in the reciprocal space: $H(k)$ and $\psi (k)$. From these forms, topological properties of mechanical systems can be well understood via topological band theories~\cite{elsevier6.3kane,rev.mod.phys.88.021004bansil,nature547.298bradlyn,annu.rev.condens.matter.phys.12.225cano}.

Before reviewing specific systems and models, we first discuss the topological classification of mechanical systems. In electronic systems, the classification of topological states is determined by the spatial dimension and symmetries, such as the time-reversal symmetry $T$, the particle-hole symmetry $C$ and the chiral symmetry $S$. In mechanical systems, these symmetries can be regarded as constraints on the Bloch Hamiltonians $H(k)$ as shown in Eqs.~\ref{time reversal symmtery},\ref{particle-hole symmtery},\ref{chiral symmtery}. Notably, the time-reversal symmetry $T$ has $T^2=-1$  due to the spin-1/2 property of electrons. However, in mechanical systems, $T^2=1$, although it can be augmented to $T^{*2}=-1$ when combined with an appropriate symmetry.

Mechanical system with $T$ symmetry follows
\begin{equation}
  {U_T H(k) - H(-k) U_T =0,}
  \label{time reversal symmtery}
\end{equation}
with some antiunitary $U_T$ and $U_T^2=1$, which represents $T$ instead of “time-reversal”, because in mechanical system it doesn’t correspond to the reversal of time. Besides, the mechanical system with particle-hole symmetry $C$ obeys the constraint that
\begin{equation}
  {U_C H(k) + H(-k) U_C =0,}
  \label{particle-hole symmtery}
\end{equation}
and $U_C^2=1$ for antiunitary $C$. Moreover, the chiral symmetry $S$ demands that
\begin{equation}
  {U_S H(k) + H(k) U_S =0,}
  \label{chiral symmtery}
\end{equation}
with a unitary $U_S$ and $U_S^2=1$. The band structures in the presence of $T$,$C$,$S$ symmetries are shown in Figs.~\ref{T,C,S}(a)-\ref{T,C,S}(c), respectively.

\begin{figure}[htbp]
\flushright
    \includegraphics[width=0.85\linewidth]{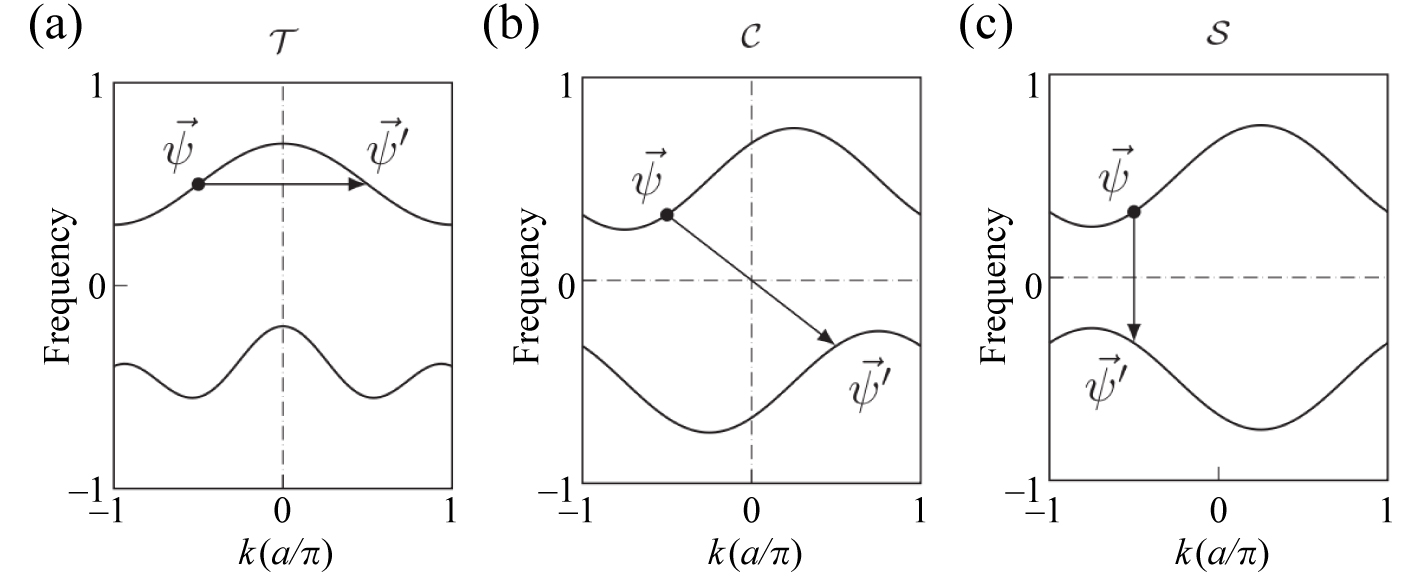}
    \caption{Visualization of $T$, $C$, $S$ symmetry on band structure. Figures adopted from Ref.~\cite{proc.natl.acad.sci.u.s.a.113.E4767susstrunk}}
    \label{T,C,S}
\end{figure}

\begin{figure}[htbp]
\flushright
    \includegraphics[width=1.0\linewidth]{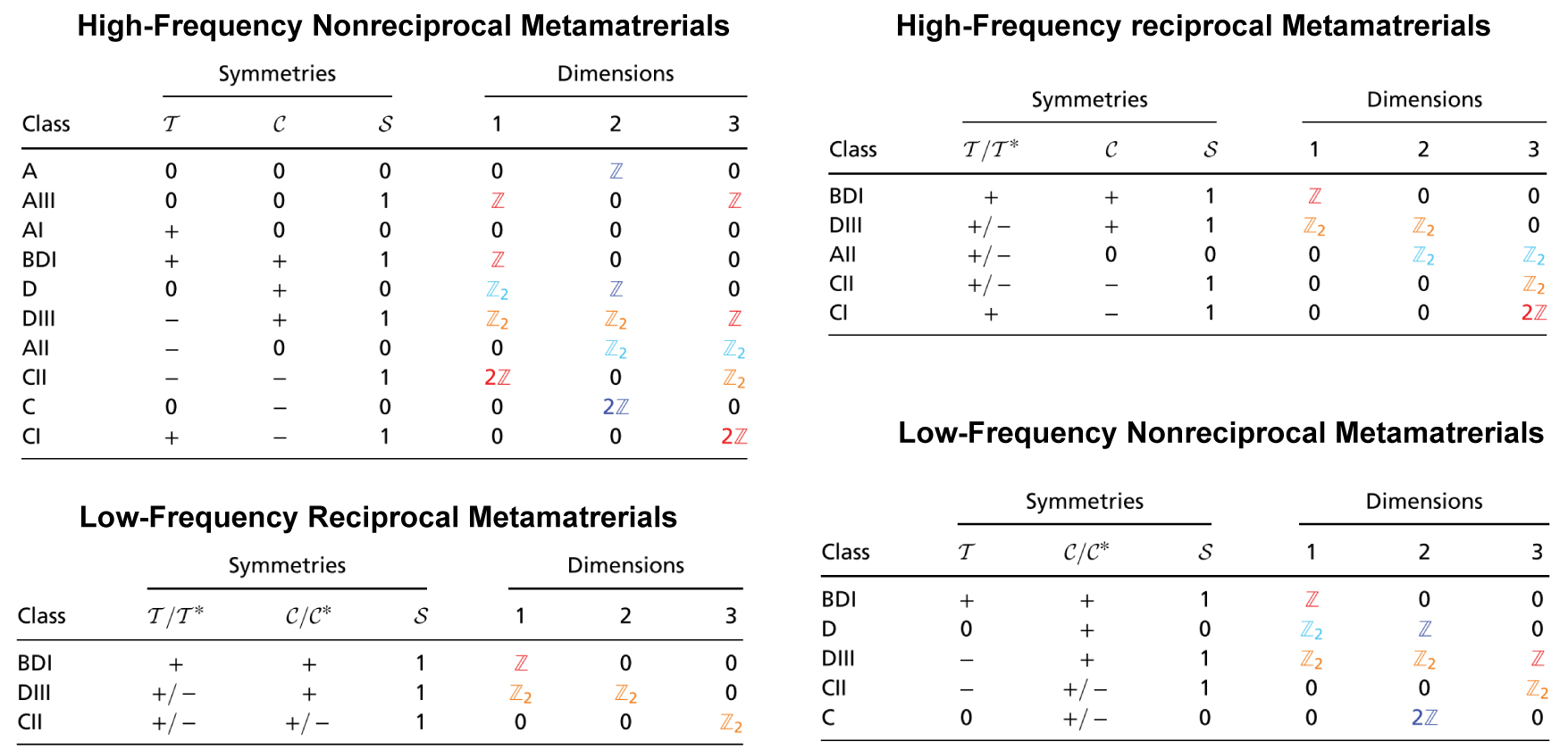}
    \caption{Topological classification of mechanical systems. The first column of one table gives the standard names of the symmetry groups. The next three columns indicate the absence $(0)$ or presence $(1, +, -)$ for a given symmetry. The $+/-$ signs the value $(+1/-1)$ of the square of symmetry, respectively. The last columns present the categories of topological numbers characterizing band structure in given dimensions. $\mathbb{Z}$ denotes an integer number; $2\mathbb{Z}$ indicates an even number; $\mathbb{Z}_2$ represents a binary index. Chern numbers in even dimensions are marked by blue color. Winding numbers in odd dimensions are labeled in red color. Additional indices can be derived from primary ones when more symmetries are present. They are marked in light blue and light red for descendants of the Chern numbers and winding numbers. Tables adopted from Ref.~\cite{proc.natl.acad.sci.u.s.a.113.E4767susstrunk}}
    \label{Topological classification of mechanical systems}
\end{figure}

Via above discussion, we have environments to implement the topological classification of mechanical systems as similar to electronic systems. Due to the formulation of topological indices is different when our target gap is located around a zero-energy line (related to thermodynamic or ground state properties) or finite frequencies, and the existence or absence of non-reciprocal elements is a significant distinction of mechanical systems, we can divide the topological classification problem into four different classes: low/high-frequency and reciprocal/non-reciprocal. The specific classification progress refers to Huber’s work~\cite{proc.natl.acad.sci.u.s.a.113.E4767susstrunk} (Prodan also provides a similar classification for classical passive metamaterials in his work~\cite{phys.rev.b98.094310barlas}), and we only show these four tables in Fig.~\ref{Topological classification of mechanical systems}, which associate the classes, symmetries, and corresponding topological indices of band structures.

Based on the different properties of low and high-frequency mechanical materials, we discuss some topological phenomena in low and high-frequency situations, respectively. For the low-frequency range of metamaterials, the topological phenomena emerging are associated with particle-hole symmetry in most cases. The intrinsic particle-hole symmetry $C$ from mass-spring models demands that spectrum is symmetric about its center point: $\left(k=0,f=0\right)$. In this case, topological edge modes arise in the gap near zero energy, which provides an excellent platform to study thermodynamic problems of bosons in metamaterials. Of course, in time reversal systems, chiral symmetry $S$ plays a similar role.

One of the most influencing work on phononic zero-energy edge modes is the work by Kane and Lubensky~\cite{nat.phys.10.39.kane}. There, they study the topological boundary states in a Maxwell frame~\cite{Phys.Rev.Lett.116.135503rocklin,Phys.RevX.6.041029meeussen,Annu.Rev.Condens.MatterPhys.9.413mao} consisting of mass points connected by rigid bonds or central-force springs. In such a system, the difference between the total number of motional freedom degrees of lattice and the number of bonds determines the number of zero-frequency modes~\cite{philosophicalmagazine27.294maxwell}. Kane and Lubensky focus on isostatic Maxwell frames, where these two numbers are balanced (zero-energy modes should not exist, but a more general Maxwell relation predicts the presence of zero-energy modes~\cite{int.j.solidsstruct.14.161calladine}) and the system is on the verge of mechanical collapse. Interestingly, in recent work, an isostatic twisted kagome lattice exhibits zero modes by rotating adjacent site-sharing triangles in opposite directions~\cite{proc.natl.acad.sci.u.s.a.109.12369sun}, which exceeds the traditional Maxwell’s count and is described by the general Maxwell’s relation---relevant for self-stress states, where springs can be under tension or compression with no net forces on the masses.

\begin{figure}[htbp]
\flushright
    \includegraphics[width=0.9\linewidth]{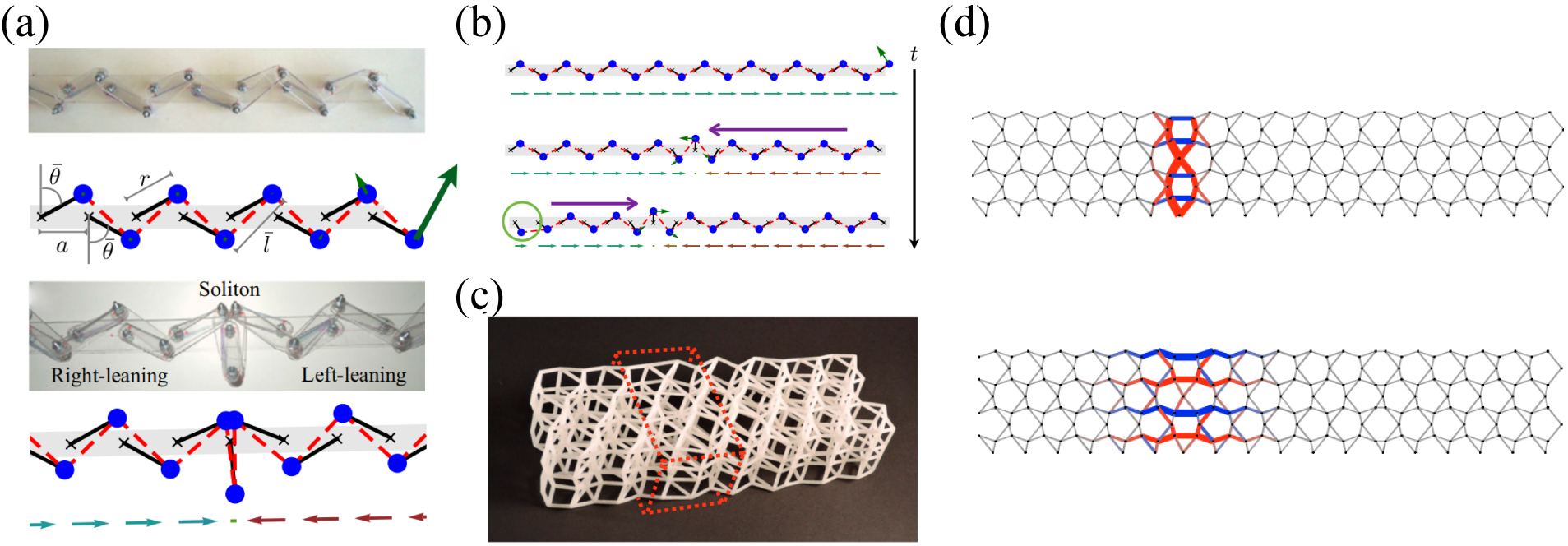}
    \caption{Floppy modes in rotors chain and self-stress modes in Maxwell frame. (a) Upper two panels give the configuration and sketch of the rotors chain with constant $\theta=\bar{\theta}$, respectively. The green arrows depict the displacement amplitude of each mass (blue circle) of the edge-localized zero mode of system. And the last panels present the case where a soliton state occurs between right-leaning and left-leaning states. The below arrows denote the $x$ projection of each rotor. (b) Evolution of the soliton state. The soliton state always localized at the domain wall moves freely in the system. (c) Maxwell frame with two domain walls. (d) Two field patterns of self-stress states in left domain wall under different structure parameters. Figures adopted from: (a) and (b), Ref.~\cite{proc.natl.acad.sci.u.s.a.111.13004chen}; (c) and (d), Ref~\cite{proc.natl.acad.sci.u.s.a.112.7639paulose}}
    \label{floppy+selfstress}
\end{figure}

For such isostatic Maxwell frames, the square root of dynamical matrix $D$ corresponds to the equilibrium matrix of a mass-spring model, where $\sqrt{D}$ connects the spring tension and displacements of the attached masses. Based on this equilibrium matrix, Kane and Lubensky propose a winding number to characterize the number of localized topological modes~\cite{nat.phys.10.39.kane,rep.prog.phys.78.073901lubensky}. For these critical Maxwell frames, a positive winding number predicts the floppy modes---parts of the system move freely. On the other hand, the dual partners of floppy modes are self-stress modes, where the external load on the material can be absorbed, predicted by a negative winding number. Both of these topological boundary modes have been confirmed in experiments. The floppy modes localized at the boundary have been verified by Vitelli in experiment, through a one-dimensional Maxwell frame mapped from Su-Schrieffer-Heeger  model~\cite{proc.natl.acad.sci.u.s.a.111.13004chen} as shown in Figs.~\ref{floppy+selfstress}(a) and ~\ref{floppy+selfstress}(b). Furthermore, they show that, these floppy modes can be transported through the bulk of system by the nonlinear mechanism. In another experiment, a self-stress mode bound on a defect is confirmed through a quasi-two-dimensional frame~\cite{proc.natl.acad.sci.u.s.a.112.7639paulose} as shown in Figs.~\ref{floppy+selfstress}(c) and ~\ref{floppy+selfstress}(d). Above discussions about low-frequency topology are appealing in both physics and applications. In physics, the low-frequency topology are relevant for thermodynamic and ground-state properties. In applications, such floppy modes and self-stress modes are useful in engineering some mechanical response instruments.

\begin{figure}[htbp]
\flushright
    \includegraphics[width=0.9\linewidth]{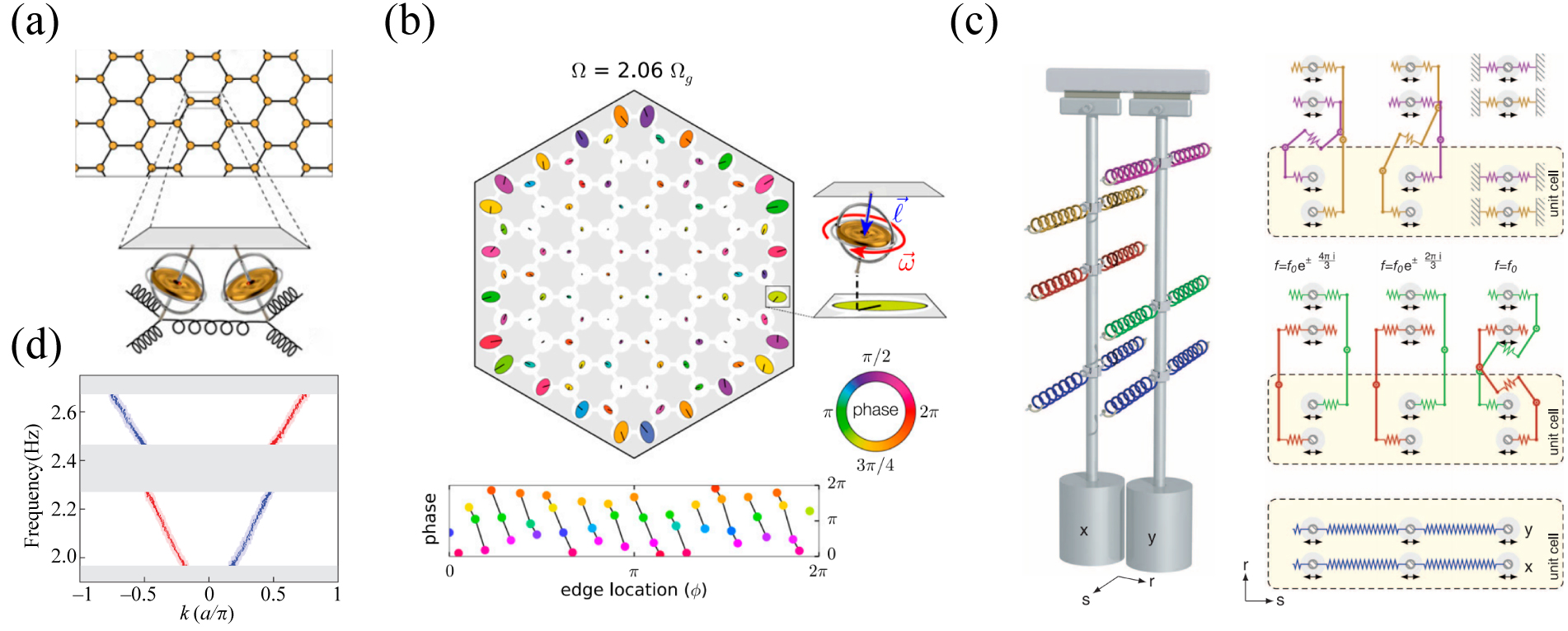}
    \caption{Quantum Hall effect and Quantum spin Hall effect in mechanical systems. (a) A honeycomb lattice with gyroscopes as sublattices coupled through effective springs. (b) The characterization of gyroscopic chiral edge mode, including the configuration of orbits labeled by ellipses and the phase vs. position of quantum Hall edge mode. (c) Schematic diagram of two one-dimensional pendula with six springs divided into three floors. Right three panels present the couplings in three layers, respectively. (d) Dispersion of helical edge states. Red and blue curves denote different polarization establishing the helical properties of quantum spin Hall edge modes. Figures adopted from: (a) and (b), Ref~\cite{proc.natl.acad.sci.u.s.a.112.14495nash}; (c) and (d), Ref~\cite{science.349.47susstrunk}}
    \label{quantum spin Hall effect+quantum Hall effect}
\end{figure}

For the high-frequency topological phenomena in mechanical metamaterials, we mainly discuss two cases with broken and preserved the time-reversal symmetry, respectively. These two cases answer the question, whether the quantum Hall effect and quantum spin Hall effect can be realized in mechanical systems. quantum Hall effect and quantum spin Hall effect were first studied in the electronic systems, where the bulk of systems has an energy gap, however their edges can be excited continuously across the gap. The difference between quantum Hall effect and quantum spin Hall effect is that the former needs to break the time-reversal symmetry and the other is protected by it. If time reversal symmetry is broken, quantum spin Hall effect will not exist, but quantum Hall effect will be supported. However, there is still a problem on implementing quantum Hall effect in mechanical systems. The dynamics of mechanical systems governed by Newton’s second law is invariant under time reversal, due to they are second order in time. Recently, a superb work solves this problem. They construct a coupled system of gyroscopes with intrinsic time-reversal symmetry breaking~\cite{Phys.Rev.Lett.115.104302wang, proc.natl.acad.sci.u.s.a.112.14495nash,Phys.RevB.98.174301mitchell} as shown in Fig.~\ref{quantum spin Hall effect+quantum Hall effect}(a). In the fast-spinning limit, they obtain a procession response of the gyroscope’s axis, which is first order in time. Using this mechanism, they organize an experiment to realize the quantum Hall effect. As a result, they overcome the inherent disorder of the gyroscopes system and successfully observe the unidirectional quantum Hall effect edge states described by Chern number~\cite{nat.phys.15.352wen} (Fig.~\ref{quantum spin Hall effect+quantum Hall effect}(b)), despite the bulk modes of the system is chaotic. This work opens a new method to construct time-reversal broken mechanical metamaterials, which may also inspire the construction of metamaterials supporting such robust topological modes in other systems.

The quantum spin Hall effect is featured with two counterpropagating edge modes that differ by the system’s spin degree of freedom. However, in spinless mechanical systems, it is difficult to build the spin degree of freedom to realize quantum spin Hall effect. Fortunately, recent works provides some methods~\cite{science.349.47susstrunk, J.Appl.Phys.119.084305pal}. They use a special pendula structure as shown in Fig.~\ref{quantum spin Hall effect+quantum Hall effect}(c) to realize a topological lattice with flux $\Phi=\pm 2\pi/3$ per plaquette, which is similar to Hofstadter model. And they replace the spin up and spin down in electronic systems with the left and right circularly polarized motions in their system. Under these conditions, they successfully observe the phononic helical edge states in a mechanical system as shown in Fig.~\ref{quantum spin Hall effect+quantum Hall effect}(d). These quantum spin Hall edge states with opposite propagating directions can’t scatter into another while time-reversal symmetry is preserved, which have potential application prospects in surface acoustic signal transmission.

Reviewing these discussions about mechanical systems, it’s not hard to find that, even though we want to discuss the linear properties of systems, there are still some inevitable nonlinear elements involved, such as self-stress. The topological classification we discuss is based on linear and non-dissipative mechanical metamaterials, whereas the nonlinearity and dissipation are very trivial in mechanical systems. On one hand, nonlinearity and dissipation are challenges for linear topology, and on the other hand, they are strong tools to enrich the topology physics of mechanical systems~\cite{proc.natl.acad.sci.u.s.a.111.13004chen,phys.rev.lett.99.084301mullin,proc.natl.acad.sci.u.s.a.109.5978shim,phys.rev.lett.115.044301coulais,proc.natl.acad.sci.u.s.a.109.5978ghatak}. For instance, establishing dissipative and non-reciprocal couplings is a good method to reach the realm of non-Hermitian topology~\cite{Proc.NatlAcad.Sci.USA.117.29561ghatak}. And nonlinearity~\cite{PhysRevLett.127.076802lo} is an excellent additive that interacts with topological phenomena, which are easily implemented on flexible mechanical~\cite{PhysRevLett.116.135501chen,nat.rev.mater.2.17066bertoldi,j.appl.phys.130.040901deng} materials with mechanical instabilities and large deformations, such as buckling-based metamaterials, frustrated and programmable metamaterials. Moreover, some recent works focusing on topological defects~\cite{nat.phys.11.153paulose,nat.mater.21.430lin} and non-Abelian~\cite{nat.phys.17.1239jiang} topology reveal more interesting topological phenomena.

\subsection{Micron- and nano-scale mechanical metamaterials}

\begin{figure}[htbp]
    \flushright
    \includegraphics[width=0.85\linewidth]{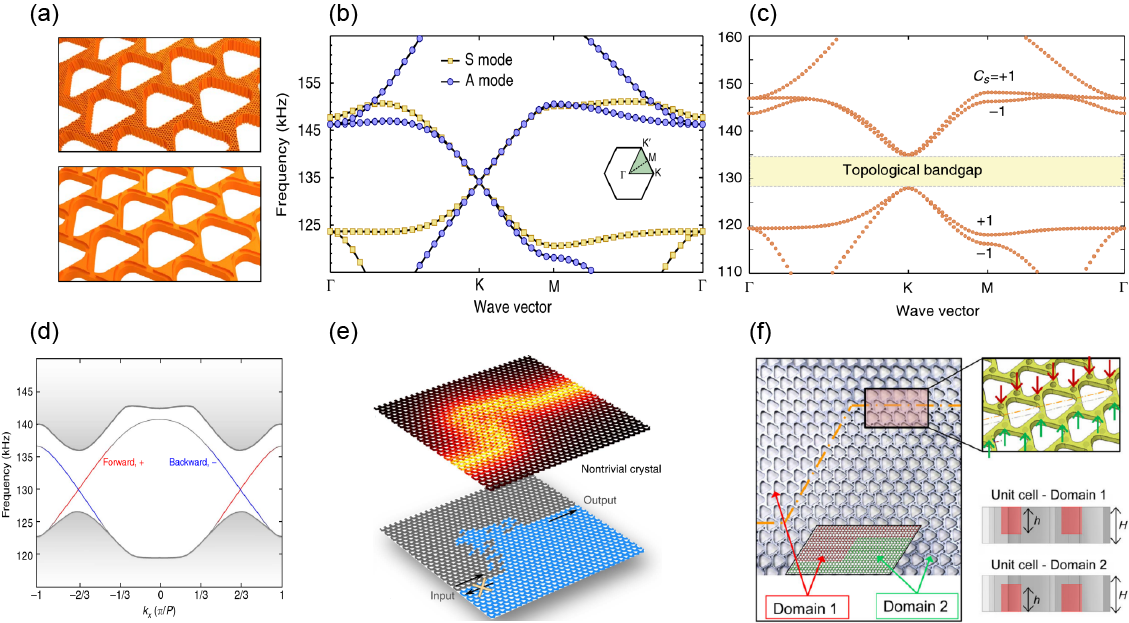}
    \caption{Topological quantum spin Hall effect in mechanical metamaterials. (a) Top (bottom) panel: the elastic membrane with (without) $z$-mirror symmetry. (b) Bulk band structure along the high symmetric lines for the membrane with $z$-mirror symmetry. A four-fold accidental degeneracy appears at the $K$ point. (c) Bulk band structure along the high symmetric lines for the membrane with broken $z$-mirror symmetry. Two two-fold accidental degeneracy appear at the $K$ point. (d) Projected band structure of a domain wall constructed by two elastic phononic crystals with opposite spin Chern numbers. (e) Robustness of the edge state along a one-dimensional boundary with random potentials. Figures (a)-(e) adopted from Ref.~\cite{nat.commun.6.8682mousavi}. (f) Experimental sample for the realization of the quantum spin Hall effect in mechanical metamaterials. Figure (f) adopted from Ref.~\cite{physrevx.8.031074miniaci}.}
    \label{xueqin.1.quantum spin Hall effect}
\end{figure}

Inspired by the topological states in electronic and photonic systems, as well as the diverse potential applications of the micron- and nano-scale mechanical metamaterials, the concept of topological physics has been extended to elastic wave systems. However, the existence of the three polarizations of elastic wave in solid structures challenges the realization. Mousavi {\sl et al.} proposed an ingenious design to exhibit the quantum spin Hall effect in mechanical metamaterials for the first time~\cite{nat.commun.6.8682mousavi}. It is known that the symmetric and anti-symmetric states usually couple with each other in a solid membrane. Mirror symmetry of the membrane makes these two states decouple, thus they can be viewed as two pseudospin states. By a deliberate design, the symmetry and anti-symmetry states are accidental degeneracy at the $K$ (or $K^\prime$) point. The patterned membrane is schematically demonstrated in the top panel of Fig.~\ref{xueqin.1.quantum spin Hall effect}(a). The associated bulk band structure along the high symmetry lines is calculated in Fig.~\ref{xueqin.1.quantum spin Hall effect}(b), where a four-fold accidental degeneracy appears at the $K$ point. Pseudospins along are not sufficient to realize the quantum spin Hall effect. Spin-orbit coupling is an essential ingredient. The breaking of the mirror symmetry in the modified membrane, as plotted in the bottom of Fig.~\ref{xueqin.1.quantum spin Hall effect}(a), plays a role of generating the coupling between the symmetric and anti-symmetric states, thus gives rise to the spin-orbit coupling. The four-fold degeneracy now reduces to two two-fold degenerate states, accompanying with a complete band gap, as shown in Fig.~\ref{xueqin.1.quantum spin Hall effect}(c). The spin Chern numbers of the two bands below the bulk gap are ±1, respectively. According to the bulk-boundary correspondence, the non-zero spin Chern number guarantees the existence of topological edge states. A domain wall formed by two phononic crystals with opposite values of spin Chern numbers is constructed. Figure~\ref{xueqin.1.quantum spin Hall effect}(d) illustrates the corresponding projected band structure. It is seen that two pairs of helical edge states present in the bulk gap. To demonstrate the robustness of the edge states, wave propagation along a domain wall with random potentials is simulated in Fig.~\ref{xueqin.1.quantum spin Hall effect}(e). Obviously, the edge state transports freely without any reflection. After this theoretical proposal, in 2018, Miniaci {\sl et al.} performed the experiment to verify the quantum spin Hall effect and the topologically protected edge states in elastic plates~\cite{physrevx.8.031074miniaci}. The experimental configuration is plotted in Fig.~\ref{xueqin.1.quantum spin Hall effect}(f).

\begin{figure}[htbp]
    \flushright
    \includegraphics[width=0.85\linewidth]{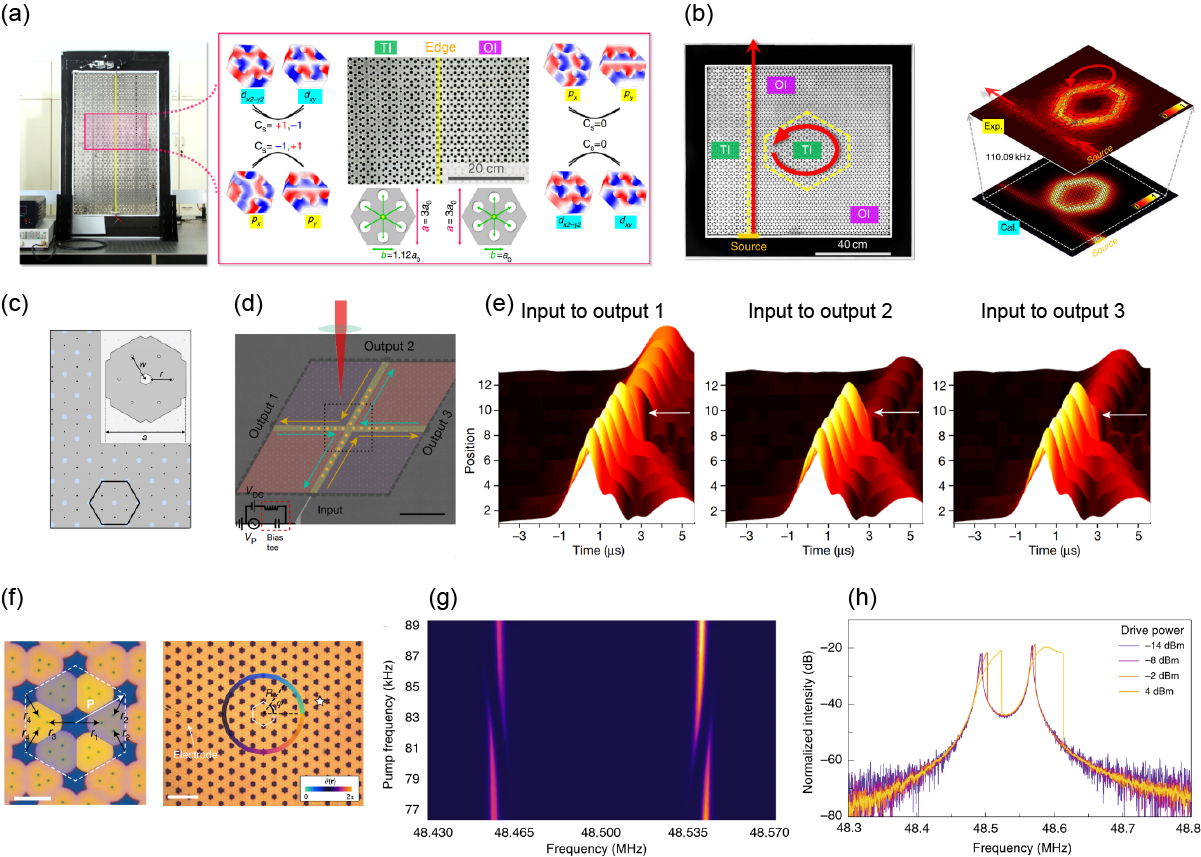}
    \caption{Elastic quantum spin Hall effect realized by the band inversion of the $p$ and $d$ orbital states on the millimeter ((a), (b)) and nano ((c)-(h)) scales. Figures (a)-(b) adopted from Ref.~\cite{nat.commun.9.3072yu}. Figures (c)-(e) adopted from Ref.~\cite{nature.564.229.cha}. Figures (f)-(h) adopted from Ref.~\cite{nat.nanotechnol.16.576.ma}.}
    \label{xueqin.2.valley}
\end{figure}

Beside the previous scheme to achieve quantum spin Hall effect in mechanical metamaterials, another approach is enlightened by that in photonic crystals~\cite{phys.rev.lett.114.223901wu}. $C_6$ symmetry protects two two-dimensional irreducible representations at the $\Gamma$ point. Considering the lowest states, these two two-dimensional irreducible representations correspond to the p and d orbital states. The combinations of these two orbital states, i.e., $p_x \pm ip_y$ and $d_{x^2-y^2} \pm id_{xy}$, function as the pseudospins. Time reversal symmetry for pseudospin-1/2 states, satisfying $T^2=-1$, is constructed by the composition of time reversal symmetry of boson and $C_6$ symmetry. Through tuning the structural parameters, the band inversion of the $p$ and $d$ orbital states arises, leading to a phase transition between the trivial and topological phase~\cite{nat.commun.9.3072yu,nature.564.229.cha,phys.rev.appl.16.044008.zhang,nat.nanotechnol.16.576.ma}. A sample fabricated by drilling perforated holes in a metallic plate, is illustrated in the left panel of Fig.~\ref{xueqin.2.valley}(a). The yellow line represents the domain wall between the trivial and topological phononic crystals. For the left phononic crystals, the frequency of the $p$ orbital states is lower than those of the $d$ orbital ones. The spin Chern numbers of the two bands below the bulk gap are $\pm 1$. While for the right phononic crystals, the frequencies of the $p$ and $d$ orbital states reverse. The spin Chern numbers become zero. Therefore, there should exist topological edge states at the boundary of these two phononic crystals~\cite{nat.commun.9.3072yu}. Take advantage of the robustness of the topological edge states, a monolithic integrated whisper-gallery resonator, consisting of a straight waveguide and a hexagon-shaped one, is designed as shown in the left panel of Fig.~\ref{xueqin.2.valley}(b). The simulated and measured field distributions are illustrated in the right panel of Fig.~\ref{xueqin.2.valley}(b). The input wave efficiently couples to the whisper-gallery resonator and then turns back to the straight waveguide with total transmission.

For the demand of the reliability of the high-frequency mechanical systems, the on-chip topological mechanical metamaterials have been implemented. In Ref.~\cite{nature.564.229.cha}, the authors fabricated a periodic array of free-standing SiN nanomembranes, as shown in Fig.~\ref{xueqin.2.valley}(c). Compared with the working frequency at a hundred Hz in Ref.~\cite{nat.commun.9.3072yu}, it is up to ten MHz in this work. To demonstrate the pseudospin-dependent propagation of the edge states, a pseudospin-filter sample was prepared, as illustrated in Fig.~\ref{xueqin.2.valley}(d). The elastic wave is injected from the bottom port, due to the pseudospin-momentum locking property, only output 1 and output 3 channels are permitted for wave propagation, while the output 2 channel is forbidden. The measured field distributions in the output1 to output 3 channels are displayed in Fig.~\ref{xueqin.2.valley}(e), which confirms this statement. Lately, the topological quantum spin Hall effect is also implemented in surface acoustic wave~\cite{phys.rev.appl.16.044008.zhang}.

Recently, the orbital degree of freedom has been introduced to study the nanomechanical topological metamaterials~\cite{nat.nanotechnol.16.576.ma}. The unit cell of the nanomechanical metamaterial is exhibited in the left panel of Fig.~\ref{xueqin.2.valley}(f). The auxiliary degree of freedom brings new advantages to modulate elastic waves. For instance, the adiabatic phase transition of different topological edge states can be achieved experimentally without closing the bulk band gap, in contrast, it cannot be realized in the case of fixed orbitals. Besides the one-dimensional edge states, the 0D vortex states have been created in a nanomechanical metamaterial with spatially varying parameters (the right panel of Fig.~\ref{xueqin.2.valley}(f)). The second-order (Fig.~\ref{xueqin.2.valley}(g)) and third-order (Fig.~\ref{xueqin.2.valley}(h)) nonlinearities have been obviously observed in the sample.

\begin{figure}[htbp]
    \flushright
    \includegraphics[width=0.85\linewidth]{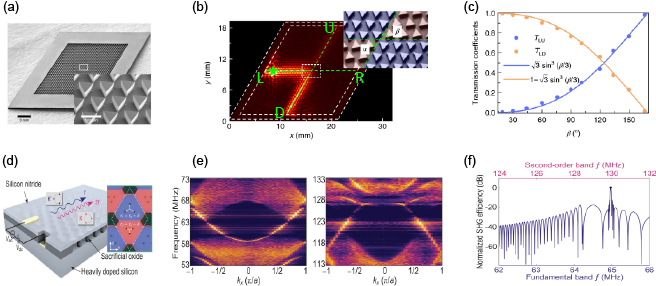}
    \caption{On-chip micron valley topological insulator (a) and abnormal partitions (b, c).  Figures (a)-(c) adopted from Ref.~\cite{nat.mater.17.993.yan}. Nano-electromechanical structure of valley topological insulator for dual bulk band gaps (d, e) and the enhanced second-harmonic nonlinearity (f). Figures (d)-(f) adopted from Ref.~\cite{adv.mater.33.2006521ma}.}
    \label{xueqin.3.on-chip valley}
\end{figure}

Orbital states, as a degree of freedom, have been studied in the above works on topological mechanical metamaterials. Valley is another a degree of freedom that can provide additional manipulation of mechanical waves. Valley topological insulator for elastic waves was firstly realized by Yan {\sl et al.}~\cite{nat.mater.17.993.yan}. The microfabricated elastic metamaterial is shown in Fig.~\ref{xueqin.3.on-chip valley}(a), which consist of triangular pillars, arranged in a hexagonal lattice, on top of a silicon plate. The elastic waves propagating on the silicon structure can be described as Lamb waves and modulated by the periodical pillars. Tuning the rotational angle of the triangular pillars, distinct valley phases arise. The demonstrations of the existence and robustness of the edge states on the boundary between different valley phases are observed experimentally. Interestingly, the energy partition of the elastic waves on the silicon chip can be modulated by the angles between different channels. A four-channel (marked as L, R, D and U) device, made up of alternative distint valley phases, is demonstrated in Fig.~\ref{xueqin.3.on-chip valley}(b). The angle between channels L and D is $\alpha$, and that between channels R and U is $\beta$. The inset in Fig.~\ref{xueqin.3.on-chip valley}(b) illustrates the enlarged view around the intersection point of the four channels. The projected edge state dispersion indicates that the edge states possessing positive group velocities in channels L, D, and U are locked at the $K$ valley, while those in channel R are locked at the $K^\prime$ valley. Because of the negligible inter-valley scattering, once the elastic wave enters from channel L, it is only allowed to channels D and U, while is prohibited in channel R. The field distribution for $\alpha=\beta=60^{\circ}$ is measured in Fig.~\ref{xueqin.3.on-chip valley}(b). At the first glance, the wave seems easily bending to channel U, rather than to channel D, because the angle between channels L and U is much smoother than that between channels L and D. Counterintuitively, the measured wave energy in channel D is much larger than that in channel U, due to the stronger coupling between channels L and D than that between channels L and U. Although the waves cannot transmit into channel R, it plays an important role on the manipulation of transmissions in channels D and U. For the fixed channels L, D and U, when increasing $\beta$, the transmissions of channels D and U can vary from 1 to 0 and 0 to 1, respectively, as illustrated in Fig.~\ref{xueqin.3.on-chip valley}(c). This abnormal partition supplies a convenient way to control the elastic wave propagation. Very recently, Ren et al. increased the working frequency of elastic valley Hall effect to several hundred MHz in an optomechanical system, which may have potential applications in implementing on-chip phononic circuits with robust topological waveguides~\cite{nat.commun.13.3476ren}.

Many previous works about valley topological insulator in mechanical metamaterials are focused on a single bulk band gap. Reference~\cite{adv.mater.33.2006521ma} experimentally demonstrates the valley Hall effect in dual bulk band gaps. The nano-electromechanical sample is illustrated in Fig.~\ref{xueqin.3.on-chip valley}(d). There are two complete bulk band gaps around 62MHz and 122MHz. The measured projected edge state dispersions of these two band gaps are given in Fig.~\ref{xueqin.3.on-chip valley}(e). The counter-propagating helical edge states are clearly observed. Because of the strong mechanical nonlinearities of the nano-electromechanical systems, the second-harmonic nonlinear process has been studied by deliberately design of the structure to satisfy the perfect phase matching condition, which results in high second-order generation efficiency, as shown in Fig.~\ref{xueqin.3.on-chip valley}(f). The valley spin-Hall effect can also be realized in a simple structure of a hexagonal lattice, with its unit cell consisting of an alternation of free and clamped holes in a metallic plate, arranged to lie at the vertices of a smaller hexagon~\cite{phys.rev.b.102.214103tang}.

\begin{figure}[htbp]
    \flushright
    \includegraphics[width=0.85\linewidth]{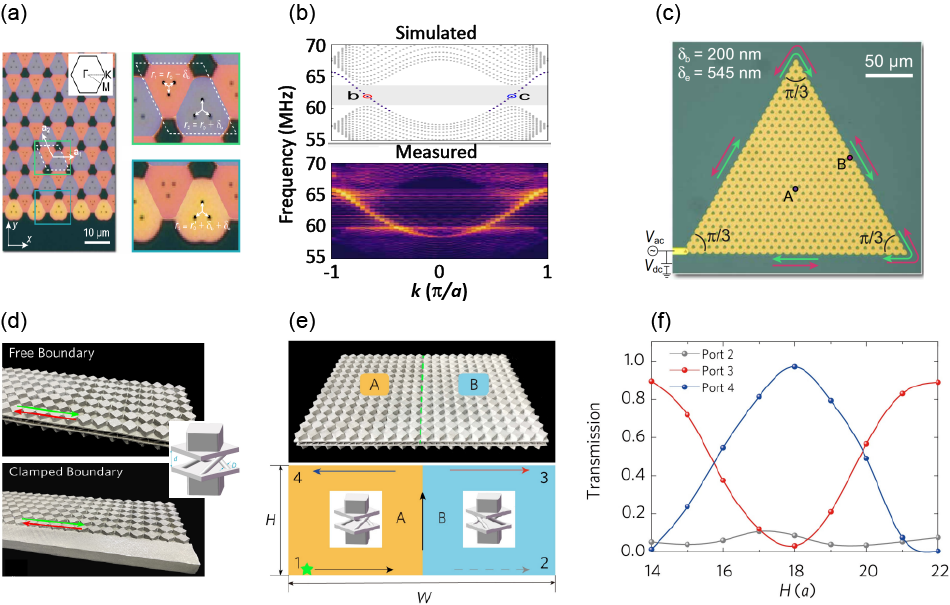}
    \caption{(a)-(c) Observation of chiral edge states in gapped nanomechanical graphene by adjusting the boundary potentials. (a) Optical microscope image of the nanomechanical graphene sample. (b) The simulated and measured gapless edge state dispersions. (c) Triangular sample for edge state propagation. Figures (a)-(c) adopted from Ref.~\cite{sci.adv.7.eabe1398xi}. (d)-(f) Chiral edge states without modulating the boundary potential. (d) Top (bottom) panel: photo of the bilayer elastic metamaterial with the free (clamped) boundary. Inset: unit cell of the structure. (e) Photo of the sample consisting of two structures with opposite chiral interlayer couplings. The height and width of the sample are $H$ and $W$ respectively. (f) Transmissions of output ports 2-4 for different $H$ at the frequency of 26.75kHz. Figures (d)-(f) adopted from Ref.~\cite{nsr.nwac203wu}.}
    \label{xueqin.4.edge states at single boundaries}
\end{figure}

All the above discussed works of the gapless helical edge states are at the domain wall between two mechanical metamaterials with distinct topology. A natural question arises: How to achieve the gapless helical edge states at the boundary of a single material. The well-known edge states at a single material are those at the zigzag boundary of graphene. However, these edge states are flat, rather than helical. Xi {\sl et al.} reported the existence of helical edge states at the single zigzag boundary of a nano-mechanical graphene~\cite{sci.adv.7.eabe1398xi}. This is constructed by adjusting the on-site boundary potentials of the edge. The sample and the zoom-in area of the edge are shown in Fig.~\ref{xueqin.4.edge states at single boundaries}(a). For a specific edge parameter, the simulated and measured edge state dispersions confirm the existence of the gapless helical edge states, as shown in Fig.~\ref{xueqin.4.edge states at single boundaries}(b). Due to the pseudospin-momentum locking, the edge states can propagate through the $\pi/3$ sharp bends without backscattering. The optical microscope image of the triangular sample is illustrated in Fig.~\ref{xueqin.4.edge states at single boundaries}(c). Another advantage of this work is its high operating frequency, up to 60MHz, which may have potential applications in elastic devices.

Very recently, a new proposal has been put forward to achieve topological insulator for elastic waves in a single boundary without delicately modulating the boundary potential~\cite{nsr.nwac203wu}. The unit cell is a bilayer structure, where each layer consists of a square block on top of a plate and the interlayer couplings are introduced by four tilted pillars. Two layers supply the pseudospin degrees of freedom, and the chiral pillars induce the spin-orbit couplings. In addition to out-of-plane polarization, the in-plane ones are indispensable to describe the properties of this system. The nonzero spin-Chern number protects the existence of the gapless helical edge states at both the free and clamped boundaries. The configurations of the samples are shown in Fig.~\ref{xueqin.4.edge states at single boundaries}(d). Beside the edge states at the boundary of a single elastic metamaterial with a certain chirality of the interlayer coupling, the edge states at the boundary between distinct topological systems give a way to manipulate the wave propagation. The anticlockwise and clockwise chiral interlayer couplings determine different topology of the system. The sample and corresponding schematic are illustrated in Fig.~\ref{xueqin.4.edge states at single boundaries}(e). The incident wave inputs from port 1, and will transport to the other ports. The transmissions of ports 2-4 are simulated in Fig.~\ref{xueqin.4.edge states at single boundaries}(f). The port 2 does not permitted for the spin-momentum locking property, whereas the transmissions of ports 3 and 4 can be flexibly modulated by the height of the structure.

The edge states confined at the boundary area restrict the efficient coupling between the surface acoustic wave and interdigital transducer. Following the idea of topological valley-locked waveguide states in acoustic heterostructures~\cite{nat.commun.11.3000wang}, Wang {\sl et al.} realized the extended topological valley-locked surface acoustic wave on a piezoelectric on-chip elastic metamaterial~\cite{nat.commun.13.1324.wang}, which can match broadband interdigital transducer efficiently. This may have promising applications for surface acoustic wave integrated circuits with high throughput and energy capacity.

The topological property of the edge state in the previous works is based on the electric polarization. Lately, the quantized polarization has been extended from dipole moment to higher multipole one~\cite{science.357.61benalcazar}. The concept of higher-order topology with extended bulk-edge correspondence greatly expands the conventional topological materials. Elastic metamaterial provides a versatile platform to exhibit the higher-order topology. The quadrupole topological insulator, featured by the zero-dimensional corner states, is firstly observed in phononic crystals~\cite{nature555.342Serra-Garcia}. In addition to the corner states protected by the higher multipole moments, there exists another type of ones characterized by the nontrivial bulk polarization~\cite{phys.rev.lett.122.204301fan}. Contents about the higher-order topology will be covered in the later section.

\subsection{Non-Hermitian topological phononic systems}
Topological properties of non-Hermitian system have attracted a lot of attention~\cite{phys.rev.appl.16.057001gu}. Different from Hermitian system, non-Hermitian system has complex spectrum so that supports more plentiful topological structures. Exceptional point is one example unique to non-Hermitian system. Its topological properties have been widely studied in coupled few modes systems~\cite{phys.rev.x.6.021007ding}. In this paper, we will mainly focus on the topological properties of band structure. As we introduced in previous section, the non-Hermitian system can support point gap and line gap. Phononic metamaterial is a feasible and versatile platform to explore those new concepts, which not only brings fundamental understanding in physics but also provides possible applications. The existing studies on non-Hermitian band topology in phononic system can be classified into three classes. The first class is about line gap topology. Gain and loss are added to a topological phononic metamaterial. The gain/loss is small enough so that it does not change the line gap topology of the system. However those gain/loss can change the transport properties and scattering properties of bulk states or topological edge states. The second class is also about line gap topology. Here the gain/loss makes a topological phase transition. Topological nontrivial states can be obtained solely by adding gain and loss to a topological trivial system. The third class is about point gap topology. Asymmetric couplings can be realized in different acoustic systems. The well-known non-Hermitian skin effect is studied.

\begin{figure}[htbp]
\flushright
    \includegraphics[width=0.85\linewidth]{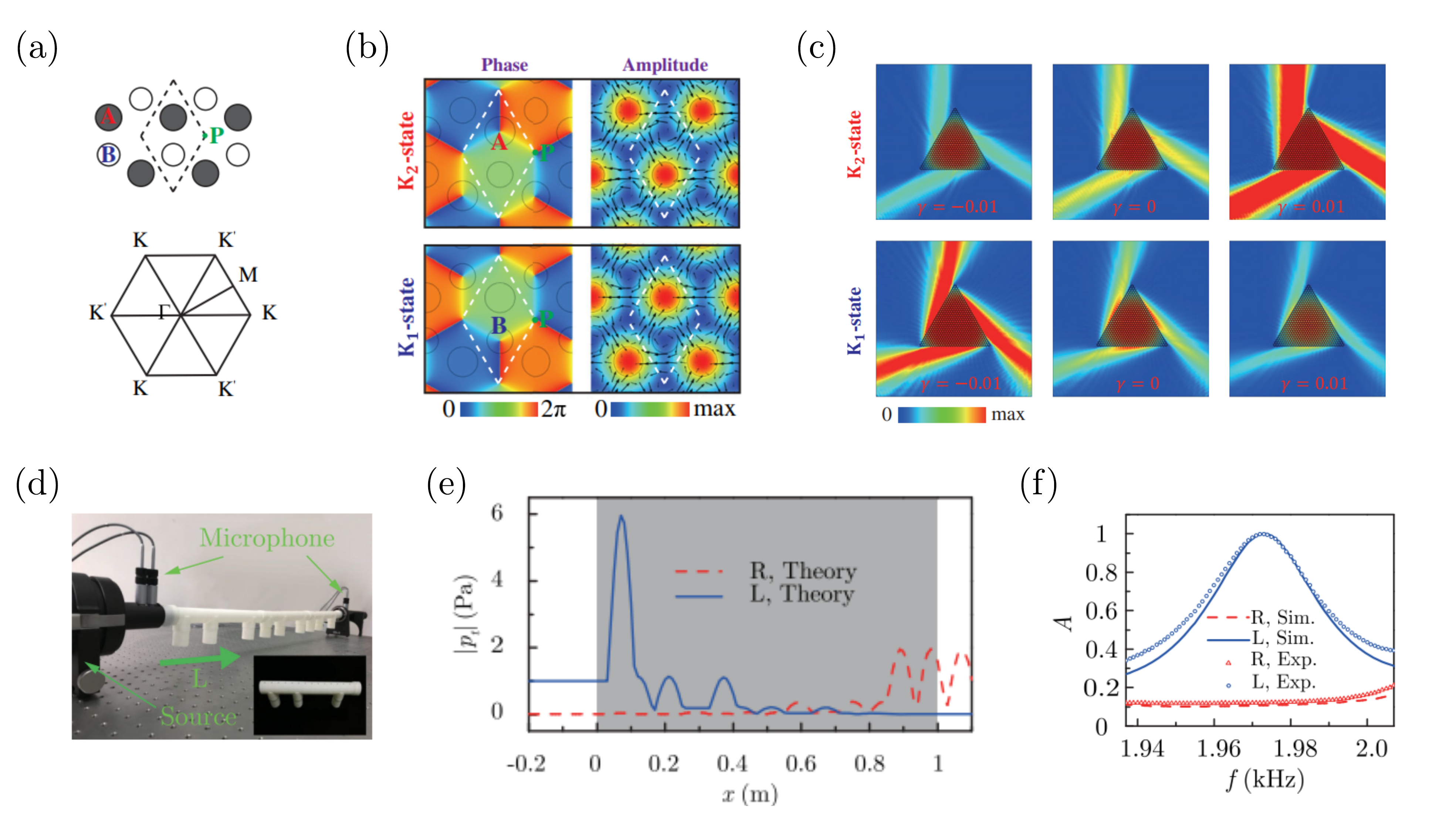}
    \caption{First class of non-Hermitian topology: non-Hermitian band in phononic metamaterials. (a) Top panel, acoustic boron nitride consisting of cylinders A and B in air. Bottom panel, the corresponding first Brillouin zone. (b) Phase and amplitude distribution for valley states. $P$ indicates the position with $C_3$ symmetry.  (c) Field profile of valley states excited by a chiral source of topological charge $m=1$ and $m=-1$, respectively. Figs.(a)-(c) adopted from~\cite{phys.rev.lett.120.246601wang}. (d) Samples of comb-like acoustic locally resonant structure with Aubry-Andre-Harper like modulation, whose profile of topological edge states is shown in (e) and sound absorption is shown in (f). Figs.(d)-(f) adopted from~\cite{phys.rev.lett.121.124501zhu}. }
    \label{nonhermitian}
\end{figure}

First class: One example is using gain and loss to amplify or attenuate valley topological states. The system is acoustic boron nitride with spatially alternating gain and loss located at two sublattices, respectively~\cite{phys.rev.lett.120.246601wang}. The unit cell of acoustic boron nitride is shown in Fig.~\ref{nonhermitian}(a), which is composed of cylinders A and B. The radius of A and B are different so the six fold rotation symmetry is broken. Without gain and loss, the system supports valley states at high symmetric momentum points $K$ and $K^\prime$, which are also the extrema of topological band gap. There are two kinds of valley states called valley $K_1$ and valley $K_2$ with different charges $+1$ and $-1$, respectively, as shown by phase in Fig.~\ref{nonhermitian}(b). Their field amplitudes are also different, where valley $K_1$ mainly locates at cylinder A and valley $K_2$ mainly locates at cylinder B. When small non-Hermitian modulation is introduced by adding gain/loss to cylinders A and B, the line gap topology of the system does not change. The topological properties of the valley states are kept. However, the transport properties of the system are dramatically changed by the non-Hermitian modulation. The non-Hermitian modulation is tuned by $\gamma$, which represents gain (loss) on cylinder A with negative (positive) value. Figure~\ref{nonhermitian}(c) shows the field amplitude of valley states excited by a chiral source. When $\gamma<0$ the $K_1$ states are amplified, and when $\gamma>0$ the $K_2$ states are amplified. Except the bulk states, the valley edge states between two acoustic boron nitride systems with different valley Chern numbers also can be amplified or attenuating.

Another example is using non-Hermitian topological edge states to realize an asymmetric absorber. The system is the comb-like acoustic locally resonant structure~\cite{phys.rev.lett.121.124501zhu}, where the positions of resonators are modulated in Aubry-Andre-Harper formula as shown in Fig.~\ref{nonhermitian}(d). Such system can be understood as locally resonators coupled by far field so that it is a nice platform to study topological states in open systems. When the positions of resonators are modulated in Aubry-Andre-Harper formula, the system supports asymmetric topological edge states in the band gap. Different from closed system, here the topological edge state is intrinsically connected with environment. We can incident a plane wave from left or right to study the scattering properties of the system. Without gain and loss, the topological edge states can be excited by the plane wave. When the topological edge states become steady states, all the waves are reflected when considering the frequency is located in the band gap.  When the loss is introduced to the resonators with a critical value, all the waves are absorbed by the topological edge states without reflection. From Fig.~\ref{nonhermitian}(e), we notice the system only supports topological edge states on left side at the specific frequency. The reflection is zero with left incident wave and one with right incident wave. So it can be used to realize a topological protected asymmetric absorber as shown in Fig.~\ref{nonhermitian}(f).

The non-Hermitian modulation can also be used to tune the eigenvalues of topological corner modes in second-order topological insulators. The topological corner modes can be tuned to be source-like or sink-like~\cite{phys.rev.lett.122.195501zhang}. By engineering non-Hermitian properties on the topological cavities, non-Hermitian topological whispering gallery with chirality can be obtained~\cite{nature.597.655hu}.

\begin{figure}[htbp]
\flushright
    \includegraphics[width=0.85\linewidth]{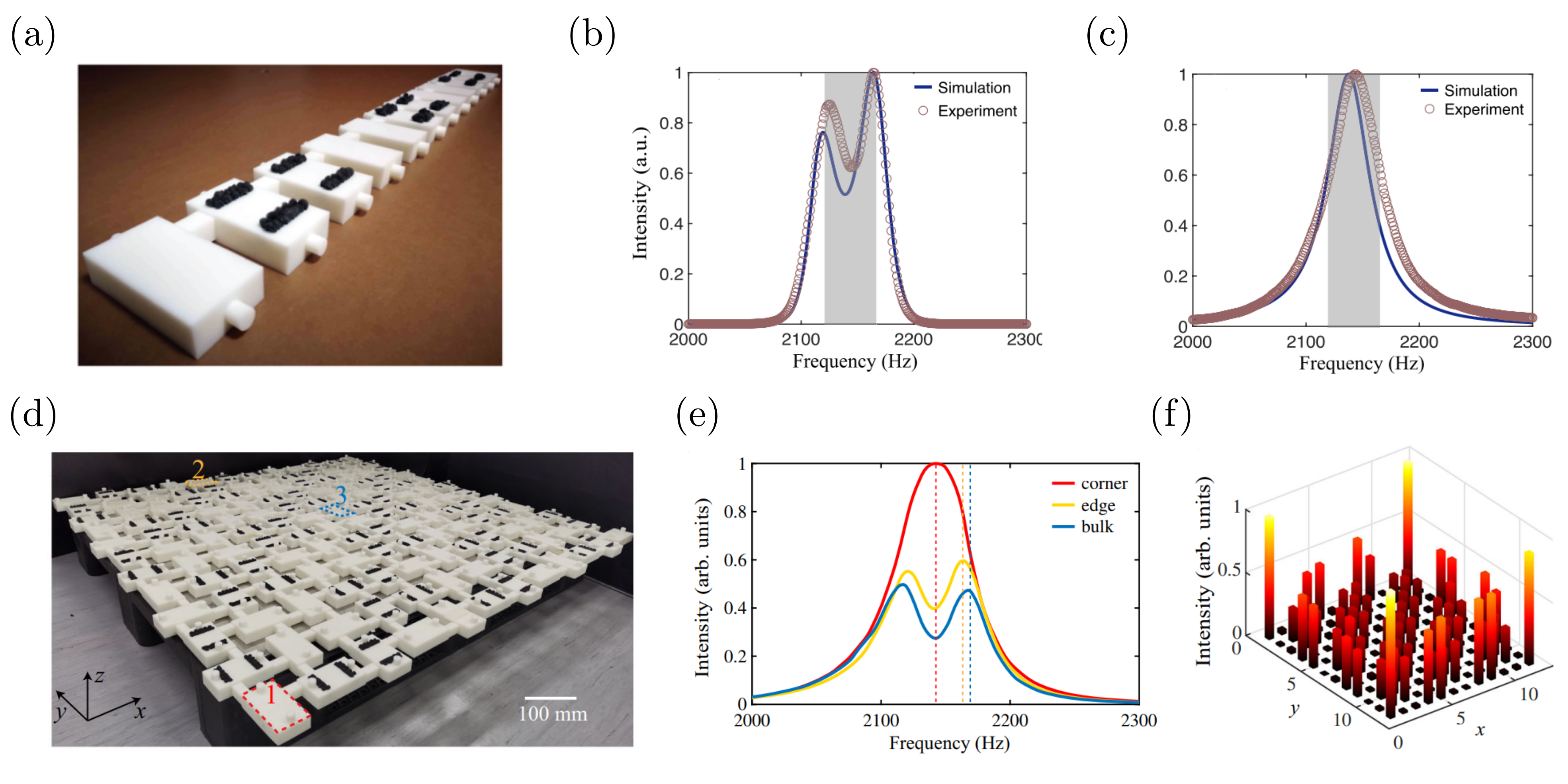}
    \caption{Second class of non-Hermitian topology: Topological insulator induced solely by gain and loss. (a) Sample of coupled acoustic resonators with spatially distributed loss. (b) Measured bulk spectrum for (a). (c) Measured edge spectrum for (a). Figs.(a)-(c) adopted from~\cite{phys.rev.b.101.180303gao}. (d) Sample of acoustic second-order topological insulator with spatially distributed loss. (e) Measured bulk, edge and corner spectrum for (d). (f) Measured field profile for the corner mode. Figs.(d)-(f) adopted from~\cite{nat.commun.12.1888gao}.}
    \label{nonhermitiansecond}
\end{figure}

Second class: In this class, topological nontrivial phase can be obtained solely by gain and loss~\cite{phys.rev.b.101.180303gao,nat.commun.12.1888gao}. Figure~\ref{nonhermitiansecond}(a) shows one-dimensional system with coupled acoustic resonators. The couplings between neighbor resonators are the identical. Without gain and loss, there is only one pass band. When loss is introduced as Fig.~\ref{nonhermitiansecond}(a), the system is dimerised into a Su-Schrieffer-Heeger-like model but a non-Hermitian version. It is proven by the measured bulk spectrum in Fig.~\ref{nonhermitiansecond}(b) where there is a band gap. Importantly, the spatially distributed loss makes the effective couplings between the two resonators at boundary weaker. It belongs to topological nontrivial phase so that the system supports topological edge states in the band gap as proven by the measured boundary spectrum in Fig.~\ref{nonhermitiansecond}(c). Gain or loss induced second-order topological insulators have also been studied in acoustic system~\cite{nat.commun.12.1888gao}. Figure~\ref{nonhermitiansecond}(d) shows the two-dimensional sample. Similar to the one-dimensional one, here the couplings between neighbor resonators have the same amplitude. Without gain and loss, there is only one pass band. When loss is introduced, one band gap opens and supports gapped edge states and topological corner states as shown in Fig.~\ref{nonhermitiansecond}(e). The fields of topological corner states are localized at four corners as shown in Fig.~\ref{nonhermitiansecond}(f).

\begin{figure}[htbp]
\flushright
    \includegraphics[width=0.85\linewidth]{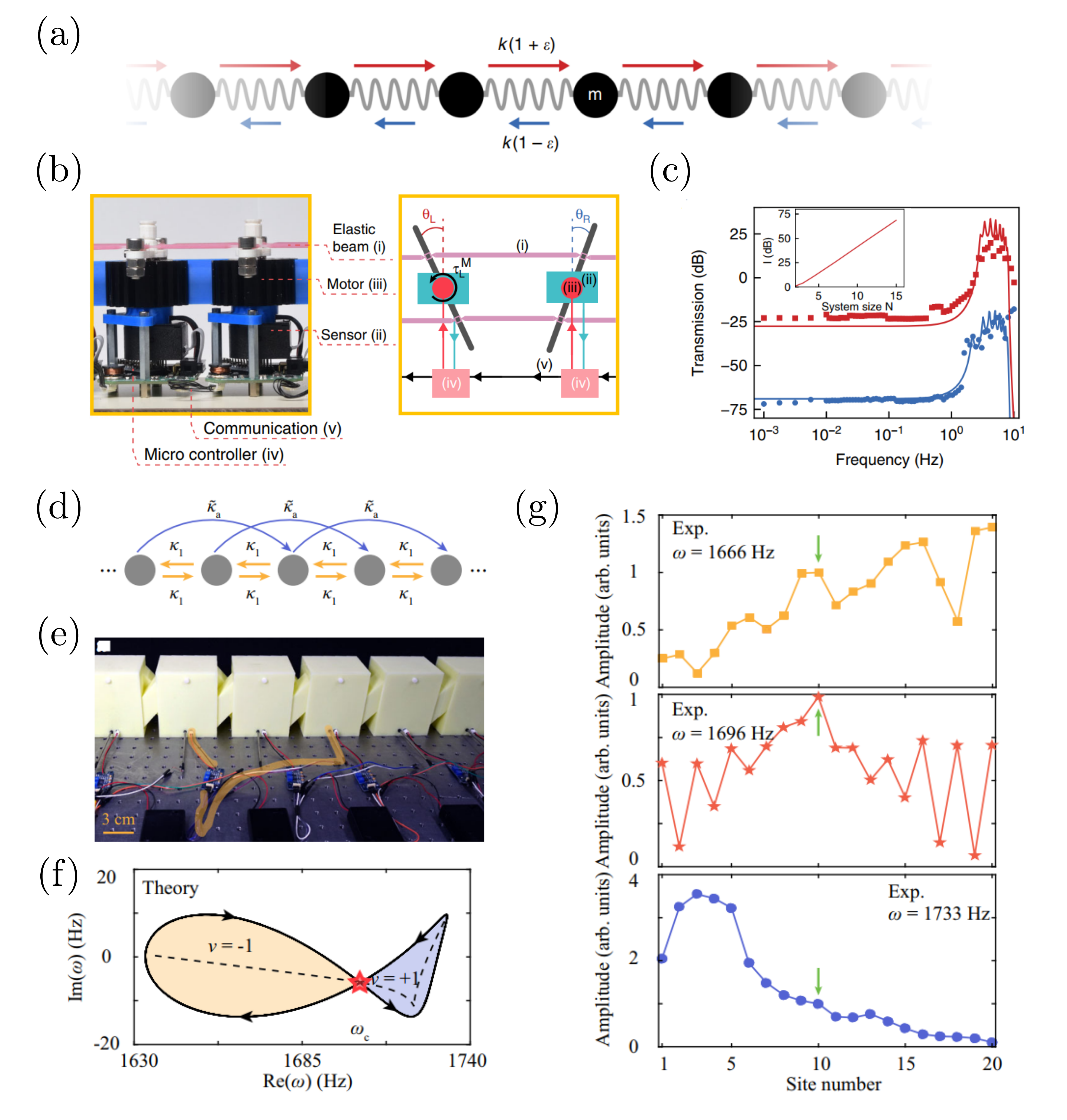}
    \caption{Third class of non-Hermitian topology: Non-Hermitian skin effect in mechanical metamaterials (a)-(c) and acoustic metamaterials (d)-(f). (a) The lattice model with asymmetric couplings. (b) Unit cell of the robotic metamaterials with nearest-neighbor asymmetric couplings. (c) Measured transmissions with excitation at the left (red squares) or right (blue dots) edge of the metamaterial. Figs.(a)-(c) adopted from~\cite{nat.commun.10.4608brandenbourger}. (d) Lattice model with nearest-neighbor asymmetric couplings and next-nearest-neighbor couplings. (e) Sample of the acoustic metamaterials. (f) The complex spectrum with twisted winding number. (g) Measured field distribution with excitation at middle site. Figs.(d)-(g) adopted from~\cite{nat.commun.12.6297zhang}.}
    \label{nhse}
\end{figure}

Third class: This class has non-Hermitian skin effect which comes from point gap topology. One important step to realize non-Hermitian skin effect is the realization of asymmetric couplings as shown in Fig.~\ref{nhse}(a). Phononic metamaterial is an ideal platform to study the non-Hermitian skin effect, due to the ease with which the couplings strength can be engineered. For example, Brandenbourger $et$ $al$. realized non-Hermitian skin effect with a robotic mechanical metamaterial, where the local control loops are used to break the reciprocity of couplings~\cite{nat.commun.10.4608brandenbourger}. The unit cell is shown in Fig.~\ref{nhse}(b). The sensors measure the oscillation of the mechanical motor and provide feedback force to the rotor by micro controller. Due to the non-Hermitian skin effect, the fields are localized at right for both left excitation and right excitation as shown in Fig.~\ref{nhse}(c). Zhang $et$ $al$. studied non-Hermitian skin effect in an acoustic metamaterial~\cite{nat.commun.12.6297zhang}. They studied one more complicated case where the next-nearest-neighbor couplings are also considered as shown in Fig.~\ref{nhse}(d). Figure~\ref{nhse}(e) provides one sample. Here the asymmetric couplings are realized by directional acoustic amplifier. The acoustic signal in one cavity measured by microphone is amplified to excite the next-nearest-neighbor cavity. Different from the nearest-neighbor asymmetric couplings, here the spectrum of the system supports twisted winding numbers in complex plane shown in Fig.~\ref{nhse}(f). In open boundary condition, the $\nu=-1$ parts accumulate to right and the $\nu=+1$ parts accumulate to left. It is confirmed by the field measurements in Fig.~\ref{nhse}(g) with excitations at middle site. Non-Hermitian skin effect was also studied in other works with the asymmetric couplings introduced by engineering non-Hermitian spatial distribution in coupled ring resonators~\cite{nat.commun.12.5377zhang,arxiv.2205.14824gao}, piezoelectric elements in mechanical metamaterials~\cite{phys.rev.lett.125.206402gao,nat.commun.12.5935chen} and non-Hermitian spatial distribution in topological active materials~\cite{phys.rev.lett.125.118001.scheibner,nat.commun.12.4691palacios}. The non-Hermitian bulk-boundary correspondence has also been studied in active mechanical metamaterials~\cite{Proc.Natl.Acad.Sci.U.S.A.117.29561ghatak}.

\begin{figure}[htbp]
\flushright
    \includegraphics[width=0.85\linewidth]{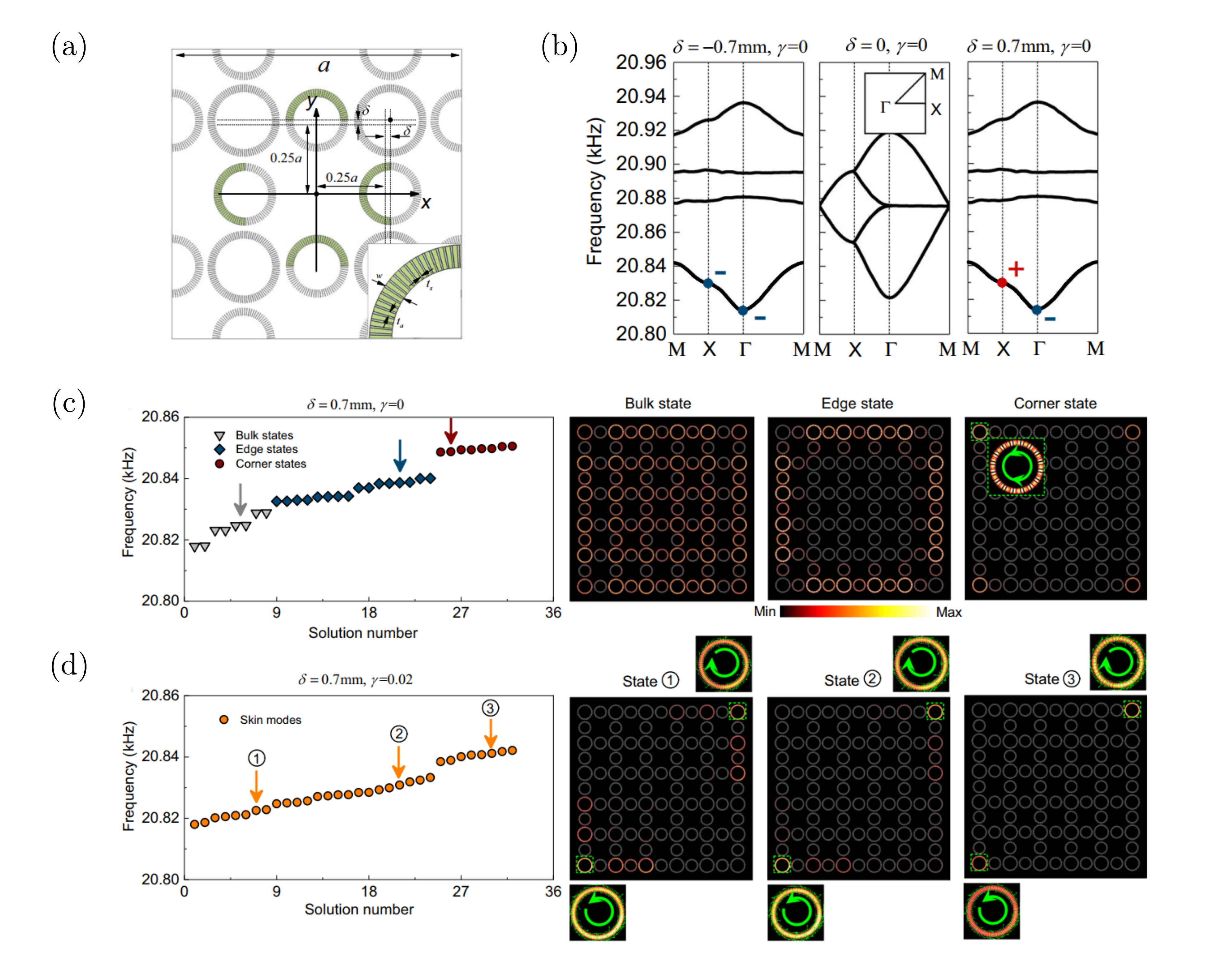}
    \caption{Non-Hermitian higher-order topology with non-Hermitian skin effect. (a) Unit cell of the acoustic metamaterial, consisting of four site rings, coupled to each other by link ring. (b) Band structure calculated for the Hermitian acoustic metamaterial with $\delta=-0.7$ mm, 0, 0.7 mm. (c) Spectrum for a box-shaped acoustic metamaterial (in the Hermitian region) with open boundary condition in  both the x and y direction. The corresponding acoustic wavefunctions for the bulk, edge and corner states are presented. (d) The same as (c) but with loss. Figures adopted from~\cite{nat.commun.12.5377zhang}.}
    \label{secondNHSE}
\end{figure}

Recently, there are also works studying non-Hermitian higher-order topology with non-Hermitian skin effect~\cite{nat.commun.12.5377zhang}. Figure~\ref{secondNHSE}(a) shows the unit cell. The loss is engineered in the link ring to realize spin-polarized asymmetric coupling. Even in the Hermitian case, such model already supports topological nontrivial states. The band structures of the system are shown in Fig.~\ref{secondNHSE}(b). As tuning the dimerization parameter $\delta$, the band gap between the second band and the third band closes and reopens, and the system undergoes a phase transition from normal insulator, to gapless states and finally to second-order topological insulator. The second-order topological phase is confirmed by the spectrum of finite structure in Fig.~\ref{secondNHSE}(c). Only the corner states are localized at corner. However when the loss is introduced, all the states are localized at the corner as shown in Fig.~\ref{secondNHSE}(d). Such localization comes from non-Hermitian skin effect. Wang $et$ $al$, realized the delocalized topological corner modes in an active mechanical lattice~\cite{nature.608.50wang,phys.rev.b.105.L201402teo}.  Besides, the second-order non-Hermitian skin effect, where only $\mathcal{O}(L)$ modes are localized at the corner and the bulk states are still extended was studied in active metamaterials~\cite{nat.commun.12.4691palacios}.

\section{Special topics}
Here we review some special topics related with topological physics, including pseudo magnetic field, synthetic dimension, acoustic Floquet topological insulators, bulk transports in acoustic heterostructures, topological pumping, topological active matter, acoustic skyrmions, twisted phononic bilayers, and fractal topological acoustic metamaterials.

\subsection{Pseudo magnetic field}
\begin{figure}[htbp]
\flushright
    \includegraphics[width=0.85\linewidth]{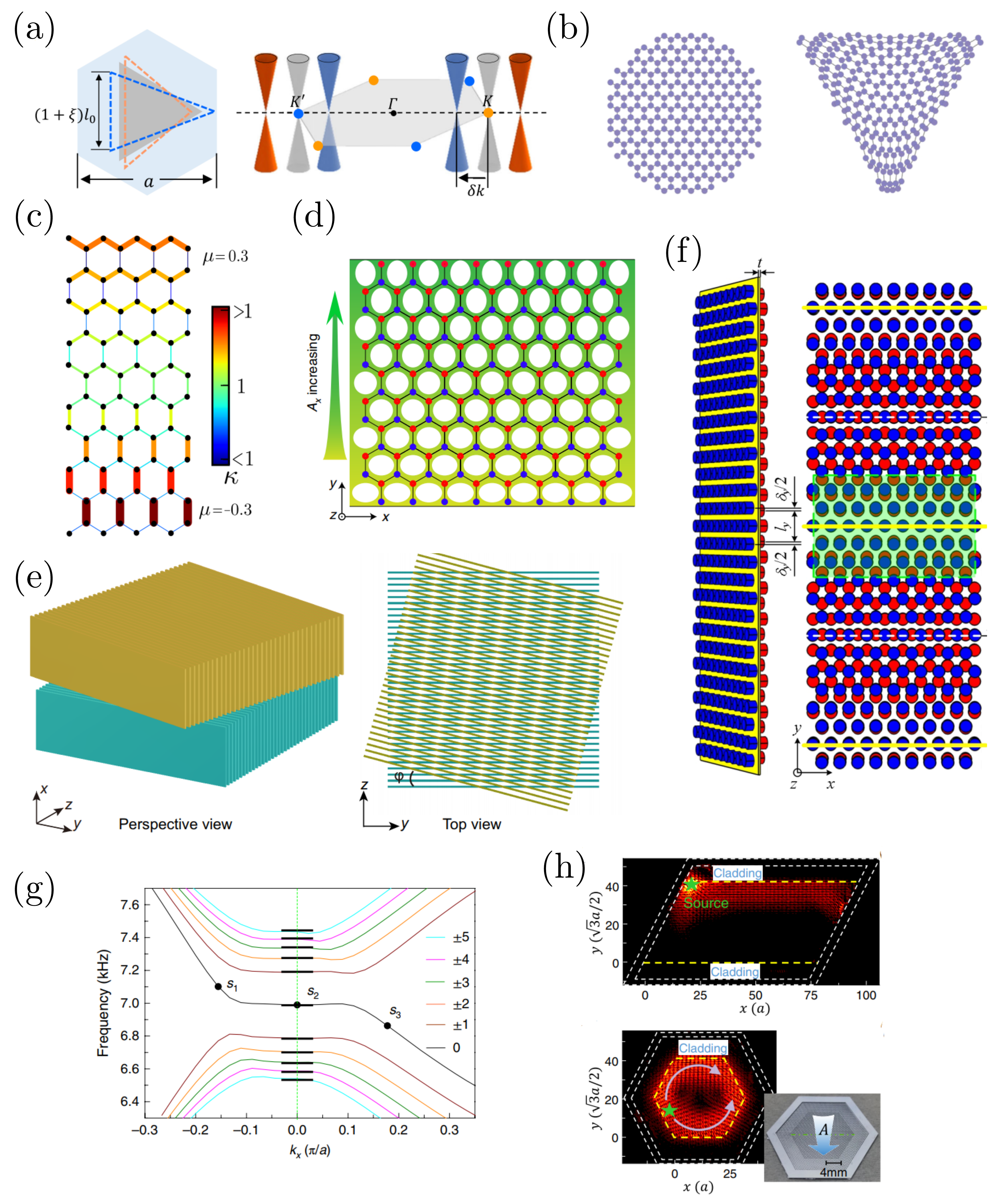}
    \caption{Pseudo gauge field in phononic metamaterials. (a) Shape deformations of triangle scatterer in one unit cell and their effects on the shift $\delta k$ of Dirac cones. (b) Schematics of the unstrained and strained acoustic lattices with triaxial strain. (c) Patterning of the local materials stiffness that leads to a constant magnetic field in mechanical metamaterials. (d) Pseudo magnetic field from the shape deformations of scatterers in acoustic metamaterials. (e) Gauge field from the twisted bilayer acoustic metamaterials. (f) Pseudo magnetic field from Moire phononic lattices. (g) Spectrum near the Dirac frequency for (d). Bold lines marks the predicted Landau levels. (h) Measured elastic edge states under psedu magnetic field for (a). Figures adopted from ~\cite{phys.rev.lett.127.136401yan,phys.rev.lett.118.194301yang,phys.rev.lett.119.195502abbaszadeh,nat.phys.15.352wen,sci.adv.7.2062yang,arxiv.2103.12265zheng}}
    \label{gaugefield}
\end{figure}

Pseudo magnetic field, which comes from the stain of the graphene-like materials, provides a nice tool to study magnetic field related phenomena specially for bosons which do not interact with the genuine magnetic field. The core of realizing pseudo magnetic field is to shift the Dirac point by adjusting the parameters of the system as shown in Fig.~\ref{gaugefield}(a). Without strain, the graphene-like structure supports effective Dirac Hamiltonian $H=v_D (k_x\sigma_x+k_y\sigma_y)$ at Brillouin zone corners $K$ or $K^\prime$. After straining the model, the momentum of Dirac point $\mathbf{k}=(k_x,k_y)$ is shift by $\mathbf{k}\rightarrow \mathbf{k}+\delta\mathbf{k}$ where the $\delta\mathbf{k}$ is equivalent to an effective vector potential $\mathbf{A}$. The pseudo magnetic field can be obtained by engineering the space dependent vector potential.

Different approaches have been proposed in phononic metamaterials to realize pseudo magnetic field. Yang have proposed to realize pseudo magnetic field in acoustic structures by applying the triaxial strain on the the acoustic honeycomb lattice as shown in Fig.~\ref{gaugefield}(b)~\cite{phys.rev.lett.118.194301yang}. In their proposal, the lattice sites are tuned away from their original positions and are connected by coupling waveguides of the proper length. A strong uniform pseudo magnetic field can be obtained. Abbaszadeh $et$ $al$. have proposed to pattern the local material stiffness of mechanical graphene [Fig.~\ref{gaugefield}(c)] to realize the pseudo magnetic field~\cite{phys.rev.lett.119.195502abbaszadeh}. The system is uniform along the $x$ direction and graded along the $y$ direction. An uniform pseudo magnetic field can be obtained to control the density states and transverse spatial confinement of sound in the metamaterials. Wen $et$ $al$. have theoretically proposed and experimentally realized the pseudo magnetic field by an uniaxial deformation of scatterers in an acoustic graphene [Fig.~\ref{gaugefield}(d)]~\cite{nat.phys.15.352wen}. Yan $et$ $al$. have theoretically proposed and experimentally realized the pseudo magnetic field by deforming the scatterers in an on-chip elastic system [Fig.~\ref{gaugefield}(a)]~\cite{phys.rev.lett.127.136401yan}. Other methods to introduce pseudo magnetic field include twisted bilayer acoustic metamaterials shown in Fig.~\ref{gaugefield}(e)~\cite{sci.adv.7.2062yang} and Moire phononic graphenes shown in Fig.~\ref{gaugefield}(f)~\cite{arxiv.2103.12265zheng}. At the Dirac cone, the quasiparticle has linear dispersion. Under the uniform pseudo magnetic field, they have the same behavior as electrons in graphene with magnetic field which supports unequally divided Landau levels. Figure~\ref{gaugefield}(g) shows one example where the Landau levels are marked. Between the Landau levels, there are topological edge states. Figure~\ref{gaugefield}(h) shows the measured elastic edge states in on-chip elastic systems. The edge states propagate along one direction for a fixed valley. Besides, the pseudo magnetic field has also been considered in three-dimensional systems, where the gauged field is induced from the shift of Weyl points~\cite{nat.phys.15.357peri}. Chiral phonons can be obtained from acoustic Weyl semimetal under the effect of pseudo magnetic field.

\subsection{Synthetic dimension} Topological states usually have different behaviors in different dimensional systems. However, higher-dimensional system is usually hard to realize in experiment. Synthetic dimension provides one way to study higher dimensional topological states in a lower dimensional system. Even system with dimension larger than three can be studied by the synthetic method. The synthetic dimension can be obtained by internal degrees of freedom of the system (frequency or angular momentum) or by parameters of the system~\cite{phys.rev.lett.108.133001boada,phys.rev.lett.111.226401kraus,phys.rev.a.82.052311tsomokos,optica.5.1396yuan}.

In phononic metatmaterials, most of the works use the parameters as additional dimensions. For example, Zhu $et$ $al$, theoretically proposed and experimentally realized asymmetric topological edge states in an acoustic locally resonant metamaterial~\cite{phys.rev.lett.121.124501zhu}. The distances between the local resonators are modulated in Aubry-Andre-Harper formula $L_n=L+\delta L\cos(2\pi b n+\phi)$, where $b$ is an rational number and $\phi$ is the modulation phase. Although the system is one-dimensional, the topological edge states are described by Chern number, which is the topological invariant of a two-dimensional system (one dimension from the space of the system and the other dimension from the modulation phase as synthetic dimension). Similar topological states can be obtained by adding Aubry-Andre-Harper-like modulation on the on-site potential or nearest-neighbor couplings of one-dimensional lattice model. Chen $et$ $al$. have studied the four-dimensional topological phase by a two-dimensional system~\cite{phys.rev.x.11.011016chen}, where each direction is coupled acoustic resonators with on-site modulations. The eigenstates of system can be treated as product states of eigenstates from two independent Aubry-Andre-Harper model. Topological corner states and topological corner states in the bulk continuum are studied there. Wang $et$ $al$. studied the three-dimensional $Z\times Z_2$ topological states by a two-dimensional system, where one direction is a Su-Schrieffer-Heeger model and the other direction is a Aubry-Andre-Harper model with modulation on the couplings~\cite{phys.rev.lett.127.214302wang}. The $b$ can also be set as an irrational number. In this case, it corresponds to quasi-crystal. Two groups have proved that the quasi-crystal also can supports topological edge states~\cite{phys.rev.lett.122.095501apigo,phys.rev.appl.13.014023xia}. And the well-known Hofstadter butterfly is measured with the help of synthetic dimension~\cite{commun.phys.2.55ni}. Other works have studied the type-II Weyl points and Weyl exceptional ring in synthetic dimension~\cite{appl.phys.express.15.037001li,phys.rev.lett.129.084301liu}.

\begin{figure}[htbp]
\flushright
    \includegraphics[width=0.85\linewidth]{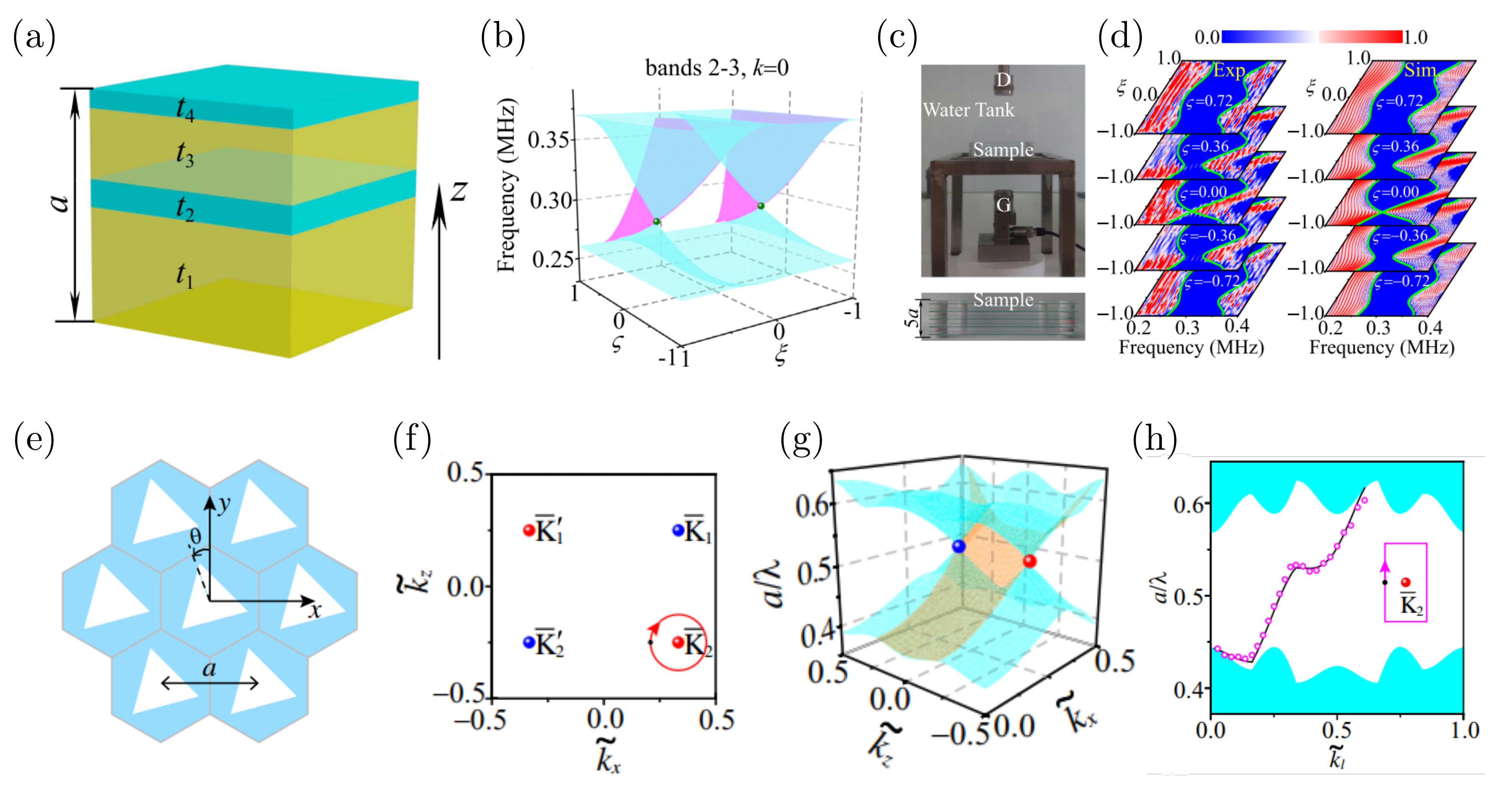}
    \caption{Synthetic Weyl point. (a) Four-layer-based unit cell, stacked alternately by water (yellow) and glass (blue) along the z direction. (b) The synthetic Weyl point degenerated at second band and third band. (c) The experiment set up. Bottom panel: an enlargement of the experimental sample. (d) Measured and simulated transmission spectra. Figs.(a)-(d) adopted from~\cite{phys.rev.lett.122.136802fan}. (e) Two-dimensional acoustic valley metamaterials made of regular triangular scatterers (white) in air background (blue). $\theta$ is the extra synthetic dimension. (f) The location of Weyl points in synthetic $k_x-k_z$ surface Brillouin zone. (g) Surface band (orange) plotted in the one half of surface Brillouin zone. (h) Measured (open circles) and simulated (black line) surface bands along the loop in the $k_x-k_z$ surface Brillouin zone (inset). Figs.(e)-(h) adopted from~\cite{phys.rev.lett.128.216403fan}. }
    \label{synthetic}
\end{figure}

The parameters can be more general. For example, Fan $et$ $al$. studied the Weyl physics in an one-dimensional acoustic metamaterial with two extra structure parameters~\cite{phys.rev.lett.122.136802fan}. The unit cell is composed of two layers of water and two layers of glass as shown in Fig.~\ref{synthetic}(a). Two extra structure parameters, $\xi=(t_1-t_3)/(t_1+t_3)$ and $\varsigma=(t_2-t_4)/(t_2+t_4)$, are introduced to control the layer thickness of the system and play the role of two additional synthetic dimensions. Figure~\ref{synthetic}(b) shows two Weyl points in the synthetic spaces. The Fermi arcs with specific boundary condition are colored in pink. The existence of Weyl points can be confirmed by a transmission measurement with setup shown in Fig.~\ref{synthetic}(c). The measured transmission confirms that the second band and the third band touch at $\varsigma=0.00$ as shown in Fig.~\ref{synthetic}(d). The experimental results are consistent with the simulation results. Besides, Fan $et$ $al$. also propose to use synthetic dimension to understand the valley topology by connecting the valley topological edge states with the Fermi arcs around Weyl points in synthetic space~\cite{phys.rev.lett.128.216403fan}. Figure~\ref{synthetic}(e) shows the structure of the acoustic valley metamaterials. The rotation angle $\theta$ of the scattering plays the role of the addition synthetic dimension. Figures~\ref{synthetic}(f) and ~\ref{synthetic}(g) show the positions of Weyl points. Surrounding one Weyl, gapless topological edge states can be obtained as shown in Fig.~\ref{synthetic}(h).

\subsection{Acoustic Floquet topological insulators} Topological insulators can be obtained by periodical driving or Floquet engineering. Fleury $et$ $al$. proposed the first realistic acoustic Floquet topological insulator~\cite{nat.commun.7.11744fleury}. They proposed a graphene lattice with trimer metamolecule on each site. The on-site potential of sub-cavities is designed to be periodically dependent on time with different phases, achieved by time-periodically modulating the bulk modulus. Such time modulation breaks the time-reversal symmetry and makes the system support topologically protected edge states. Such method was extended to elastic system and experimentally realized the first elastic Floquet topological insulator~\cite{sci.adv.6.8656darabi}.

\begin{figure}[htbp]
\flushright
    \includegraphics[width=0.85\linewidth]{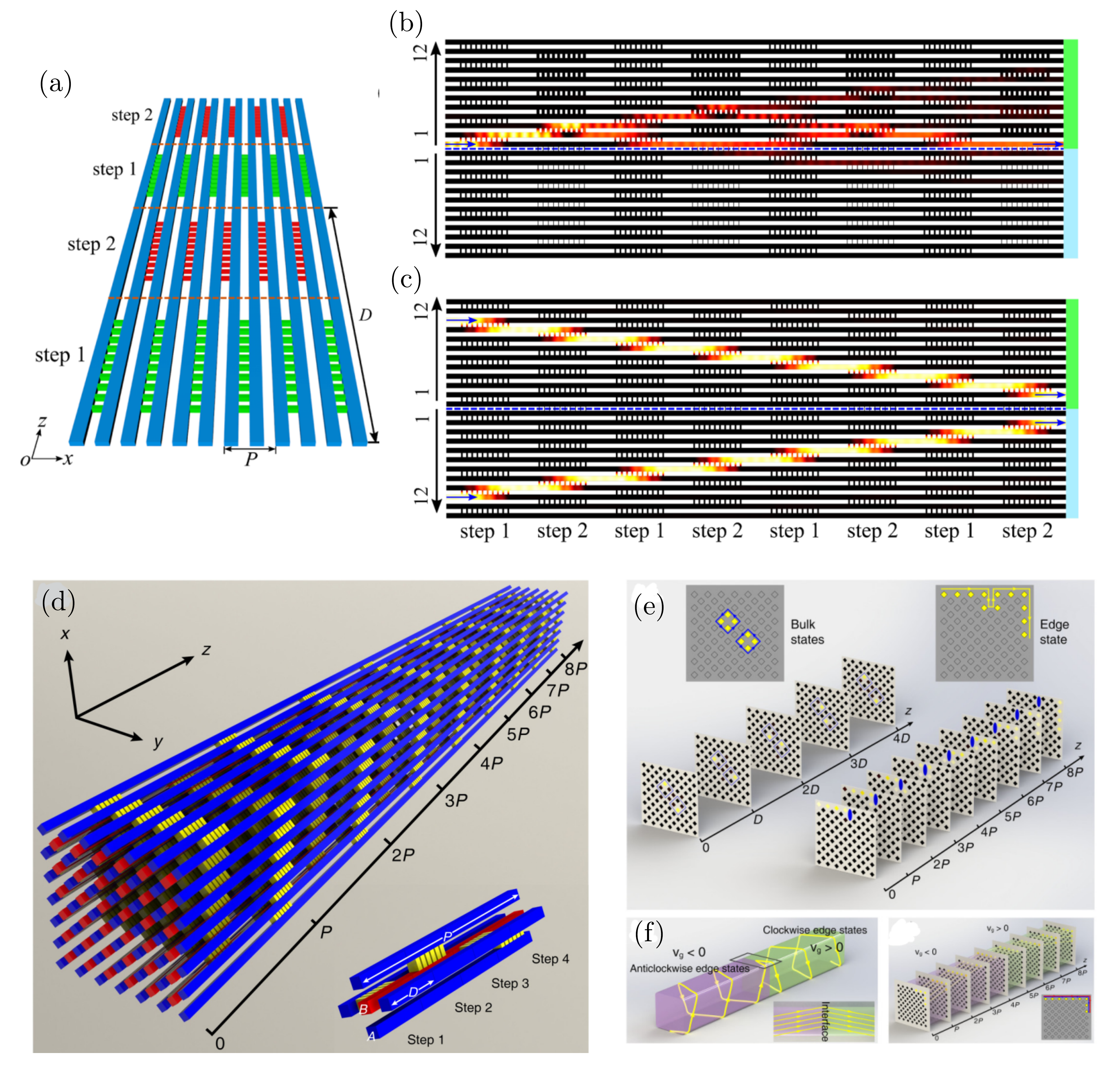}
    \caption{Acoustic Floquet topological insulator in coupled waveguide arrays. (a) Schematic diagram of one dimensional coupled waveguide arrays. (b) Field evolution of topological interface state. (c) Field evolution of bulk state. Figs.(a)-(c) adopted from~\cite{j.appl.phys.123.091716peng}. (d) Two dimensional coupled waveguide arrays. (e) Field evolution of bulk state and topological edge state. (f) Topological negative refraction. Figs.(d)-(f) adopted from~\cite{phys.rev.res.1.033149peng}.}
    \label{acousticFloquet}
\end{figure}

The time modulation method has high experimental requirements. There are other alternatives that can be used to achieve acoustic Floquet topological insulators. One method is to treat additional space dimension as time. Such method is borrowed from photonics, where they use the helical waveguide arrays to realize a photonic Floquet topological insulator~\cite{nature.496.196rechtsman}. In acoustics, Peng $et$ $al$. used the coupled acoustic waveguide arrays to realize an one-dimensional Floquet topological insulator~\cite{j.appl.phys.123.091716peng}. The schematic diagram of one-dimensional coupled waveguide arrays is shown in Fig.~\ref{acousticFloquet}(a), which are composed of regularly distributed waveguides and coupling blocks. Here, the $z$ direction plays the role of time. The couplings are modulated along the $z$ direction. Specially, such system supports two band gaps, one at quasienergy 0 and the other at quasienergy $\pi/D$. The topological properties of the system are described by a $Z_0\times Z_\pi$ topological invariant. By designing the couplings in step 1 and step 2, two different phases ($0\times0$ and $1\times1$) can be obtained. Figure~\ref{acousticFloquet}(b) shows the field evolution of topological interface state along $z$ and Fig.~\ref{acousticFloquet}(c) shows the field evolution of bulk states. Such method has also been used to study two-dimensional acoustic Floquet topological insulator~\cite{phys.rev.res.1.033149peng}. The structure is shown in Fig.~\ref{acousticFloquet}(d). The chiral modulation of couplings along the $z$ direction makes the system support chiral edge states as shown in Fig.~\ref{acousticFloquet}(e). Such novel states are different from the usual Chern insulator in static systems. It was known as anomalous Floquet topological insulator which supports topological chiral edge states with zero Chern number. They also use the structure to realize a topological negative refraction as shown in Fig.~\ref{acousticFloquet}(f). More recently, the coupled acoustic waveguides have also been used to realize an acoustic Floquet higher-order topological insulator and acoustic Floquet $\pi/2$ mode~\cite{nat.commun.13.11zhu,arxiv.2207.09831cheng}. Different from the static system, the topological boundary states in Floquet system usually have anomalous dynamical evolution. For example, the time-periodic corner states from Floquet higher-order topology were reported in Ref.~\cite{nat.commun.13.11zhu}.

\begin{figure}[htbp]
\flushright
    \includegraphics[width=0.85\linewidth]{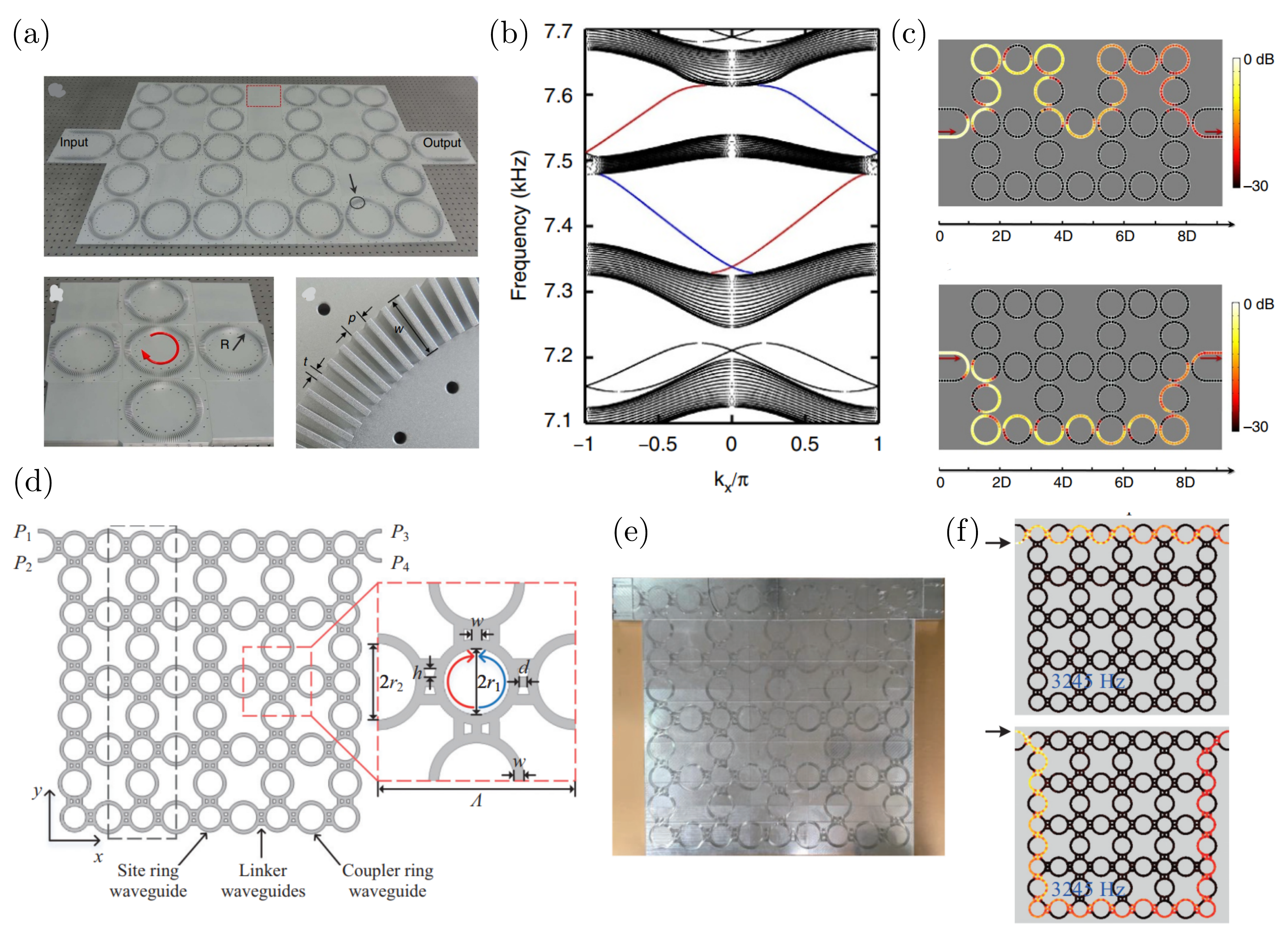}
    \caption{Acoustic Floquet topological insulator in coupled ring resonators. (a) Two dimensional coupled metamaterial ring lattice composed of $3\times4$ unit cells. Left-down is one unit cell. Right-down is the details of waveguide. (b) Projected band structure. (c) Measured topological edge states for spin up and spin down, respectively. Figs.(a)-(c) adopted from~\cite{nat.commun.7.13368peng}. (d) Schematic diagram of waveguide networks. (e) Photograph for the sample. (f) Measured pressure amplitude for topological edge states. Figs.(d)-(f) adopted from~\cite{phys.rev.b.95.094305wei}.}
    \label{CRRs}
\end{figure}

Another method to realize acoustic Floquet topological insulator is using coupled ring resonators which are described by scattering matrix~\cite{phys.rev.b.89.075113pasek}. Given that both scattering matrix and Floquet operator are unitary matrices, their eigen equation problems are equivalent. Peng $et$ $al$. used acoustic waveguide to realize the ring resonators~\cite{nat.commun.7.13368peng}. The configuration is shown in Fig.~\ref{CRRs}(a). The system is composed of site ring resonators coupled by link ring resonators which are all made up of waveguides. Right-down shows the details of the waveguide which confines the acoustic wave on the microstructure. Waves on one waveguide can propagate to the near waveguide by near field couplings. By designing parameters, they can realize perfect couplings. At those parameters, the system can support topological edge states as shown by projected band structure in Fig.~\ref{CRRs}(b). The measured pressure amplitudes of topological edge states are shown in Fig.~\ref{CRRs}(c) for spin up and spin down excitations. Wei $et$ $al$. proposed a simpler structure to realize the acoustic anomalous Floquet topological insulator~\cite{phys.rev.b.95.094305wei}. The schematic diagram is shown in Fig.~\ref{CRRs}(d) and the photograph of sample is shown in Fig.~\ref{CRRs}(e). Here the waveguide is realized by hard boundaries and the couplings between two waveguides are realized by link waveguides. Thanks to the simplicity of the structure, the loss of the system is small. The measured pressure amplitude of topological edge states is shown in Fig.~\ref{CRRs}(f). The coupled ring resonators have also been used to study non-Hermitian skin effect by adding designed loss~\cite{nat.commun.12.5377zhang,arxiv.2205.14824gao} and topological dislocation modes by extending to three-dimensional system~\cite{nat.commun.13.508ye}.

\subsection{Bulk transports in acoustic heterostructures}

\begin{figure}[htbp]
    \flushright
    \includegraphics[width=0.85\linewidth]{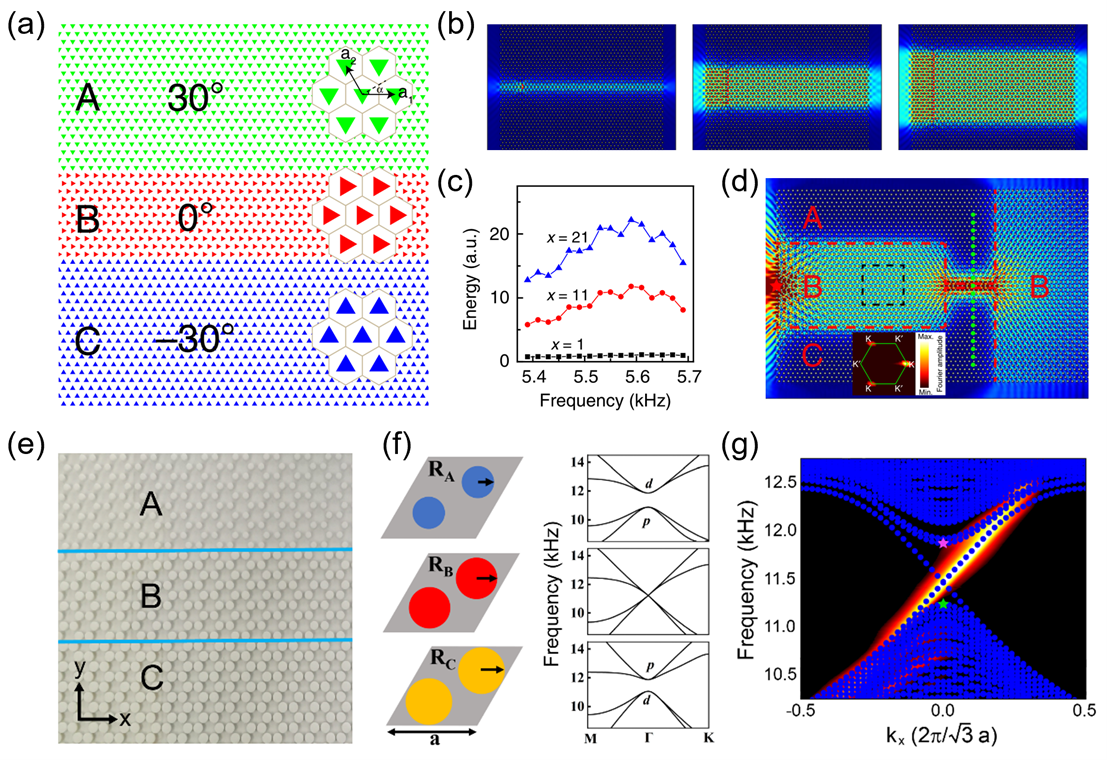}
    \caption{(a) Schematic of heterostructure (${\rm A}|{\rm B}_x|{\rm C}$) consisting of three phononic crystals domains, A, B, and C. Here phononic crystals in domains A and C are of different valley phases, while phononic crystals in domain B have Dirac dispersions. (b) Simulated pressure fields at a frequency of 5.55 kHz in heterostructure (${\rm A}|{\rm B}_x|{\rm C}$) with the number of layers in domain B, $x$, being 1, 11, and 21, respectively. Red lines denote the excitation sources. (c) Total transmitted energy versus frequency for the three structures. (d) Simulated pressure fields at a frequency of 5.45 kHz in heterostructure with a stepped domain B with x sharply changing from 21 to 1, showing the valley-locked converging. Dashed red lines delineate the boundaries of domains A, B, and C. Inset shows is the Fourier spectrum of the acoustic field in the black dashed rectangle. Figures (a)-(d) adopted from~\cite{nat.commun.11.3000wang}. (e) Photograph of a heterostructure similar to that in (a), but phononic crystals in domains A and C are of different pseudospin Hall phases, while phononic crystals in domain B have double Dirac dispersions. (f) Left panel: Schematics of unit cells in domains A, B and C with the same lattice constant but different radii for the scatterers. Right panel: The corresponding bulk dispersions. (c) Projected dispersions of the heterostructure of (${\rm A}|{\rm B}_9|{\rm C}$) as in (e). The color map and blue dots represent the measured and simulated results. Figures (e)-(g) adopted from~\cite{phys.rev.appl.18.034066liu}.}
    \label{jiuyang_Picture6}
\end{figure}

As discussed in Section 3.1, acoustic topological edge transports protected by valleys and pseudospins have been observed in the domain walls of phononic crystals, where the gapless valley-locked and pseudospin-locked edge states localize in and decay away from one-dimensional interfaces~\cite{nat.phys.12.1124he,nat.phys.13.369lu,Phys.Rev.Lett.118.084303zhang}. Based on the interface transports, phononic crystals with gapless bulk bands can be further inserted into the domain walls to construct heterostructures and forming topological waveguides~\cite{nat.commun.11.3000wang,nat.commun.13.1324.wang, phys.rev.appl.18.034066liu, phys.rev.appl.18.054073yin}. Inherited from the domain wall edge states, the waveguide states possess gapless dispersions and are immunity against defects. More importantly, the waveguide states have a high capacity for energy transport, and thus are more flexible for interfacing with the existing acoustic devices. The heterostructures based on valley topology are shown in Fig.~\ref{jiuyang_Picture6}(a), which is a sandwich structure consisting of three domains~\cite{nat.commun.11.3000wang}. Within a hexagonal lattice array, domain B is a typical Dirac phononic crystals constructed by regular triangular rods~\cite{phys.rev.b.89.134302lu}, and domains A and C are phononic crystals of valley phases realized by deviating the orientations of the rods. This heterostructure can host valley-locked guiding states, which, different from those in domain wall structures, extend into the entire domain B. Benefited from the spatial expansion from an interface to a domain, these waveguides have more space to transmit energy. Moreover, the energy transport capability of the heterostructure can be adjusted by the width of domain B. As illustrated in Fig.~\ref{jiuyang_Picture6}(b), line sources of different width (marked in red) are used to excite the topological valley waveguide states, and the acoustic guiding waves propagate dominantly inside domain B and the associated pressure fields are almost uniformly distributed. Obtained by collecting at the right side of the waveguides, the total transmitted energy increases with increasing width of domain B [Fig.~\ref{jiuyang_Picture6}(c)], exhibiting more flexible application potentials compared with the domain wall structure. The heterostructures also serve as versatile new devices for acoustic wave manipulation, such as acoustic splitting, reflection-free guiding, and converging~\cite{nat.commun.11.3000wang}. For acoustic converging, as shown in Fig.~\ref{jiuyang_Picture6}(d), excited by a point source on the left, the topological valley waveguide states are almost not reflected when the states transport through a stepped domain B with its width sharply dropping from 21 to 1. The reflection immunity arises from the valley-locking properties and is due to the presence of only the $K$ valley in domain B, which is evidenced by Fourier transforming the acoustic fields in domain B [inset of Fig.~\ref{jiuyang_Picture6}(d)]. It is the finite width of the topological waveguide that makes the reflection-free converging possible, which may contribute to acoustic enhancement or energy harvesting. This waveguide scheme is also proven to be suitable for acoustic topological insulators with pseudospins~\cite{phys.rev.appl.18.034066liu}. As shown in Figs.~\ref{jiuyang_Picture6}(e) and ~\ref{jiuyang_Picture6}(f), inserting an acoustic graphene with double Dirac dispersions~\cite{appl.phys.lett.105.014107li} into the domain wall of phononic crystals with inverted bands~\cite{nat.phys.12.1124he}, a topological heterostructure can similarly be constructed with a pair of gapless helical waveguide modes [Fig.~\ref{jiuyang_Picture6}(g)]. The helical waveguide modes gain new abilities of adjustable transmission capacity and beam collimations from the freedom of the width of domain B~\cite{phys.rev.appl.18.034066liu}, paving a new way to explore the topological transports in versatile designs of heterostructures for acoustic waves.

\subsection{Topological pumping} Topological pumping which supports unidirectional wave transfer is another hot topic about band topology. Thouless pump which has quantised transport of wave and particle is one example~\cite{phys.rev.b.27.6083thouless}. Long $et$ $al$. studied the acoustic Thouless pumping in the acoustic analogue of Aubry-Andre-Harper model~\cite{j.acoust.soc.am.146.742long}. The treat one additional space dimension as time and the on-site potential of the system is modulated in Aubry-Andre-Harper formula. For a localized bulk wave packet, the center of position of the wave packet will change in a quantised unit cell in a pumping cycle. More recently, the non-Abelian thouless pumping, where the order of two evolution loops with same base point makes the pumping different, was proposed and realized in an coupled acoustic waveguide arrays~\cite{phys.rev.lett.128.244302you}.

\begin{figure}[htbp]
\flushright
    \includegraphics[width=0.85\linewidth]{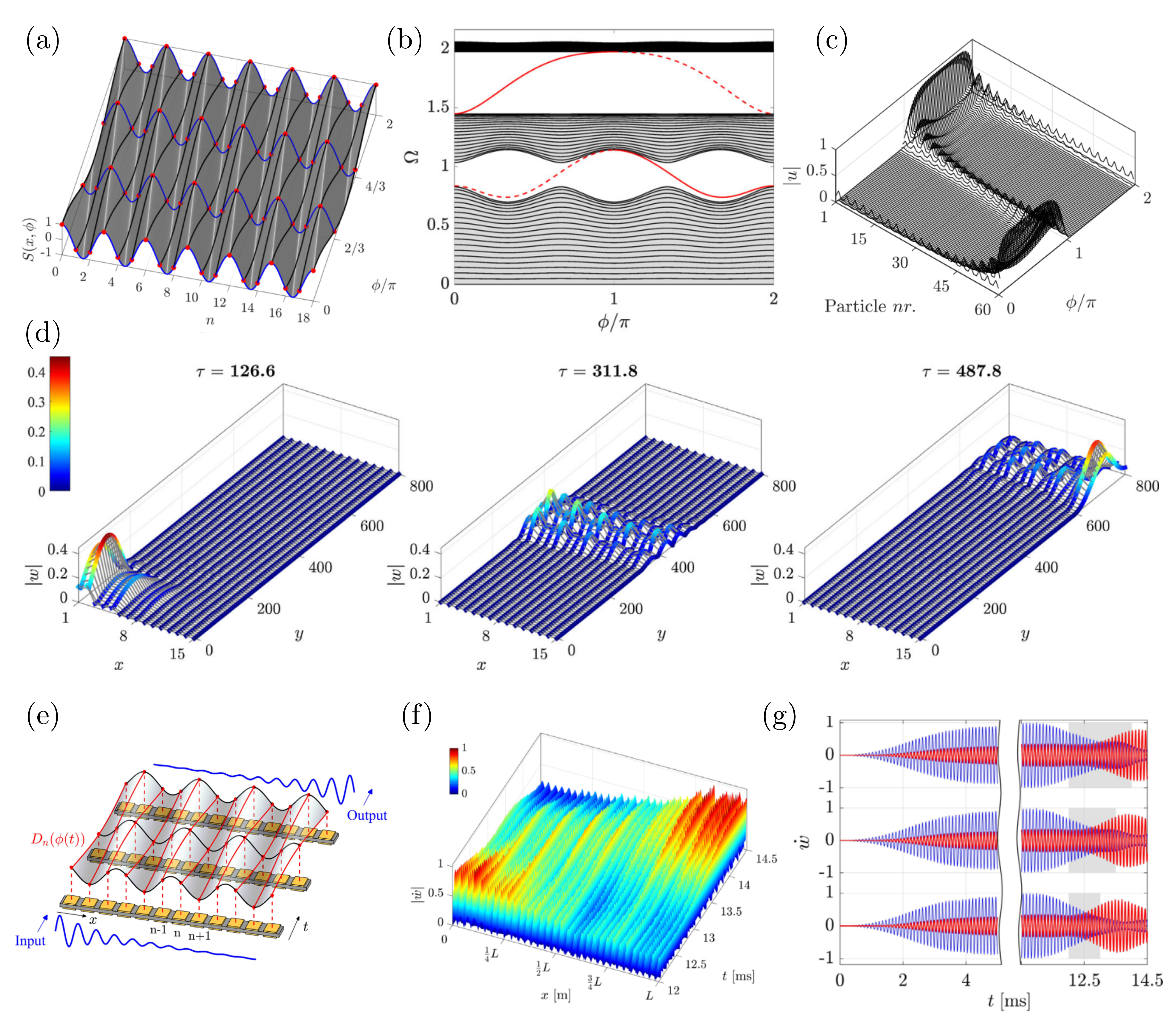}
    \caption{Acoustic topological pumping. (a) Surface $S(x,\phi)=\cos(2\pi\tau x+\phi)$ generating stiffness constants by sampling at $x_n=n$. (b) Band structure of finite system as a function of $\phi$. (c) Adiabatic pumping of topological edge states from right to left. (d) Temporal pumping of topological edge states. Figs.(a)-(d) adopted from~\cite{phys.rev.lett.123.034301rosa}. (e) Illustration of the input and output electromechanical beams. (f) Measured temporal pumping of topological edge states. (g) Signals at left (blue) and right (red) boundaries of the beam for temporal pumps induced within different modulation windows (shaded gray regions). Figs.(e)-(g) adopted from~\cite{phys.rev.lett.126.095501xia}.}
    \label{pumping}
\end{figure}

Another type of topological pumping is about adiabatic pumping of waves from one topological edge state to another~\cite{phys.rev.lett.109.106402kraus}. They were first studied in optical waveguide arrays and were recently introduced to acoustics and mechanics. Rosa $et$ $al$. proposed to realize topological pumping of edge states in spatially modulated elastic lattices~\cite{phys.rev.lett.123.034301rosa}. The stiffness of the one-dimensional lattice is modulated by a function shown in Fig.~\ref{pumping}(a). By changing the phase of the function, the topological edge states of the one-dimensional finite structure can be tuned across the band gap. Fig.~\ref{pumping}(b) shows the band structure of the finite structure as function of $\phi$, where the gapless edge states are clear. These topological edge states can be described by gap Chern number of the corresponding band gap. The field distribution of these topological edge states are plotted in Fig.~\ref{pumping}(c). For $0<\phi<\pi$ ($\pi<\phi<2\pi$), they are localized at right (left) edge. They do a simulation with time varying $\phi$ from $-0.25\pi$ to $0.25\pi$. The left topological edge state slowly transfers to the bulk and finally is pumped to the right as shown in Fig.~\ref{pumping}(d). Xia $et$ $al$. designed an electromechanical waveguide to realize the temporal pumping of topological edge states~\cite{phys.rev.lett.126.095501xia}. The structure of the system is composed of a waveguide and piezoelectric patches as shown in Fig.~\ref{pumping}(e). An effective modulation of stiffness as $D_n(\phi)=D_0[1+\alpha\cos(2\pi\theta+\phi)]$ can be realized. The temporal pumping of topological edge states as theory predicted is experimentally measured as shown in Fig.~\ref{pumping}(f). Fig.~\ref{pumping}(g) shows different cases with different modulation window. In the initial 12 ms, steady-state vibration of left topological edge states are induced. When the modulation window is turned on, it is topologically pumped to the right. Different mechanics have been used to realize topological edge mode pumping, like the incommensurate bilayer acoustic metamaterials~\cite{phys.rev.lett.125.224301cheng}, waveguide with dynamical boundary~\cite{phys.rev.lett.125.253901xu}, and Landau-Zener transition between two coupled topological edge states~\cite{phys.rev.lett.126.054301chen}. Most recently, the adiabatic transport of coherent phonon is realized in a dynamically modulated nanomechanical system by exploiting three coupled topological modes ~\cite{phys.rev.lett.129.215901tian}. Such work is valuable for phononic quantum information processing.

\subsection{Topological active matter}Topological active matter is another hot topic. Different topological states including topological defects, topological protected sound waves and non-Hermitian skin effect have been proposed or realized in active matter system. The experimental platforms for active matter contains active-liquid metamaterials~\cite{nat.phys.13.1091souslov,phys.rev.x.7.031039Suraj,phys.rev.lett.122.128001Anton}, active nematic cells~\cite{arxiv.2008.10852Yamauchi} and robotic metamaterials~\cite{nat.commun.10.4608brandenbourger}, etc. About this topic, there is a nice review~\cite{nat.rev.phys.4.380shankar}. Here we only discuss some advantages of active matter when used to realize backscattering-free topological sound. As we know, it is important to break time reversal symmetry to realize the topological chiral edge modes. In electrons, it is realized by magnetic field. In acoustics, an effective magnetic field can be obtained by rotating fluids. In usual acoustic system, we use an external drive to realize it. However, it brings a lot of difficulties to the experiment. On the one hand, the external drive usual makes the system unstable. On the other hand, the velocity of rotating fluids, which is important to tune the band gap width of the system, is hard to control. Active matter provides us a nice platform to solve these problems~\cite{nat.phys.13.1091souslov}. Active matters, whose individual components motion at the microscales, naturally break the time reversal symmetry. Besides, its velocity can be easily tuned to be comparable with the velocity of sound so that wide-frequency topological chiral edge states can be obtained~\cite{nat.phys.13.1038alu}.

\subsection{Acoustic skyrmions}
Skyrmions characterized by a real-space topological number have been realized in different systems including helimagnetic materials and photonic system~\cite{SKYRME1962556,nature.442.797Bogdanov,science.323.915Muhlbauer,science.361.993tsesses,phys.rev.lett.124.106103guo}. One of the keys to realize it is the vector field. Acoustic waves are described by the acoustic field pressure which is a scalar quantity, so that it is generally believed that skyrmions can not be realized in the acoustic systems. Recently, Ge $et$ $al$. found a new way to use acoustic velocity as a vector field to realize skyrmions in the acoustic system~\cite{phys.rev.lett.127.144502ge}. Using the spoof surface acoustic waves , whose z-component and in-plane component of acoustic velocity have a $\pi/2$ phase difference, they realize the acoustic skyrmions distributed in triangle lattice. The skyrmions can be moved by controlling the phase of the sources. They also show the skyrmions are robust against defects. The acoustic skyrmions may be used to transport information in a stabler way.

\subsection{Twisted phononic bilayers}
The rise of twistronics and twist-photonics~\cite{adv.mat.32.1903800liang,nature.577.42wang,nature.582.209hu} has inspired the lots of studies on twisted phononic bilayers recently~\cite{2d.mater.gardezi,Phys.Rev.Lett.125.214301lopez,Phys.Rev.B.104.L020301kuan,Phys.Rev.Applied.17.034061wu,Phys.Rev.B.102.180304deng,Phys.Rev.B.105.184104zheng}. The study of twisted phononic systems started from the analog with twisted bilayer graphene with an aim of tuning the phononic bands via a the twisting degree of freedom. For instance, a mechanical analog of twisted bilayer graphene made of two vibrating plates was studied in Ref.~\cite{Phys.Rev.Lett.125.214301lopez} in the aim of find the magic angles for quasiflat bands of phonons (specifically, the flexural waves) with vanishing group velocity. Similar purposes were pursued in other studies in acoustic and mechanical systems~\cite{2d.mater.gardezi,Phys.Rev.B.104.L020301kuan,Phys.Rev.B.102.180304deng}. Later, the topological properties of twisted bilayers of phononic systems were studied~\cite{Phys.Rev.Applied.17.034061wu,Phys.Rev.B.105.184104zheng} focusing on the phononic band gaps emerging due to the twisting and interlayer couplings. One special merit of these twisted phononic bilayers is that the interlayer couplings are directly tunable, giving access to strong and ultrastrong interlayer couplings that are unavailable in the electronic counterparts. In particular, in Ref.~\cite{Phys.Rev.Applied.17.034061wu}, the ultrastrong interlayer couplings yield an unconventional higher-order topological band gap with unique topological indices and layer-hybridized corner states that was not found before in electronic twisted graphene bilayers. The ultrastrong interlayer couplings also yield a large band gap and nearly flat band around the band gap. These studies demonstrate the richness of phononic properties in twisted phononic bilayers.

\subsection{Fractal topological acoustic metamaterials}
Going beyond the conventional Euclidean geometry that constrains the study of topological physics is a frontier in the field. In electronic systems, this aim was partly realized in a Sierpiński-triangle geometry in Bi thin films where the fractal geometry is found to turn off the topology of electrons~\cite{Phys.Rev.Lett.126.176102liu}. Later, topological physics of fractal geometry was studied in photonic systems~\cite{Light.Sci.Appl.9.128yang,science.376.1114biesenthal}, demonstrating the existence of topological edge states in both the outer and inner edges and showing reshaped topological phase diagrams compared to the model defined in an Euclidean geometry. For acoustic systems, topological states in fractal geometry were studied recently in the Chern insulator phase~\cite{arXiv.2205.05297li} and higher-order topological insulator phases~\cite{science.bulletin.67.2069zheng,arxiv2205.05298li}. For the Chern insulator phase which was realized in acoustic metamaterials with synthetic gauge flux, the fractal geometry leads to considerable squeeze of the topological phase region but with one-way edge states that are protected by a robust mobility gap. For the higher-order topological states, it was found that fractal geometry fundamentally changes the higher-order topological phenomena, leading to massive number of edge and corner states that are comparable with the number of the bulk states. In fact, it was found~\cite{science.bulletin.67.2069zheng,arxiv2205.05298li} that the numbers of the edge and corner states scale polynomially with the system size according to the fractal dimension of the system, giving a direct manifestation of the fractal dimension in topological physics.

\section{Potential applications}

\subsection{Acoustic double-zero-index medium at Dirac-like point} Dirac cone where two bands are degenerate at Dirac points has been widely studied in two-dimensional phononic crystals. Experimental results show that the three-dimensional double-zero-index medium can be used to bend acoustic waves in three-dimensional space with high efficiency.

Dirac cones often exist at the Brillouin zone boundary and is protected by the lattice symmetry. In two-dimensional phononic crystal, there is another kind of degenerated points, called Dirac-like points, has attracted a lot of attention. They usually happens at the zone center and is accidental degeneracy. Different from the Dirac cone, Dirac-like points are three fold degenerated. One important properties of the phononic crystal at Dirac-like point is that their effective mass density and effective reciprocal of bulk modulus are zero, which are called double-zero-index medium. Different phononic crystal systems have been proposed to realize the double-zero-index medium and show they can be used for supercoupling, bending and cloaking of acoustic waves~\cite{phys.rev.b.84.224113liu,appl.phys.lett.100.071911liu,phys.rev.lett.111.055501fleury,appl.phys.lett.104.161904zheng,appl.phys.lett.105.014107li,j.appl.phys.121.135105dai}. The experimental realization of Dirac-like points in acoustic system is very challenging for its realization requires the refractive index of the scattering is larger than the surrounding medium, air, whose refractive is almost the largest for airborne acoustics. Dubois $et$ $al$. provides a clever way to change of the effective sound velocity by tuning the thickness of the acoustic waveguides~\cite{nat.commun.8.14871dubois}. They experimentally show the double-zero-index medium can bend a point source into a plane wave. Another works shows the Dirac-like cone exist at the interface of two higher-order acoustic topological insulators~\cite{adv.sci.9.202201568meng}. For a more general discussion on different Dirac cones, see a recent review~\cite{Europhysics.lett.137.15001lin}. Recently, the Dirac-like points were also realized in three-dimensional system~\cite{phys.rev.lett.124.074501xu}. In three-dimensional system, the Dirac-like points are four fold degenerate and the waves can be bent into any direction in three dimension.

\subsection{Topological analog signal process for equation solving and image processing} 
\begin{figure}[htbp]
\flushright
    \includegraphics[width=0.85\linewidth]{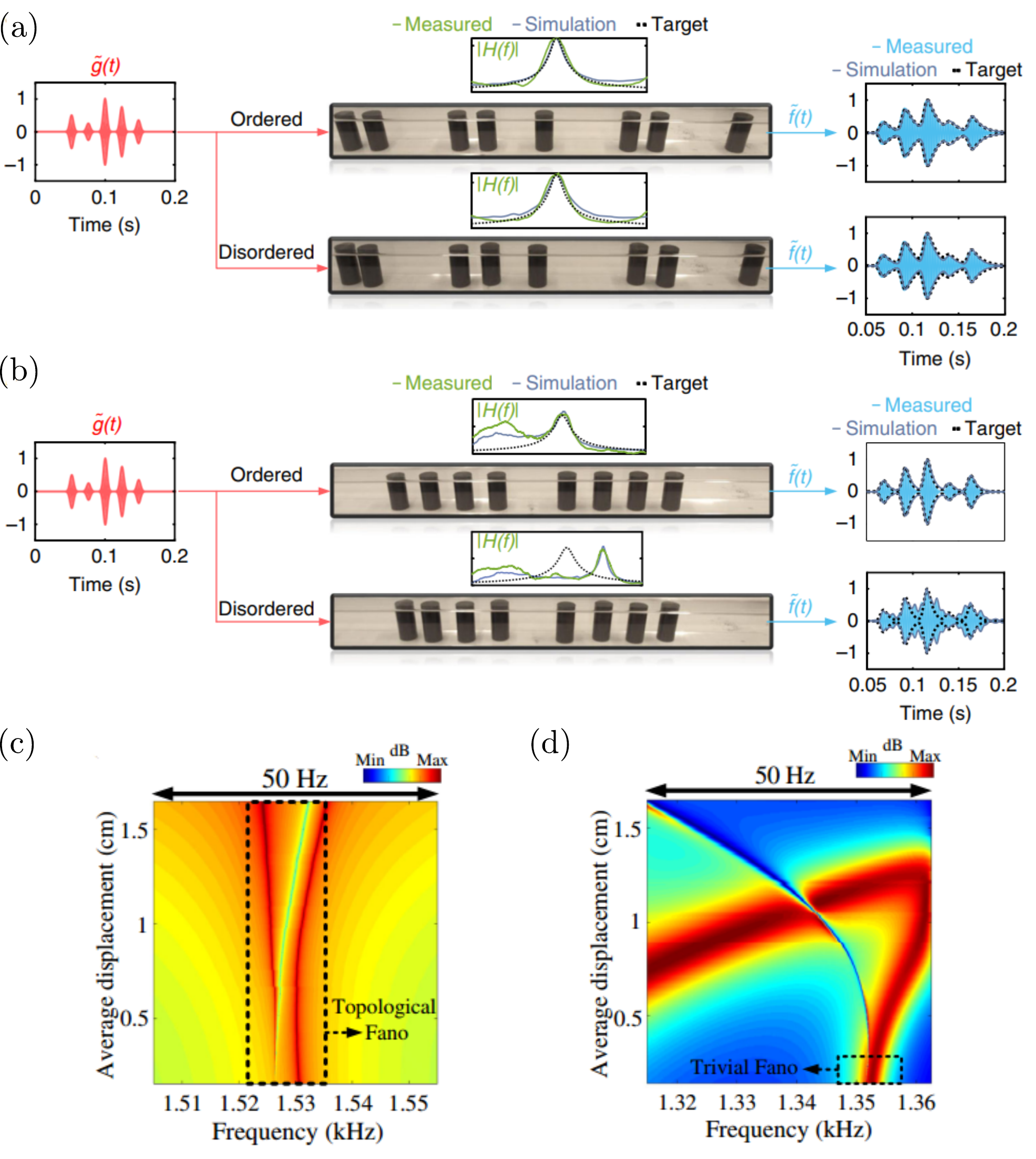}
    \caption{Numerical and experimental results for topological equation solver. (a) Results for ordered and disordered topological waveguide. (b) Results for ordered and disordered trivial waveguide. Figs.(a) and (b) adopted from~\cite{nat.commun.10.2058.zangeneh-nejad}.(c) Transmission spectrum of topological Fano resonance as function of frequency and disorder strength. (d) Same as (c) but for trivial Fano resonance.Figs.(c) and (d) adopted from~\cite{phys.rev.lett.122.014301.zangeneh-nejad}.}
    \label{analogsignal}
\end{figure}

Recently, analog signal process based on metamaterials has attracted a lot of attention for its high efficiency and low power consumption. The usual methods based on cavities suffer from disorders which may change the line shape of the system so that make the analog signal process unstable. Topological acoustic states robust against disorders provide us one way to solve this problem. Zangeneh-Nejad $et$ $al$. used a Su-Schrieffer-Heeger like topological interface states to solve a differential equation~\cite{nat.commun.10.2058.zangeneh-nejad}. It works for the Lorenz shape transmission $H(f)=\frac{A}{i(f-f_0)+f_0/2Q}$. If we consider input signal source $\tilde{g}(t)=g(t)\cos(2\pi f_0t)$, its relation with out put signal $\tilde{f}(t)=f(t)\cos(2\pi f_0t)$ can be obtained by the inverse Fourier transform of $H(f)$, leading to the differential equation $f'(t)+\alpha f(t)=\beta g(t)$ where $\alpha=\pi f_0/Q$ and $\beta=2\pi A$. So the differential equation can be solved by measuring the transmitted signal in time domain. Need to say, such Lorenz spectra line-shape is not specific to topological interface mode and also exist in usual defect mode or cavity mode. The advantage of topological interface mode is its resonant frequency is robust against disorders. Fig.~\ref{analogsignal} compared the results for topological equation solver and trivial equation solver. The topological equation solver is almost immune to disorder. Transmission with different kind of line-shape, like the Fano shape~\cite{phys.rev.lett.122.014301.zangeneh-nejad}, can be used to solve more complex differential equation. Figs.~\ref{analogsignal}(c) and ~\ref{analogsignal}(d) show the transmission spectrum of topological Fano resonance and trivial Fano resonance. We notice the line-shape of topological Fano resonance is robust against disorders. Topological acoustic states can also be used for image processing. Zangeneh-Nejad $et$ $al$. used an acoustic topological Anderson insulator to realize a disorder-induced analog filtering of an imagine~\cite{adv.mater.32.2001034.zangeneh-nejad}.

\subsection{Topological acoustic energy harvest and topological acoustic tweeters}
Compared with the conventional defect states or waveguide states, topological edge states are robust against disorder which give them an advantage in many applications. Next we will provide some examples for their possible applications. Acoustic energy, which can automatically provide power for low-power-consumption devices, has attracted a lot of attention. Given the enhancement to acoustic field, topological acoustic edge states can be used for acoustic energy harvest. Compared with the usual defect modes or cavity modes, the frequency of the topological acoustic edge states is more stable which is benefit for energy harvest. Fan $et$ $al$. proposed to use topological interface states in acoustic crystal to harvest energy~\cite{appl.phys.express.13.017004.fan}. They put piezoelectric cantilever beam at the interface between topological trivial region and topological nontrivial region. The field at the interface was enhance 66 times and an optimal output power of 135 mW can be obtained with the incidence sound of 1 Pa. Lan $et$ $al$. proposed to use locally resonant topological metamaterials for acoustic energy harvest~\cite{j.appl.phys.129.184502.lan}. Compared with the acoustic crystal, the frequency of the system can be tuned lower and the unit cell of the system is in subwavelength region.

Another application direction of topological acoustics is used to control microparticles, known as topological acoustic tweeters. Dai $et$ $al$. realized a topological acoustic tweeters in one dimensional arrays of on-chip Helmholtz resonant cavities~\cite{phys.rev.appl.15.064032.dai}. A 20-$\mu$m microparticle was rotated by the acoustic radiation force and streaming. Liu $et$ $al$. used the pseudospin topological edge states to control microparticle~\cite{appl.phys.lett.120.222202.liu}. By choosing pseudospin up or speudospin down states, the microparticle can be rotated clockwise or anticlockwise. Besides, the topological interface states can also be used for microparticle separation. Dai $et$ $al$. propose a topological interface states based on the Su-Schrieffer-Heeger model~\cite{appl.phys.lett.119.111601.dai}. By calculating the flow distribution and acoustic radiation force, they show the topological interface states can be used to separate two particles with same size and density but different compressibility.

\subsection{Topological acoustic antennas}
\begin{figure}[htbp]
\flushright
    \includegraphics[width=0.85\linewidth]{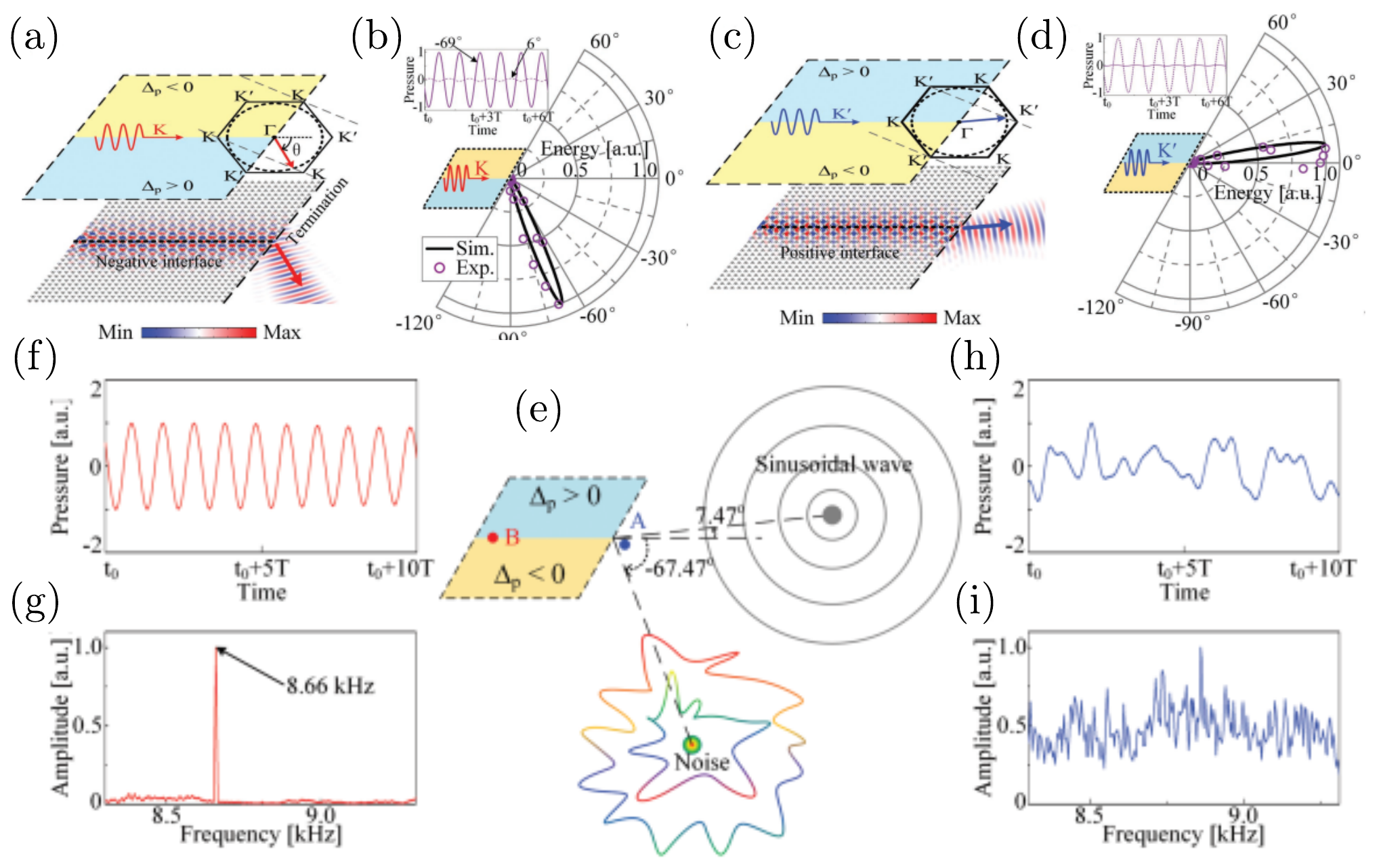}
    \caption{Directional acoustic antennas based on valley topological edge states. (a)Simulated field distribution and out-put coupling of $K$ valley topological edge states. (b) Simulated and measured field pattern of (a). (c) Simulated field distribution and out-put coupling of $K^\prime$ valley topological edge states. (d) Simulated and measured field pattern of (c). (e) Experimental schematics for anti-interference acoustic communication. (f) Measured time domain signal and (g) corresponding Fourier transform spectra at B point with topological acoustic materials. (h) and (i) are similar to (f) and (g), but without topological acoustic materials. Figures adopted  from~\cite{adv.mater.30.1803229.zhang}.}
    \label{topologicalantennas}
\end{figure}

Topological edge states, specifically the valley topological edge states, can be used as topological acoustic antennas. We know that the valley topological edge states are momentum locked. At the interface of valley topological edge states and free space, the valley topological edge states will radiate along specific direction. Zhang $et$ $al$. realized directional acoustic antennas in a Kagome lattice phononic crystal~\cite{adv.mater.30.1803229.zhang}. By choosing $K$ valley or $K^\prime$ valley topological edge states, the acoustic field can emit to different direction as shown in Figs.~\ref{topologicalantennas}(a) and ~\ref{topologicalantennas}(c). The measured field patterns in Figs.~\ref{topologicalantennas}(b) and ~\ref{topologicalantennas}(d) are consistent with the simulated results. Fig.~\ref{topologicalantennas}(e) provides an application scenarios showing the directional acoustic antennas can be used for anti-interference acoustic communication. There are sound sources, one source emitting sinusoidal waves with frequency $f=8.66$ kHz and the other one radiating broadband white noise. The fields at B point are measured with and without topological acoustic materials. We notice with the topological acoustic materials, the target signal can be extracted without the effect of noise as shown in Figs.~\ref{topologicalantennas}(f) and~\ref{topologicalantennas}(g). In free space without the help of topological acoustic materials, the target signal is hidden by the noise as shown in Figs.~\ref{topologicalantennas}(h) and~\ref{topologicalantennas}(i). Song $et$ $al$. designed two rotatable phononic crystals to realize a switchable directional emission~\cite{appl.phys.lett.117.043503.song}. By changing the interface to zizag formula or armchair formula, the emission direction can be tuned. Zheng $et$ $al$. have applied such method for underwater acoustic positioning~\cite{int.j.mech.sci.174.105463.zheng}.

\subsection{On-chip phononic topological devices.} 
\begin{figure}[htbp]
\flushright
    \includegraphics[width=0.85\linewidth]{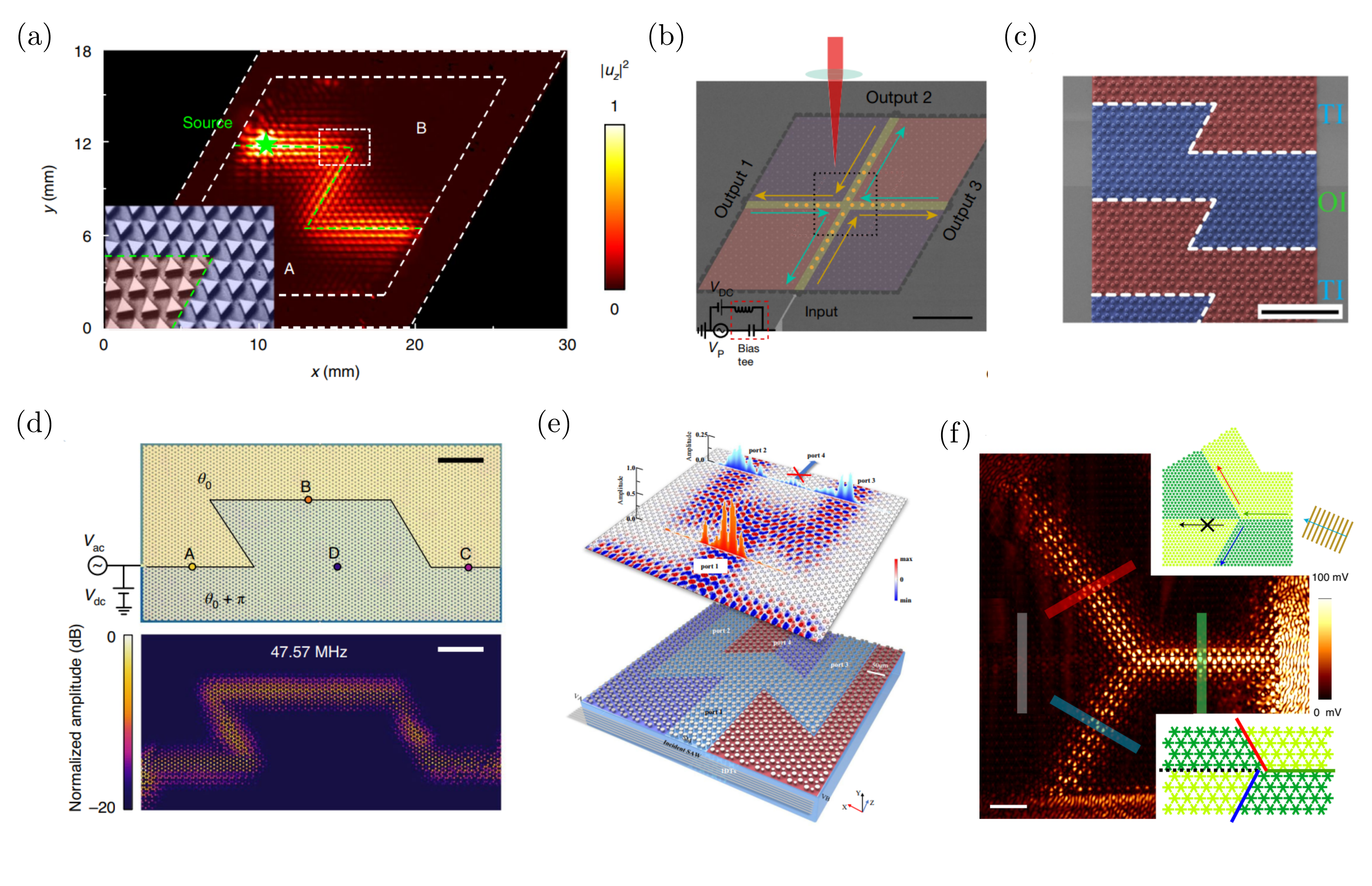}
    \caption{On-chip phononic topological devices. (a) Bending waveguide based on valley topological edge states in silicon chip. (b) Pseudospin filter based on pseudospin topological edge states in a membrane structure. (c) Bending waveguide based on pseudospin topological edge states of surface acoustic waves. (d) Bending waveguide based on auxiliary orbital topological edge states in a nanomechanical membrane structure. (e) Surface acoustic wave splitter based on the valley topology. (f) Gigahertz topological beamsplitting in nanoelectromechanical phononic crystal. Figures adopted from ~\cite{nat.mater.17.993.yan,nature.564.229.cha,phys.rev.appl.16.044008.zhang,nat.nanotechnol.16.576.ma,nat.commun.13.1324.wang,nat.electron.5.157.zhang}.}
    \label{devices}
\end{figure}

Phononic waves, specially those are compatible with electronic or photonic integrated circuits, plays an important role in loading, modulating or processing signals for its relative slower group velocity and smaller size. An important challenge in combining many discrete devices to form a phononic networks is to deal with impedance matching problems. Topological phononics where the topological edge states are robust against disorders provide us a platform to solve the problem so that the on-chip phononic topological devices are desirable. Different platforms, like those supports surface acoustic waves, silicon chips with designed geometry and lattice of nanomechanical membranes, have been used to realize on-chip phoononic topological devices with different functionality. Yan $et$ $al$. fabricated a valley topological materials on silicon chips~\cite{nat.mater.17.993.yan}. Based on the valley topological edge states, they realized a bending waveguide as shown in Fig.~\ref{devices}(a). They also observed the energy partition for valley transport. Cha $et$ $al$. reported a pseudospin topological insulators in a lattice of nanoelectromechanical membranes~\cite{nature.564.229.cha}. Given the propagation direction of the topological edge states is determined by the pseudospin, they realized a pseudospin filter as shown in Fig.~\ref{devices}(b). The devices are composed of four domains with different topological properties. The signal from input transports to output 1 and output 3 as shown by yellow arrows. Zhang $et$ $al$. studied the surface acoustic waves topological states by preparing periodical microstructure on lithium niobate~\cite{phys.rev.appl.16.044008.zhang}. They realized the speudospin topological edge states by shrinking and expanding the six microstructures in unit cell. Based on the pseudospin topological edge states, they realized the bending waveguide for surface acoustic wave as shown in Fig.~\ref{devices}(c). They also realized the topological beamsplitter in the same system. Ma $et$ $al$. proposed a new mechanics to realize topological states~\cite{nat.nanotechnol.16.576.ma}. Different from previous works using valley or pseudospin degree of freedom, they used the auxiliary orbital degree of freedom. They realized their proposal in a lattice of nanomechanical membrane structure. The bending waveguide can be realized as shown in Fig.~\ref{devices}(d). Wang $et$ $al$. realized topological beamsplitter~\cite{nat.commun.13.1324.wang} based on the extended topological valley-locked surface acoustic waves shown in Fig.~\ref{devices}(e). While most of the works are focusing on the frequency region in megahertz, Zhang $et$ $al$. has realized the gigahertz topological valley states in an nanoeletromechanical system~\cite{nat.electron.5.157.zhang}. Topological beam splitter can be realized as shown in Fig.~\ref{devices}(f). They also realized a bending waveguide in the same system.

\subsection{Topological sasers.} 
\begin{figure}[htbp]
\flushright
    \includegraphics[width=0.85\linewidth]{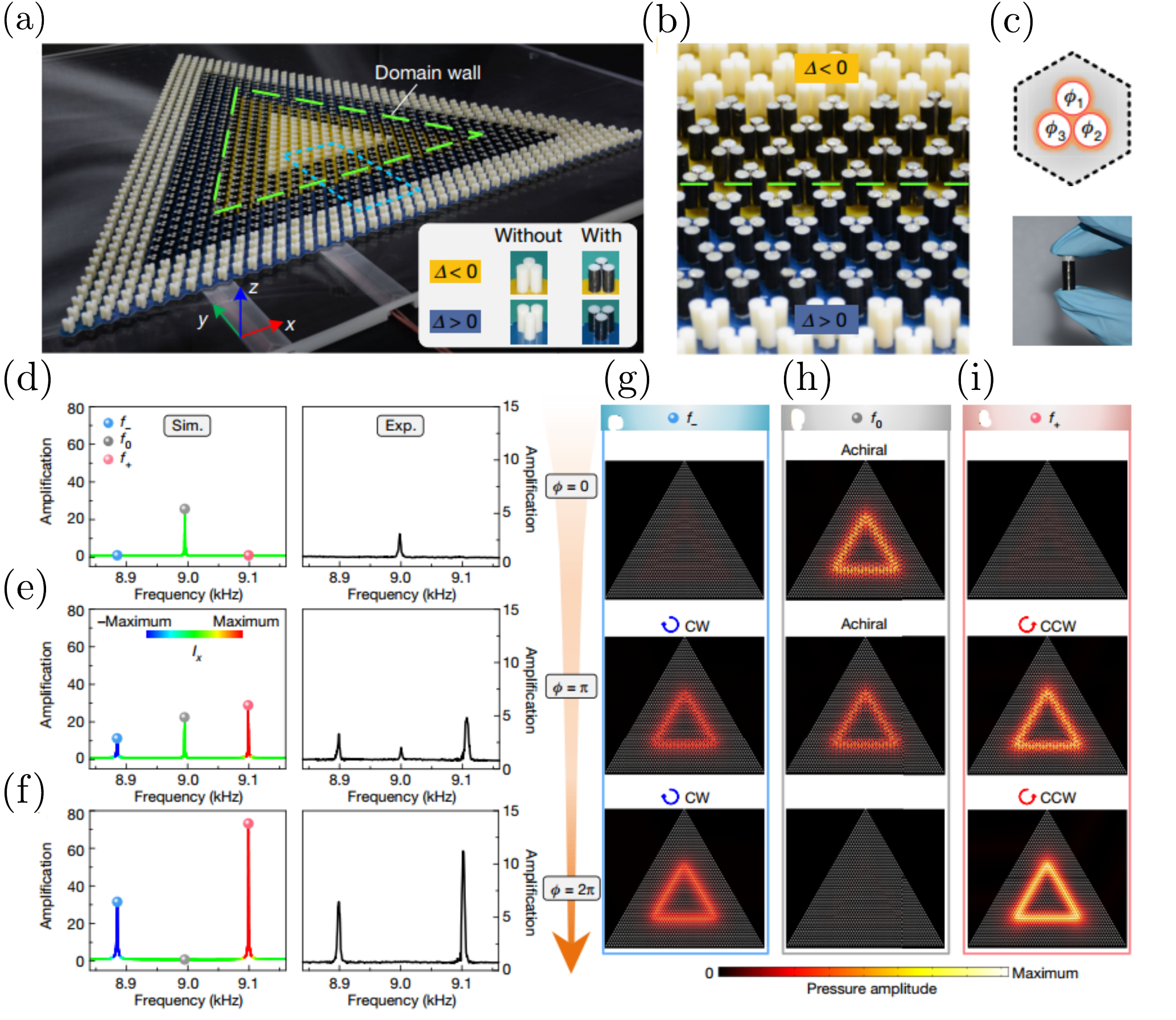}
    \caption{Topological sasers. (a) Photograph of the topological cavity. Insets show the cylinder trimers wrapped without or with carben nanotube films. (b) The cyan zone in (a). The cylinders around the domain wall are coated with carben nanotube films. (c) The phase for three gain-cylinders can be controlled. (d)-(f) The amplification factor for three different gain-phase textures, $\phi=0$ for (e), $\phi=\pi$ for (e) and $\phi=2\pi$ for (f). (g)-(i) Chirality and field for three resonance frequencies $f_{-}(g)$, $f_{0}(h)$ and $f_{+}(i)$ at three different gain-phase textures. Figures adopted from~\cite{nature.597.655hu}.}
    \label{saser}
\end{figure}

Topological lasers, realizing lasers based on the topological boundary states which are robust against fabrication imperfections and defects, are appealing for applications as they would yield lasers outperform conventional lasers~\cite{nat.photon.16.279yang}. When the concept is extended to acoustic system for sound waves, as the topological saser, gain is hard to be obtained. Recently, Hu $et$ $al$. realized a topological seser by using the electro-thermoacoustic coupling in carbon nanotube as gain, and combining with valley topological edge state in a phononic crystal~\cite{nature.597.655hu}. Compared with other methods to realize gain like loudspeaker, the electro-thermoacoustic method has negligible vibration displacement of samples. Figure~\ref{saser}(a) shows the photograph of the topological cavity constructed by valley topological edge states at domain wall of two kinds of phononic crystals with different valley topologies. At the domain wall, the cylinders are coated by carbon nanotube films as shown in Fig.~\ref{saser}(b). The carbon nanotubes are biased by external alternating electrical signals. The gain-phase of different cylinders also can be precisely controlled as shown in Fig.~\ref{saser}(c). The lasing modes can be controlled by a phase $\phi=3(\phi_{j+1}-\phi_j)$ with $j=1,2$. Figures~\ref{saser}(d)-\ref{saser}(f) show the amplification factors for three different gain-phase textures, $\phi=0$ for (d), $\phi=\pi$ for (e) and $\phi=2\pi$ for (f), respectively. The color bar represents the intensity of clockwise modes (+) and counterclockwise modes (-). Figures~\ref{saser}(g)-\ref{saser}(i) show the chiralities and fields. We notice, when the phase $\phi$ is nonzero, chiral lasing modes can be obtained, a clockwise mode at $f_{-}$ and a counterclockwise mode at $f_{+}$. A chirality routing of topological acoustic signals also can be realized. The topological sasers may be used to improve acoustic communication system.

\subsection{Phononic topological logic gates} 
\begin{figure}[htbp]
\flushright
    \includegraphics[width=0.85\linewidth]{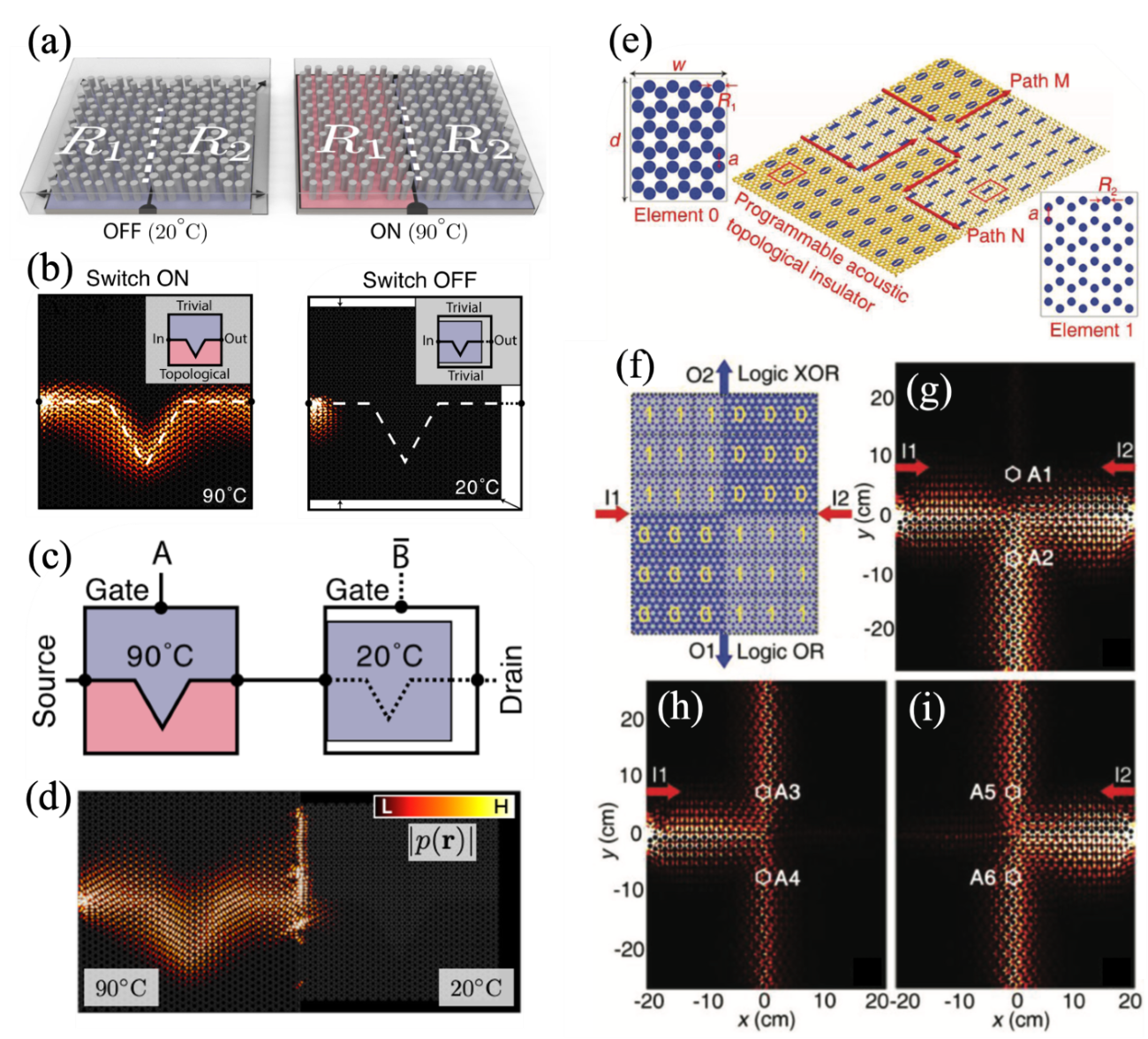}
    \caption{Acoustic topological logic gates. (a) A temperature-controlled topological switch, which combines two sizes of steel pillars (radii $R_1$ and $R_2$) in a honeycomb lattice anchored to a high-thermal-expansion base plate. When the system is cooled (left panel), both the topological phases with radii $R_1$ and $R_2$ are trivial, and no edge states emerge at the interface, as the off state of the switch. When the system is heated (right panel), the topological phase with radius $R_1$ transits to nontrivial and gives rise to edge states at the interface, as the on state of the switch. (b) Field distributions of the on (left panel) and off (right panel) states of the topological switch, revealing the topological feature robust against the bends. (c) Acoustic topological AND gate constructed by two switches in series. Only both control signals (A and B) to be high can register an output. (d) Field distribution for the case that the control signal B is low. The acoustic wave scatters at the interface, and cannot propagate to its output, as the off state of the topological AND gate. (e) Programmable acoustic topological insulator consisted of two digital elements of “0” and “1”. (f) Acoustic topological OR and XOR gates. The system is composed of a $6\times6$ element array. I1 and I2 denote the input ports, and O1 and O2 represent the output ports. (g)-(i) Field distributions for input both in I1 and I2 (g), only in I1 (h) and only in I2 (i), demonstrating that output O1 behaves as the OR gate while O2 is the XOR gate. (a)-(d) adopted from~\cite{phys.rev.lett.128.015501Pirie}, and (e)-(i) adopted from~\cite{adv.mater.30.1805002Xia}.}
    \label{logic}
\end{figure}

Acoustic topological insulators can be used to construct the topological logic gates. In Ref.~\cite{phys.rev.lett.128.015501Pirie}, Pirie $et$ $al$. constructed the topological AND and NOT gates by designing the temperature-controlled topological switch, based on the acoustic topological insulator. As shown in Fig.~\ref{logic}(a), the temperature-controlled topological switch is designed by using the edge states, i.e., turning on with the edge states while off without the edge states. Figure~\ref{logic}(b) shows the topological feature of the switch, since the edge states are robust against the weak disorder and bends. The topological AND gate can be constructed by two topological switches in series, as depicted in Fig.~\ref{logic}(c). A and B, as the control signals, should be both high to heat the two switches and allow the acoustic waves (information) propagate from the source to the drain. While one or all of the control signals is low, the information cannot propagate to its output. Figure~\ref{logic}(d) shows the case that signal B is low, in which the acoustic wave scatters at the interface, stops to its output, corresponding to the off state of the topological AND gate. The topological NOT gate can be design by utilizing the high-thermal-expansion base plate with negative coefficient. When the control is off, the device cools and expands, giving rise to nontrivial topological phase with edge states to allow information to propagate. Another work in Ref~\cite{adv.mater.30.1805002Xia}, Xia $et$ $al$. realized the acoustic topological OR and XOR gates.
Figure~\ref{logic}(e) shows the programmable acoustic topological insulator consisted of two digital elements of “0” and “1”. By constructing a $6\times6$ element array, as shown in Fig.~\ref{logic}(f), the OR and XOR gates can be achieved at the same system. As calculated in Figs.~\ref{logic}(g)-\ref{logic}(i), as long as the acoustic waves come from the input ports, no matter from I1 and I2, they will propagate to the output port O1, corresponding to the OR gate. However, the output port O2 behaves as the XOR gate, since there is no information when the acoustic waves are excited at I1 and I2 at the same time.

\section{Methodology}

\subsection{Fabrication of phononic metamaterials}

The sample for acoustic waves is fabricated by 3D printing of plastic stereolithography material~\cite{nat.phys.14.30li,nat.mater.20.812wei,nat.mater.20.794luo,nat.mater.18.113ni,nsr.nwac203wu}. In the 3D printing technology, firstly, the three-dimensional digital model is designed by computer into several layers of plane slices. Then, the 3D printer superimposes power, liquid, plastic or metal materials layer by layer according to the slice pattern. Finally, the entire object is fabrica·ted. Another way to fabricate acoustic samples is by using mechanical machining~\cite{nat.mater.18.108xue}.

The on-chip sample for elastic waves is prepared by microfabrication technology on silicon wafers~\cite{nat.mater.17.993.yan,phys.rev.lett.127.136401yan}. The fabrication procedure is as follows. Firstly, the wafer with double sides polished is cleaned up by sulfuric acid solution and deionized water. Secondly, a desired pattern is etched after lithography and developing. Thirdly, the silicon dioxide on top of the etching area is removed by hydrofluoric acid. Fourthly, deep reactive ion etching process is applied to etch the wafer to a certain depth. Finally, the photoresist and silicon dioxide are removed by hydrofluoric acid.

\subsection{Measurements}

For the measurement of acoustic waves, a sub-wavelength headphone is put inside the sample or at the boundary for bulk or edge state excitation. A sub-wavelength microphone probe attached to a steel rod is inserted into the sample or the boundary to record the acoustic signals. The probe is fixed on a 3D scanning platform for automatic scanning. Acoustic signals are sent and received by a network analyzer. The bulk and edge states dispersions can be obtained by Fourier transforming of the measured fields~\cite{nat.phys.14.30li,nat.mater.20.812wei,nat.mater.20.794luo}. For the measurement of the acoustic response spectra of the corner states of the higher-order topology, both the headphone and probe are placed at the same position simultaneously~\cite{nat.mater.18.108xue,phys.rev.lett.126.156801yang}.

For the measure of elastic waves, a tungsten needle with a sharp tip, fixed on a piezoelectric ceramic disc, is applied to excite elastic waves. To identify the exact excitation location, a CCD camera is used to monitor the position of the needle. Moreover, to avoid the backscattering of the elastic waves, photoresist and silica gel are coated at the border of the sample to absorb them. The displacement field of the sample is scanned by a laser vibrometer, which is mounted on an automatic scanning platform. All the displacement signals are collected by a network analyzer, and then used to Fourier transform for the bulk and edge states dispersions ~\cite{nat.mater.17.993.yan,phys.rev.lett.127.136401yan}.

\section{Summary and outlooks}

Topological phononics is an interdisciplinary field at the interface between topological physics (an important branch of condensed matter physics) and phononics. Its basic elements include discovering topological mechanisms and phenomena in phononics as well as using them to achieve unconventional manipulation of phononic waves. In many cases phononic metamaterials are designed and used to realize novel topological phenomena that are not yet observed before---thanks to the remarkable controllability of phononic metamaterials and the versatile phononic measurements in a wide range of temperatures and scales. At the application levels, topological mechanisms and phenomena can be used to achieve unprecedented functions in controlling the energy and information carried by phonons. It is indeed marvelous that topological mechanisms can make some acoustic and phononic phenomena resilient to disorders and imperfections, making applications based on them more feasible. In some cases, such as topological Fano resonances~\cite{phys.rev.lett.122.014301.zangeneh-nejad}, devices based on topological mechanisms are much superior than the conventional ones.

Beyond its importance in materials science and applications, topological phononic metamaterials open a new pathway toward experimental topological physics. Many topological phenomena are discovered/observed first in phononic metamaterials. For instance, topological negative refraction~\cite{nature560.61he}, topological sasers~\cite{nature.597.655hu}, pseudospin- and valley-polarized edge states~\cite{nat.phys.12.1124he,nat.phys.13.369lu}, higher-order topological insulators~\cite{nature555.342Serra-Garcia,nat.phys.15.582zhang,nat.mater.18.108xue,nat.mater.18.113ni}, higher-order Weyl semimetal phases~\cite{nat.mater.20.794luo,nat.mater.20.812wei}, Dirac vortex states~\cite{PRL.123.196601gao}, topological Wannier cycles~\cite{nat.mater.21.430lin}, spectral flows in fragile topological insulators~\cite{science.367.797peri}, non-Abelian braiding of band nodes~\cite{nat.phys.17.1239jiang,jiang2022experimental,qiu2022minimal}, non-Hermitian morphing of topological modes~\cite{nature.608.50wang,phys.rev.b.105.L201402teo}, and topological exceptional nexus~\cite{science.370.1077tang} were first discovered in phononic metamaterials.

Looking into the future, topological phononic metamaterials will continue to have important impacts on research in metamaterials, as they could provide new mechanisms for robust wave manipulations that may lead to superior functional devices and systems. This is particularly important as topological phononic metamaterials go smaller and smaller and merging with micromechanical systems,  nanomechanical systems, and optomechanical systems~\cite{nat.commun.13.3476ren,nature.612.666youseffi}. Developing high-quality on-chip phononic topological systems and their applications is still a prior target in future research. Such systems may enable fine control of phonons in the linear, nonlinear, non-Hermitian, and even quantum regimes. Applications and devices based on topological phononic metamaterials can range from topological Fano devices, topological sasers, topological sensors, topological transducer, topological surface acoustic wave devices, topological wave guiding and concentration, topological on-chip phononic transmission lines, topological on-chip resonators, electrically tunable topological phononic devices, etc.

Meanwhile, from the perspective of fundamental science and materials, there are lots of opportunities for future exploration, such as non-Abelian topological states (e.g., Euler class topological phases), non-Hermitian topological states, fragile topological states, topological semimetals (including non-Abelian topological semimetals and other exotic phenomena), nonlinear topological phononic metamaterials (including topological sasers and other intriguing topics), phonon-phonon interactions and nonclassical phonon states as well as their interplay with topological phenomena, topological optomechanical systems, emergent phenomena at topological defects, tunable phononic systems, topological phonons in active matters, artificial gauge fields of phonons, and synthetic dimensions in topological phononic systems, etc. Much remains to be done to explore these topics and directions. One of the encouraging signs is that these topics may have lasting impacts on both fundamental physics and applications (e.g., in ultrasonic technologies, micro- and nano-mechanical applications, sensing, detection, imaging, topological transducers, topological SAW devices, and on-chip communication devices). Yet there are still many challenges in this field, including the fine control of non-Hermitian effects, physical realization of non-reciprocal effects, tunable nonlinear effects, high-quality phonon resonators, nano-fabrication and integrations, controlling phonons in micro- and nano-mechanical systems. These expectations and obstacles will shape future researches in topological phononic metamaterials.

\section*{Acknowledgements}
W.~Z. thanks support from the National University of Singapore. W.~D., J.~L., X.~H. and Z.~L. were supported by the National Natural Science Foundation of China (Grant Nos. 11890701, 11974120, 11974005, and 12074128), the National Key R$\&$D Program of China (2018YFA0305800), the Guangdong Basic and Applied Basic Research Foundation (Grant Nos. 2019B151502012, 2021B1515020086, and 2022B1515020102). Y.~L., Z.-K.~L. and J.-H.~J. were supported by the National Natural Science Foundation of China (Grant Nos. 12125504 and 12074281), the Priority Academic Program Development (PAPD) of Jiangsu Higher Education Institutions, and the Key Lab of Advanced Optical Manufacturing Technologies of Jiangsu Province. H.-X.~W. was supported by the the National Natural Science Foundation of China (Grant No. 11904060).

\section*{References}
\bibliographystyle{iopart-num}
\bibliography{References_new}

\end{document}